# Studies on $R_K$ with Large Dilepton Invariant-Mass, Scalable Pythonic Fitting, and Online Event Interpretation with GNNs at LHCb

Dissertation

zur

Erlangung der naturwissenschaftlichen Doktorwürde
(Dr. sc. nat.)
vorgelegt der

Mathematisch-naturwissenschaftlichen Fakultät

der

Universität Zürich

von

Jonas N. Eschle

von

Entlebuch LU

**Promotionskommission**

Prof. Dr. Nicola Serra (Vorsitz)

Prof. Dr. Gino Isidori

Dr. Patrick Owen

Zürich, 2024

## Abstract

The Standard Model of particle physics is the established theory describing nature's phenomena involving the most fundamental particles. However, the model has inherent shortcomings, and recent measurements indicate tensions with its predictions, suggesting the existence of a more fundamental theory. Experimental particle physics aims to test the Standard Model predictions with increasing precision in order to constrain or confirm physics beyond the Standard Model.

A large part of this thesis is dedicated to the first measurement of the ratio of branching fractions of the decays $B^+ \to K^+ \mu^+ \mu^-$ and $B^+ \to K^+ e^+ e^-$, referred to as $R_K$, in the high dilepton invariant mass region. The presented analysis uses the full dataset of proton-proton collisions collected by the LHCb experiment in the years 2011-2018, corresponding to an integrated luminosity of $9 \, \text{fb}^{-1}$. The final result for $R_K$ is still blinded. The sensitivity of the developed analysis is estimated to be $\sigma_{R_K}^{(\text{stat})} = 0.073$ and $\sigma_{R_K}^{(\text{syst})} = 0.031$. Applying all analysis steps to a control channel, where the value of $R_K$ is known, successfully recovers the correct value.

In addition to the precision measurement of $R_K$ at a high dilepton invariant mass, this thesis contains two more technical topics. First, an algorithm that selects particles in an event in the LHCb detector by performing a full event interpretation, referred to as DFEI. This tool is based on multiple Graph Neural Networks and aims to cope with the increase in luminosity in current and future upgrades of the LHCb detector. Comparisons with the current approach show at least similar, sometimes better, performance with respect to decay reconstruction and selection using charged particles. The efficiency is mostly independent of the luminosity, which is crucial for future upgrades.

Second, a PYTHON package for likelihood model fitting called ZFIT. The increasing popularity of the PYTHON programming language in High Energy Physics creates a need for a flexible, modular, and performant fitting library. The ZFIT package is well integrated into the PYTHON ecosystem, highly customizable and fast thanks to its computational backend TENSORFLOW.

# Acknowledgements

First and foremost, I would like to thank Prof. Dr. Nicola Serra for his guidance since my Bachelors until the end of my PhD. and for all the opportunities he gave me. His way of thinking that I encountered in my first particle physics lecture sparked my interest for the field; his passion for machine learning and goals beyond the reasonable have always been a great inspiration. I would like to thank all the supervisors, from my Bachelors to my PhD, for their guidance and support. These are Prof. Dr. Rafael Silva Coutinho, who introduced me to the physics and analysis in my bachelor thesis and with whom I will be working with again in the near future; Dr. Albert Puig who introduced me to the world of software in physics, papers and navigating politics during my master thesis. But most importantly, I would like to thank Dr. Patrick Owen, my supervisor during my PhD, who taught me not only about the physics analysis but guided me through roadblocks and helped to progess in my PhD, including in the whole process of writing the thesis.

I would also like to thank Gino Isidori as a member of the Ph.D committee, and Eduardo Rodrigues for the external review report of the thesis.

There is a variety of colleagues that I was able to discuss and share a great time with, both in and outside the office; more than I can mention here. A huge thanks also goes to my family, my brothers who are always here for me, and my parents, who brought me on this path.

And finally, I would like to thank Anja, for way more than I can even imagine right now. Dadruf, das mer zruglued, "weisch no, damals, ohni Doktor?"

# Contents

# List of Figures

# List of Tables

# 1

# Introduction

In the year 2012, the ATLAS and CMS collaborations announced that they have both found statistically significant evidence of a particle consistent with the Brout-Englert-Higgs-Guralnik-Hagen-Kibble particle - often shortened to "Higgs particle". The Higgs discovery completed the Standard Model of particle physics (SM), the prevailing theory describing fundamental particle interactions, after over 50 years of development. Due to its highly successful predictions of a plethora of observations, peaking with the aforementioned discovery, the SM is the widely accepted theory describing the laws of physics governing the most fundamental particles in our universe.

Despite its success, the SM is not a complete theory of nature, as it neither includes gravity nor accounts for the existence of dark matter or dark energy. Following the lack of explanations for many observations and the tensions between some measurements and their SM predictions, it is often assumed that the SM is only an effective theory, valid up to a certain energy scale. In order to develop a more complete theory, the shortcomings of the SM need to be mapped out precisely. This requires theoretically clean observables as well as high-statistics data samples with a thorough understanding of the detector. The measurement of $R_K$, the ratio of branching fractions of two decays that only differ in the flavor of the final-state leptons, represents such a test which is contained in this thesis.

At the energy frontier of High Energy Physics (HEP), Large Hadron Collider (LHC) has been running throughout the last decade and provides proton-proton or heavy-ion collisions at four interaction points, populated by the ATLAS, ALICE, CMS and LHCb detectors. The LHCb experiment is at the center of this thesis. Its design is optimized for the study of $b$ and $c$ decays. During the years 2011-2018, it collected a dataset from proton-proton collisions at center-of-mass (CoM) energies from 7 to 13 TeV, corresponding to an integrated luminosity of about $9\,\mathrm{fb}^{-1}$. During a long shutdown of the LHC, the LHCb detector was upgraded and is currently collecting data at a world-record CoM energy of 14 TeV and a higher instantaneous luminosity than previously. Further upgrades of the LHC and the LHCb detector are planned for the next decades. The future increase in instantaneous luminosity holds unprecedented challenges for the detector as well as for the software, requiring solutions that reach beyond an increase in computing power. In this thesis, a deep learning algorithm for triggering and event reconstruction, called Deep Full Event Interpretation (DFEI), designed to tackle the increase in luminosity, is presented.

Large-scale data analysis requires a variety of tools involving physics knowledge as well as technical expertise. Alongside the general success of machine learning algorithms over the past decade, machine learning has become a powerful toolbox in HEP where especially the PYTHON ecosystem provides highly optimized libraries. As a consequence, many HEP analyses employ the PYTHON programming language. However, for a long time, the PYTHON ecosystem has not included a fitting framework suitable for HEP analyses. This thesis presents the zfit project that aims to bridge this gap and enjoys a growing user base.

The thesis starts with an overview of the current state of knowledge about fundamental particles and forces in Chapter 2. With the theoretical knowledge at hand, the LHCb experiment is introduced in Chapter 4. In Chapter 3, an overview of the statistical tools used to interpret the collected data is given. An approach to interpret events inside the LHCb detector based on machine learning is provided in Chapter 5. The main analysis of this thesis is the measurement of $R_K$, a decay ratio with two different leptons in the final state, which is presented in Chapter 6. The last of the three main chapters, Chapter 7, presents a PYTHON package for likelihood fitting.

*"All that remains to do in physics is to fill in the sixth decimal place"*

Albert Michelson, 1894

<div align="right">

# 2

</div>

# Particles, Forces and Interactions

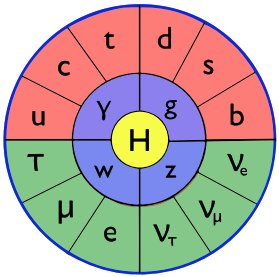

## 2.1 Overview

At the end of the 19th century, scientists accepted the atom as the indivisible unit of matter. They considered physics to be complete — apart from minor "details". With the turn of the century, the picture started to change. Einstein revolutionized the understanding of space and time with his theory of relativity. Radioactivity was discovered by accident. Rutherford and his colleagues fired alpha particles at gold foils and discovered the inner structure of atoms. And about 30 years later, the theory of quantum mechanics (QM), initially introduced as an "act of desperation" by Max Planck, was widely accepted as the descriptor of particles on the atomic scale.

These discoveries lay the basis of modern physics and constitute essential building blocks of the theoretical description of particle physics, whose journey started in the 1930s with the discovery of new particles such as the positron, the muon, and the pion. Initially, the understanding of new particles was based on the idea of a "particle zoo", where each particle was considered to be a fundamental building block of matter. This changed in the 1960s, when the existence of quarks was postulated through theoretical models that describe the interactions between them and the following discovery in deep inelastic scattering experiments. These turned out to be the founding stones of the modern understanding of particle physics, the SM. In the decades that followed, the discovery and measurement of new particles and interactions confirmed the predictions of the SM, up to the discovery of the before-mentioned Higgs boson in 2012.

The SM is a gauge quantum field theory (QFT) that describes the fundamental particles and their interactions. There are quarks and leptons, both fermions, constituting the elementary particles. Quarks come in two types, up-type and down-type quarks, and exist in three generations, called "quark flavors". Leptons follow the same structure of three generations — "lepton flavors" — and have two types, massive particles and their corresponding electrically neutral neutrinos. For all fermions, an antiparticle with the same mass but opposite charge.

The three fundamental forces included in the SM are the electromagnetic, the weak, and the strong force. In the SM, the forces are mediated by gauge bosons between particles that carry the relevant charge. The electromagnetic force carrier is the massless photon, which can act on particles with electric charge, *i.e.*, all fermions except the neutrinos. The weak force is mediated by the massive $W$ and $Z$ bosons, which interact with all *left-handed* fermions. Due to the large mass of the weak gauge bosons, the force appears weak at low energies, hence the name historically. Weak interactions are the only interaction able to change the flavor of a fermion, ultimately causing the decay of the heavier generations into the lighter ones. The strong force is mediated by the massless gluons that couple to the color charge. Quarks and the gluons themselves carry color charge and can hence interact strongly and form hadrons. The picture is completed by the Higgs boson, causing the fermions and the $W$ and $Z$ bosons to be massive. Fig. 2.1.1 contains an illustration of the SM particles including their properties.

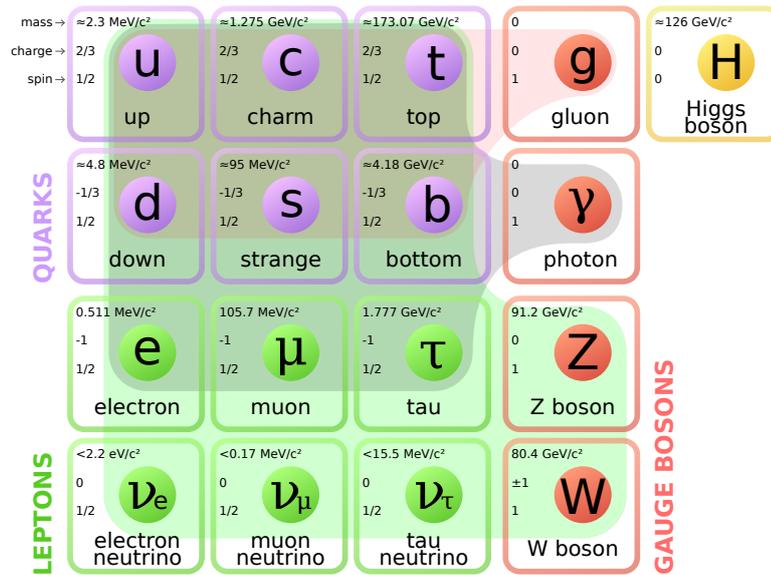

**Figure 2.1.1:** Overview of the particles and forces in the SM with the respective charges, spins and masses [1]. For each fermion, there exists an anti-particle with the same mass but opposite charges. The gauge bosons are "force carriers" and are responsible for the interactions. The possible interaction of particles through a specific force carrier is drawn with a thin line containing the respective particles, non-exclusively. The Higgs boson causes most of the particles to be massive.

The SM accurately predicts a plethora of experimental observations but ultimately also explains the physical world around us. Nuclei and atoms consist of quarks that are bound together by the strong force, shaping the picture at the nuclear scale. Electrons orbit the nucleus and are bound by the electromagnetic force, dominating the interaction on the scale of atoms and molecules. The weak force is responsible for crucial, yet more subtle effects, such as the decay of radioactive elements and a certain matter-antimatter asymmetry. The SM is today, for good reason, celebrated as one of the most successful theories in physics. Yet, despite its success, the actual theory already contains minor modifications beyond its clean nature. Even with those, disagreements in measurements and major open questions remain. Nothing big, just some minor "details".

## 2.2 Field theoretical description of the SM

Before the development of particle physics theories, the physics of particles was governed by classical field theory for electromagnetism and gravity, special relativity (SR) for speeds $\approx c$ and QM for scales $\approx \hbar$. QFT incorporates all of them, automatically reproducing the results of previous theories in their respective limits. This made QFT a natural choice for the SM. Developing the SM was also motivated by *symmetries*, characterizing the invariance of a system under a transformation. According to Noether's theorem [2], every symmetry is associated with a conserved quantity - classically explaining conservation of energy, momentum and charge, among others.

The SM is a gauge QFT with symmetry group $SU(3)_C \times SU(2)_L \times U(1)_Y$, where $C$ denotes the color charge, $L$ the left-handed chirality and $Y$ the hypercharge. By introducing the Higgs field with a nonzero vacuum expectation value, the group is spontaneously broken. Requiring local gauge invariance gives rise to extra terms representing gauge bosons in the SM. The SM Lagrange density can be split into four terms

$$\mathcal{L} = \mathcal{L}_{\text{fermion}}^{\text{kin+int}} + \mathcal{L}_{\text{gauge}} + \mathcal{L}_{\text{Higgs}} + \mathcal{L}_{\text{Yukawa}} \tag{2.1}$$

which are the kinetic and interaction term of the fermions, the gauge bosons or force carriers, the Higgs field and the Yukawa coupling, respectively.

The SM relies *explicitly* on experimental measurements for the value of specific parameters. In total, there are about 19 free parameters - mostly the masses of the fermions, coupling strengths, the Cabibbo-Kobayashi-Maskawa (CKM) parameters and Higgs characteristics - that are not predicted by the theory itself. When extending the SM to include neutrino masses, as discussed in Section 2.3 this number even increases to at least 26.

The following sections discuss the different components of the SM. The focus is to provide an overview to motivate the SM and the coming chapters concerned with experimental aspects. A thorough mathematically correct SM description is nowadays well-established knowledge and can be found in standard textbooks and lecture notes about particle physics, such as [3–5], which the following sections are based on. Section 2.2.1 discusses quantum electrodynamics (QED), the first and highly successful gauge QFT covering concepts that also appear in other parts of the SM. Then, quantum chromodynamics (QCD) is introduced in Section 2.2.2, governing quarks and gluons. Section 2.2.3 introduces the weak force and unifies it with QED into the electroweak (EW) theory. Thereafter, the Higgs mechanism is described in Section 2.2.4 and the effects of Spontaneous Symmetry Breaking (SSB). Finally, the coupling to fermions in Section 2.2.5 and the resulting effects on flavor are discussed.

### 2.2.1 A quantum theory for electromagnetism

In the 1920s, Maxwell's equations were well established in describing the electromagnetic fields in a relativistic correct way, implying the existence of a photon. At that time, Dirac formulated the quantized, relativistic description of spin-1/2 particles, initially aimed at describing the electron. This motivated the development of QED, historically the first QFT, describing the electromagnetic force between elementary particles in a consistent manner. QED served historically and serves in this chapter as an example for the formulation of other forces in the SM.

Taking inspiration from the formulation of classical mechanics, the Lagrangian density $\mathcal{L}$ is the starting point. A complete description requires three components: the fermions, the gauge bosons, and the interaction between them. Thus, the $\mathcal{L}_{\mathrm{QED}}$ takes the form

$$\mathcal{L}_{\mathrm{QED}} = \mathcal{L}_{\mathrm{fermion}}^{\mathrm{kin}} + \mathcal{L}_{\mathrm{fermion}}^{\mathrm{int}} + \mathcal{L}_{\mathrm{gauge}} \tag{2.2}$$

with each of the terms discussed in the following sections.

#### 2.2.1.1 Fermions

Using the Dirac equation, the Lagrangian density for a free particle is given by

$$\mathcal{L}_{\mathrm{fermion}}^{\mathrm{kin}} = i\bar{\psi}\slashed{\partial}\psi - m\bar{\psi}\psi \tag{2.3}$$

where $\psi$ is a Dirac spinor, $m$ is the fermion mass and, using the slash notation, $\slashed{\partial} \equiv \gamma^\mu \partial_\mu$ with $\gamma^\mu$ the Dirac matrices.

#### 2.2.1.2 Local gauge invariance

The Lagrangian in Eq. 2.3 is invariant under global phase transformations $\psi \to e^{iQ\alpha}\psi$, with $Q$ and $\alpha$ two arbitrary constants, that will be identified as the charge of the particle and the phase, respectively. Furthermore, physical processes should not depend on an arbitrary *local* phase either. Imposing invariance under the phase transformation $\psi \to e^{iQ\alpha(x)}\psi$ results in an additional term in Eq. 2.3:

$$\frac{1}{g_e}\partial_\mu\alpha(x) \tag{2.4}$$

with the coupling constant $g_e$. This imposed local gauge invariance belongs to the symmetry group $U(1)_{EM}$, resulting in the conserved quantity of the electric charge $Q$.

The extra term in Eq. 2.4 can be compensated by the introduction of a gauge field $A_\mu$. Following local gauge invariance, the new field must transform as $A_\mu \to A_\mu + \frac{1}{g_e}\partial_\mu\alpha(x)$.

By introducing the *covariant derivative*

$$\partial_\mu \to D_\mu = \partial_\mu - i g_e A_\mu \ , \tag{2.5}$$

the ordinary derivative generalizes to incorporate invariance under local gauge transformations.

As the treatment of other forces will show, this is a general procedure to obtain the Lagrangian of a gauge field and the covariant derivative will be extended throughout this section.

Inserting transformation Eq. 2.5 into the $\mathcal{L}_{\text{fermion}}$ given by Eq. 2.3 yields

$$\mathcal{L}_{\text{fermion}} = \underbrace{i\bar{\psi}\partial\!\!\!/\psi - m\bar{\psi}\psi}_{\mathcal{L}_{\text{fermion}}^{\text{kin}}} + \underbrace{g_e Q A_\mu \bar{\psi}\gamma^\mu\psi}_{\mathcal{L}_{\text{fermion}}^{\text{int}}} . \tag{2.6}$$

where the first two terms correspond to the Lagrangian of a free spin-1/2 fermion, see Eq. 2.3. The third term represents the electromagnetic interaction between a fermion with charge $Q$ and the electromagnetic gauge field $A_\mu$.

### 2.2.1.3 A massless propagator

The Proca equation, describing spin-1 particles, adds the propagator term for the new gauge field

$$\mathcal{L}_{\text{gauge}} = -\frac{1}{4}F^{\mu\nu}F_{\mu\nu} + \underbrace{m_\gamma^2 A^\mu A_\mu}_{\text{needs to vanish}} . \tag{2.7}$$

The $m_\gamma$ is the photon mass and $F^{\mu\nu}$, the electromagnetic field tensor, is defined as

$$F^{\mu\nu} = \partial^\mu A^\nu - \partial^\nu A^\mu \ . \tag{2.8}$$

The second term in Eq. 2.7 is not invariant under gauge transformations and vanishes by requiring a massless boson with $m_\gamma = 0$. Eq. 2.7 is the free Lagrangian of the electromagnetic field, describing the gauge boson of the electromagnetic interaction, the photon.

As the photon exists in Maxwell's equations, applying the Euler-Lagrange equations on the four-potential $A^\mu$ recovers the former in covariant form

$$\partial_\mu F^{\mu\nu} = J^\nu \tag{2.9}$$

where $J^\nu = (c\rho, \vec{j})$ is the four-current density, with $\rho$ the electric charge density and $\vec{j}$ the electric current density. Combining Eq. 2.6 and Eq. 2.7 results in the full Lagrangian of the electromagnetic interaction,

$$\mathcal{L}_{\text{QED}} = \underbrace{-\frac{1}{4}F^{\mu\nu}F_{\mu\nu}}_{\mathcal{L}_{\text{gauge}}} + \underbrace{i\bar{\psi}\slashed{D}\psi - m\bar{\psi}\psi}_{\mathcal{L}_{\text{fermion}}^{\text{kin+int}}} \,. \tag{2.10}$$

#### 2.2.1.4 Interaction probabilities

The transition rate of an event $\Gamma_{i\to f}$ depends on the square amplitude of the process $\mathcal{M}_{i\to f}^2$. In order to calculate the amplitude, a perturbative procedure can be used in QED, where Feynman diagrams of the process need to be evaluated, thereby taking into account all possible contributions. An example of electron-muon scattering is shown in Fig. 2.2.1, where every vertex is a three-body vertex with one propagator and two fermions, carrying a coupling constant (*i.e.*, factor) $\alpha_e \approx 1/137$. Higher order diagrams are therefore suppressed by powers of $\alpha_e$, allowing for arbitrary precision calculations and reliable uncertainty estimates.

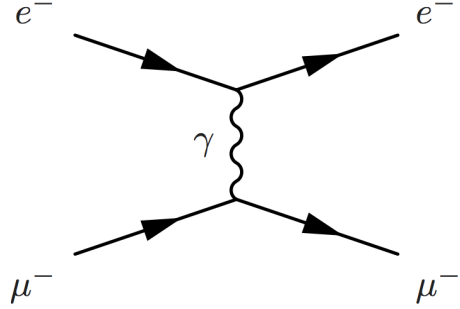

**Figure 2.2.1:** Feynman diagram of electron-muon scattering via photon exchange. It is read from left to right, with an electron and a muon coming in, exchanging a photon and going out as an electron and a muon again. The x-axis represents time, the y-axis space.

#### 2.2.1.5 Renormalization

Looking closer to the possible Feynman diagrams of a given process, higher order contributions from fermion loops as depicted in Fig. 2.2.2 enter the calculation. Evaluating a diagram involves an integral over all possible momenta of the loop particles, which is divergent. Renormalization adds a momentum transfer ($q^2$) dependent counter-term to the coupling constant, keeping the whole theory finite:

$$g_e^2 \to g_{e,R}^2(q^2) \propto g_e^2(1 - g_e^2 C) \tag{2.11}$$

where $C$ is a $q^2$ independent term. The coupling thereby becomes a "running coupling constant" due to the $q^2$ dependence.

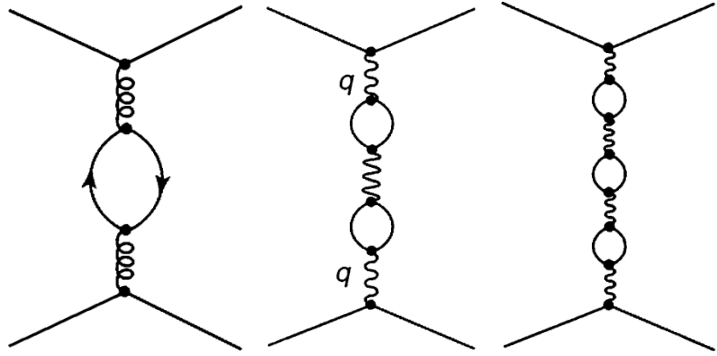

**Figure 2.2.2:** Depiction of a fermion loop (left) and higher order contributions with multiple loops (middle, right) extending the diagram in Fig. 2.2.1, also called "vacuum polarization". These loops effectively shield the interaction and create an effective coupling constant that depends on the momentum transfer $q^2$.

The effect is called "vacuum polarization" as the virtual electron-positron dipoles screen the charge. Implicitly, the $g_{e,R}(q^2)$ is the electric elementary charge *actually* measured in the laboratory, whereby $g_e$ is the hidden, underlying coupling constant of nature. In QED, correction terms to $g_e$ are in the order of permil and thus usually negligible, leaving $\alpha_e = \frac{g_e^2}{4\pi} \approx \frac{1}{137}$ a valid approximation for most processes.

## 2.2.2 The Strong Interaction

In the 1960s, a range of newly discovered particles, such as $\pi^{+,-,0}$, $K^{+,-,0}$ and $\Lambda^{+,-,0}$, gave way to the idea that they are composed of smaller constituents, quarks, held together by the strong force. The force carriers are massless gluons that interact with color charged particles, resulting in the name QCD. The color charge introduces a new quantum number, a three-dimensional vector $\vec{\psi}_c$, as there are three types: red, blue, and green.

Obtaining the Lagrangian is done analogously to the electromagnetic interaction. Given the three colors, the required local gauge invariance is now under the symmetry group $SU(3)_C$ instead of $U(1)_{EM}$. The Lagrangian for a free quark from Eq. 2.3 needs to be extended by a sum over the three colour charges $c$

$$\mathcal{L}_{\text{quark}} = \sum_f \bar{\psi}_c (i\slashed{\partial} - m_c)\psi_c \ . \tag{2.12}$$

The transformation, following Section 2.2.1, is given by

$$\psi \to e^{ig_s t^a \Theta^a(x)} \psi \tag{2.13}$$

where $g_s$ is the coupling strength and $t^a = \frac{1}{2}\lambda^a$ the eight generators of the $SU(3)_C$ group, with $\lambda^a$ the Gell-Mann matrices. They are analogous for $SU(3)$ what the Pauli matrices are for $SU(2)$.

The covariant derivative is given by

$$D_\mu = \partial_\mu + i g_s t^a G_\mu^a \tag{2.14}$$

Due to the non-Abelian, *i.e.*, a group that does not commute $[A, B] \neq 0$, (non-Abelian) nature of the symmetry group $SU(3)_C$, the field strength tensor has an additional term *w.r.t.* Eq. 2.8:

$$G^{\mu\nu} = \partial_\mu G_\nu^a - \partial_\nu G_\mu^a + g_s f^{abc} G_\mu^b G_\nu^c, \tag{2.15}$$

where $G^\mu$ is the gluon gauge field and $f^{abc}$ are the structure constants of the group $SU(3)_C$, defined by the commutator $[t^a, t^b] = i f^{abc} t^c$.

The Lagrangian of the strong interaction can be written using the previously established Lagrangian of the quarks as

$$\mathcal{L}_{\mathrm{QCD}} = -\frac{1}{4} G_a^{\mu\nu} G_{\mu\nu}^a + \bar{\psi}_c (i \slashed{D} - m) \psi_c \ . \tag{2.16}$$

#### 2.2.2.1 Asymptotic freedom

Expanding the additional term in the field strength tensor of Eq. 2.15 results in purely gluonic interaction terms

$$\mathcal{L}_{\mathrm{QCD}}^{self} = f_{abc} f^{abc} G_\mu^a G_\nu^b G^{\mu\nu c} \ . \tag{2.17}$$

Fig. 2.2.3 contains example Feynman diagrams of such gluon-gluon interactions.

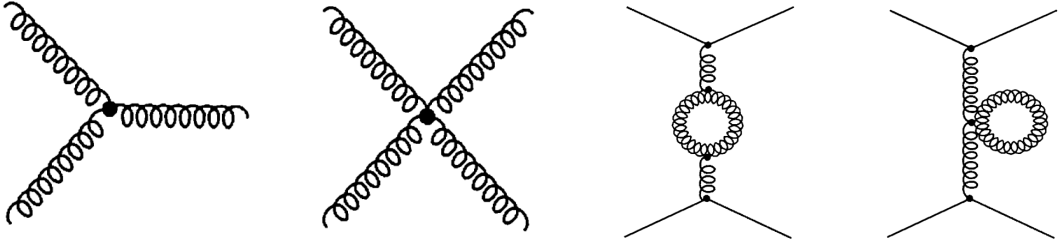

**Figure 2.2.3:** Due to the non-Abelian nature of the $SU(3)_C$ symmetry group, the QCD Lagrangian shows terms of gluon self-interactions (left diagrams). Thereby, vertices with three or four gluons are possible. These allow for gluon loops (right diagrams) that are responsible for the effects of asymptotic freedom and confinement.

QCD allows not only for the same fermion loops, here restricted to quarks, as in Section 2.2.1.5, but also gluon loops as depicted in Fig. 2.2.3. The presence of the latter has the opposite effect than quark loops have, so that the renormalized QCD coupling *decreases* with increasing momentum transfer $q^2$. This effect leads to "asymptotic freedom" of quarks in the SM. The asymptotic freedom has profound implications for the observation — or rather lack thereof — of free quarks: when pulling two quarks apart, the force between them increases until it is so large that it is energetically favorable to create a quark-antiquark pair from the vacuum. Besides forbidding the observation of free quarks, this also explains the lack of any free particles with color charge, called "(color) confinement".

The strong coupling is measured to be $\approx 0.1$ at an energy corresponding to the $Z$ mass $M_Z \approx 91\,\text{GeV}/c^2$ and becomes $\approx 1$ at low energies of about $1\,\text{GeV}$. Perturbative calculations using higher order diagrams are therefore not possible in the way it works for QED. This is of particular importance, as low energy processes involve the hadronization of quarks. But even for higher energies, the coupling constant is still an order of magnitude larger than for the electromagnetic force. Overall, this renders theory predictions for QCD processes notoriously difficult and subject to large uncertainties.

### 2.2.3 Electroweak unification

In 1933, Fermi successfully formulated a theory of the weak force that describes the beta decay $n \to pe\bar{\nu}_e$ as a point-like interaction with Fermi's coupling constant $G_F$. Unbeknownst to Fermi at that time, the theory is an effective field theory (EFT) that worked so well at low energies due to the heaviness of the actual weak force carriers of $80\,\text{GeV}/c^2$ ($W^\pm$) and $91\,\text{GeV}/c^2$ ($Z$). After the successful formulation of QED in the 1950s, attempts were made to formulate a weak theory in a similar fashion but in $SU(2)$ to include the three gauge bosons $W^\pm$ and $Z$. In the 1960s, Glashow, Weinberg and Salam proposed the electroweak theory unifying the electromagnetic and weak interactions into a single one. They based the formulation on the combination of the $SU(2)_L$ and $U(1)_Y$ groups. The $L$ subscript denotes the left-handed chirality of the fermions, which is a consequence of the weak interaction only coupling to left-handed particles and the $Y$ denotes the hypercharge[1].

The fermion fields are grouped into doublets under the $SU(2)_L$ symmetry group

$$
\begin{aligned}
Q_L^i &= \begin{pmatrix} u_L^i \\ d_L^i \end{pmatrix} = P_L \begin{pmatrix} u^i \\ d^i \end{pmatrix} \\
L_L^i &= \begin{pmatrix} \nu_L^i \\ e_L^i \end{pmatrix} = P_L \begin{pmatrix} \nu^i \\ e^i \end{pmatrix}
\end{aligned}
\tag{2.18}
$$

and singlets

$$
\begin{aligned}
u_R^i &= P_R u^i \\
d_R^i &= P_R d^i \\
e_R^i &= P_R e^i
\end{aligned}
\tag{2.19}
$$

---

[1]It does not denote the same as $U(1)_{EM}$ in the context of the electromagnetic interaction.

where $P_{L/R} = \frac{1}{2}(1 \mp \gamma^5)$ are the projection operators onto the left-handed (minus) and right-handed (plus) chirality and $i$ denotes the generation.

### 2.2.3.1 Electroweak Lagrangian

The derivation goes by proceeding as before with QED in Section 2.2.1 by requiring local gauge invariance, here under $SU(2)_L \times U(1)_Y$. The covariant derivative and the field strength tensor are

$$D_\mu = \partial_\mu + ig_2 t^a W_\mu^a + ig_1 \frac{Y}{2} B_\mu \tag{2.20}$$

and

$$\begin{aligned} W^{\mu\nu} &= \partial^\mu W^\nu - \partial^\nu W^\mu + g_2 t^a W^\mu W^\nu \\ B^{\mu\nu} &= \partial^\mu B^\nu - \partial^\nu B^\mu \ . \end{aligned} \tag{2.21}$$

Here, $g_2$ and $g_1$ are the coupling constants of the $SU(2)_L$ and $U(1)_Y$ groups, respectively, and $W^\mu$ and $B^\mu$ are the four-potentials of the gauge bosons. $t^a$ are the generators of $SU(2)_L$ that are the Gell-Mann matrices when acting on the doublets and zero on the singlets. Eq. 2.20 and Eq. 2.21 contain the same structure as in QCD, Eq. 2.14 and Eq. 2.15. However, this time the $SU(2)_L$ and $U(1)_Y$ groups mix in Eq. 2.20.

The electroweak Lagrangian for the fermion fields is

$$\mathcal{L}_{\text{fermion}} = i\bar{Q}_L^i \slashed{D} Q_L^i + i\bar{L}_L^i \slashed{D} L_L^i + i\bar{u}_R^i \slashed{D} u_R^i + i\bar{d}_R^i \slashed{D} d_R^i + i\bar{e}_R^i \slashed{D} e_R^i \tag{2.22}$$

Inserting Eq. 2.20 and Eq. 2.21 gives rise to three terms

$$\mathcal{L}_{\text{EW}} = \mathcal{L}^{\text{kin}} + \mathcal{L}_{\text{CC}} + \underbrace{(\mathcal{L}_{\text{NC}}^\gamma + \mathcal{L}_{\text{NC}}^Z)}_{\mathcal{L}_{\text{NC}}} \tag{2.23}$$

where the second term describes charged currents, the third terms neutral currents. The conserved quantities are the weak isospin $T_a$ with $a = 1, 2, 3$ the three components, and the one-dimensional weak hypercharge $Y$. Both are related to the electric charge $Q$ by the Gell-Mann-Nishijima formula $Q = T_3 + \frac{Y}{2}$.

### 2.2.3.2 Finding the gauge bosons

The fields $W_\mu^a$ and $B_\mu$ are not the physical gauge bosons, but a linear combination of them. Rotating the fields by the weak mixing angle $\theta_W$ reveals the physical gauge bosons

$$\begin{pmatrix} A_\mu \\ Z_\mu \end{pmatrix} = \begin{pmatrix} \cos\theta_W & -\sin\theta_W \\ \sin\theta_W & \cos\theta_W \end{pmatrix} \begin{pmatrix} B_\mu \\ W_\mu^3 \end{pmatrix} \ . \tag{2.24}$$

They correspond to the $\gamma$, $A_\mu$, and the $Z^0$, $A_\mu$, bosons.

Similarly, the charged gauge bosons $W^{\pm}$ appear from mixing the fields $W^1_\mu$ and $W^2_\mu$ as

$$W^{\pm}{}_\mu = \frac{W^1_\mu \mp iW^2_\mu}{\sqrt{2}} \ . \tag{2.25}$$

The mixing angle $\theta_W$ relates the electromagnetic and the weak force through

$$g_e = g_1 \sin\theta_W = g_2 \cos\theta_W \ . \tag{2.26}$$

Combining Eq. 2.20 and Eq. 2.25, the charged Lagrangian can be written as

$$\mathcal{L}_{\mathrm{CC}} = \frac{g_2}{2\sqrt{2}}(W^+_\mu(\bar{u}^i\gamma^\mu \underbrace{(1-\gamma^5)}_{P_L} d^i + \bar{\nu}^i\gamma^\mu \underbrace{(1-\gamma^5)}_{P_L} e^i) + \mathrm{h.c.}) \tag{2.27}$$

Here, $h.c.$ denotes the hermitian conjugate and only left-handed doublets are involved. The projection term $(1-\gamma^5)$ mixes a vector and an axial-vector, which is characteristic of the weak interaction. In this stage, the $W^{\pm}$ coupling is the same for all fermion generations and no parity violation is observed.

The neutral Lagrangian for $Z$ can be written using Eq. 2.20 and Eq. 2.24 as

$$\begin{aligned}
\mathcal{L}^Z_{\mathrm{NC}} &= \frac{g_2}{2\cos\theta_W} Z_\mu (T_3 - Q\sin^2\Theta_W)\bar{f}^i_H\gamma^\mu f^i_H \\
&= \frac{g_2}{2\cos\theta_W} Z^0_\mu \bar{f}^i_H\gamma^\mu (C^H_V - C^H_A\gamma^5)f^i_H \ ,
\end{aligned} \tag{2.28}$$

where $H = L, R$ denotes the chirality of the fermion. The vector ($C_V$) and axial-vector ($C_A$) couplings that were $1/2$ for the charged Lagrangian in Eq. 2.27 are for the neutral current

$$\begin{aligned}
C^H_V &= T_3 - Q\sin^2\Theta_W \ , \\
C^H_A &= T_3 \ ,
\end{aligned} \tag{2.29}$$

Due to $T_3 = 0$ for the right-handed singlets, the coupling is different from the left-handed doublets and notably $C^R_A = 0$.

Using again Eq. 2.20 and Eq. 2.24, the Lagrangian for the photon is

$$\mathcal{L}^\gamma_{\mathrm{NC}} = g_e A_\mu \bar{f}^i_H\gamma^\mu Q_i f^i_H \ , \tag{2.30}$$

where notably no axial-vector term appears conserving parity as was expected from QED.

### 2.2.4 Higgs field

The gauge bosons $W^{\pm}$ in Eq. 2.27 and $Z$ in Eq. 2.28 appear massless, contrary to expectations. As seen in Eq. 2.7, terms that are quadratic in $\psi$ of the form of $-m\psi\bar{\psi}$ are not invariant under local gauge transformations. This contradicting puzzle was solved in 1964 by the introduction of the Higgs mechanism and SSB, named after one of the proponents, Peter Higgs.

This mechanism relies on the introduction of a new field $\phi$ that is a doublet of complex scalar fields of the group $SU(2)_L$ with hypercharge $Y = 1$ as

$$\phi = \begin{pmatrix} \phi^+ \\ \phi^0 \end{pmatrix} \tag{2.31}$$

where $\phi^+$ and $\phi^0$ are complex fields with electric charge $+1$ and $0$ respectively.

The Lagrangian term for the Higgs field respects the $SU(2)_L \times U(1)_Y$ symmetry and is given by

$$\mathcal{L}_{\text{Higgs}} = (D_\mu \phi)^\dagger (D^\mu \phi) - V(\phi^\dagger \phi) \tag{2.32}$$

where $D_\mu$ is the covariant derivative of the Higgs field and $V(\phi^\dagger \phi)$ is the potential of the Higgs field. The potential $V(\phi)$ is arbitrary and chosen to be

$$V(\phi) = \mu^2 \phi^\dagger \phi + \lambda (\phi^\dagger \phi)^2 \; , \tag{2.33}$$

which has minima for any $\phi^\dagger \phi = \frac{\mu^2}{2\lambda} \equiv \nu$, defining the vacuum expectation value $\nu$. At the unstable equilibrium $\phi = (0, 0)$, the system is perfectly rotationally invariant. Moving to any ground state at $\phi^\dagger \phi = \frac{\mu^2}{2\lambda}$ breaks the symmetry of the system. As the breaking appears only below a critical temperature $160\,\text{GeV}$ [6] and preserves the symmetry above it, it is called SSB.

A common choice is the "unitary gauge"

$$\phi_0 = \frac{1}{\sqrt{2}} \begin{pmatrix} 0 \\ \nu + h(x) \end{pmatrix} \tag{2.34}$$

containing the real, massive scalar field of the Higgs boson $h(x)$.

### 2.2.4.1 Effects of spontaneous symmetry breaking

After SSB, the residual symmetry is $U(1)_{EM}$, as seen in Section 2.2.3, broken down from the $SU(2)_L \times U(1)_Y$ group. The massless gauge bosons from the electroweak theory acquire mass terms that appear in $\mathcal{L}_{\text{Higgs}}$

$$D_\mu \phi^\dagger (D^\mu \phi) \propto \underbrace{(\partial_{mu} h)(\partial^\mu h)}_{\mathcal{L}_{\text{Higgs}}^{kin}} + \underbrace{(\frac{g_2 \nu}{2})^2 W^+{}_\mu W^{-\mu}}_{m_W^2} + \underbrace{\frac{(g_1^2 + g_2^2)\nu^2}{4}}_{m_Z^2} Z_\mu Z^\mu \ . \quad (2.35)$$

Using Eq. 2.26, the masses of the $W^\pm$ and $Z$ bosons can be expressed as

$$\begin{aligned} m_W &= \frac{1}{2} g_2 \nu \approx 80 \text{ GeV/c}^2 \\ m_Z &= \frac{1}{2} \sqrt{g_1^2 + g_2^2}\, \nu = \frac{m_W}{\cos \theta_W} \approx 91 \text{ GeV/c}^2 \ . \end{aligned} \quad (2.36)$$

The photon remains massless as no term of the form $A_\mu A^\mu$ appears in Eq. 2.35.

### 2.2.5 Yukawa coupling

Yukawa interactions are responsible for the coupling of fermions to the Higgs field and ultimately for the ability to differentiate between flavors.

Using the doublets from Eq. 2.18 and singlets from Eq. 2.19, the Lagrangian can be written as

$$\mathcal{L}_{\text{Yukawa}} = \underbrace{-Y_u \bar{Q}_L \tilde{\phi} u_R - Y_d \bar{Q}_L \phi d_R}_{\mathcal{L}_{\text{Yukawa}}^{\text{quarks}}} \underbrace{-Y_l \bar{L}_L \phi l_R}_{\mathcal{L}_{\text{Yukawa}}^{\text{leptons}}} + h.c. \quad (2.37)$$

Here, $Y_f$ are the Yukawa couplings, generic complex matrices in the flavor space and

$$\tilde{\phi} = i\sigma_2 \phi^* = \frac{1}{\sqrt{2}} \begin{pmatrix} \nu + h(x) \\ 0 \end{pmatrix} \quad (2.38)$$

is the conjugate Higgs doublet.

Expanding the Lagrangian in Eq. 2.37 and using the SSB from Section 2.2.4.1 yields

$$\mathcal{L}_{\text{Yukawa}} = -\left(1 + \frac{h}{\nu}\right) \left(\frac{Y_u \nu}{\sqrt{2}} \bar{u}_L u_R + \frac{Y_d \nu}{\sqrt{2}} \bar{d}_L d_R + \frac{Y_l \nu}{\sqrt{2}} \bar{l}_L l_R\right) + h.c. \quad (2.39)$$

The mass terms of the fermions arise as

$$M_f^{ij} = \frac{Y_f^{ij} \nu}{\sqrt{2}} \ . \quad (2.40)$$

The fermion fields in Eq. 2.39 are in their flavor eigenstates, a linear combination of the mass eigenstates. An appropriate bi-unitary rotation of the fields diagonalizes the mass matrices

$$\Lambda^f = (U_L^f)^\dagger M_f U_R^f \ , \quad (2.41)$$

where $\Lambda_f^j$ are the fermion masses and $U^f$ are unitary matrices that transform the fermions between the flavor and mass eigenstates as

$$f'_H = U_f f_H \ , \quad H = L, R \ , \quad f = u, d, l \ . \tag{2.42}$$

### 2.2.5.1 CKM matrix

The interaction in charged currents in Eq. 2.27, and neutral currents, in Eq. 2.28 and Eq. 2.30, are written in terms of the flavor eigenstates. Using the transformation in Eq. 2.42 on both quark fields yields an extra term of the form, for example

$$\bar{u}_L d_L \rightarrow \bar{u}_L \underbrace{U_u^\dagger U_d}_{V_{CKM}} d_L \ , \tag{2.43}$$

Both fields can be connected through the unitary CKM matrix [7] $V_{CKM} = U_u^\dagger U_d$ as

$$\begin{pmatrix} u_L \\ c_L \\ t_L \end{pmatrix} = V_{CKM} \begin{pmatrix} d_L \\ s_L \\ b_L \end{pmatrix} \tag{2.44}$$

with the approximate values from [7]

$$\begin{aligned} V_{CKM} &= \begin{pmatrix} V_{ud} & V_{us} & V_{ub} \\ V_{cd} & V_{cs} & V_{cb} \\ V_{td} & V_{ts} & V_{tb} \end{pmatrix} \\ &\approx \begin{pmatrix} 0.97 & 0.23 & 0.004 \\ 0.22 & 0.97 & 0.04 \\ 0.009 & 0.04 & 0.999 \end{pmatrix} \end{aligned} \tag{2.45}$$

The CKM matrix describes the mixing of the quark generations and is close to identity, yet notably not diagonal. It can be parametrized by three angles $\theta_{12}$, $\theta_{23}$ and $\theta_{13}$ and a complex phase $\delta$ responsible for charge and parity ($CP$) violation in the SM. The mixing is hierarchical, with the first generation being of $\mathcal{O}(1)$, reduced coupling between the first and second generation of $\mathcal{O}(0.1)$ and minimal coupling between the first and the third generation of $\mathcal{O}(0.01)$.

The charged and neutral current interactions from Eq. 2.27, Eq. 2.28 and Eq. 2.30 after the SSB can be written as

$$\begin{aligned} \mathcal{L}_{CC}^{quarks} &= g_2 \bar{u}_L V_{CKM} d_L W^+ + h.c. \\ \mathcal{L}_{NC} &= \frac{g_2}{\cos \theta_W} \bar{f} \gamma^\mu (g_V - g_A \gamma^5) f Z_\mu \ . \end{aligned} \tag{2.46}$$

The mixing of different quark generations via the CKM matrix appears only in the charged current interaction, but not in the neutral current interaction. The consequence of this is that flavor changing interactions can occur at tree level in the charged current interaction, but flavour changing neutral current (FCNC) only appear at loop-level, strongly suppressing the latter by the Glashow-Iliopoulos-Maiani (GIM) mechanism [8]. The implications for the search for new particles will be discussed in further detail in Section 2.3.3.

The Lagrangian after SSB is still invariant under continuous $U(1)$ transformations of all quark field phases simultaneously. This results in an accidental symmetry of the Lagrangian, not required by the theory but observed in nature, the baryon number conservation.

### 2.2.5.2 Lepton masses

The Lagrangian in Eq. 2.37 only contains the charged leptons

$$\mathcal{L}_{\text{Yukawa}}^{\text{lepton}} = -Y_l \bar{L}_L \phi l_R - \underbrace{Y_\nu \bar{L}_L \tilde{\phi} \nu_R}_{\text{Neutrino term (hypothetical)}} + h.c. \tag{2.47}$$

The neutrino term vanishes in the SM and is not actually present in the Lagrangian. Under the assumption of massless neutrinos, the coupling is $Y_\nu^{ij} = 0$. This strongly simplifies the transformations in Eq. 2.42 as we can set the mass eigenstates equal to the flavor eigenstates. The transformation matrix $U_\nu \dagger U_l$ is simply the identity matrix.

As for the quarks, an accidental symmetry is also present in the lepton sector, the invariance under *each* lepton field phase transformation, known as lepton flavour universality (LFU). This implies the lepton number conservation in each flavor, $L_e$, $L_\mu$ and $L_\tau$.

## 2.3 Beyond the Standard Model

The predictions of the SM are in remarkable agreement with experimental data and have been confirmed to a high degree of precision. And yet, it is certainly not the end of the story, as minor and major open questions remain. In this section, firstly, some of the more pronounced open questions and inconsistencies with measurements are discussed in Section 2.3.1. Section 2.3.2 elaborates on different methods of probing the SM for possible extensions. Finally, Section 2.3.3 discusses LFU and possible violations, motivating the measurements in Chapter 6.

### 2.3.1 Open questions

There are two types of disagreements with the SM, depending on their size: small discrepancies in specific measurements that probe the SM and hint towards a crack. They may accumulate to a bigger picture, but are limited now. They help in the search, for *what* is broken and will be further discussed in Section 2.3.2. But there are large problems that pose *fundamental* issues, making clear that the SM, in today's form is not the end of the story. Those open questions will be discussed in the following.

**Neutrino Oscillations**

In the early 2000s, experiments started to measure the flavors of neutrinos and found that neutrinos would change their flavor while propagating [9]. This is compelling evidence that the mass eigenstates of neutrinos do not coincide with the flavor eigenstates. A difference in the masses *squared* of different neutrinos explains a possible way to allow them oscillating back and forth, but thereby implying that neutrinos have a mass. It is — in contrast to, say, photons — *not* a fundamental requirement of the SM for them to be massless. In Eq. 2.47, a hypothetical neutrino term is added to the Lagrangian and setting $Y_\nu^{ij} > 0$ would allow for neutrinos to have a mass. The current extension describes the mixing of mass and weak eigenstates by the Pontecorvo-Maki-Nakagawa-Sakata (PMNS) matrix, connecting the mass and flavor eigenstates similar to how the CKM matrix works for quarks. From measurements, the mixing angles of the PMNS seem to be significantly larger than those in the CKM matrix [10] and currently at the percent level of precision.

**Strong and electroweak unification**

The SM is also by itself not a fully satisfying theory. It describes the unification of the electromagnetic and weak interactions that behave only so differently in the "cold" environment of today's universe, below the Planck energy scale of about $10^{19}$ GeV. This leaves the strong interaction as an independent force and begs the question whether the SM is itself a low energy EFT approximation of a more general theory that unifies these forces at some energy scale.

**Gravity**

Gravity is described by general relativity (GR) and is not part of the SM at all. As it is about 40 orders of magnitude weaker than the electromagnetic force, it is not relevant for particle interactions at the scales of elementary particles. It stands separated from the SM, without even a quantized description available that *could* potentially be unified. As it stands, GR blows up for high energies, rendering it a non-normalizable theory.

**Dark matter and dark energy**

From astrophysical observations and simulations, it is well established that the detectable matter only makes up about 5 % of the total energy density of the universe [11]. The rest is made up of so-called — in lack of a better name — dark matter and dark energy, none of which is part of the SM. Gravity itself thereby brings up the question about the nature of dark matter and dark energy. But this poses a fundamental problem: as dark matter interacts through the gravitational force, it is possible that it interacts *at least* via one of the other forces. As such it should be visible, in one way or another, in interactions studied at particle colliders.

**Baryon genesis**

The universe holds another stunning question, the matter-antimatter asymmetry. According to the SM, the same amount of matter and anti-matter should have been produced in the big bang, yet the latter seemed to have disappeared. There is a possible mechanism in the SM [12] that allows for this, the *CP* violation [13], where a decay of a particle and its anti-particle do not occur at the same rate. It originates from the complex phase in the CKM matrix, although at the known values, it is far too small to explain the observed asymmetry.

### 2.3.2 Probing the Standard Model

Answering the open questions from Section 2.3.1 is not as straightforward as confirming some prediction, as no prevalent theory that solves all problems has come forward yet successfully. In fact, the most promising theory for the last decades, supersymmetry, a theory that suggests a symmetry between fermions and bosons, (SUSY), has not shown any success at the LHC experiments [14]. To find where the SM is incomplete, a search for New Physics (NP) contributions is necessary by probing the SM. But with no clear path forward, the question of *how* to find what lies beyond the SM becomes more prevalent. More specifically, by using collider experiments such as the ones situated at the LHC.

Different NP models that aim to solve existing problems predict effects that usually involve new, often heavy or feebly interacting particles. And whatever NP the models involve, *some* effect should be measurable at the energy scales of the LHC. Effects mostly mean new particles — as elementary particles themselves that extend the SM, as fundamental constituents of currently known fundamental particles or even new forces that are only relevant at higher energy scales. One way to look for new particles is to look for directly produced particles in a decay, so-called *direct searches*. As particles can only be created with high enough accelerator beam energies, this strongly limits the mass of the particles that can be produced and detected. Whatever the particle, even very feebly interacting ones, they are detectable through unknown signatures in the detector, prominently some that miss energy or momentum. Due to the latter requiring a hermetic detector such as the ATLAS and CMS experiments, such searches are often not well suited for the LHCb experiment and only a subset of the possible NP models can be probed in direct searches.

Following the main rule of QM, any process that can happen, will happen. Hence, if yet unknown particles and interactions exist, they contribute to the cross-section of any process that contains particles, which interact with such new physics. This can enable otherwise prohibited decays that violate SM imposed conservation laws or that *alter* the decay rate of a process by contributing — constructively as well as destructively — to the amplitude of the process. This is called *indirect searches* and are effectively precision measurements of already known quantities that are believed to be in agreement with the SM predictions.

Given the overwhelming success of the SM, any NP process is expected to be relatively small. Therefore, rare decays provide a particularly useful tool for indirect searches as small NP contributions may be more prominent *w.r.t.* their small SM branching fractions. Precision measurements of the SM also pose a theoretical challenge as the SM predictions need to be calculated to at least the same degree of precision as the experimental measurement is sensitive to.

### 2.3.3 Flavor anomalies

On the search for NP, decays with an underlying $b \to s\ell^+\ell^-$ transition with denoting a lepton, FCNC processes, offer a promising avenue to explore. On the one hand, this is due to the heavy $b$ quark: many NP models build on the EW symmetry breaking and, given the $b$'s strong Yukawa couplings to the Higgs field, predict particles preferentially coupling to the third quark generation. The $b$ is not as well studied as lighter quarks, given their abundance and possibility to easily measure in lower energy experiments.

On the other hand, FCNC decays are highly sensitive to contributions from NP: as elaborated in Section 2.2.5.1, in the SM, FCNC decays are forbidden at tree level and only occur at loop level. This means the amplitude incorporates only higher order processes whose contributions are significantly smaller, making them rare in the sense of a tiny branching fraction.

$b \to s$ transitions, such as $b \to s\ell^+\ell^-$ decays, are a type of FCNC decays as depicted in Fig. 2.3.1. Apart from the loop-level suppression, they are strongly suppressed by the GIM mechanism. The dominant term involves a product of two CKM off-diagonal matrix elements $V_{bc}V_{cs}^*$, which are small, as the CKM matrix is almost diagonal.

Many NP models predict new possible contributions to $b \to s\ell^+\ell^-$ decays [15–17], with two possible examples shown in Fig. 2.3.2.

In recent years, different measurements involving mostly $b \to s\mu^+\mu^-$ processes have shown deviations from the SM predictions, referred to as *B* or *flavor anomalies*. Direct measurements of differential branching fractions for $B^+ \to K^+\mu^+\mu^-$ and $B^0 \to \phi\mu^+\mu^-$ decays were both found to be below their theoretical predictions in the $q^2$ region between 1.1 and 6.0 GeV$^2/c^4$ (central $q^2$ region) [18, 19]. In $B^0 \to K^{*0}\mu^+\mu^-$, an optimized angular observable, $P_5'$, shows a deviation from the SM prediction at the $\approx 3\sigma^2$ level [20], as shown in Fig. 2.3.3.

Another approach is to use the EFT formalism, which describes physics up to a given energy scale and absorbs effects originating from higher energy scales into effective coupling constants, the Wilson coefficients (WCs). This integrates out the effect of high energy contributions, making an interaction point-like above the energy scale of interest[3], here $m_B$. This provides an approximation at the desired energy scale and allows to study physics in this energy domain despite a lack of knowledge about the physics at larger energies. A number of measurements related to $b \to s\mu^+\mu^-$ processes and to $b \to s\gamma$ processes measured $C_7'$, $C_9'$, $C_{10}'$ appearing in the Hamiltonian describing the underlying transitions [21–23]. They seem to show a similar tendency as the other $B$ anomalies, namely that the fit favors a shift of $C_9$, or of a combination of $C_9$ and $C_{10}$, from their SM expectations, as illustrated in Fig. 2.3.4.

---

[2]Depending on the exactly chosen theoretical model and whether to compare $P_5'$ in individual bins or other variables, but the order of magnitude is the point.

[3]Similar to Fermi's theory of the weak interaction as briefly mentioned in Section 2.2.3.

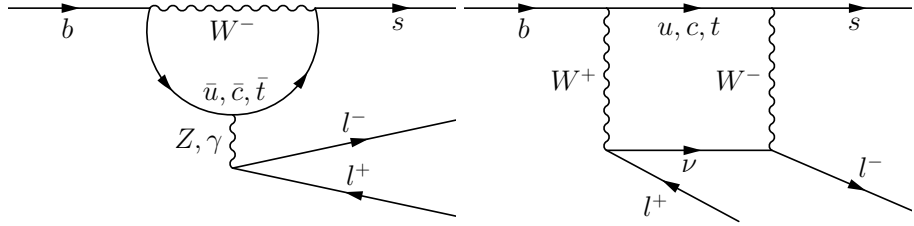

**Figure 2.3.1:** Feynman diagrams of FCNC decays of $b$-hadrons featuring general $b \to s\ell^+\ell^-$ decays. The left diagram is often referred to as "penguin" diagram, the right one as "box" diagram.

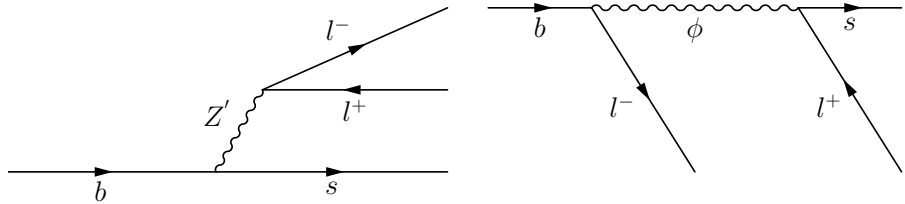

**Figure 2.3.2:** Feynman diagrams of $b \to s\ell^+\ell^-$ decays featuring tree-level contributions of beyond the Standard Model (BSM) particles. The left diagram shows a $Z'$ boson as a mediator [17]. The right diagram contains a leptoquark-like particle that converts leptons to quarks and vice versa [16].

The quoted significances of these measurements originate from experimental as well as theoretical uncertainties, increasing the significance implies reducing these uncertainties. On the experimental side, the measurements depend on a thorough understanding of the detector response and the efficiency of the selection, both mainly through simulation and the limited amount of data. For theoretical calculations, hadronic interactions are notoriously difficult to compute, as discussed in Section 2.2.2.1. The difficulty arises in non-perturbative scenarios, occurring at low energies that appear in the hadronisation of particles, for example in charm loops, and in the calculation of form factors, thereby leading to large uncertainties. These shortcomings add more variables to the equation: not only is the experimental measurement affected by uncertainties, but also the theoretical predictions are, further lowering the overall significance of a possible deviation. Thereby, they paint a consistent, yet inconclusive, picture.

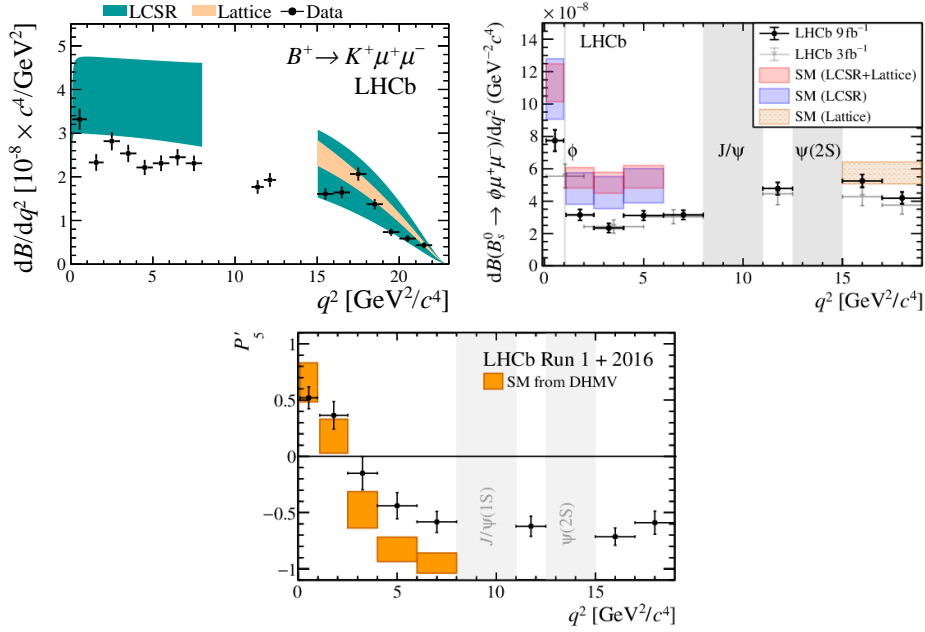

**Figure 2.3.3:** Recent LHCb measurements of $b \to s\mu^+\mu^-$ observables. The top left plot [18], and the top right plot [19] show the differential branching fractions for $B^+ \to K^+\mu^+\mu^-$ and $B^0 \to \phi\mu^+\mu^-$ decays, respectively. The LHCb data points are shown in black and the theory predictions are shown with blue bands. The bottom plot shows the $P_5'$ angular observable from the $B^0 \to K^{*0}\mu^+\mu^-$ decay [20], with the data points shown in black and the theory prediction in orange. All three plots show a discrepancy with respect to the SM in the central $q^2$ region.

### 2.3.4 Lepton flavor universality

"*Who ordered that?*" — I. I. Rabi, 1936

According to the SM, the three lepton flavors electron, muon and tau are identical, except for their masses. Different particles that are essentially the same already raised eyebrows in the early days of particle physics, when the muon was discovered as the heavier sibling of the electron. As seen in Section 2.2.3, the different lepton flavors have the same coupling strength to the electroweak gauge bosons, known as LFU. This is only accidentally preserved in the SM after SSB, as the Yukawa couplings are diagonal in lepton flavor space, as stated in Eq. 2.47.

A cleaner way to test $b \to s\ell^+\ell^-$ transitions is therefore to measure LFU directly, as LFU implies that, whatever the individual branching fraction of two decays,

$$X \to h\ell^+\ell^- \text{ , with } \ell = e, \mu, \tau \tag{2.48}$$

where $X$ and $h$ are hadrons, the *ratio*, referred to as $R_h$, of two such branching fractions has to be unity, under the assumption that the leptons have the same mass. Specifically, for the ratio of electrons and muons, the actual lepton masses differ, but minimally compared to the overall energy transfer in the process. A difference arises in the final state radiation (FSR), which is proportional to the lepton mass and needs to be taken into account experimentally in simulation. The difference in the branching fractions is less than a percent and is well under control, using simulation with PHOTOS[4], in LHCb analyses [25].

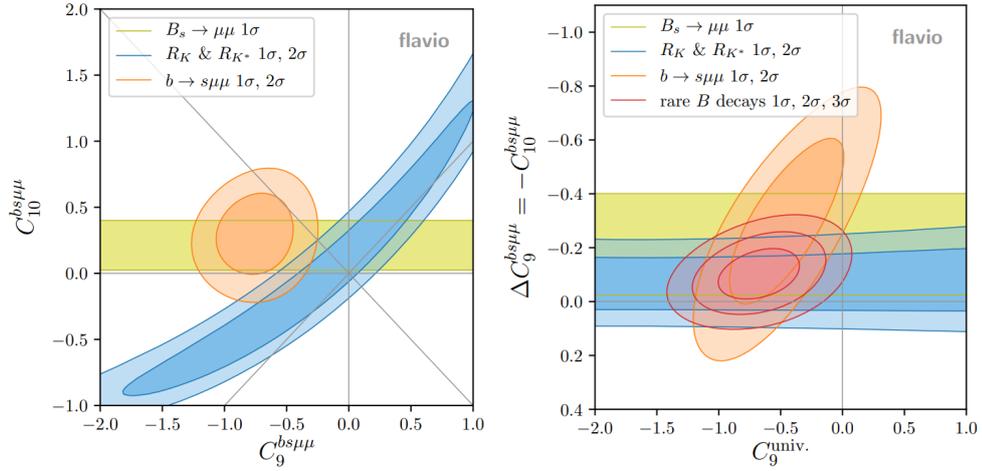

**Figure 2.3.4:** Results of the global fits to $C_9$ and $C_{10}$ using $b \to s\mu^+\mu^-$ data from [24]. The left-hand plot shows the 1 and 2 $\sigma$ contours for the global fit to $C_9$ and $C_{10}$ in orange, the 1 $\sigma$ contours from the angular data in yellow, the combined $R_K$ and $R_{K^{*0}}$ data fit in blue. The right-hand plot shows the equivalent contours of measurements but now assuming a lepton universal contribution to $C_9$ and a $V - A$ (*i.e.*, $\Delta C_9 = -C_{10}$) contribution to the muon modes. The experimental measurements are in good agreement and show a slight deviation from the SM expectation.

Measuring a ratio of two branching fractions has the advantage that many uncertainties cancel out, both theoretical and experimental. On the theory side, hadronic non-perturbative, contributions entering in the $X \to h$ transition — according to the SM — appear in both decays and therefore cancel exactly, allowing for an extremely precise theoretical prediction. A similar argument goes for experimental uncertainties, as most of them are common between different decays as in Eq. 2.48 and are expected to cancel in the ratio.

Several of these ratios have been measured and are given, together with the previous measurements in an overview in Fig. 2.3.5. The current picture shows no major deviation in $R_K$-like measurements, yet all of them have been done in either the $q^2$ region between 0.1 and 1.1 GeV$^2$/$c^4$ (low $q^2$ region) or the central $q^2$ region and none was performed in the $q^2$ region between 14.3 and 23.0 GeV$^2$/$c^4$ (high $q^2$ region). The latter provides a complementary measurement of a statistically independent sample and accesses a new phasespace, which differs experimentally. A measurement of the LFU through the ratio $R_K$ in the channels $B^+ \to K^+ e^+ e^-$ and $B^+ \to K^+ \mu^+ \mu^-$ in the high $q^2$ region will be discussed in Chapter 6.

---

[4]PHOTOS is used to generate bremsstrahlung in the simulated decay of particles.

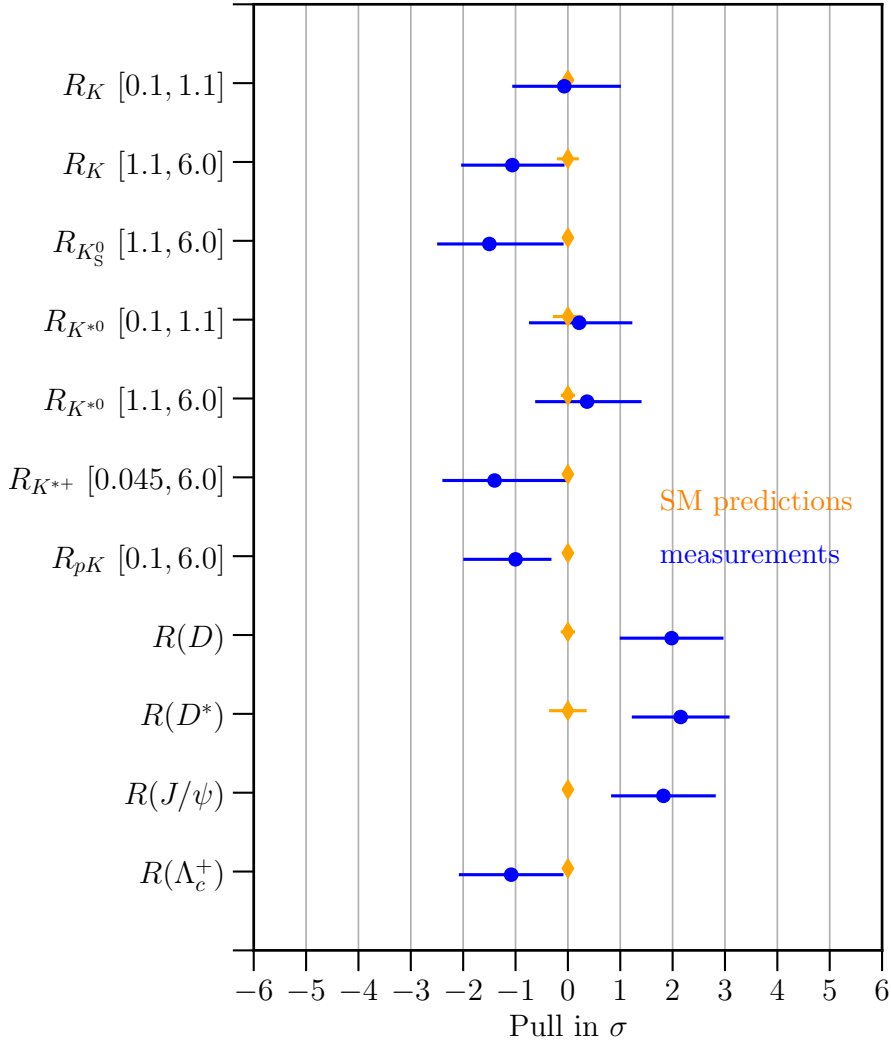

**Figure 2.3.5:** Overview of recent $R_h$LFU measurements [26]. The type of measurement is denoted on the left, with $[x, y]$ indicating the $q^2$ region in GeV$^2/c^4$. SM prediction of each measurement (orange) is moved to zero. The significance of an experimental measurement (blue) is shown relative to zero and scaled, by requiring that both, the theoretical and experimental, uncertainties are summed in quadrature sum up to unity. If a significance is given in the relevant publication, this value is used instead of the computed significance.

# 3

# Applied Statistical Methods

Randomness is a fundamental aspect of nature. It is baked into the laws of quantum mechanics and can be seen in the macroscopic world due to the complexity of systems. Statistical tools are needed to describe such stochastic processes in order to make any inference about the underlying laws of nature. In HEP, stochastic processes appear in decay probabilities, detector effects, simulation generation, and many more.

Statistical statements are usually exactly computable in only three cases: tiny samples, infinitely large samples, or normally distributed processes. In reality, the actual statistical inference is often diluted by practicalities such as computationally expensive calculations or the lack of exact descriptions for most processes. As a consequence, statistical methods used in reality rely on necessary simplifications and assumptions, often resulting in field-specific adoptions of certain methods. In HEP experiments, with a large amount of available data and simulation, the methods tend to be sophisticated and a high degree of correctness is expected. In the following, a few applied statistical methods that are used throughout this thesis will be introduced and motivated; a detailed derivation and explanation can be found in textbooks such as Refs. [27–29] , whose parts of this chapter are based on.

The first section introduces some concepts of stochastics and statistics, see Section 3.1. The second and third sections present the field of machine learning more generally, Section 3.2, and with a focus on deep learning, Section 3.2.2. Section 3.2.3 discusses the validation of different inference algorithms. Section 3.3 introduces the likelihood and its application in inference. Section 3.4 provides an overview over optimizations in the context of likelihood maximization and deep learning training. Finally, the computational aspects of all these methods are considered in Section 3.5.

## 3.1 Randomness

The most frequently spoken words in natural language are words that are anticipated, such as *the, a, to, of, etc.*, and therefore add very little information. This reasoning is captured in the Shannon entropy, which describes the amount of "surprise" or "information content" of a random outcome. Some formal definitions related to entropy and randomness are provided in the following.

The sample space $\Omega$ is the set of all possible outcomes of an experiment.

**Definition 1** *The `probability distribution` of a random variable (RV) X is a function p that maps the sample space $\Omega$ to the real numbers $[0, 1]$. The probability distribution p is a probability measure, i.e., it satisfies the following properties:*

1. *$p(\Omega) = 1$: The probability over the whole sample space $\Omega$ is 1.*

2. *$p(\emptyset) = 0$: The probability over the empty set $\emptyset$ is 0.*

3. *$p(A \cup B) = p(A) + p(B)$: The probability of the union of two subsets A and B of $\Omega$ is the sum of the probabilities of the two sets.*

Equivalently, a continuous distribution with similar properties can be defined, replacing the sum with an integral and the probability of a single event being zero, but the probability of an interval being non-zero. The definition of a probability density function (PDF) given a non-negative function $f$ is

**Definition 2**
$$PDF_f(x) = \frac{f(x)}{\int_\Omega f(x)dx}$$

In the following, Def. 1 of the discrete distribution will be used, but the results are basically equally valid for Def. 2.

The general definition of entropy is given by

**Definition 3** *The `entropy` of an RV X is defined as*
$$\mathrm{H}(X) = -\sum_{x \in X} p(x) \log(p(x))$$

This can be dissected into the information content of an event $x$, which is defined as $-\log p(x)$, and the expectation value of the information content, the sum over all events weighted by their probability. The entropy is thus the average information content of an event and can be reformulated as

$$\mathrm{H}(X) = \mathbb{E}[-\log p(X)] \tag{3.1}$$

The expression for cross-entropy $H(p,q)$ can be motivated by taking the expectation value *w.r.t.* a different distribution in Def. 3, yielding

$$\begin{aligned}
\mathrm{H}(p,q) &= -\sum_{x \in X} p(x) \log q(x) \\
&= \mathrm{H}(p) + \sum_{x \in X} p(x) \log \frac{p(x)}{q(x)}
\end{aligned} \tag{3.2}$$

The cross-entropy is thus the entropy of the true distribution $p$, a constant for a given dataset, plus the *Kullback-Leibler divergence*[1] [30] between the two distributions. It is widely used in deep learning and will be applied in Section 5.4. The Kullback-Leibler divergence is a measure of how different two probability distributions are, which is nonnegative and zero if and only if the two distributions are identical. This result can be interpreted as self-information of the true distribution plus the difference between the distribution $p$ and $q$. Both distributions $p$ and $q$ can depend on a set of parameters $\theta$, written as $p(x|\theta)$ and $q(x|\theta)$, which will be used later in this chapter.

## 3.2 Machine learning

The core concept of machine learning is to approximate a transformation by data, rather than explicitly formulating it, with a highly flexible function. This involves obtaining a set of parameters $\Theta$ for a function $f(x; \Theta)$, which minimizes some metric of error with the data; in machine learning, this is called the *training*. Deep learning is a subset of machine learning algorithms that use artificial neural networks, which are distinct from general machine learning algorithms due to their specific structure. Deep learning has become a huge field itself, however, not always are neural networks the right approach. They tend to be the most flexible algorithms and have the power to adapt to the most complex problems; this is called the "capacity" of an algorithm and should correspond to the complexity of the problem at hand.

---

[1] A disparity measure between two probability distributions based on information.

### 3.2.1 Trees and ensembles

**Decision trees**, and ensemble versions, mainly boosted decision trees (BDTs) [31], have a special place in machine learning (ML). They have existed for a long time, with the first proposal of a decision tree (DT) in 1963 [32], and are probably the most intuitive algorithms that still deliver close to or even state-of-the-art performance for some problems. As such, they are used within the LHCb experiment, see Section 4.7.2, and in analyses, as discussed in Section 6.2.5.

A DT is a simple form of decision making. It is a binary tree structure where each node represents a decision and each branch a possible outcome of that decision. The tree is constructed by recursively splitting the data into subsets based on a feature, as illustrated in Fig. 3.2.1. Hereby, the feature to split in and the split value itself are chosen such that the resulting subsets are as pure as possible. The purity of a subset is measured using an appropriate metric.

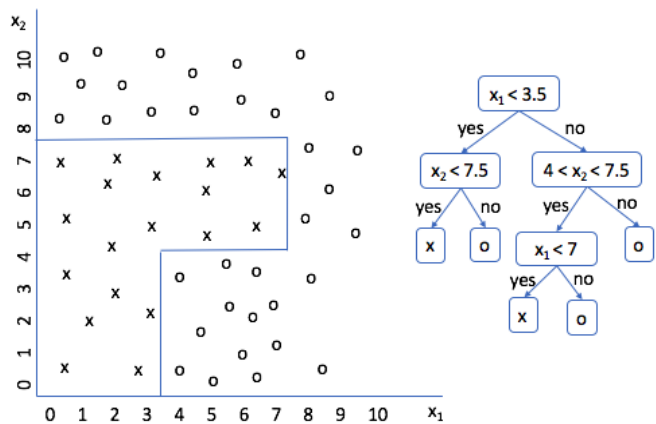

**Figure 3.2.1:** Illustration of a DT. The data points are depicted on the left side with the decision boundaries of the tree, with the latter depicted on the right side. This splits the input space into rectangular hypercubes and allows to account for non-linear correlations.

A decision is made by going from the top of the tree to the bottom where the actual outcome is located, *i.e.,* in the leaves of the tree. Factually, this splits the input space into rectangular hypercubes. Although a single deep tree can already be powerful, they are prone to overfit and hardly ever generalize well. They are, however, an extremely powerful tool when used with ensemble methods as a weak learner.

The algorithm of **boosting** [33] approaches the learning task with an iterative method, based on "weak learners", and can be extended to other tasks, see Section 6.3.3.2. A weak learner is considered an algorithm that has a low capacity on its own *i.e.,*, that is significantly below the problems complexity, and underfits the original problem. In each iteration, a new weak learner is trained on the data. Then, the data is weighted so that the weak learner focuses on the data that was misclassified by the previous weak learners. Finally, the algorithm prediction is a weighted sum of the predictions of all weak learners, whereby the weights are determined by the performance of the weak learners. The better a weak learner performs, the higher the weight of its prediction. This is repeated until a stopping criterion is reached, typically a maximum number of weak learners or a minimum performance of the weak learners.

A **Boosted Decision Tree** BDT combines these two parts, a DT and the boosting technique. When talking about BDTs, the weak learner is a shallow decision tree, with a typical depth — the longest path from top to bottom — of 2 to 10. Their simplicity promises a fast training time and a stable output compared to deep learning (DL) algorithms. They are the prevalent choice for supervised classification on high-level variables in HEP analysis.

### 3.2.2 Deep learning

A computer mimicking neurons in 1943 represents the first neural network prototype. The groundwork for most modern deep-learning algorithms was laid in the 1960s and 1970s leading to the invention of convolutional neural networks in 1980 [34]. Several constraints, most importantly the lack of suitable data and computational power, prevented the field from taking off at the time. Only in the 2000s, BDTs became a common tool and the first successful classifications using convolutional neural networks (CNNs) were achieved [35]. The following introduces the basics of neural networks and will be built upon in Section 5.2 with the introduction of geometric learning.

At the heart of neural networks are matrix multiplications and nonlinear functions, called *activation functions*. The input vector corresponds to the input data size, and the output corresponds to the desired predictions. In between are so-called *hidden layers*. Mathematically, the structure of a neural network is described as

$$
\begin{aligned}
\vec{y} &= f(\vec{x}; \Theta) \\
&= f^{(L)} \circ f^{(L-1)} \circ ... \circ f^{(1)}(\vec{x}; \Theta) \\
&= f^{(L)} \left( f^{(L-1)} \left( ... f^{(1)} \left( \vec{x}; \Theta^{(1)} \right); \Theta^{(2)} \right); ... ; \Theta^{(L)} \right) \\
&= f^{(L)} \left( f^{(L-1)} \left( ... f^{(1)} \left( \vec{x} \cdot \Theta^{(1)} + \vec{b}^{(1)} \right) \cdot \Theta^{(2)} + \vec{b}^{(2)} \right) ... \right) \cdot \Theta^{(L)} + \vec{b}^{(L)}
\end{aligned}
\tag{3.3}
$$

where $\Theta$ is the set of all parameters, which are called weights in deep learning, $\Theta^{(l)}$ the weights of the $l$-th layer and $\vec{b}^{(l)}$ the biases, an additive weight, of the $l$-th layer. The function $f^{(l)}$ is the activation function of the $l$-th layer. One of the most common activation functions is the rectified linear unit (ReLU) defined as

$$
f^{(l)}(x) = \max(0, x)
\tag{3.4}
$$

but also other functions, such as the `sigmoid` function defined as

$$f^{(l)}(x) = \frac{1}{1 + e^{-x}} \tag{3.5}$$

are commonly used, the only requirement being that they are non-linear.

The weights are arbitrary in the beginning but are optimized during the training, see Section 3.4. The loss function plays a central role, as it measures the disagreement between the output of the network and depends on the task at hand. For classification, it is often based on the cross-entropy, as defined in Eq. 3.2.

A deep neural network (DNN) as defined in Eq. 3.3 is a general formulation of a function. Using different activation functions, number of layers and nodes, and different connections between them, all summarized under the term "architecture", can lead to vastly different performances. Theoretically, a one-layer neural network with enough nodes can approximate any function [36]. However, "enough" is usually not feasible, and deep networks, *i.e.,* with more layers, were found to perform better than shallow, wide networks, hence the name "deep learning"[2]. In practice, deep learning has proven to be a general purpose function approximator as theoretically suggested, given enough data *and* computational power. However, being general comes at a cost, and not all forms of generality are always needed. Instead, a function tailored to the problem at hand is a better choice, as it removes the need to infer the problem-specific structures from data. This greatly reduces the demand for data and computational power, making certain problems actually feasible to be learned. This was what ultimately brought "AlexNet" [35] to its success; CNNs are a restriction of the general function that is tailored to image recognition. CNNs encode in the architecture that only neighboring pixels matter to extract basic geometric features in an image, such as diagonal or horizontal lines.

Different architectures matter as restrictions of more general functions, baking problem-specific knowledge into the function itself. This is the place where actual improvements are made: Using expert knowledge to choose the most restricted architecture that can still encode the complexity of the problem. An example of such a task, interpreting an event from a particle collision with deep learning, is discussed in Chapter 5.

### 3.2.3 Validation

Machine learning models, and especially deep learning models, can be powerful tools as they are able to capture higher-order correlations that are practically inaccessible using human-engineered algorithms. It is often desirable to understand the internals of an algorithm to derive a conclusion and an intuitive understanding. However, as machine learning algorithms inhibit a high degree of complexity *by design*, this makes them a black box. Nevertheless, for all algorithms, it is crucial to actually understand their response to inputs by testing the algorithm on data. This will be relevant for a variety of analysis tools in Chapter 6 and in Section 5.6.

---

[2]The definition of "deep" changes by the year; generally, anything with multiple hidden layers counts.

### 3.2.3.1 Bias and overfitting

When studying the response of an algorithm, the interesting property is the performance, *i.e.,* how well the algorithm performs on a given task. Having an *unbiased* performance estimate of the algorithm is key to understanding its behavior and comparing it to other algorithms. This is a necessity with deep learning models, as the typically large capacity makes them prone to overfitting, whereby the algorithm picks up statistical fluctuations of the training dataset. The term overfitting is used to describe a machine learning model whose parameters have been tuned to remember the specific *dataset* instead of generalizing to learn the *distribution* that the dataset was drawn from. It manifests itself mainly in a different performance of the algorithm on the training dataset compared to the validation dataset. The opposite of this effect is called underfitting, and implies that the parameters have not yet sufficiently been tuned, or the model has not enough capacity, resulting in a suboptimal but equal performance on both training and validation data. Unbiased means that when a performance estimate is repeated, the average performance will be the same as the performance on the whole dataset or on new, unseen data; a biased estimate systematically over or underestimates[3] the performance when repeated many times.

While underfitting seems generally more accepted, overfitting seems sometimes to be (mis)understood as something worse. This is not the case: both are a different expression of the same issue, namely that the algorithm was not perfectly designed, here through either over- or under-optimizing the parameters in the training. It implies that the training dataset can be used more efficiently, easily improving the performance. Therefore, a model can overfit, leading to non-optimal performance, while a performance estimate can be biased. Bias may be thought of as "wrong" and overfit as "not optimal".

### 3.2.3.2 Unbiased validation

Two commonly used methods to obtain an unbiased performance estimation are splitting and $K$-folding. Splitting divides the dataset into two parts, the training and testing datasets, as mentioned before. This is comparably simple, but bears the inherent consequence that the training dataset is smaller by the events taken in the testing dataset, thereby possibly lowering the statistical power of the training dataset.

To minimize the loss of statistical power, a $K$-folding technique can be used. The dataset is split into $K$ folds. The algorithm is then trained on $K-1$ folds, acting as the training dataset, and tested on the remaining fold, the test dataset. This is repeated $K$ times, each time with a different test fold and a set of training folds. The *K*-folding technique has the advantage that the entire dataset is used: After *K* iterations, the algorithm has evaluated each event in an unbiased way - the algorithm has not seen this event in the training dataset. Additionally, the dataset used for training can be as close to the original size as desired by choosing *K* appropriately. An ambiguity remains, which classifier to use for the actual prediction, as there are now *K* similar classifiers. A common choice is to pick one randomly, as any choice is equally good; alternatively, the average prediction can be used.

---

[3]It overestimates if the training dataset is used.

To better equalize the response of the different classifiers when using *K*-folding or splitting, these techniques are often combined with a stratification procedure. This ensures that the same amount of each class is in each fold, creating a balanced output. Throughout the analysis, described in Section 6, the stratified *K*-folding technique is used.

## 3.3 Likelihood

Cross-entropy, as defined in Eq. 3.2, can be approximated by the empirical cross-entropy with $x_i$ drawn from $q$, where the expectation is taken with regard to $\hat{q}$ instead of $q$

$$\mathbb{E}[-\log p(x; \theta)] = -\frac{1}{n} \sum_{i=1}^{n} \log p(x_i; \theta)$$

To find the minimum, which implies $p(x; \theta) \approx q(x)$, the constant $\frac{1}{n}$ can be ignored and only the sum is considered. Performing the following transformations, involving an exponentiation of the term to remove the logarithm, as

$$argmin_\theta - \frac{1}{n} \sum_{i=1}^{n} \log p(x_i|\theta) = argmin_\theta - \prod_{i=1}^{n} p(x_i|\theta)$$
$$= argmax_\theta \, \mathcal{L}(\theta) \tag{3.6}$$

and correspondingly, the likelihood can be defined:

**Definition 4** *The* `likelihood` *of a model $\mathcal{M}$ given the data $x$ is defined as*

$$\mathcal{L}(\theta) = \prod_{i=1}^{n} p(x_i|\theta)$$

Maximizing the likelihood or minimizing the cross-entropy between the model and the data is, therefore, equivalent. The likelihood can be understood from information theory, but it can also be seen from a more intuitive understanding: given all probabilities of the data points, the total probability of the data is the product of the probabilities of the individual data points. Making the likelihood a function of the parameters of the model, as opposed to the PDF which is a function of the data, the expression Def. 4 appears.

### 3.3.1 Maximum Likelihood method

The maximum likelihood estimate (MLE) is a method to estimate the true parameters $\theta$ using the likelihood $\mathcal{L}(\theta)$ as defined in Def. 4. This method is used throughout HEP to infer parameters, as seen in the analysis in Chapter 6.

$$\hat{\theta} = \arg\max_{\theta} \mathcal{L}(\theta). \tag{3.7}$$

The MLE offers a lot of desirable properties:

**Consistency** Being a consistent estimator means that the estimates $\hat{\theta}$ go to the true value $\theta$ in the limit of the number of measurements $N$ going to infinity.

**Efficiency** An efficient estimator has a variance $V(\hat{\theta})$ that is equal to the Cramér-Rao bound, the theoretical lower bound of the variance of any unbiased estimator.

**Asymptotic normality** In the limit of $N$ going to infinity, the estimator is normal distributed with mean $\theta$ and variance $V(\theta)$.

**Invariance** The estimator is invariant under transformations of the parameters $\theta$.

The solution is well defined with Eq. 3.7, yet finding the set of parameters $\hat{\theta}$ usually requires numerical methods, as the likelihood function is often not analytically solvable. For practical reasons, the likelihood is often transformed through a log transformation and a negation into the negative log-likelihood (NLL):

**Definition 5** *The negative log-likelihood is defined as*

$$\text{NLL}(\theta) \equiv -\log \mathcal{L}(\theta) = -\sum_{i=1}^{n} \log p(x_i|\theta) \tag{3.8}$$

Minimizing this term results in the same estimate $\hat{\theta}$ as maximizing the likelihood

$$\hat{\theta} = \arg\max_{\theta} \mathcal{L}(\theta) = \arg\min_{\theta} \text{NLL}(\theta) \ . \tag{3.9}$$

### 3.3.2 Customizing the likelihood

The likelihood is defined as the *product* of PDFs $p(x_i|\theta)$ and the data $x_i$. However, some terminology became common in HEP to describe the individual terms of a likelihood; all terms appear in the HEP likelihood fitting library ZFIT in Chapter 7. The basic term of a "simple" fit refers to the likelihood as defined in Def. 4 with one, potentially multidimensional, PDF $p(x|\theta)$ and multiple observations $x$.

**Simultanous** A likelihood that is the product of multiple PDFs $p^j(x_i^j|\theta^j)$, where $j$ is the index of the PDF and $i$ the index of the observation; each PDF is associated with a set of exclusive observations $x_i^j$, whereas $\theta^j$ are sets of parameters with arbitrary overlap. This is often understood as a fit to multiple datasets, also referred to as "auxiliary" measurement; mathematically, it is a product of uncorrelated PDFs.

**Shared parameter** A parameter is called "shared" if it appears in a simultaneous likelihood in multiple parameter sets $\theta^j$ of different PDFs .

**Extended** The likelihood contains an additional term that takes into account the number of observations $N$ as a measurement on its own. This can be used to fit the number of events in a sample.

**Constraint** A constraint is a term in the likelihood that approximates prior or auxiliary knowledge about a parameter. This can be an uncertainty from an ad hoc estimate, such as theoretical uncertainties, or the approximation of a potentially large likelihood term.

**Fixed from simulation** If the likelihood of a parameter given a simulation sample is narrow enough, *i.e.,* the uncertainty is sufficiently small, the uncertainty is neglected, and the parameter is assumed to be known as indefinitely precise.

In the end, all this is just terminology and does not change the likelihood itself. The only exception arises from weighting each observation with a weight $w_i$ that has various applications in HEP. Mathematically, a weighted likelihood is defined as

$$\mathcal{L}_{\text{weighted}}(\theta) \equiv \prod_{i=1}^{N} p(x_i|\theta)^{w_i}. \tag{3.10}$$

The maximum likelihood estimate of a weighted likelihood is still valid and allows for the same conclusions. However, while the maximum is still the same, the uncertainties obtained through the profile no longer have the same meaning and dedicated methods can be used to correct for the weights; such an implementation is mentioned in Section 7.4.5.

## 3.4 Optimization

Many applications, such as in the maximum likelihood method in Section 3.3.1, require the knowledge of a function's minimum value or position.[4] Finding the minimum can be complicated especially when there is no analytical solution.

Mathematically, a minimum is a point where the first derivative is zero and the second derivative is positive. Knowing the first derivative, also called *gradient*, of the function at a point gives the direction of the steepest descent. Following the descent in a convex function is a promising strategy to find a minimum. In the context of likelihood fits, functions are mostly convex in the vicinity of the global minimum. The functions considered in deep learning, on the other hand, are often highly nonconvex with many local minima. Non-differentiable functions require general black-box optimization and are only of limited interest in deep learning and likelihood fitting, where loss functions are generally differentiable.

### 3.4.1 Stochastic gradient descent

Minimizing a function by moving in the direction of the negative first derivative is called *gradient descent*. Mathematically, a gradient descent step of size $\eta$ at position $x_n$ is described by

$$x_{n+1} = x_n - \eta \boldsymbol{\nabla} f(x_n) \ .$$

The size of the (initial) step, in the context of machine learning, is often called the *(initial) learning rate*, and chosen ad hoc. Most algorithms have built-in heuristics to adapt the step size during the minimization process, removing some dependency on the initial choice.

The stochastic gradient descent (SGD) method combines the gradient descent with a stochastic component. The gradient provides the direction towards the minimum. Evaluating the function only on a randomly chosen subset of the data helps to avoid overfitting as every subset of data has a slightly different minimum. For a wide variety of deep learning applications, the optimal batch size varies from a few dozens to a few hundreds, see also the different training configurations in Section 5.4. Due to the randomness involved, the algorithm will settle in a minimum that is comparably flat. This is a desirable property for deep learning algorithms, as the output becomes insensitive to minor changes in the input, resulting in a more general and robust tool.

---

[4]Sometimes the maximum, but every maximum can be transformed to a minimum with $f(x) \rightarrow -f(x)$

### 3.4.2 Second-order methods

Optimizing a likelihood differs from the previous method, as the global minimum is needed; furthermore, significantly fewer parameters are trained, opening the possibility for more sophisticated algorithms. The following serves as an introduction, different implementations in libraries will be used in Section 7.4.4.

Newton's method is based on the local second-order approximation of the objective function. The expansion of a function $f(\vec{x})$ around a point $\vec{x}_0$ is given by

$$f(\vec{x}) = f(\vec{x}_0) + \boldsymbol{\nabla} f(\vec{x}_0) \cdot (\vec{x} - \vec{x}_0) + \frac{1}{2}(\vec{x} - \vec{x}_0)^T \cdot \text{Hess } f(\vec{x}_0) \cdot (\vec{x} - \vec{x}_0) + ... \quad (3.11)$$

where $\text{Hess } f(\vec{x}_0)$ is the Hessian matrix, a square matrix of second-order partial derivatives of a scalar-valued function, of the function $f$ at the point $\vec{x}_0$. The minimum of the function, neglecting terms above order three, is found by setting the gradient to zero, which yields

$$\vec{x}_{min} = \vec{x}_0 - \text{Hess } f(\vec{x}_0)^{-1} \cdot \boldsymbol{\nabla} f(\vec{x}_0) \quad (3.12)$$

This is called a Newton step. It effectively consists of a starting point $\vec{x}_0$ and a step in the direction of the negative gradient, scaled by the inverse of the Hessian matrix. For a polynomial of order two, this method will solve the problem in one step. For real-world problems, this method becomes an iterative procedure, which is repeated until the minimum is found.

Newton's method requires the calculation of the inverse Hessian, an $n \times n$ matrix, where $n$ is the number of free parameters in the function, an extremely costly step. To reduce this computational cost, the inverse Hessian can also be approximated - a valid procedure as the solution in Eq. 3.12 is also based on the approximation of the actual function. The approximation makes assumptions about the Hessian matrix when updating it only through the gradient at each step in the minimization, thus these methods are called *quasi*-Newton methods. These methods are computationally less expensive than Newton's method, but still require the approximation of the Hessian matrix and are useful for functions with a small number of parameters; small in the sense that the Hessian matrix can be stored in memory.

## 3.5 Computational aspects

The success of deep learning was only partially caused by the availability of data and the development of algorithms. The third ingredients are advances in the computational resources, on the hardware and software side, that could cope with the high demands of deep learning training. These aspects are not only relevant for the training and deployment of deep learning algorithms, see Chapter 5, but enable statistical inference frameworks for HEP in PYTHON, see Chapter 7.

### 3.5.1 Hardware acceleration

The basic structure of computers is called *von Neumann architecture* and consists of a memory, a central processing unit central processing unit (CPU) and an input and output, such as a storage disk, with all elements connected by buses. A CPU is a self-contained independent unit[5]. Hardware accelerators are devices optimized for specific computations. Their ability to perform these specific simple tasks efficiently and in a parallelized manner results in a reduced ability to deal with complex instructions compared to a CPU. This is sometimes referred to as vectorized computation and single instruction, multiple data (SIMD).

The optimal usage of hardware accelerators or distributed computing on multiple CPUs depends on the problem. Choosing when to use accelerators is a complex task and predicting the performance is difficult, especially in real-world applications. Deep learning algorithms with their parallel nature and simplicity are a perfect task for graphics processing units (GPUs). Modern deep learning frameworks usually have different registered computing kernels that allow seamless execution of code on CPUs as well as GPUs.

### 3.5.2 Software advances

The software side is as crucial to the success of deep learning as the hardware side. On the one hand, the user-friendliness of PYTHON and its ecosystem made deep learning accessible to a broader audience, making prototyping and experimentation easier. On the other hand, the availability of efficient libraries and frameworks made the training of large models feasible as they take care of the low-level details of the hardware and software. Gradients can be an extremely useful tool in the optimization of functions as seen in Section 3.4. Apart from trivial cases where the analytic expression is known beforehand, the gradient has to be computed by the program, usually using one of three methods:

**Symbolic** If a closed-form mathematical expression is available for the entire function, the derivative can be calculated using symbolic math libraries like Mathematica [37] and Sympy [38]. While useful for insight and human interpretation, those expressions can be computationally inefficient and cause numerical problems.

**Automatic** If a closed-form expression is only available for an individual operation at a time, the automatic gradient can be calculated using the chain rule $f(g(x))' = f'(g(x)) \cdot g'(x)$ consecutively. Apart from the reduced knowledge required about the function, this generally provides greater numerical stability than symbolic evaluations and is the method of choice for most deep learning frameworks.

---

[5] With the age of parallelism and specific examples, such as Apple's M and Intel's Skylake series, these lines start to be blurry again.

**Numeric** Regardless of the knowledge about the function, a numerical derivative using finite differences, $f'(x) = (f(x + \Delta h) - f(x))/\Delta h$, can be calculated if a gradient exists. However, choosing the right $\Delta h$ is not trivial as too large steps ignore the local shape, while too many steps lead to numerical instabilities; numerical gradients are still highly needed even with other options available, as discussed in Section 7.2.

Moreover, programs can be either compiled, like C++, interpreted, like pure PYTHON, or a mixture of both. The former allows significantly faster code execution following optimal allocation of memory and the elimination of redundant code[6]. The advantage of interpreted code is flexibility, which allows for more dynamic manipulation of objects and code execution. Ahead-of-time (AOT) compiled libraries that can be called from an interpreted language marry the two concepts.

To reduce the PYTHON interpreter invocation even more, just-in-time (JIT) compilation of code can be used. Contrary to AOT compilation, code is only compiled during run-time when it is actually needed and possibly cached. In JIT compilation, a function is called for the first time with some arguments and, before executed, compiled to create a version of itself that is specialized on the input arguments. The specifics of the signature that are used to create specializations can vary depending on the use case and range from data types, such as different float types, to specific memory layout of objects. Once a specialized version is created, it is cached, and subsequent calls to the function with the same input signature will use this cached version. This specialized version is, in general, as fast as an AOT compiled one. The advantage of a JIT-compiled function in comparison to AOT is that this usually allows even better specializations on specific arguments. The cost of this is apparent: the compilation time adds directly to the run time. If a function is called relatively often and is computationally expensive, this initial overhead of the first function call is negligibly small. Given that machine learning models are trained on the same type of input data, JIT compilation is a perfect fit for this use case. Most deep learning frameworks use JIT compilation to optimize the execution of the model, while allowing the flexibility and user-friendliness of a language such as PYTHON for the rest of the code.

The advantages of JIT compilation for the execution of specialized functions have a more prominent place in another, newer computing language, JULIA [39]. It combines an accessible high-level syntax similar to PYTHON with the before-mentioned JIT functionality and a powerful typing system *built into the core language.* With an ecosystem purely focused on scientific and numerical computing capabilities, JULIA is becoming a promising alternative language for HEP applications [40]; it is, as of now, not yet on the same level of maturity and adaption as PYTHON.

---

[6]Fully optimized and compiled code has typically little resemblance with the user-generated code.

*"A physicist is just an atom's way of looking at itself"*
Niels Bohr

# 4

# LHCb Experiment

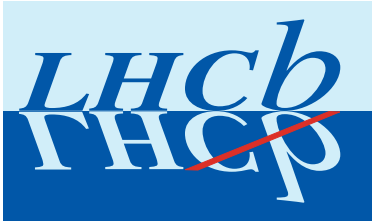

## 4.1 Large Hadron Collider

Right outside Geneva, on the Swiss-French border, the Conseil Européen pour la Recherche Nucléaire (CERN) is located. The origins of CERN can be traced back to 1949, when a scientific institution was envisioned to unite postwar Europe. Nearly a decade later, in 1957, the first accelerator started up at CERN, with a CoM energy of 600 MeV and was 16 meters in circumference. It would not be the last accelerator at CERN.

Today, one hundred meters underground and 27 kilometers in circumference, is the LHC, the current high-energy frontier, situated in a tunnel that was built in the mid-1980s. Initially, the tunnel hosted the Large Electron-Positron Collider (LEP) with different experiments at interaction points along the beamline, along with them was the DELPHI experiment. In the early 2000s, the accelerator and experiments were dismantled and replaced by the LHC, the new energy frontier with a designed CoM energy of 7 to 8 TeV in the first run period and up to 13 TeV in the second. In the cavity of the DELPHI experiment, the LHCb experiment was built [41], whose design was studied in the 1990 and received its final approval in 1998. The LHCb experiment is a single-arm forward spectrometer designed to excel in studying *b*- and *c*-hadrons.

Proton beams at the LHC start their journey as hydrogen atoms in a pressurized bottle of hydrogen gas, where an electric field strips the electrons from the protons. What follows is a chain of accelerators, from a linear accelerator to several circular accelerators that increase the energy of the protons consecutively to the desired level. During the acceleration, the protons are grouped into separate bunches. Each bunch of protons consists of about $10^{11}$ protons, and the bunches are separated in time by 25 ns. Two beams traverse the collider, one in each direction. They are kept in their circular path by superconducting dipole magnets and are focused by quadrupole magnets. Inside the beam pipe, an ultrahigh vacuum of $10^{-11}$ mbar is maintained.

At four interaction points along the accelerator, the bunches are focused to a diameter of 16 μm and brought to collision. The shape of the crossing bunches is tuned to achieve a higher overlap, resulting in more collisions per bunch crossing, leading to more interactions, tracks, and ultimately to more interesting events. However, this leads to a plethora of challenges for the detection, reconstruction, and online selection of the events, as distinguishing between different vertices and correctly assigning hits to tracks becomes exponentially difficult with more combinatorial possibilities. To mitigate these effects, the LHCb experiment was designed to operate at a lower luminosity than other experiments, resulting in one interaction per bunch crossing on average.

## 4.2 LHCb detector

The LHCb detector is built to cover the region of predominant *b*-hadron production, the forward angular region between 10 mrad and 300 mrad in the bending plane and 250 mrad in the non-bending plane [42]. It is built of several subdetectors, each with a specific purpose to infer properties of traversing particles that are combined to reconstruct the whole event in the detector acceptance. In the period after 2018, the LHCb experiment underwent a major upgrade, Upgrade I, which is described in Section 4.7, and is mainly relevant for Chapter 5. In the following sections, the LHCb detector is described as it was operating during Run 1 and Run 2[1] in the years 2011 a to 2012 and 2015 r to 2018 respectively, which is when the data analyzed in Chapter 6 was taken. The content, numbers and figures are based, absent any other reference, on the LHCb upgrade [42] and performance [43] reports.

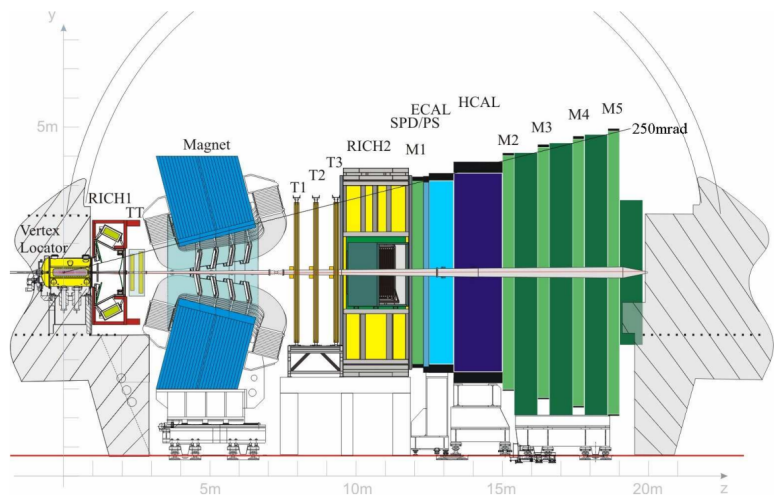

**Figure 4.2.1:** Schematic side view of the LHCb detector with its several subdetectors. The gray areas outside the detector are the cavern walls.

The subdetectors are arranged around the beam pipe ase shown in Fig. 4.2.1. When the bunches collide in the interaction region and a *b*-hadron is produced, it travels a distance of about 1 cm on average before decaying into other particles. Closely around this point is the Vertex Locator (VELO) that will be explained in more detail in Section 4.3.1. The VELO has a high spatial resolution, enabling to track the *b*-hadron and the decay products, which can be used to find the collision primary vertex (PV). Outside the VELO, the particles traverse the first of two Ring Imaging Cherenkov detectors (RICH s), RICH1, both of which are explained in Section 4.4.1 and then the first planar tracking station, Tracker Turicensis (TT) as explained in Section 4.3.3. After passing the LHCb dipole magnet described in Section 4.3.2, the particles traverse the planar tracking stations T1,T2, and T3 and then RICH2. After these detectors, which are only capable of detecting charged particles, the particles enter the Electromagnetic Calorimeter (ECAL), depleting predominantly electromagnetically interacting particles of their energy. The ECAL is followed by the Hadronic Calorimeter (HCAL). Both calorimeters are described in Section 4.4.2. The only particles that regularly penetrate the detector beyond this point are muons, which are tracked and identified in the muon stations described in Section 4.4.3.

---
[1]A "run" refers to a data-taking period.

All subdetectors are read out and processed by electronics that are located in situ - partially right on the detector readout itself and partially in the detector cavern in a radiation-protected area. The data is then sent to the LHCb computing infrastructure for further processing and storage. The reconstruction assigns particle tracks to different hits in the detector, as described in Section 4.3.3, and particle identification assigns a probability for a certain particle type to these tracks as described in Section 4.4. During these stages, a trigger system reduces the data throughput coming from the detector readout system to a manageable level. It selects interesting events and finally determines which ones are stored on disk, as described in Section 4.5.

A sound understanding of the detector response is needed to interpret the collected data. This is achieved by using a detailed simulation of the detector to propagate particles through it and study the response *w.r.t.* different physics decays and detector conditions. The production of simulated events in LHCb is described in Section 4.6.

## 4.3 Tracking

### 4.3.1 Vertex locator

In the heart of the LHCb detector, as close as 8 mm to the beam axis, the VELO subdetector is located. The purpose of the VELO is to precisely locate PVs and secondary vertices (SV), which are a distinct feature of *b*-hadron, as they travel a distance of about 1 cm before decaying.

The sensor modules of the VELO are arranged in halves on both sides of the beam, each side consisting of 21 modules. On each side of a detector module, the silicon micro strip detectors are placed, on one side radial to have $\phi$-angle resolution, with some skewing to improve position determination, and on the other side of the detector module in a concentric circle to have radial resolution. Combining this information with the detector position along the beam axis allows determining the three-dimensional coordinates of a hit inside the VELO volume. Each module had an average occupancy of 1 % per bunch crossing, a quantity that guided some of the overall design. The arrangement of the modules is shown in Fig. 4.3.1.

The ambitious design of having a detector in such proximity to the interaction point came with two main challenges: An unstable beam requires the detector to remain at significantly more distance from it, and the need to cope with a harsh radiation environment. Unstable beams have a larger circumference than the planned minimum distance of the VELO. Therefore, the design needed to incorporate a detector that moves closer to the beam once it was stable, and in fact so close that the detector is placed inside the actual beam pipe volume. Therefore, the two-half design with each sensor placed within a thin-walled aluminum box, was chosen. This meets the ultra-high vacuum requirements of the LHC, which is higher than $10^{-8}$ mbar, while the detector vacuum is expected to be around $1 \times 10^{-4}$ millibar minimum from outgassing and leaks through cables and surfaces. The aluminium box also shields the detector from electromagnetic effects from the beam and the beam pipe. Operating inside a vacuum requires the VELO to be actively cooled. This is achieved by a $CO_2$ cooling system that is routed through the detector and attached to the modules, keeping the temperature below $-5\,°C$ at all times. Cooling is also crucial for the radiation hardness of the silicon sensors, required in this environment, and reduces the leakage current.

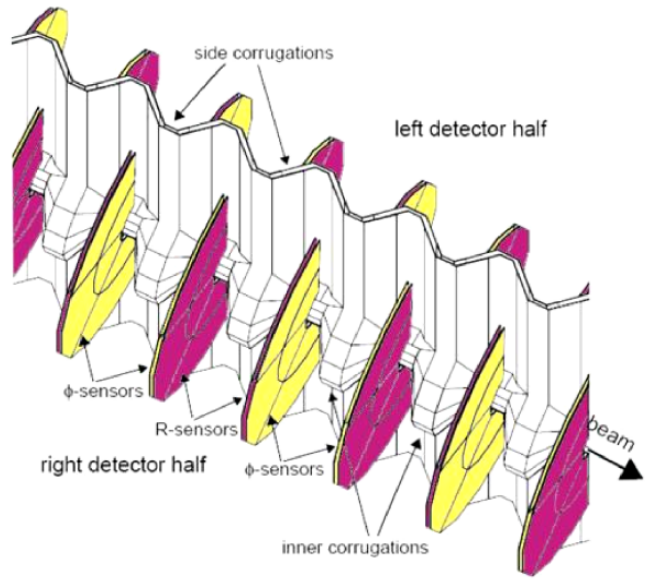

**Figure 4.3.1:** Enlarged view of a few VELO detector modules in the closed state. The two silicon strip arrangements that lead to radial (pink) and $\phi$-angle (yellow) measurements are shown in different colors. The construction apart from the detector is a thin-walled aluminum vessel that separates the beam and detector vacua.

### 4.3.2 Magnet

In the middle of the LHCb detector sits the dipole magnet. It bends the trajectories of charged particles, allowing to determine their momentum from the curvature of the tracks. The magnet had to overcome a couple of contrasting needs and constraints. On the one hand, the magnetic field has to be as high as possible to allow for a good momentum resolution even for high-momentum particles. On the other hand, the field strength within the RICH detectors, situated in the vicinity of the detector as seen in Fig. 4.2.1, must not exceed $2\,\mathrm{Tm}$. Furthermore, since LHCb was built in an already existing cavern, the dimensions were limited and proved challenging for the assembly phase. The chosen design was a warm magnet that uses iron yokes and coils with two identical but mirrored magnets with a field strength of 4 Tm for tracks of about 10 m. A schematic representation of the magnet is shown in Fig. 4.3.2.

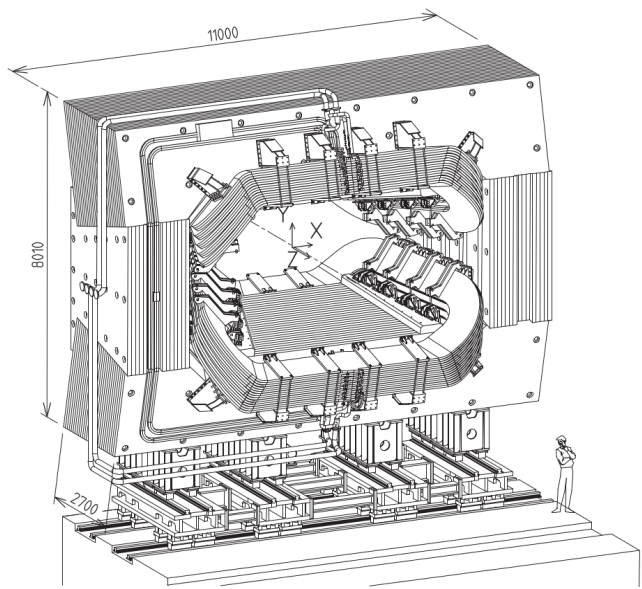

**Figure 4.3.2:** Schematical view of the LHCb dipole magnet with connectors and the holding structure. Both magnets are identical but mirrored and consist of iron yokes and coils. The magnetic polarity is regularly flipped to account for any asymmetric effects in the detector.

To understand the particle behavior and to simulate the detector response, the magnetic field has to be known comparably well. To map the field strength, an array of Hall probes was used to measure the magnetic field on a grid, spanning from the interaction point up to RICH2. In most regions, the measurement was in agreement with the simulated field strength expected from the magnet design within below 1 %. Even with a good understanding of the magnetic field, there is still ample room for biases in terms of one-sided detector inefficiencies. To account for this, the magnetic field is flipped periodically, taking into account the hysteresis effect. With the right procedure, the magnet strength was reproduced beyond the measurement uncertainty, so that switching the magnet polarity regularly helped to counter any asymmetric effects that may have been present in the detector.

### 4.3.3 Tracking stations

The LHCb detector consists of multiple tracking detectors, besides the aforementioned VELO, the TT, the Inner Tracker (IT), and the Outer Tracker (OT). Before the particles pass through the magnet, referred to as upstream, they encounter the TT. Passing the magnet and having been bent, referred to as downstream, the particles leave a trace in the IT, consisting of T1, T2, and T3. The IT extends close to the beam axis as this is where the particle flux is the highest; outside the beam axis the OT is used to reconstruct tracks as depicted in Fig. 4.3.3.

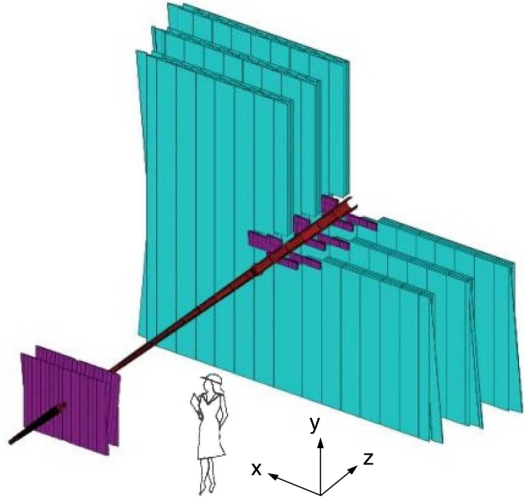

**Figure 4.3.3:** Schematic view of the LHCb tracking stations. The red station close to the interaction point, *i.e.,* with the lower $z$ coordinate, is the TT, the three stations downstream in red close to the beampipe are the T1, T2, and T3, and the outermost station in blue is the OT.

Both the TT and IT stations, summarized named Silicon Tracker (SI), are made up of silicon microstrip detectors with a strip pitch of $200\,\mu m$ and built with four detection layers. The layers have small angles between the strip alignment, allowing for spacial information in both the $x$ and $y$ directions. Silicon sensors have a short pulse time of $25\,ns$ and are read by a custom-designed ASIC chip. They provide a spatial resolution of $50\,\mu m$.

The OT greatly extends the acceptance area of the downstream tracker. The purpose of the OT is to cover the area of tracks that are not close enough to the beam axis to be reconstructed by the IT. Each module consists of two staggered layers of straw tube detectors of $4.9\,mm$ in diameter. The layers of straw tubes are arranged in a similar way as the silicon strips in the IT to reconstruct the $x$ and $y$ coordinates of a hit. The tubes are filled with a gas mixture of $70\,\%$ Ar and $30\,\%$ $CO_2$. The gas was chosen to keep the drift time to a manageable $50\,ns$, long enough for two collisions to occur, and to allow for a good spatial resolution of $200\,\mu m$.

### 4.3.4 Vertexing and tracking

The hits of the tracking stations are used to reconstruct the trajectories of charged particles, combining the information from the VELO, the TT, the IT, and the OT. The main algorithm starts with a seed in regions of a low magnetic field and extrapolates the trajectory to the next tracking station. An inherent trade-off arises from ghost tracks, which are spurious tracks that are not associated with a real particle, and the efficiency to reconstruct real tracks. To consider a track to be successfully reconstructed, it must have a minimum number of hits in the considered subdetectors and at least 70 % of the hits associated with it originating from a single particle. In each $b\bar{b}$ event, there are on average about 72 successfully reconstructed tracks, of which about one third are long tracks, *i.e.,* tracks that have hits in all tracking stations. The reconstruction efficiency of long tracks with a momentum greater than $10\,\text{GeV}/c$ is about 96 % and is shown against the number of PVs in Fig. 4.3.4 at a ghost rate of about 9 %, of which most have a low transverse momentum. With the curvature of the magnet and the precision on the track position, the momentum resolution is about $\sigma_{p_\text{T}}/p_\text{T}^2 = 0.5\%$ at low momentum, up to 1.0% at $200\,\text{GeV}/c$, where the trajectory is less bent for higher energies, making uncertainties on the track position more impact the momentum resolution. This allows the determination of the impact parameter ($IP$), denoting the closest approach between a reconstructed track and the true origin of the particle of $(15 \pm 29)\,\text{µm}$, as shown in Fig. 4.3.4.

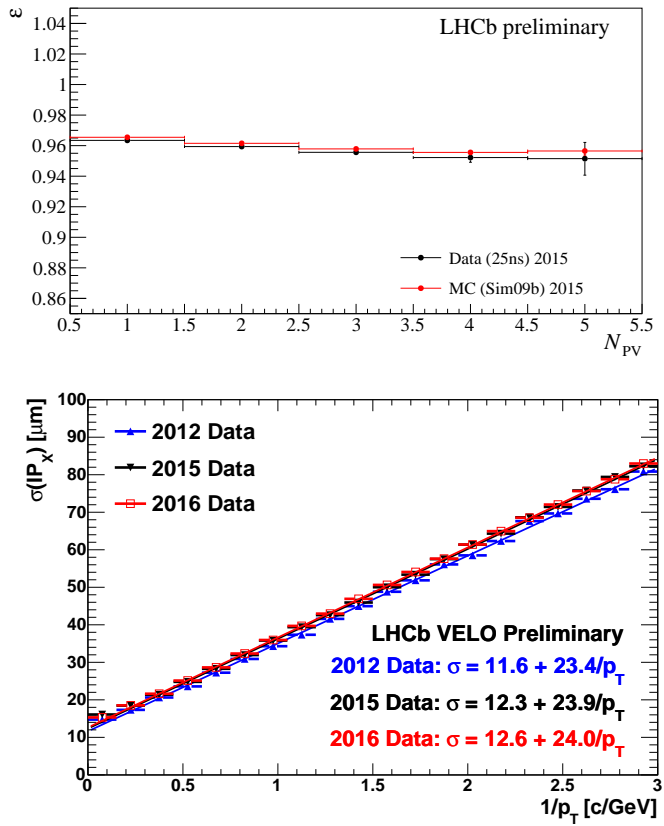

**Figure 4.3.4:** Track reconstruction efficiencies for data and simulation in 2015 conditions for muons (top) and $IP$ resolution in the $x$ component in 2012 data (blue), 2015 data (black) and 2016 data (red) using events with one PV (bottom).

## 4.4 Particle identification

Within the LHCb detector, every detected particle is reconstructed with an identity. Some of the most challenging separations of particle species are the kaons and pions, yet they differ crucially in their physical behavior, as the kaon contains a $s$-quark. To achieve an optimal distinction between different particle species, LHCb deploys a dedicated particle identification (PID) system, consisting of the two RICH s , the two calorimeters ECAL and HCAL and the muon stations at the end of the detector.

### 4.4.1 RICH

From the tracking system, the momentum of the particle is determined. Combining the momentum with the velocity measured by the RICH detectors, the mass of the particle, its identification, can be inferred. There are two RICH detectors in LHCb, RICH1 situated upstream and RICH2 downstream. They make use of the Cherenkov effect, which is the emission of electromagnetic radiation when a charged particle travels through a medium with a velocity greater than the local velocity of light in that medium, the equivalent of a sonic boom. Both RICH detectors are composed of a radiator, a mirror-based optical system, and photon detectors. The radiator is the medium through which the particle travels and emits Cherenkov radiation in a cone-like shape. The angle of the radiation cone is directly related to the velocity of the particle, but also depends on the refractive index of the medium.

For the two RICH detectors, two different media are used to accomplish optimal angle resolution - and therefore velocity resolution. RICH1 has a wide acceptance covering the full LHCb acceptance and specializing in the low momentum region, $\sim$1 GeV/c to 60 GeV/c, by using aerogel and $C_4F_{10}$ in Run 1, with the former removed in Run 2. RICH2 specializes in the high-momentum region, $\sim$15 GeV/c to 100 GeV/c, by using $CF_4$ as radiator gas. It spans a smaller acceptance region than the full acceptance of LHCb, which is the typical range for the high-momentum region. The expected cone angle of different particles as a function of momentum and as produced by RICH1 is shown in Fig. 4.4.1. These photons are reflected by a set of spherically shaped mirrors to focus them on the plane containing the photon detectors. There, the photons are registered by hybrid photon detector, a vacuum tube with a photocathode and a silicon pixel detectors (HPD s), which are situated outside the LHCb acceptance. The closeness of RICH1 to the magnet raises the problem that the magnetic field can distort the functioning of the HPD s and is therefore placed in large iron boxes to be shielded.

### 4.4.2 Calorimeters

After the tracking stations and RICH detectors, particles enter two separate calorimeters, the ECAL and the HCAL, a design chosen by many experiments. The ECAL stops the particles that predominantly interact through the electromagnetic force, which have a significantly shorter radiation or interaction length than most remaining particles. The energy of the latter is measured in the HCAL, which is situated behind the ECAL. In front of the calorimeters is a pre-shower detector situated with a scintillator pad detector to optimally distinguish between electrons and neutral pions. In contrast to other experiments, the calorimeters are used primarily for neutral particles and in the hardware first-level trigger, and less for PID of charged particles.

Both calorimeters follow a similar and well-established design in that they are composed of alternating layers of dense and active material. The dense material does not directly detect deposited energy but has a relatively low radiation- and interaction-length to deplete particles of energy and induce showers, and is significantly more cost-effective compared to the active material. The active material, which is for both calorimeters scintillating crystals, creates photons when traversed by a particle shower. These photons are guided via optical cables that are attached to the end of the scintillators to photomultiplier tubes (PMT s). The ECAL uses a design with longitudinal alternating layers and lead as the dense material. The HCAL has vertically alternating layers, *i.e.,* the rather unusual design choice of layers running *parallel* to the beam pipe, using iron as the dense material, resulting in a radiation depth of about 5.6 interaction lengths. Both calorimeter cell designs are illustrated in Fig. 4.4.2.

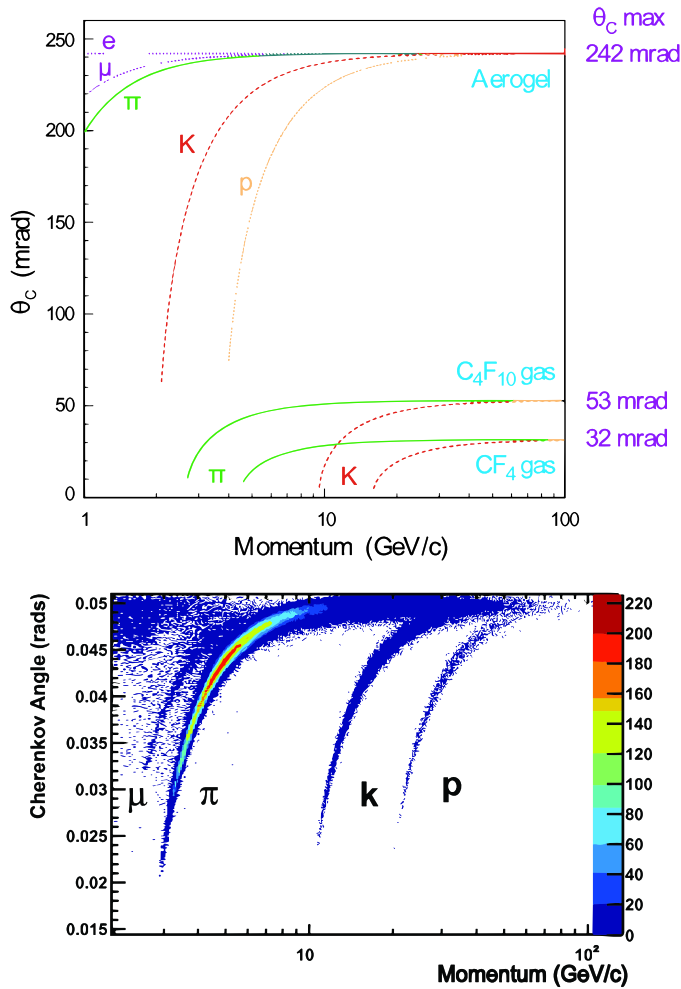

**Figure 4.4.1:** Cone angle of different particles as a function of momentum and as expected with different gases (top) and as measured by RICH1 (bottom) [44].

These sampling calorimeter designs were chosen as a balance between the cost of the active material, the radiation resistance, the reliability, and the energy resolution. The particle flux density is highest close to the beam pipe and decreases with the distance from it. To follow this trend, the calorimeters are built with different sections, the ECAL with three and the HCAL with two, which have higher granularity closer to the beam pipe and a coarser granularity further away. The ECAL achieves a nominal energy resolution

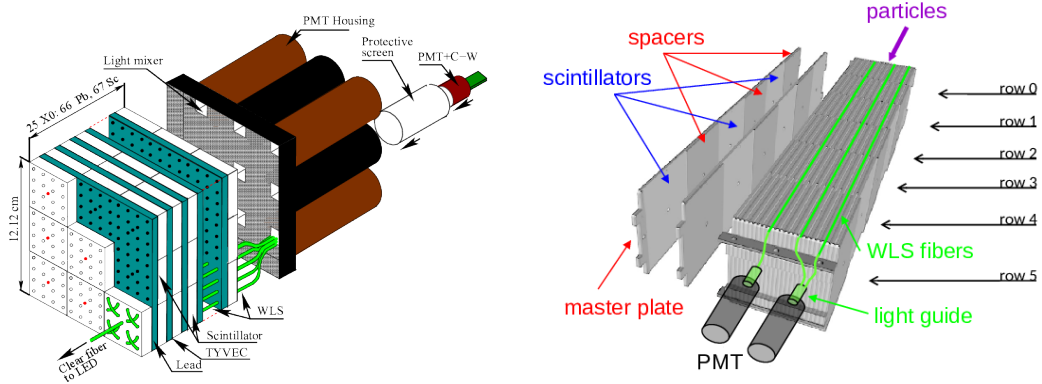

**Figure 4.4.2:** Design of the ECAL cell (left) and HCAL call (right), both with alternating layers of dense and active material. The orientation with respect to the beam pipe is perpendicular for the ECAL and parallel for the HCAL.

of about $1\% + 10\%/\sqrt{E\,[\,\text{GeV}\,]}$.

### 4.4.3 Muon stations

Muons are deeply penetrating particles that are usually not stopped by the detector modules. To measure the muons, LHCb has a total of five dedicated muon stations with M1, the first station, placed before the calorimeters and M2 to M5 placed behind. The first station, M1, is a tracking station that measures muon tracks and helps determine their transverse momentum. The other four stations are interleaved with iron absorbers to stop other particles and only allow highly penetrating muons with a momentum of at least 6 GeV to pass.

The inner region of M1 is made up of a triple-GEM detector to cope with the high particle flux density, while the outer region of the station and all other muon stations consists of multi-wire proportional chambers (MWPC s). Hereby, M2 and M3 have been designed with a resolution high enough to accurately measure the muon tracks and, therefore, the momentum, while M4 and M5 are designed to help positively identify muons.

### 4.4.4 Particle identification performance

The subdetectors described in this section are used to infer the identity of the particles through dedicated PID variables. These were initially based on the delta log-likelihood of the particle hypothesis against an alternative hypothesis, and have been extended by an additional set of PID variables from the output of a neural network that uses the different subdetectors as input. The muon identification has an additional, dedicated boolean variable, called isMuon, which is solely based on the information from the muon stations. For the electron identification, a 90 % efficiency is achieved with a 5 % misidentification rate for hadrons, while for kaons, a 95 % efficiency is achieved with a 5 % misidentification rate for pions. The muons achieve an even higher PID efficiency of 97 % with a misidentification rate of 2 % for pions, with the caveat that muon identification is only available for muons with a momentum of at least 6 GeV [45], and is depicted in Fig. 4.4.3.

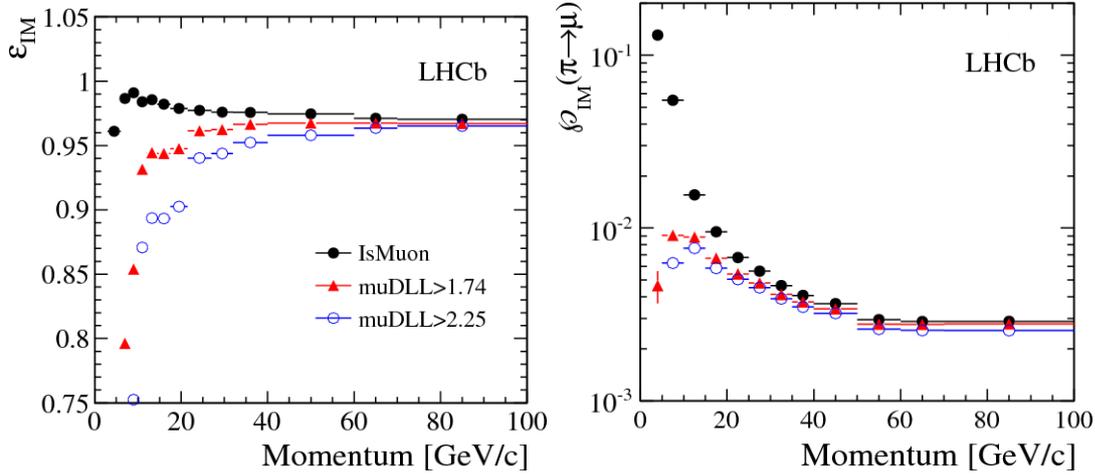

**Figure 4.4.3:** Muon identification efficiency as a function of momentum for muons (left) and pion misidentification as muon (right) using the isMuon variable only (black), with additionally muDLL> 1.74 (red) or muDLL> 2.25 (blue) selection requirements [45]. The muDLL variable is the delta log-likelihood of the muon hypothesis against the pion hypothesis. isMuon is a dedicated, binary variable based on the muon stations alone.

## 4.5 Trigger system

The LHCb experiment collects data at a rate of about 40 MHz. This rate is too high to be fully stored for offline analysis, from bandwidth to storage capacity considerations, and requires a trigger system to reduce the rate to a manageable level. The selected and obtained data should be within the physics program of LHCb, focusing mainly on *b*-hadron decays, but also including *c*-hadron decays and other rare processes.

The LHCb trigger system is responsible to decide, *which event*, and sometimes *what part of* an event, is stored for further analysis. Given that a collision happens every 25 ns, the trigger system must be able to make a first decision within that approximate amount of time; due to the possibility of parallel processing, multiplied by the available number of computing cores. The trigger is composed of multiple stages, starting with the simple yet fast level-zero trigger (L0), followed by HLT1 and HLT2, which reduce the data stream to about 1 kHz that is written to disk for offline analysis.

### 4.5.1 Level zero

The L0 trigger is implemented in dedicated hardware, field-programmable gate arrays (FPGA s) with a high radiation tolerance, and works in sync with the 40 MHz of the bunch crossing to reduce the rate of data to 1 MHz. The total time for a decision is 4 µs with half of it available as computing time. This severely constrains the complexity of the computations to simple selection requirements without elaborate reconstruction and takes information from only a dedicated set of detectors. Since *b*-hadrons are heavy, their decay products usually carry large transverse momentum and transverse energy. The L0 trigger makes use of inputs from both the ECAL and HCAL, which have been described in Section 4.4.2, as they provide information about the transverse energy of hadron, photon and electron candidates. It also uses the muon system, which provides direct information about the transverse momentum of muon candidates. The decision to keep or discard an event at this stage is based on a threshold on the transverse energy or momentum of the candidates. A pile-up system in the VELO, which estimates the number of interactions per bunch crossing, complements the information of specific events. This can reject events with too many interactions, as they would otherwise take a disproportionate fraction of the readout bandwidth and disk storage capacity.

### 4.5.2 High-level trigger

The high-level trigger (HLT) is divided into two stages, HLT1 and HLT2, which are implemented in software and run asynchronously on a dedicated computing farm, with all the information from the detectors available for events that passed the L0 trigger. Since 1 MHz coming from the L0 trigger is still a considerable high rate and, given the limited computing resources, not all information was used by HLT1 in Run 1. Instead, only a primitive track fit was performed using information from the VELO and the tracking stations. If the fitted track passed the fit quality and $p_{\mathrm{T}}$ thresholds, it was sent to HLT2 and the track reconstruction was redone offline. HLT2 makes use of all information available from the subdetectors to reconstruct the event and takes the decision to persist or discard it.

With Run 2, the available computing resources were expanded by increasing the number of computing cores and adding GPUs. The whole HLT1 computation was moved to the new GPUs, freeing up even more resources for the HLT2. Furthermore, having learned from Run 1 that the actual time during which the detector actively takes data averages to 30%, a more sophisticated strategy was implemented. The buffer storage that keeps the events before they are processed in HLT was systematically used: during the data taking, the buffer was filled with events, creating a backlog that was processed by the HLT during the off-time of the LHC. These improvements drastically increased the computing time available per event from approximately 40 ms to almost 1 s and allowed a complete offline quality event reconstruction in HLT2. Additionally, automated online alignment and calibration allowed the reconstructed variables to be on-pair with offline quality. The trigger efficiency after HLT2 is shown in Fig. 4.5.1 for a three-body decay with an average efficiency of about 30 %; muons have an average detection efficiency of about 90 %.

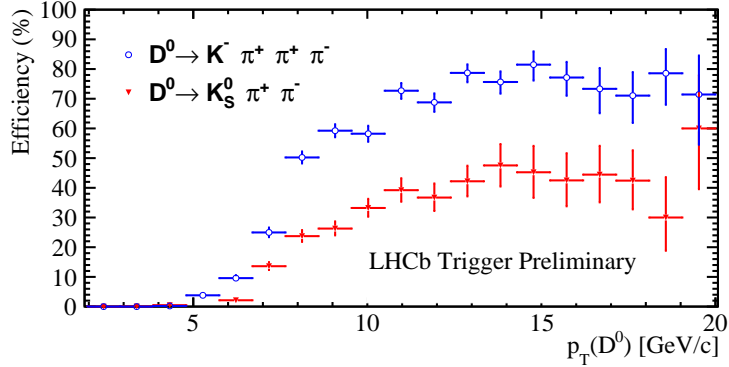

**Figure 4.5.1:** Trigger efficiency of HLT2 for the three-body decay $D^0 \to K_S^0 \pi^+ \pi^-$ (red) and four-body decay $D^0 \to m\pi^+\pi^+\pi^-$ (blue) decays as a function of the transverse momentum of the $D^0$ meson.

### 4.5.3 Turbo stream

With the Run 2 trigger strategy, all variables and tracks are available in offline quality from the HLT2 stage, making an offline reconstruction de facto obsolete [46]. The relevant tracks that are reconstructed, the high-level objects that describe the whole event, take less disk space than storing the whole low-level detector information, such as individual hits in the tracking stations. With many analyses limited by the size of the collected data and with an ever-increasing data rate expected for future upgrades, as described in Section 4.7, a new data processing was introduced in Run 2, called `Turbo stream`. The idea of the `Turbo stream` takes the efficiency of disk space usage and computing to the next level: it only stores the high-level objects that come from the relevant decay, discarding not only the low-level information but also any other particle in the event that is not relevant to the decay, without additional offline reconstruction performed. This reduces the event size by an order of magnitude, in turn allowing to store more events and increasing the statistical power of the collected data. Such a strategy will not be optional for the upgrades of the LHCb experiment, but will instead serve as the only way to store data. In Run 2, about 70 % of the events were processed with the `Turbo stream`, with the remaining 30 % fully stored.

There are multiple drawbacks to this strategy:

- The data is not available for other analyses than the one that triggered the event;

- It also does not allow to repeat the reconstruction of tracks or the inference of variables in case of a bug in the software or a change in the calibration;

- Finally, it does not allow to use the whole available knowledge of an event for the inference, such as the presence of unrelated tracks or other physical backgrounds that potentially interfere with the reconstruction of the signal decay.

These drawbacks can be mitigated by storing additional tracks, as required for some decays, but thereby increasing the size of the data stream again. Another approach, mostly related to the last point, is to store the relevant discarded information in compressed form, as discussed below Section 4.5.4. A more general approach is the interpretation of the whole event, which will be introduced in Chapter 5.

### 4.5.4 Isolation variables

The different analyses using Run 1 data lead to the development of more sophisticated tools to improve the purity of the signal in Run 2. One of the main challenges in the offline analysis is posed by combinatorial background events, which are random combinations of other tracks in the event, and by partially reconstructed background events, where a particle is not assigned to the correct decay or entirely escaped detection and is therefore missing in the reconstructed final state. Therefore, in Run 2, a new set of variables was introduced, *isolation variables*, that describe the kinematic and spatial isolation of a track in the event with respect to the other tracks, thus implicitly preserving a part of the full event information. These variables became more crucial to be available with the `Turbo stream` as any information not inferred using only the decay particles is not available for offline analysis. A more sophisticated approach to reconstruct events by taking into account all tracks in the event and inferring all required variables online is discussed in Chapter 5.

## 4.6 Simulation

Simulation is an essential tool in HEP for many purposes. It is critical to build a detector and develop subdetectors that do not yet exist. With simulation, different designs can be tested and optimized, their interplay and outputs can be merged, and a realistic dataset can be generated to perform a pseudoanalysis. For analysis, it provides access to events that look like real data yet have the underlying true information available. This provides information on the expected shapes of variables and events that are not detected directly. The latter makes it possible to infer the actual number of events that were produced in collisions given a data sample that underwent detector inefficiencies, geometric boundaries, and selection requirements.

The simulation framework in LHCb consists of several software components that are stacked on top of each other. The first stage is the generation of the proton-proton collision, which is usually carried out by PYTHIA [47]. The PYTHIA generator takes into account various aspects of the physics in a collision and produces a realistic description of the final-state particles. To propagate the decay, a package specialized on *b*-physics, EvtGen [48], takes the output of Pythia and simulates the decay of the particles, according to the known branching fractions and lifetimes. The resulting events contain true physical quantities and are often referred to as "generator level" events. These are propagated through the LHCb data collection and processing chain, taking into account the effects on the particles, such as multiple scattering and bending by the magnetic field, and simulating all subdetector responses, implemented in Geant4 [49]. An additional digitization step is performed by the Boole software on the output of Geant4, which is then processed by the same reconstruction software used for real data.

This chain produces simulation samples that are close to the real data, referred to as "(full) simulation". In contrast to real data, simulation events still contain the true variables they were generated with, such as true kinematics or true PID information. These allow studying the actual efficiency of detector parts, algorithms, analysis selection requirements, and inference sensitivity. However, the simulation quality differs for different decays and different regions of the phasespace due to uncertainties in the physics generators and the imperfect simulation of the detector response. Known issues exist, for example, for the distributions of *b*-hadron kinematics or the emulation of certain PID and trigger variables. Dedicated procedures are often used in offline end-user analysis to correct these variables using data-driven methods in order to avoid any biases.

## 4.7 Upgrades

The LHCb detector has had two very successful runs, starting with Run 1 and improving in several aspects in Run 2, including operating at nearly double the CoM energy, leading to a total of 9 fb$^{-1}$ collected data at CoM energies from 7 to 13 TeV. The LHC is not yet near its end of lifetime, which is projected to be in 2040, nor is the LHCb experiment. The LHCb detector has just undergone its first major upgrade, Upgrade I, in the long shutdown after Run 2 and started collecting new data during Run 3, with Run 4 planned to follow and run until approximately 2030 [50]. The LHCb detector will then undergo a second major upgrade, Upgrade II, which will be the last major upgrade of the LHCb detector and enter the High Luminosity LHC (HL-LHC) era [51]. In both upgrades, the collision energy is planned to remain the same, but the instantaneous luminosity will increase by a factor of five in Upgrade I and ten in Upgrade II. This increase will result in a total integrated luminosity of 50 fb$^{-1}$ for Upgrade I and up to 300 fb$^{-1}$ for Upgrade II. The increase in available data will provide major improvements for physics analysis at LHCb, as many current results are limited by the statistical uncertainty. The higher luminosity brings a set of challenges for the detector hardware, electronics, and software.

### 4.7.1 Subdetector upgrades

With Upgrade I, most of the subdetectors have been completely or partially renewed [50]. The upgraded LHCb detector is schematically depicted in Fig. 4.7.1 and the changed subdetectors are described in the following.

**VELO**   A completely new VELO is installed based on pixel silicon modules, which cope significantly better in a high luminosity environment. The modules are placed closer to the beam at 5 mm to 3.5 mm to improve the $IP$ resolution.

**Tracking**   With the Upstream Tracker (UT) and the Scintillating Fibre Tracker (SciFi), two completely new tracking detectors are installed; the other planar tracking stations have been removed. The UT is based on silicon microstrip modules and based upstream, before the magnet. It inhibits finer granularity and is closer to the beam. The SciFi is located after the magnet and based on scintillating fibers, read out by PMT s at the end. This marks the first use of scintillating fiber technology on this scale and in such a radiative environment.

**PID**   The RICH detector received a new mirror system that is able to cope with the higher track density.

**Calorimeters & muon system**   These detectors are the only parts that remain partially the same, as only the electronics readouts for both calorimeters, ECAL and HCAL, and for the muon system, have been completely overhauled. Furthermore, the Scintillating Pad Detector (SPD) pre-shower detector, which was installed in front of the ECAL, and M1, the first muon station, have both been removed.

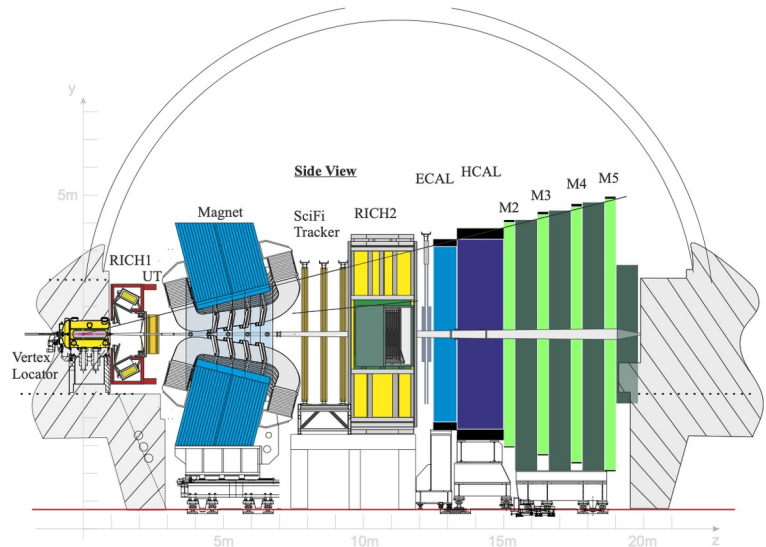

**Figure 4.7.1:** Illustration of the LHCb detector with Upgrade I subdetectors [50].

### 4.7.2 Trigger and reconstruction

A higher instantaneous luminosity will lead to an increase in visible collisions in LHCb from 1 to 5 and up to 50 for Upgrade I and Upgrade II, respectively [52]. Most notably for Upgrade II, this will increase the number of visible *b*-hadron events per bunch crossing from  1% to about  0.5 per event and *c*-hadrons from  25% to > 1 per event. The higher occupancy will require a faster readout, reconstruction and a more efficient trigger system, moving along the direction of the `Turbo stream` as explained in Section 4.5.3 and only keeping the relevant information.

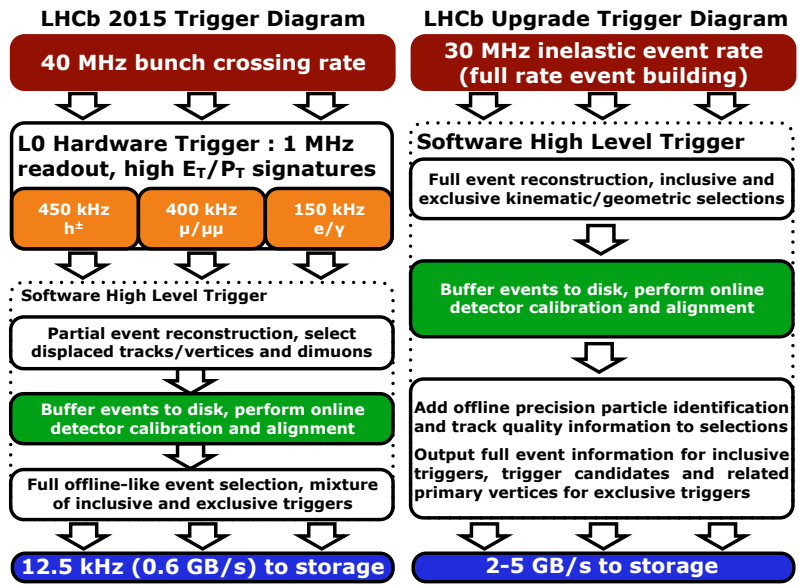

**Figure 4.7.2:** Trigger strategy for Run 2 (left) in comparison to the strategy planned for the upgrade in Run 3 (right) [53]. 30 MHz correspond to the expected, visible interaction rate for Run 3. The main difference is in the removal of the hardware-based level zero trigger leaving a fully software-implemented trigger system in place. This allows the exploration of more sophisticated trigger algorithms such as DL based tools.

For Upgrade I, crucial steps towards coping with the large number of tracks have been taken not only on the detector side but most notably also on the software side. The trigger system moved away from the L0, hardware implemented, trigger to a system that is purely implemented in software and runs asynchronously on a computing farm, with HLT1 implemented on GPUs and HLT2 on CPUs. A comparison of the Run 2 and Run 3 trigger strategies is shown in Fig. 4.7.2. This allows testing and verifying new, ambitious trigger strategies that can select events and optimally condense all event information online, such as DFEI explained in the following Chapter 5.

# 5

# **Deep Full Event Interpretation**

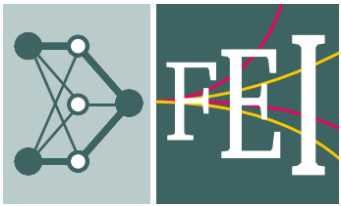

For every interesting event inside the LHCb detector, the tracks and clusters from neutral particles are reconstructed and, together with other detector information, particle candidates are built, as described in Section 4.3. To reduce the amount of data bandwidth and data stored on disk, a multistage trigger is used to determine which events to keep, as described in Section 4.5. With the upgrades of the LHCb detector, as described in Section 4.7, and the increase in luminosity, the same event selection strategy will no longer work. In this chapter, an approach to solve this challenge, to select and interpret events online, the machine learning-based DFEI algorithm, is presented.

First, the problem of event selection and interpretation is introduced in Section 5.1. Then, graph neural networks (GNNs) are introduced in Section 5.2 as a tool to learn the structure of the event. The dataset used for training the DFEI algorithm is described in Section 5.3. In Section 5.4, the DFEI algorithm is presented with its different stages. Further simplifications of the event preprocessing are described in Section 5.5. The performance of the DFEI algorithm is evaluated in Section 5.6. Finally, the findings are discussed and an outlook on future developments is given in Section 5.7.

## 5.1 From pulses to decays

The advent of Run 3 marked the successful completion of LHCb Upgrade I and is expected to be surpassed by Upgrade II in a decade. During the data taking periods in the years 2011 to 2018, the average number of protons interacting with each other during a bunch-crossing at the LHCb interaction point was about one, producing tens of tracks per event, of which the relevant ones are stored on disk, as previously described in Section 4.5. With Upgrade I and Upgrade II, the number of interactions per bunch-crossing will increase by a factor of five and ten, leading to hundreds and thousands of tracks per event, respectively, substantially changing the event topology as summarized in Tab. 5.1.1 [51].

**Table 5.1.1:** Approximate average quantities per event for the different LHCb run conditions, as estimated from the simulated dataset as described in Section 5.3. Only objects within the LHCb geometrical acceptance are considered. The runs 3 to 4 correspond to the Upgrade I, runs 5 to 6 to the Upgrade II.

| LHCb period | # vis. $pp$ collisions | # tracks | # $b$ hadrons | # $c$ hadrons |
|---|---|---|---|---|
| Runs 1 to 2 | $\sim 1$ | $\sim 50$ | $\ll 1$ | $\ll 1$ |
| Runs 3 to 4 | $\sim 5$ | $\sim 150$ | $\ll 1$ | $\sim 1$ |
| Runs 5 to 6 | $\sim 50$ | $\sim 1000$ | $\sim 1$ | $\sim 5$ |

The current algorithm and computing infrastructure will not be able to cope with the increase in demand, as previously discussed in Section 4.7, as these conditions bring many technical challenges:

- the number of hits and signals recorded in the detectors increase with the luminosity, which increases the number of possible track combinations approximately as $\mathcal{O}(n^2)$. This non-linear increase directly translates to demand of computing power needed for the tracking and reconstruction, under the assumption that the same algorithms from Run 2 were to be used;

- pile-up in the detector leads to ghost tracks and increased overlap of tracks in the same points, increasing the noise;

- an increase in the number of tracks will proportionally increase the size an event takes to be stored on disk;

- due to the increase in collisions, an increase in decays containing a *b*-hadron and an overall interesting signature is expected. In Upgrade II, approximately one interesting decay per event is expected. A trigger strategy based on selecting *which event to store* will not work anymore, as nearly every event would be triggered.

The last two points together are independent, and their effects accumulate[1]. As an increase in event size with $\mathcal{O}(n)$ also increases the chance of a *b*-hadron being present and thereby the chance of triggering by a factor of $n$, this will require an increase in disk space scaling as $\mathcal{O}(n^2)$: There will be $n$ times as many events containing a *b*-hadron, each one taking up $n$ times the disk space. Summarizing, storing the whole event and applying the current algorithms and computing, the requirement for bandwidth and disk space will be far exceeded. Instead, a new strategy is needed that will not save the whole event to cope with this new reality.

The future strategy, which has already been partially addressed in LHCb, relies on online reconstruction and selection of events, within a limited amount of time and computing resources. With the particle tracks reconstructed from the hits, the particle decay topology has to be correctly identified. Due to random overlaps of tracks that do not originate from the same parent particle, referred to as combinatorial background, or the omission of a track that actually belongs to a specific decay vertex, referred to as partially reconstructed background, events that mimic the actual signal decay appear in the collected data. The approach partially used in Run 2 with the `Turbo stream` as described in Section 4.5.3, and currently partially in use, relies on different variables associated with the decay, such as the quality of the vertex reconstruction or the associated tracks themselves, to remove parts of these backgrounds in the offline analysis. Additionally, a few dedicated variables describe the isolation of the events using the closest tracks, which are not part of the decay itself; apart from these, the approach does not take into account the rest of the event. This limitation mainly comes from the lack of powerful algorithms that could actually make use of additional information provided by the event, a task already complicated by the increase in luminosity. This represents a trade-off: the more information is discarded, the harder it is to later interpret the whole event, even bearing the risk to make it worthless if crucial information was not stored; yet storing the full event, which allows for full interpretation, brings an unmanageable bandwidth and disk storage requirement. While this is going in the right direction, at present there is no nominal strategy in LHCb to systematically select which parts of the event may be interesting for all different physics analyses.

---

[1] This argument saturates with more than one *b*-hadron per event, but at this point, the discussion on *which* event to store is meaningless.

The question of designing a trigger algorithm can be largely divided into *what part of an event* to store and *how best to interpret the event*. They are not independent questions: For an application in LHCb and, for that matter, in any other LHC experiment, the time when all the information from the event is available will be only in the online selection. Discarding information in the online selection implies that offline analysis of the event is not possible, and the problem shifts to how best to interpret the event *at the time when all information is available*. Minimizing the average event size will directly translate into maximizing the number of events LHCb can record. Given the general power and scalability of ML based approaches and their application in similar scenarios, this is the most promising one. The goal of the DFEI algorithm is to fill this gap: to take the entire event into account for an online selection of the tracks and an inference of relevant variables using GNNs, mostly the event decay topology reconstruction. Although this is not the first algorithm to solve either of the two problems, the combination of both is novel and has not been explored before in a hadron-collider environment. The following approaches have been used to solve similar tasks.

### 5.1.1 Full Event Interpretation

A similar approach that takes into account all the available information is the Full Event Interpretation (FEI) algorithm [54] used in the Belle II experiment, situated at the electron-positron collider SuperKEKB, which was developed for exclusive tagging of $B$-decays. The algorithm differs in that the environment is significantly cleaner and the possible decays are more limited at Belle II than in LHCb. The Belle II experiment has a hermetic detector situated at an electron-positron collider; thus the event is a fully reconstructible system with known initial states and significantly fewer tracks, making the task of inference less challenging. In addition, only two species of $b$-hadrons are studied, $B^0$ and $B^+$ mesons, while LHCb has access to and is interested in all $b$-hadron species, such as $B_s^0$, $B_c^+$ mesons, $\Lambda_b^0$ baryons and $c$-hadron decays.

From an algorithmic point of view, the FEI algorithm is based on a fixed set of different boosted decision tree classifiers, one for each decay type considered. A direct adoption of this approach would be unfeasible for LHCb, given the much greater variability in terms of different signal decay topologies, further enhanced by the fact that a fraction of the particles produced in the decays fall outside the LHCb geometrical acceptance. And while the accumulation of BDTs remains a possible strategy, more powerful and flexible ML algorithms have been established in the meantime, such as neural networks and specifically GNNs. Recently, an extension to the FEI algorithm based on a GNN was proposed [55, 56], showing better performance than the previous implementation.

This resembles the DFEI approach more closely, but is still designed for a very different environment.

## 5.1.2 General approaches based on ML

Some projects have used ML techniques for online processing, yet it is still a relatively new occurrence, especially within LHCb. The first and so far only use of a ML based approach on the full set of reconstructed tracks within LHCb is a probabilistic model based on DT for the inclusive flavor tagging of signal *b*-hadrons [57]. The combined processing of all event information demonstrated better results compared to the classical tagging algorithm, which uses only a subset of the reconstructed particles in the event. In comparison, the task of flavor tagging is much simpler than the explicit decay chain reconstruction attempted by DFEI. Another approach that compresses information about an event with ML are isolation tools as mentioned in Section 4.5.4. However, these efforts are based on a combination of features of the signal candidate and a few extra particles, disregarding any correlation with the other particles in the event. In other LHC experiments, a type of full event reconstruction is performed in CMS [58] and ATLAS [59], using an algorithm called particle flow. Particle flow reconstructs all stable particles in a decay, allowing them to be taken into account in further, derived objects, mostly jets. It uses all the final state particles for a global event description, significantly improving the performance of jet reconstruction with respect to the previous baseline that used basic geometric cones to cluster particles. Although the original algorithm uses classical methods, an approach based on GNNs [60] was proposed in CMS, which takes as input all particles from an event and predicts variables such as particle identification and transverse momentum of each particle. Although similar to DFEI at a technical level, the particle flow algorithm differs crucially in that it does not attempt to explicitly reconstruct the decay chains for all relevant decays of interest.

The task of decay chain reconstruction is conceptually close to the hierarchical reconstruction of jets, for which a variety of algorithms based on GNNs have been developed [61–75]. The ultimate goal of those algorithms, however, is typically focused on inferring quantities of the jet overall, for example, flavor tagging of the jet to determine the initial particle, and reconstruction of the jet to infer its kinematic properties. The jet substructure, which comes closest to the goal of DFEI, is only studied to a limited extent. The limitations of these algorithms for the task of reconstructing all ancestors in particle decay chains are reviewed in detail in Ref. [56].

## 5.1.3 Design goals

To address the challenges in online selection and reconstruction for the LHCb upgrades with increased event multiplicity, the DFEI algorithm is being developed. There are two main goals of DFEI: to select the interesting parts of an event and to reconstruct the full decay chains of all relevant *b*-hadrons in the event. The algorithm is based on GNNs that are well suited to take into account the topology of the data, such as the connections between the different particles in the event. As this kind of algorithm in a hadronic environment like LHCb is the first of its kind, the goal was to have a prototype that is capable of performing the task, albeit neither fully optimized nor providing all possible information about the event. For DFEI, only decays with charged particles are considered, as they carry the most information from the detector, with the extension to include neutral particles planned for future work. To break down the complexity, the algorithm is split into three parts where the first two layers serve to reduce the complexity of the problem and the third layer is the actual inference layer.

## 5.2 Graph Neural Networks

The workings of machine learning was introduced in Chapter 3, and the importance of choosing adequate architectures that represent the underlying structure of the data was discussed in Section 3.2.2. As for any ML problem, the motivation for a specific algorithm should not come from the algorithm itself, but from the problem that it is supposed to solve. In this section, graph structures and how they ideally represent an event will be introduced, followed by GNNs that provide a natural extension, or generalization, of other neural network architectures to work with graph structures.

### 5.2.1 Graphs

When there are many input features with the same input substructure in close spatial proximity, this substructure is ideally encoded in the network architecture, as discussed in Section 3.2.2. An example is images consisting of pixels, which are usually best tackled by CNNs that use a convolutional filter on a subset of pixels to extract features. A CNN thus naturally encodes the knowledge about images in general, that neighboring pixels are relevant to each other while distant ones are not, at least not for a first-level interpretation[2]. Particle tracks in a detector show a similar substructure: Particles have a relation to each other, often physically meaningful, such as the angle between them. However, potentially all particles are connected, and there is no a priori ordering of the different objects, as there is for pixels in an image. To further complicate matters, the number of objects can vary from event to event and cannot be expressed as a vector of fixed length.

This is where graph representations come into play. A graph is an abstraction that consists of nodes, particles in this case, which are interconnected with edges, reflecting the relationship of the different objects to each other. As illustrated in Fig. 5.2.1, an image can be thought of as a special case of a graph, where the nodes are the pixels

---

[2]This interpretation is usually forwarded to a DNN, connecting the whole picture.

and the edges are the connections between the pixels, or as words in a text are the one-dimensional case of a graph, where the nodes are the words and the edges are the connections between the words. Furthermore, a graph representation can have global

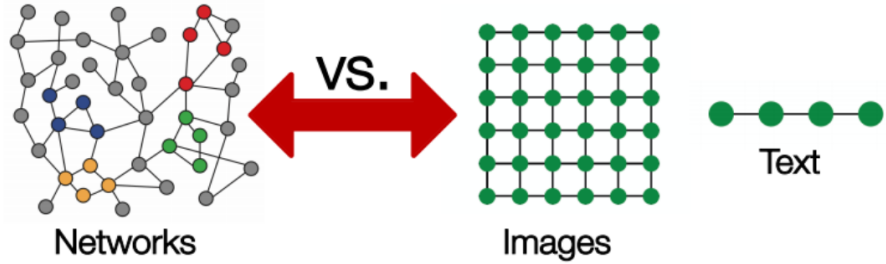

**Figure 5.2.1:** Comparison of a graph or generalized network structure to an image and text. The image and text show a specific, rectangular and only locally connected structure, while the graph is more general and can have arbitrary connections between the nodes.

attributes, too.

With the graph structure at hand, a decay consisting of particle tracks in a detector can be fully described.

- The nodes are the tracks and contain properties such as PID, momenta, charge, *etc.*;

- The edges are the connections between the tracks and describe how they relate to each other, such as the angle between the tracks, the distance between the tracks, *etc.*;

- The global attributes are the event properties, such as the number of tracks in the event, the number of vertices, *etc.*.

So far, this is only a conceptual abstraction that does not yet have a mathematical representation. While there are many ways to map a graph, one commonly used is the adjacency matrix. The adjacency matrix is a square matrix with the number of nodes as the number of rows and columns. The entries in the matrix are 0 or 1, depending on whether there is a connection between the nodes or not, or can, more generally, be a weight $v_{ij}$ that describes the strength of the connection between node $u_i$ and $u_j$.

### 5.2.2 Message passing GNNs

In a general sense, GNNs refer to neural networks that can deal with graph structures. The basic architecture of typical GNNs is that there are layers that perform graph-to-graph transformations using neural networks, as illustrated in Fig. 5.2.2. These can be chained together and return a graph, which can be the target $y$ itself. Such tasks are referred to as *node* or *edge* prediction, where the target graph contains information about a node or edges that should be learned. However, information about the whole graph can be inferred by performing graph classification and putting the transformed graph through a DNN, similar to the typical architecture of a CNN.

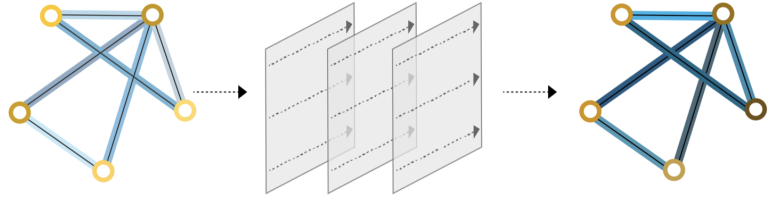

**Figure 5.2.2:** General workings of a GNN. The input graph is transformed by multiple layers, each consisting of a neural network, and the final output graph is used for the task at hand.

Most parts of working with GNNs, such as activation functions, layers, regularization, loss functions, and optimizers, are the same as for other neural networks. What differs is the update steps of the individual parts and the aggregation of arbitrary neighbors, which is specific to GNNs and presents in a way a generalized version of the convolutional layers of CNNs. One of the most general methods of a GNN is *message passing*, and many others can be derived from it as a special case. In message passing, nodes send messages to their neighbors and neighbors aggregate the messages to update their state. The aggregation is the crucial part: As each node can have a different number of connections, the messages need to be aggregated in a way that is invariant under the number of neighbors and, more importantly, invariant under a change of the ordering of the neighbors. To implement this invariant aggregation, there are some network architectures with varying degrees of sophistication. The simplest approach uses operations such as a sum or mean over the nodes while others use neural networks, such as recurrent neural networks (RNNs), to learn the aggregation function. The message passing is then implemented as a neural network that takes the current state of the node and the current state of the neighbors and returns the new state of the node.

In DFEI, the main building blocks consist of message passing blocks, called Graph Networks (GNs) as described in [76]. These blocks take as input a graph and transform it, returning another graph. Multiple GN blocks are serially stacked to build a more capable transformer. A block is made up of three update functions $\phi^v$, $\phi^e$ and $\phi^u$, and three information aggregation functions $\rho^{v \to e}$, $\rho^{e \to u}$ and $\rho^{v \to u}$. Each of the three update functions is implemented using a multilayer perceptron (MLP), and the aggregation functions are piecewise summations. The workings of a single block and its update strategy are illustrated in Fig. 5.2.3 and explained in the following. $v_i^k \in R^F$ denote the node features of node $i$ in layer $k$ and $e_{j,i} \in R^D$ the edge features from node $j$ to node $i$, and $u^k \in R^G$ the global features, where $F$, $D$ and $R$ are the dimensionality of the respective input vectors.

The GNN functions by first transforming the edges using the edge features, the neighboring nodes, and the global features. The updated edge features are computed as

$$e_{j,i}^{k+1} = \tilde{\phi}^e \left( v_i^k, v_j^k, e_{j,i}, u^k \right) \tag{5.1}$$

where $\tilde{\phi}^e$ denotes a differentiable function, usually, and as is the case for DFEI, an MLP.

For the node update, an aggregation of the updated edge features is performed, as well as the node features and the global features. The updated node features are obtained through

$$v_i^{k+1} = \tilde{\phi}^v \left( v_i^k, \bigoplus_{j \in \mathcal{N}i} e_{j,i}^{k+1}, u^k \right) \tag{5.2}$$

where $\bigoplus$ denotes a differentiable, permutation invariant function, such as sum or mean, over all the neighboring nodes $\mathcal{N}_i$, and $\tilde{\phi}^v$ is the node update function, which is as before an MLP. Note that the updated edges $e_{j,i}^{k+1}$, as obtained from Eq. 5.1, were used.

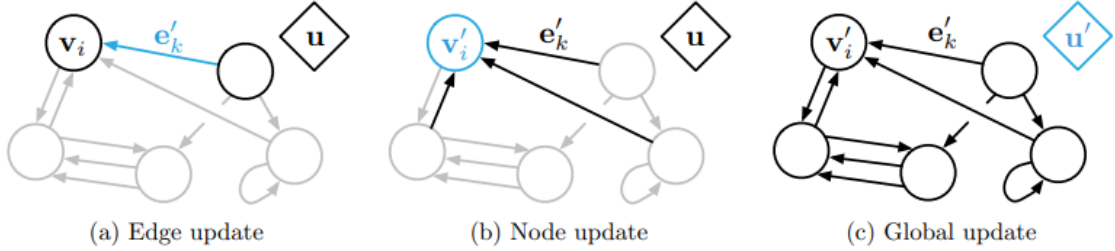

(a) Edge update      (b) Node update      (c) Global update

**Figure 5.2.3:** Illustration of the update of a GNN layer. First, the edge in a) gets updated, follwed by the node in b) and finally the global attributes in c). In each step, the updated component is shown in blue while the information that is used to update it is shown in black. All other components that are irrelevant in an update step are grey.

Finally, the global features are updated using all the updated node features, the updated edge features, and the global features. The updated global features are obtained through

$$u^{k+1} = \tilde{\phi}^u \left( u^k, \bigoplus_{i \in \mathcal{N}u} v_i^{k+1}, \bigoplus_{i \in \mathcal{N}e} e_{j,i}^{k+1} \right) \tag{5.3}$$

where $\mathcal{N}_u$ denotes the set of all nodes in the graph and $\mathcal{N}_e$ denotes the set of all edges in the graph. $\tilde{\phi}^u$ is the global update function. In this step, the updated edges from Eq. 5.1 and nodes from Eq. 5.2 were again used. This returns a fully transformed graph.

## 5.3 Dataset

To develop DFEI, the training dataset itself is crucial: in the end, any ML algorithm is a *data-driven* density approximation and can only be as good as the data sample and the variables used to train it. One decision to make in this regard is whether to train an ML algorithm on simulation or data. Simulation provides access to the true variables in the event, but it does not always represent the data well enough, to which the algorithm is designed to be applied to, as described in more detail in Section 4.6. As the task is supervised training, the ground truth is needed, therefore, the simulation is the most convenient place to start; possible refinements through data in the training can later be added.

As the algorithm is a novelty, benchmarking and comparisons with other approaches should be encouraged, and therefore a publicly available dataset is desirable. The LHCb simulation samples are restricted to internal member access only, and there is no publicly available dataset that would be suitable. Thus, a new dataset was created using openly available tools and published [77]. The data samples are generated with PYTHIA 8 [47] and EVTGEN [48], both introduced in Section 4.6, with the particle collision conditions manually replicated as expected in LHCb Run 3. This includes a number of steps, starting with simulation parameters set to the expected conditions, adding and smearing vertices, and applying the efficiency of geometric and reconstruction effects.

### 5.3.1 Simulation conditions

First, single proton-proton collisions at a CoM energy of 13 TeV are generated with PYTHIA, using an inclusive `softQCD` interaction model. Multiples of these collisions are then combined into an event, such that the number of collisions follows a `Poisson` distribution with an average of 7.6, corresponding to the average multiplicity expected in Run 3 conditions as described in Section 4.7. It was further required that each event contain at least one collision that has produced a *b*-hadron for the studies done on inclusive *b*-hadron decays. For studies done on exclusive decays, the dataset generated in the previous configuration is reused with a collision that produced an inclusive *b*-hadron decay replaced by a collision that produced the specified exclusive decay, which was generated again by combining both PYTHIA and EVTGEN generators. At this stage, the events have a description of the decay kinematics but lack an origin and detector effects.

### 5.3.2 Geometric constraints

The collisions are placed inside the LHCb detector with the point of origin of the coordinate system considered here to be the center of the VELO, as described in Section 4.3.1. A right-handed Cartesian coordinate system was adopted to describe the spatial components where the $z$ axis coincides with the beamline, the $y$ axis is aligned with the vertical pointing upward, and the $x$ axis coincides with the horizontal. To emulate the bunch crossing, the true position of each $pp$ collision is sampled from `Gaussian` distributions centered at the origin, with widths of $0.05\,\mathrm{mm}$ along the $x$ and $y$ and $10\,\mathrm{mm}$ along the $z$ axes.

The LHCb detector is a forward arm spectrometer and misses particles that go outside of its acceptance, as described in Section 4.2, as well as particles outside of the VELO coverage, which was described in Section 4.3.1. Therefore, charged stable particles, namely pions, kaons, protons, electrons, and muons, are only kept if they are within the LHCb acceptance, *i.e.,* their pseudorapidity $\eta$ must be within $1.9 \leq \eta \leq 4.9$. Furthermore, its origin along the $z$ direction is required to be within a distance of $500\,\mathrm{mm}$, which is the approximate coverage of the VELO.

### 5.3.3 Detector effects and reconstruction

The events currently are still close to the truth and neither contain detector effects due to noise and limited resolution nor limitations arising from the reconstruction. To emulate these effects, a few ad-hoc steps are taken. As a first step, the reconstruction of the PV is considered. Decays with fewer than four charged particles are discarded as they are considered to be not reconstructable. For all remaining PVs, a `Gaussian` smearing is applied to their position in each of the three dimensions.

The resolution of this smearing is taken from the LHCb Run 2 conditions as a function of the total number of charged particles in the collision taken from Fig. 5 of Ref. [78], as discussed in Section 4.3.4. The resolution for the $x$ and $y$ coordinates are assumed to be identically distributed.

In a second step, the origin of the particles is adjusted. In the real detector, this would correspond to the position measurement of the first hit from the associated track inside the VELO, which is segmented into 48 measurement planes along the $z$ direction as described in Section 4.3.1. To approximate this discretization, an equally spaced grid was chosen, and the actual $z$ coordinate of the track origin is assigned to that of the plane closest to the true origin position. For the $x$ and $y$ coordinates, the actual coordinates are determined by projecting the track on the previously determined $z$ plane and applying a `Gaussian` smearing with a resolution of $8.5\ \mu m$ in both the $x$ and $y$ directions, approximating the expected resolution in Run 3.

The last step concerns the kinematic variables by emulating the measurement of the three-momentum of the particles. Both the direction of momentum and the magnitude are smeared by a `Gaussian` distribution, approximating the performance as described in Section 4.3. The momentum slopes in the beam orthogonal axes $x$ and $y$ are smeared relative to the $z$ axis, while for the magnitude, a relative resolution of $0.4\%$ was assumed.

Combining these emulation steps and requirements, around $40\%$ of the events are observed to contain more than one $b$-hadron decay, while the observed maximum number of *b*-hadron decays is five. With this procedure, a total of 100 000 events was generated, and their usage in the training of DFEI is described in Section 5.4.

### 5.3.4 Features

The relevant variables of the dataset, which are used as input features in the DFEI GNN modules, are described below. Variables that describe properties of particles, the node features, are:

**transverse momentum** component of the three-momentum transverse to the beamline.

**impact parameter ($IP$)** distance of closest approach between the trajectory of the particle and its associated PV, defined as the one with the smallest $IP$ for the given particle among all primary vertices in the event.

**Pseudorapidity ($\eta$)** spatial coordinate describing the angle of a particle relative to the beam axis, computed as $\eta = \mathrm{arctanh}(p_z / \|\vec{p}\|)$.

**Charge ($q$)** electric charge of the particle. Since only charged reconstructed particles are considered, the charge can take only the values 1 or -1.

$O_x$, $O_y$, $O_z$ cartesian coordinates of the origin point of the particle.

$p_x$, $p_y$, $p_z$ cartesian coordinates of the three-momentum of the particle.

$PV_x$, $PV_y$, $PV_z$ cartesian coordinates of the position of the PV that was associated to the decay.

The following variables describe the relationship between two particles. They are used as edge features in the GNN modules.

**Opening angle ($\theta$)** angle between the three-momentum directions of two particles.

**Momentum-transverse distance ($d_{\perp \vec{P}}$)** distance between the origin points of the two particles projected onto a plane that is transverse to the combined three momentum of the two particles.

**Distance along the beam axis ($\Delta_z$)** difference between the $O_z$ of both particles.

### 5.3.5 Post processing

With the emulated physics of the LHCb detector applied to the dataset, a large fraction of the possible decay chains are only partially reconstructable. On the one hand, parts of a decay in an event can be outside the acceptance of LHCb. On the other hand, only charged reconstructed particles are considered valid in the DFEI prototype. As a consequence, these decays cannot be used in the training as-is, strongly reducing the amount of available events. To avoid this limitation, decay chains are custom versions of decays instead of the ones output by the PYTHIA generator. Valid decay chains within an event have been taken to create new decays by removing the ancestors that either correspond to very short-lived resonances or do not have enough charged-particle descendants to allow the proper formation of a vertex.

Using the previously mentioned setup, a total of 100 000 simulated inclusive events have been produced. In addition to this dataset, several other samples on the order of a few thousands, each with exclusive decays, have been generated. They represent specific decay channels that are some of the most common signal topologies studied in physics analyses at LHCb and are utilized to precisely evaluate the performance on these common use cases. The generated decay types and their performance evaluation are listed in Tab. 5.6.2.

## 5.4 Algorithm

A typical event that DFEI is supposed to interpret contains in the order of a hundred charged particle tracks. As DFEI is based on GNNs, each event is transformed into a graph structure. The charged particle tracks are represented as nodes and the relations between them as edges. This implies that there will be not only hundreds of nodes, but also tens of thousands of edges as input, each of them with a couple of features. Although technically a GNN should still be able to find the desired function, as explained in Section 3.2.2, this can require a significant amount of training, and is not well suited for the exploration of a prototype. Furthermore, building a graph of this size is computationally expensive and possibly not feasible for online applications, especially when considering the Upgrade II conditions.

Hence, the nominal DFEI algorithm consists of three stages with increasing difficulty of the target functions compensated with decreasing event sizes, keeping the overall size, complexity, and scalability of the algorithm manageable. The first two stages are event prefiltering stages that are applied to reduce the number of candidates before the decay chain reconstruction is performed. The first stage, called Node Pruning (NP), reduces the data sample to particles that actually originate from a *b*-hadron decay. It also doubles as a potential standalone event size reduction stage; therefore, it is further discussed with alternative lower latency bearing implementations in Section 5.5. The second stage, called Edge Pruning (EP), reduces the edges between particles that do not originate from the same *b*-hadron and are therefore not directly related. The actual event reconstruction inference, called Lowest Common Ancestor Inference (LCAI), is the final stage, the most training intense of the three stages. The final output of the DFEI processing chain in the LCAI module can be directly translated into a set of selected charged reconstructed particles and their inferred ancestors, with the predicted hierarchical relations among them. A description of the algorithm with its three stages is illustrated in Fig. 5.4.1.

Using three stages has one main disadvantage, namely that no direct end-to-end optimization can be performed, as hard selection requirements on the events after each filtering stage are applied. Additionally, this requires a fixed threshold for the selection requirement. As the goal of prefiltering is to reduce complexity, a high signal efficiency of 99 % is targeted, resulting in a sizable background reduction of about 70 %. This ensures that the LCAI training uses nearly all originally available signal events. All GNNs in the stages require a graph built using the input events for each stage. Each of the different modules uses independent MLPs for the node, edge, and global update functions, as introduced in Section 5.2.2; the latent size of all MLPs is the same with differing numbers of layers. The number of GN block iterations or message-passing steps is also configured separately for each stage. The modules are built using the `graph_nets` library [79], a library for GNN specific tools built on top of the TENSORFLOW framework, described in Section 7.2.

The training is carried out in multiple stages, following the sequence of the algorithms. To equalize the output, class weights are used in all loss functions to counter the imbalance between classes present in the dataset. The training itself is done using the Adam minimizer, which is an implementation of stochastic gradient descent as described in Section 3.4.1. For all stages, the data sample as described in Section 5.3 is used, where the entire data sample is divided into a training dataset of about 40 000 events, a test dataset with about 10 000 events and an evaluation dataset consisting of 50 000 events. The variables in the data set, as explained in Section 5.3.4, are used and expanded for the purpose of supervised training with the following two features:

- *FromSamePV*: boolean variable indicating whether the two particles share the same associated PV.

- *IsSelfLoop*: boolean variable indicating whether the edge connects a particle to itself or whether the edge connects two different particles.

In the following, each of the stages, their layouts and the training will be described in detail.

### 5.4.1 Node pruning

The first GNN has the objective of removing particles, represented as nodes, that have not been produced in the decay of any $b$-hadron. To distinguish particles with a true $b$-hadron origin from other backgrounds, the GNN mainly exploits the fact that particles produced in the decay of a $b$-hadron typically have large $\chi^2_{\text{IP}}$ and $p_{\text{T}}$ values. Since the main contributing factor to the prediction of each node in this case comes from the same node's features, self-loop connections are included in the graphs. The architecture of the model and the input features are listed in Tab. 5.4.1; the *number of layers* refers to the depth and the *latent size* to the width of the neural network used, the *number of GN blocks* describes the number of transformations of the graph through the network, *batch*

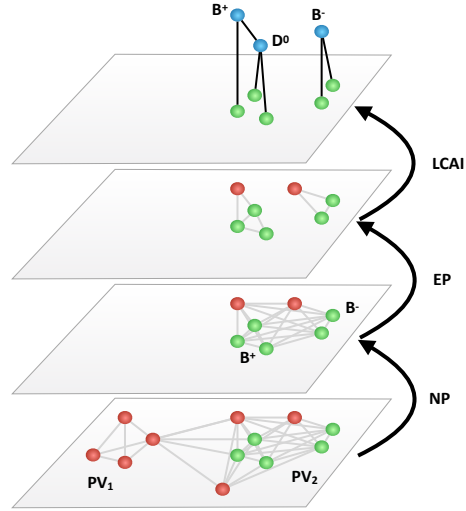

**Figure 5.4.1:** Schematic representation of an event processing by the algorithm. Green (red) graph nodes represent particles originated in the decay chain of a *b*-hadron (from the rest of the event). The reconstructed ancestors are represented in blue.

*size* and *training steps* refer to the training sample size on each update and the number of training iterations through the whole data set. The training loss is a binary cross-entropy loss with the goal of predicting the true origin from a *b*-hadron. After the training of the network, this module is used to reduce the event size by requiring that the selection keep about 99 % of the particles that originate from the same *b*-hadron decay.

| # layers | Latent size | # GN blocks | Batch size | # training steps |
|----------|-------------|-------------|------------|------------------|
| 3 | 50 | 3 | 32 | 500 |

**Table 5.4.1:** Configuration of the NP module. The top table lists the hyper-parameters used for the module architecture. The bottom table shows all the input features used in the graph. For nodes and edges, the total number of stable particles per event ($NumParts$) is included also as a global input variable. Edges connecting two different particles are differentiated from edges connecting a particle with itself (self-loops) by the boolean variable $IsSelfLoop$.

| Node variables | Edge variables |
|----------------|----------------|
| $p_T$, $IP$, $\eta$, $q$ | $FromSamePV$, $\theta$, $d_{\perp \tilde{P}}$, $\Delta_z$, $IsSelfLoop$ |

## 5.4.2 Edge pruning

Although the number of nodes has been reduced and therefore also the number of edges, which is quadratically correlated, the transformed graph from the previous stage still has a large number of edges. However, not all of them are meaningful and therefore are further reduced by the second GNN, the EP module. Its goal is to remove edges between particles that do not share the same *b*-hadron ancestor and therefore belong to a different decay. The model architecture and input features are described in Tab. 5.4.2. It is trained using binary cross-entropy to predict whether an edge connects two particles that originate from the same *b*-hadron decay. The edges in the output are removed from the event, again keeping a signal efficiency of 99 %, as for the NP module, and nodes that are no longer connected to other nodes are completely removed from the graph.

| # layers | Latent size | # GN blocks | Batch size | # training steps |
| --- | --- | --- | --- | --- |
| 4 | 100 | 5 | 32 | 500 |

**Table 5.4.2:** Configuration of the EP module. The top table lists the hyper-parameters used for the module architecture. The bottom table shows all the input features used in the graph. For nodes and edges, the total number of stable particles per event (*NumParts*) is included also as a global input variable.

| Node variables | Edge variables |
| --- | --- |
| $p_T$, $IP$, $\eta$, $q$ | $FromSamePV$, $\theta$, $d_{\perp \bar{P}}$, $\Delta_z$ |

## 5.4.3 Topological reconstruction

The most important layer of DFEI is the last layer, the LCAI, which takes the reduced graph of the two previous filtering stages and aims at inferring the so-called Lowest Common Ancestors (LCAs) of each pair of particles. This technique is similar to the recently proposed LCA-matrix reconstruction in the Belle II experiment [55] and allows to encode the topological structure of a decay in an adjacency matrix with only the lowest generation of decay products available.

The LCA matrix is constructed by first building a virtual decay tree of particles in that decay, all of them in different generations. Next, all of them are pulled down to the lowest generation, *i.e.,* starting with the children and stacking particles from the ground upwards, so that all stable, observed final-state particles are at the bottom, the lowest generation. This describes the relation between all particles in terms of how far away the lowest common ancestor is. Particles that share an ancestor at the lowest level will have first order LCA, particles that share an ancestor at the next-to-lowest level will have a second order LCA value, *etc.*. The LCA value does not directly relate to the number of intermediate resonances. For example, an LCA value of two does not necessarily imply that the particles share a *grandparent*, but could also mean that they share a parent. This parent in turn is the *grandparent* of two other particles, which have a parent in common and are connected with an LCA of one, *i.e.,* an intermediate resonance; this "pushes" the LCA value of the first particles up from the expected LCA value of one to two. As intermediate resonances are currently not part of the graph, this approach allows to encode the entire topological structure as illustrated in Fig. 5.4.2. The goal of the module is a multilabel supervised classification on the edges, where the output score is a topological description, encoded in a LCA matrix using a softmax[3] function. The architecture and the input variables for the LCAI module are shown in Tab. 5.4.3.

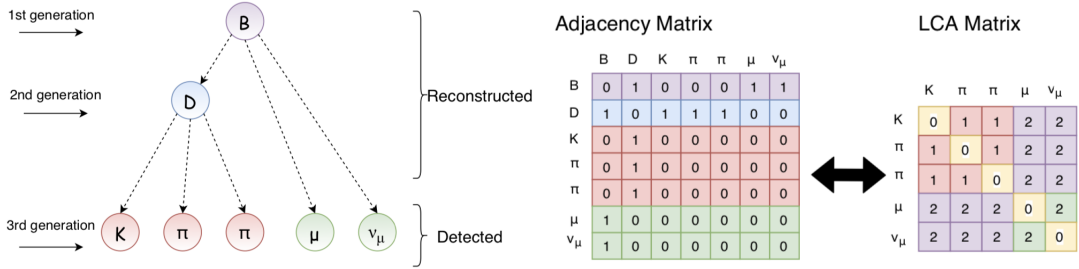

**Figure 5.4.2:** Example of a decay that is encoded in a LCA matrix, which only contains the final state particles. The corresponding true adjacency matrix contains intermediate resonances, not encoded in the input graph.

The training dataset used for the creation and study of DFEI contains basically only decays with a true order of up to and including three. As the fraction of higher-order decays in the simulation is tiny, these higher-order targets are neglected. In addition, an LCA value of 0 is included, labeling the case in which the two particles do not originate from the same decay chain. This implicitly provides an additional stage of node filtering, as fully disconnected particles, those whose edges are all predicted to have a LCA value of 0, can be safely removed from the resulting dataset.

| # layers | Latent size | # GN blocks | Batch size | # training steps |
|----------|-------------|-------------|------------|------------------|
| 5 | 100 | 5 | 128 | $2 \cdot 10^5$ |

**Table 5.4.3:** Configuration of the LCAI module. The top table lists the hyperparameters used for the module architecture. The bottom table shows all the input features used in the graph. For nodes and edges, the total number of stable particles per event (*NumParts*) is included also as a global input variable.

| Node variables | Edge variables |
|----------------|----------------|
| $O_{(x,y,z)}$, $p_{(x,y,z)}$, $PV_{(x,y,z)}$, $q$ | $FromSamePV$, $\theta$, $d_{\perp \bar{P}}$, $\Delta_z$ |

---
[3] Function that normalizes an input vector into a probability.

## 5.5 Event filtering

The nominal approach chosen in the different DFEI modules NP, EP and LCAI as described in Section 5.4 all use a message-passing GN block with node, edge, and global features. This architecture is one of the most versatile and powerful architectures, but comes at the cost of being computationally expensive, both in terms of computing time and memory demands. Such a powerful architecture is necessary to achieve maximum performance in the LCAI stage, where a comparably small graph makes training and inference feasible. This is possible due to the strong reduction in complexity achieved by the prefiltering stage NP, which takes the bulk of computational resources and identifies as the bottleneck in the algorithm. The main difference from the LCAI stage is that there is no intrinsic need to over-optimize at this stage: a certain percentage of signal particles should be kept, chosen high enough so that the filtering does not noticeably lower the signal content.

To optimize this bottleneck, weaker but faster algorithms were considered, meaning that more background particles will be present. In the case where filtering the events in an online trigger-like application is the goal, such as selecting the relevant particles for storage persistence, having more backgrounds implies a slightly larger space taken on disk. It allows to still remove or take into account the additional particles with other means in the offline analysis. In the case of complexity reduction of the input graph to LCAI, the LCAI stage will still be able to perform the classification, but will have to deal with more background particles. As the LCAI algorithm is meant to be constructed for maximum performance and as the additional backgrounds are the easiest to separate, this also does not pose a problem and requires a slightly higher demand in computing resources and training iterations.

### 5.5.1 Alternative algorithms

To reduce the computational complexity of the NP stage, some alternative algorithms have been investigated. The power of the GN blocks comes when they take into account the whole actual problem and fully encode the information in the graph. Reducing the complexity means to reduce either the input data or the complexity of the network. The latter quickly brings up two other challenges: the variability of the input size and the invariance of the ordering. The invariance is required, as there is no natural ordering for the particles in the event. Therefore, the output of an algorithm needs to be invariant under the exchange of particles in an input vector. Both the variable input size and the ordering invariance are challenges that the current GNN used in the DFEI modules can handle, although many other simpler ML algorithms cannot. While there are techniques to deal with parts of the problem, it implies that more hand-engineering is needed for the input and assumptions are required.

#### 5.5.1.1 Reducing the input data

There are mainly two approaches to reduce the complexity of the input data, either by reducing the information `per particle` or by reducing the information `about the event`. The current approach uses the whole event, providing additional discrimination power through the inclusion of surrounding particles in an event.

One possibility is to only take into account each track separately, ignoring the rest of the event, *i.e.*, using only information from the node features and global event variables to classify each track as signal or background. This is a comparably simple approach that allows to use arbitrary machine learning algorithms, having a small, fixed, and well-defined input vector.

An improvement to this approach can take some neighboring particles into account. This boils down to filtering by the $K$ nearest neighbors, where an appropriate metric must be chosen to measure the distance between particles. Although the distance of closest approach to the beamline (DOCA) seems like a feasible candidate, spatial distance is only a special case here, and distance, in general, is an arbitrary quantity that the current GNNs in DFEI encode in the transformed edge features. This approach can limit the input data to an arbitrary number of relevant decays while still preserving the most relevant correlations and was tested in Section 5.5.2.1.

### 5.5.2 Simplified algorithms

Two different architectures have been tested to perform the first stage NP with reduced capacity: a standard DNN and a GravNet [80].

#### 5.5.2.1 Fully connected network

A fully connected DNN is used with each track separately as input and possibly with nearest neighbors using the DOCA as a metric. Different architectures of a DNN were tried, mainly by varying the number of hidden layers from one to five and the number of neurons, a similar performance being observed regardless of the capacity of the network. It turned out that an architecture with two hidden layers and about 30 neurons each is sufficient. The model with two additional neighbors, ordered by the DOCA, reached a Receiver Operator Characteristic (ROC) Area Under the Curve (AUC) (ROC AUC) of about 0.91, as can be seen in Fig. 5.5.1. This corresponds to a background rejection of about 35% at 99 % signal efficiency and is approximately a factor of two worse than the nominal model described in Section 5.6.1. As DNNs are highly optimized in terms of deployment performance, this approach is a decent fallback option to guarantee a fast pruning stage.

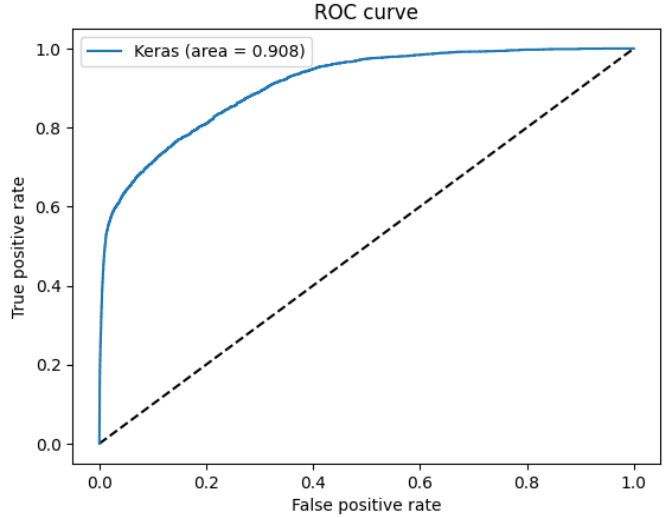

**Figure 5.5.1:** ROC curve for a DNN to predict the origin of a particle from a *b*-hadron, evaluated on a validation set.

### 5.5.2.2 GravNet

The GravNet layers have been utilized before in the CMS experiment for tracking, and also in Belle II. This comes with the advantage that a fast implementation is available and that the timing performance is already known. The GravNet layers are a simplified version of the GNN layers that only use the distance between particles as input features where the distance is calculated as the DOCA between two particles. Contrary to the DNN, described in Section 5.5.2.1, the GravNet layers can take into account the surrounding particles without specific ordering but do not use the relations between them. It uses only the $K$ nearest neighbors as input features and dynamically builds a graph, resulting in a fixed number of edges per node. The performance for $K$ equal to five is shown in Fig. 5.5.2 and reaches a ROC AUC of 0.91, the comparison with different $K$ is shown in Tab. 5.5.1.

**Table 5.5.1:** Background rejection power at 99 % signal efficiency in percent. The more neighbors the GravNet algorithm uses, the better the performance. However, the gain of including more than 10 neighbors becomes negligible.

| Number of neighbors | 3 | 5 | 10 |
|---|---|---|---|
| Background rejection at 99 % signal efficiency | 0.25 | 0.28 | 0.29 |

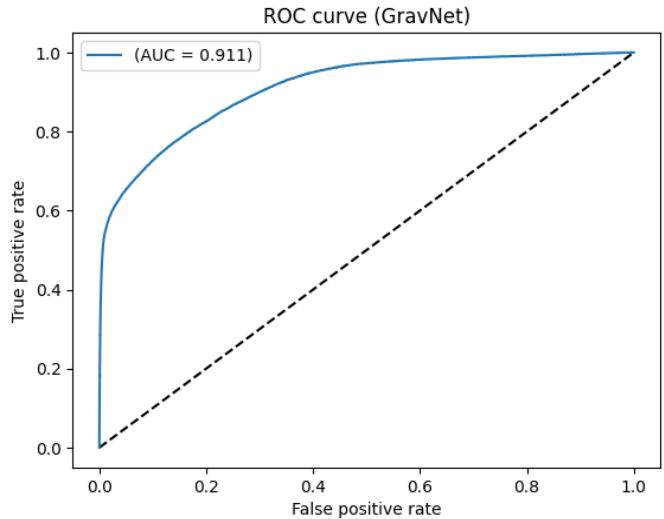

**Figure 5.5.2:** ROCs curve of the GravNet architecture with *K* equal to five reaches a ROC AUC of 0.91; in comparison, the default DFEI architecture that uses all the particles and the GN blocks reaches about 0.95.

## 5.6 Performance

As discussed in Section 3.2.3, investigating the response of any ML algorithm is crucial to understanding the algorithm that was created. The performance of DFEI was measured with respect to its main goals, the reduction of an event to the relevant particles using the NP filtering, as well as the topological event reconstruction using the whole DFEI algorithm. But other aspects are also important, namely the scaling of the algorithm with the total number of particles in the events in terms of bandwidth and efficiency. The different performance and possible problems will be assessed below.

### 5.6.1 Nominal event reduction

In this section, the nominal architecture as described in Section 5.4 is used, with which the maximum possible performance is achieved. The signal efficiency here is defined as the fraction of correctly selected particles over the number of particles that all truly come from a *b*-hadron decay in an event with the background efficiency defined accordingly. A selection requirement that allows a comparison with existing algorithms yields an average signal efficiency of 94 % and an average background rejection of 96 %. As a remarkable result, an almost flat response is found as a function of the total number of particles in the event, contrary to naive expectations that the distinction between signal and background events becomes more ambiguous with more particles, as shown in Fig. 5.6.1. As this stage can select which particles will actually be saved for offline analysis, another quantity to study is the total number of particles selected in the event, with the goal of including all relevant particles. The confusion matrix showing the true number of particles and the ones actually chosen is shown in Fig. 5.6.2.

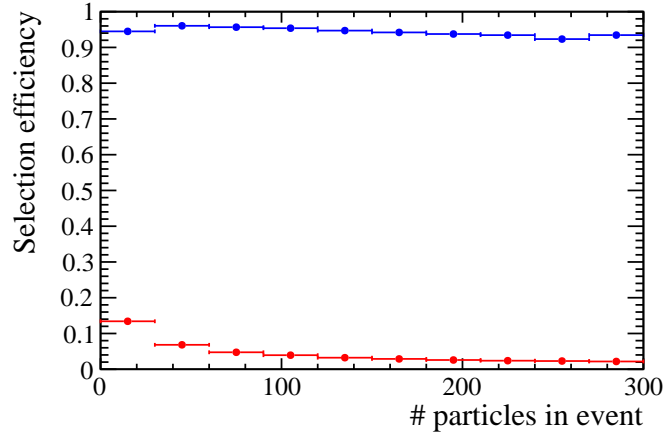

**Figure 5.6.1:** DFEI event selection efficiency and background rejection as a function of the number of particles in the event. The response is predominantly flat as a function of the number of particles in the event.

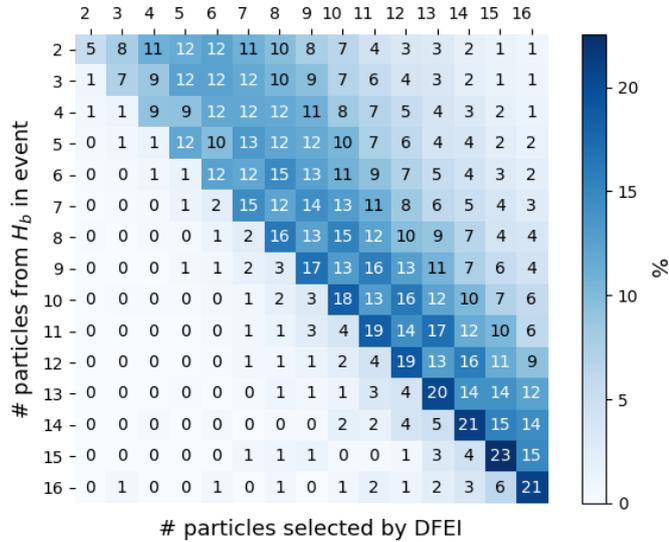

**Figure 5.6.2:** Confusion matrix of the number of selected particles per event. The problematic region is the lower triangle, as it means that relevant particles have been discarded. The upper triangle stores too many particles, which are not problematic as they can be taken care of in the offline analysis.

Overall, the average number of selected particles per event is $\approx 10$, from the initial number of $\approx 140$, and an average signal purity of $60\,\%$ is found. Purity is the relative fraction of the number of selected particles that truly originate from *b*-hadron decays over the total number of selected particles. Note that these numbers currently only apply to charged final states and do not take into account the neutral particles, making for an easier task. Therefore, the performance in event size reduction for these cases cannot be *directly* translated into an overall and general event size reduction; the overall order of reduction is still expected to be accurate, as the vast amount of information collected in an event comes from charged particles and their reconstruction.

## 5.6.2 Graph simplification

A simple preselection requirement on $\chi^2_{\mathrm{IP}}$ and $p_{\mathrm{T}}$, with $100\,\%$ signal efficiency, and the first two stages aim to reduce the complexity of the problem by removing irrelevant nodes and edges. The former can also be used as a generic particle removal, but the latter, edge removal, has no corresponding use case. The zeroth stage, edge preselection, removes about $11\,\%$ of the edges. The NP stage is the most important: It reduces the number of nodes by about $1/3$ and the number of edges correspondingly by a squared difference, that is, about $5/9$. The EP stage works directly on the edges, reducing them by about $1/3$. Due to some isolated nodes that are no longer connected to the rest of the particles after the EP stage, a few nodes are also removed. The overall reduction in the size of the graph is significant, as shown in Tab. 5.6.1. This simplification allows training and inference of the LCAI using reasonable computational resources and time.

**Table 5.6.1:** Cumulative average efficiencies on the total number of nodes and edges in the graph after each pre-filtering step. In each of the steps, a high signal efficiency of approximately $99\,\%$ per step or higher is aimed at preserving the relevant particles in the decay.

| Filtering step | Node eff. | Edge eff. |
|----------------|-----------|-----------|
| Preselection   | ~$100\,\%$ | ~$89\,\%$ |
| NP             | ~$29\,\%$ | ~$6\,\%$  |
| EP             | ~$27\,\%$ | ~$2\,\%$  |

## 5.6.3 Signal reconstruction

The inclusive reconstruction of decays is at the heart of DFEI but it may not be the most common use case at LHCb. Therefore, the performance of DFEI on specific, exclusive decay chains has also been tested. This means that the true decay chains of a certain type inside an event are compared with the predictions of DFEI and classified into the following exclusive categories that determine the performance:

**Perfectly reconstructed (PER)** these decays have all the particles originating from the same *b*-hadron, the reconstruction is disconnected from any other part of the event, and the particles in the decay are in the correct topological order. It is to note that this category is the most useful, but not necessarily needed for physics analysis, as additional tracks from other parts of the event can be removed in the offline analysis. The presence of additional tracks would not fit the reconstructed decay into this category.

**Wrong hierarchy (HIE)** decays that fulfill the same condition as before, but there is at least one mistake in the reconstruction of the "topological" ancestor decay chain.

**Not isolated (NOISO)** chains in this category contain not only all reconstructed particles originating from the *b*-hadron in the same connected decay, but at least one extra particle from the event was incorrectly included. This category is still very useful, as it allows dealing with additional particles in the offline analysis.

**Partially reconstructed (PRC)** not all reconstructed particles originating from the *b*-hadron decay have been correctly identified as belonging to the *same decay*. This does not exclude the possibility that the missing particles have been included in another reconstructed decay chain, such as in a non-isolated decay. Nevertheless, this category contains events that are mostly not useful for physics analysis.

**Table 5.6.2:** Performance of the DFEI algorithm. The response is measured for inclusive $H_b$ decays as well as for specific exclusive decay types. If the fraction is measured to be zero, the frequentist Wilson upper limit [81] for a 68 % coverage is provided.

| Decay mode | PER (%) | HIE (%) | NOISO (%) | PRC (%) |
|---|---|---|---|---|
| Inclusive $H_b$ decay | $4.6 \pm 0.1$ | $5.9 \pm 0.1$ | $76.0 \pm 0.2$ | $13.4 \pm 0.1$ |
| $B^0 \to K^{*0}[K\pi]\mu^+\mu^-$ | $35.8 \pm 0.7$ | $19.2 \pm 0.6$ | $44.9 \pm 0.7$ | $<0.02$ |
| $B^0 \to K^+\pi^-$ | $38.0 \pm 0.7$ | $-$ | $54.7 \pm 0.7$ | $7.2 \pm 0.4$ |
| $B_s^0 \to D_s^+[K^-K^+\pi^-]\,\pi^+$ | $32.8 \pm 0.7$ | $7.1 \pm 0.4$ | $53.7 \pm 0.8$ | $6.4 \pm 0.4$ |
| $B^0 \to D^0[K^+\pi^-]D^+[K^-\pi^+\pi^+]$ | $22.7 \pm 0.6$ | $22.4 \pm 0.6$ | $54.9 \pm 0.8$ | $<0.02$ |
| $B^+ \to K^+K^-\pi^+$ | $35.7 \pm 0.7$ | $10.2 \pm 0.4$ | $46.4 \pm 0.7$ | $7.7 \pm 0.4$ |
| $\Lambda_b^0 \to \Lambda_c^+[pK^-\pi^+]\,\pi^-$ | $21.7 \pm 1.0$ | $8.9 \pm 0.7$ | $36.8 \pm 1.2$ | $32.6 \pm 1.1$ |
| $B_s^0 \to J/\psi[\mu^+\mu^-]\,\phi[K^+K^-]$ | $26.9 \pm 0.6$ | $20.5 \pm 0.5$ | $52.5 \pm 0.6$ | $<0.02$ |

The performance of DFEI is measured for inclusive $H_b$ decays, by taking the average efficiency of all *b*-hadron decays as well as for specific exclusive decay types. To focus on the decays that are generally studied at LHCb, exclusive decays have been specifically measured. Both performances are listed in Tab. 5.6.2 and show some striking differences in the performance:

- The performance of PER events differs greatly, with the inclusive case being at 4.6 % and the exclusive cases being in the range of 20 % to 40 %. This is because the inclusive case contains many more complicated topologies that are not of primary interest to LHCb. The high rate for exclusive decays is a very promising result, as it shows the discriminative power of DFEI to distinguish between different decay topologies.

- The largest category for the exclusive cases and, most importantly, the inclusive ones is NOISO, which includes all particles from the decay of the *b*-hadron and, therefore, is well suited to be used in a trigger application.

- The difference in performance for exclusive decays is mostly due to the complicated topologies of the decays. The most complicated cases in terms of the shear number of reconstructed tracks is $B^0 \to D^0[K^+\pi^-]D^+[K^-\pi^+\pi^+]$, which shows low performance in all categories and has a notably high fraction of HIE events.

- The $\Lambda_b^0 \to \Lambda_c^+[pK^-\pi^+]\pi^-$ decay performs considerably worse since the $\Lambda_c^+$ baryon has a long lifetime, and therefore, the decay vertex is far away from the primary vertex, making the reconstruction of the decay chain more difficult.

## 5.7 Discussion

The performance of DFEI is especially interesting in comparison to the current LHCb trigger strategy. However, the LHCb trigger is still undergoing final commissioning and a nominal strategy, and therefore performance, is not yet available. For a direct comparison of the trigger selection efficiency with the expected nominal LHCb approach, the closest study that was found is reported in Ref. [82]. It considers the subset of reconstructed particles that have been selected by a standard LHCb inclusive trigger algorithm as described in Section 4.5. The algorithm combines different requirements around the vertexing quality and the output of a multivariate algorithm trained on individual particle features. With this approach, an approximate signal selection efficiency for particles from the same *b*-hadron decay of 90 % and an approximate background rejection power of 90 % is expected. Compared to the performance obtained by DFEI, as described in Section 5.6.3, this study is based on official LHCb simulation. The full simulation is more difficult to select, with fake tracks, and more sophisticated effects from material interactions. The DFEI approach has at least a similar performance level, for some charged decays even clearly above, compared to the current approach for the Run 3 conditions in LHCb; for higher multiplicities as expected in Upgrade II, DFEI is expected to scale significantly better given its flat response.

The reconstruction of the decay topology is low for inclusive events with a 2 % efficiency for perfect reconstruction, but significantly higher in the range of 20 % to 40 % for specific decay modes of interest, as described in Section 5.6.3. With the caveat of limitations using the ad hoc simulation, as presented in Section 5.3, a comparison of the performance of DFEI with the FEI algorithm in an electron-positron environment as encountered in Belle II can be made. The FEI algorithm is able to perfectly reconstruct the full decay chain of semileptonic decays in a few percent of cases, while for purely hadronic decays, the efficiency is in the permille region, as described in [54]. Finding the same order of magnitude in the DFEI performance in a hadronic environment clearly shows the feasibility of this approach.

The DFEI algorithm at its current stage has still multiple limitations that are planned to be addressed in the future:

- Neutral particles are not accounted for, which is a significant fraction of the particles in the event, and are planned to be added in the future; the difficulty herein lies with the fact that neutral particles do not leave a track in the detector, making the reconstruction of their origin vertex impossible.

- Not all available variables, notably the PID information, are used in the training of the algorithm, which could further improve performance or allow for more inferred variables.

- The three disconnected stages of DFEI are suboptimal from an algorithmic point of view, in that not all information is used in the final stage. Using the whole input graph is large and the training computationally expensive; studies are underway to solve both of the issues. Expanding on the idea of GravNet, a smooth self-learned distance can potentially replace manual cuts in the first two stages NP and EP. An incremental training, first pretraining on a strongly reduced sample and continuing with the full sample, can help to train the full graph at once.

- Timing studies under trigger-like conditions, with an optimized implementation in a language such as C++ that can realistically be used, are a crucial component for the possible application of DFEI and are planned to be performed in the future.

In summary, deep learning has been driving innovations in the field of HEP in recent decades, and the relatively new field of GNNs is no exception to this. It is on the daily agenda for an experiment like LHCb to squeeze the most statistical power out of the available data. This becomes even more important when considering upgrades at LHCb, where the event complexity will increase by orders of magnitude and the entire event is only available online at the trigger stage. DFEI is the first prototype of an algorithm that can interpret a whole event at once and significantly reduce the event size while remaining scalable. With an average performance of perfectly reconstructed decays on pair with the FEI algorithm in Belle II and a comparable performance in event selection as current LHCb trigger strategies, DFEI has been shown to be potentially a feasible candidate for the LHCb Upgrade I and, most importantly, for the future Upgrade II. The current stage of the DFEI development allows the integration of DFEI into the LHCb trigger to test and calibrate its response in real data, continuously improving the algorithm over the next decade.

# 6

# Studies on $R_K$ with Large Dilepton Invariant-Mass at LHCb

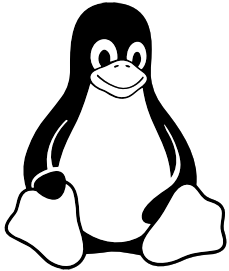

Decays involving $b \to s\ell^+\ell^-$ transitions are ideal to probe for NP as they are FCNC processes and therefore highly suppressed, as discussed in Section 2.3.3. As elaborated further in Section 2.3.4, the ratios $R_h$ of $b \to s\ell^+\ell^-$ branching fractions have many advantages over conventional measurements, as theoretical and experimental uncertainties each cancel to a high degree. This chapter presents a measurement of $R_K$, the ratio of $B^+ \to K^+\mu^+\mu^-$ and $B^+ \to K^+e^+e^-$ branching fractions, defined as

$$R_K \equiv \frac{\int_{q^2_{\text{min}}}^{q^2_{\text{max}}} \frac{\mathrm{d}\mathcal{B}(B^+ \to K^+\mu^+\mu^-)}{\mathrm{d}q^2}\mathrm{d}q^2}{\int_{q^2_{\text{min}}}^{q^2_{\text{max}}} \frac{\mathrm{d}\mathcal{B}(B^+ \to K^+e^+e^-)}{\mathrm{d}q^2}\mathrm{d}q^2} \tag{6.1}$$

where $q^2$ is the invariant mass of the di-lepton pair, $q^2_{\text{min}}$ and $q^2_{\text{max}}$ are the lower and upper boundaries of the $q^2$ range, respectively. Integrated over all $q^2$, the theoretical prediction is $R_K = 1.00 \pm 0.01$ [25, 83], as previously discussed in Section 2.3.4, a precision that goes far beyond the current experimental precision of a few percent.

The decay $B^+ \to K^+\ell^+\ell^-$ signal of interest is the "rare" mode, where the $b \to s\ell^+\ell^-$ transition occurs at loop level. The same initial and final state can arise at tree level in different "resonant" modes through $b \to sc\bar{c}$ transitions, where a charmonium ($c\bar{c}$) resonance is produced, mostly $J/\psi$ and $\psi(2S)$, which later decays into two leptons. Both decays are shown in Fig. 6.0.1.

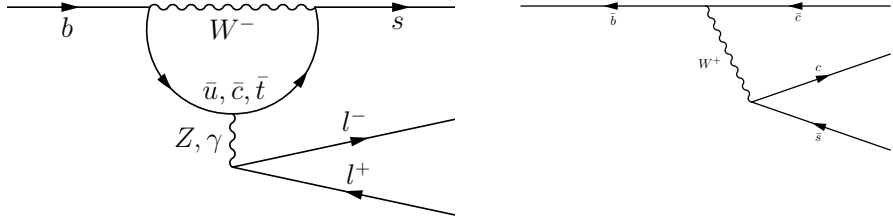

**Figure 6.0.1:** Loop level Feynman diagrams of the $B^+ \to K^+\ell^+\ell^-$ decay. The diagram on the left shows the "rare" mode, where $Z$ or $\gamma$ couples to the quark loop and the two leptons. The diagram on the right shows the "resonant" mode, where a charmonium ($c\bar{c}$) state is produced, which will later decay into two leptons.

The different decay modes are distinguished by the $q^2$ momentum transfer in the analysis, and the exact definitions used in the following are listed in Tab. 6.2.3. Fig. 6.0.2 shows the $q^2$ spectra of $B^+ \to K^+\ell^+\ell^-$ decays, where the resonant modes are clearly visible as peaks in the spectra. The central $q^2$ region and high $q^2$ region are "rare" modes, which are of interest for probing the effects of NP.

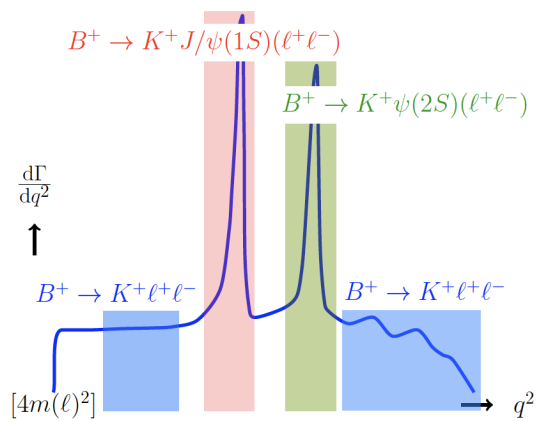

**Figure 6.0.2:** Spectra of $B^+ \to K^+ e^+ e^-$ decay channels in $q^2$. The central $q^2$ region (blue, left) was measured by the LHCb collaboration, above are two regions dominated by the resonances $J/\psi$ (red) and $\psi(2S)$ (green) with two orders of magnitude higher branching fractions. The high $q^2$ region (blue, right) above the resonances is the target of this analysis.

In the past, $R_K$ has been measured by different collaborations, notably BaBar [84] and Belle [85], and found to be consistent with the SM. Analysis [86] was recently published by the LHCb collaboration and dominates the precision on both $R_K$ and $R_{K^{*0}}$, which were measured simultaneously in the central $q^2$ region. The values obtained, $R_K = 0.949^{+0.042}_{-0.041}(\text{(stat)})^{+0.022}_{-0.022}(\text{(syst)})$ and $R_{K^{*0}} = 1.027^{+0.072}_{-0.068}(\text{(stat)})^{+0.027}_{-0.026}(\text{(syst)})$, are consistent with the SM prediction of unity. The results are depicted in Fig. 6.0.3, additionally showing the measurement in the low $q^2$ region which was expected to be unity for $R_{K^{*0}}$ due to the photon-pole and served as a cross-check.

This motivates the analysis in this chapter of $R_K$ at LHCb in the unmeasured high dilepton mass region as a direct test of LFU. The high $q^2$ region provides an independent dataset, which also differs experimentally given the higher di-lepton momentum. Due to previous work on the central $q^2$ region in multiple analyses, extensive studies have been done on efficiencies and backgrounds and lay a large part of the groundwork for analyzing the high $q^2$ region above the $J/\psi$ and $\psi(2S)$ resonances. However, the high $q^2$ region comes with some additional challenges, all of which are discussed alongside and compared to the procedures established previously in the following sections.

First, the overall analysis strategy is outlined in Section 6.1. In Section 6.2, the different selection criteria are discussed, involving trigger, offline, and multivariate analysis (MVA) requirements. Section 6.3 introduces different correction procedures to the simulation, derived from data, to account for the imperfections in the simulation. Following this, the selection efficiency is discussed in Section 6.4. In Section 6.5, different fit models that are used to extract $R_K$ and perform cross-checks are discussed. The fits to the normalization modes are shown in Section 6.6. Section 6.7 discusses the uncertainties on the measurement. A series of cross-checks validating the analysis strategy are presented in Section 6.8. Finally, the results and expected sensitivity on $R_K$ are presented in Section 6.9.

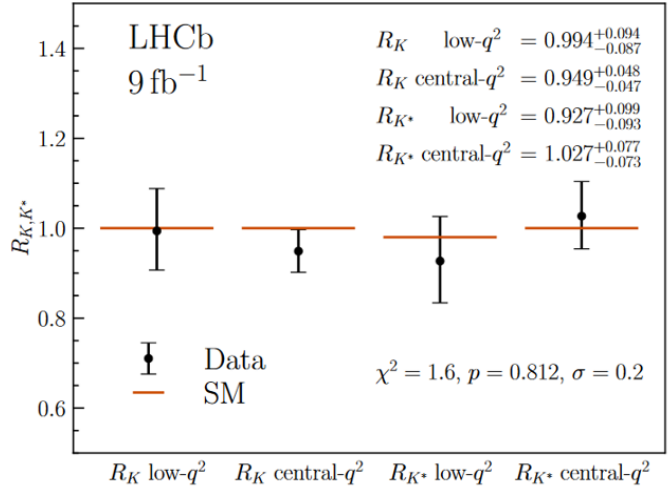

**Figure 6.0.3:** Measurements of $R_K$ and $R_{K^{*0}}$ performed by the LHCb collaboration in the central $q^2$ region $1.1 < q^2 < 6.0$ GeV$^2/c^4$ and in the low $q^2$ region $0.045 < q^2 < 1.1$ GeV$^2/c^4$. As the low $q^2$ region in $R_{K^{*0}}$ is dominated by the electromagnetic interaction, it is expected to be in agreement with the SM.

## 6.1 Strategy

The general strategy in the high $q^2$ region is based on the analyses of $R_K$ that have been performed in the central $q^2$ region at LHCb. Changing the $q^2$ region introduces a few potential problems that must first be understood. This led to an initial feasibility study of the whole analysis with the most prominent background components and only some of the corrections to understand the challenges and possible gain in precision. The entire analysis plans to incorporate data from Run 1 and Run 2, *i.e.*, for 2011, 2012, and 2015 to 2018, with a total of $9\,\text{fb}^{-1}$ data collected at the LHCb. The thesis reports nominally on the years 2016 to 2018 and mainly uses 2018 to showcase procedures, as the years differ minimally, and where they do, the differences are explicitly mentioned.

The analysis framework is built on a mixture of C++ code, mainly on top of ROOT [87], including RooFit [88], and PYTHON, with the use of the SCIKIT-HEP [89] ecosystem including the uproot [90], hist [91] and boost-histogram [92], particle [93], mplhep [94], hepstats [95], hepunits [96], iminuit [97] packages and the SCIKIT-HEP associated ZFIT [98] library.

The following sections establish the overall strategy and challenges. First, the different final states in the context of the LHCb detector are explained in Section 6.1.1, with a special focus on the electrons. The strategy and challenges of the $q^2$ selection are described in Section 6.1.2. Furthermore, Section 6.1.3 elaborates on the normalization channels $B^+ \to K^+ J/\psi(\ell^+\ell^-)$ and the control channel $B^+ \to K^+ \psi(2S)(\ell^+\ell^-)$. Finally, the different background components and strategies to deal with them are discussed in Section 6.1.4.

### 6.1.1 $K^+e^+e^-$ and $K^+\mu^+\mu^-$ final states

The measurement $R_K$ is made with the final states of the decays $B^+ \to K^+e^+e^-$ and $B^+ \to K^+\mu^+\mu^-$. Topologically, they consist of three reconstructed tracks, a kaon and two leptons, all of which are combined into a single decay. Given that the two leptons, electrons and muons, have *relative* to each other significantly different masses, the final states pose separate challenges. The electrons of $B^+ \to K^+e^+e^-$ radiate bremsstrahlung photons when passing through the LHCb detector as they interact with the detector material. Due to their light mass, this effect is large and crucially affects the momentum resolution, as further discussed in Section 6.1.2. Electrons are not penetrating and are difficult to identify correctly; therefore, they are difficult to trigger on. In stark contrast, $B^+ \to K^+\mu^+\mu^-$ muons are significantly heavier than electrons. They do not emit bremsstrahlung photons and are highly penetrating, even reaching the muon stations, as explained in Section 4.4.3.

The comparison of the momentum resolution highlights this in the two-dimensional distribution of the invariant mass of the three tracks $K^+\ell^+\ell^-$ ($m(K^+\ell^+\ell^-)$) and the four-momentum transfer squared, also called di-lepton mass squared ($q^2$), as shown in Fig. 6.1.1. The plots correspond to the entire Run 1 and Run 2 datasets, to which the preselection and trigger requirements, as discussed later in Section 6.2, have been applied. Various features are visible in these plots: Both datasets in Fig. 6.1.1 present $B^+ \to K^+\ell^+\ell^-$ decays in the $q^2$ range up to $25\,\text{GeV}^2/c^2$, yet the distributions look strikingly different with the muon dataset having a much better $m(K^+\ell^+\ell^-)$ and $q^2$ resolution than the electron dataset. Apart from the resolution, since the physics between both the muon and electron modes is the same, both datasets inhibit the same features that can be identified, mostly where $m(K^+\ell^+\ell^-) \approx$ invariant $B$ mass $\approx 5.28\,\text{GeV}/c^2$:

- A peaking structure centred at $m(K^+\ell^+\ell^-) \approx 5.28\,\text{GeV}/c^2$ and $q^2 \approx 9.6\,\text{GeV}^2/c^2$, corresponding to the resonant $B$-decay $B^+ \to K^+ J/\psi(\ell^+\ell^-)$. The radiative tails from this decay appear as a diagonal elongation;

- Another peaking structure centered at $m(K^+\ell^+\ell^-) \approx 5.28\,\text{GeV}/c^2$ and $q^2 \approx 13.6\,\text{GeV}^2/c^2$, corresponding to the resonant $B$-decay $B^+ \to K^+ \psi(2S)(\ell^+\ell^-)$;

- To the left of these peaks, elongated horizontal structures corresponding to partially-reconstructed decays;

- The non-resonant rare decay mode $B^+ \to K^+\ell^+\ell^-$, corresponding to a vertical band at $m(K^+\ell^+\ell^-) \approx 5.28\,\text{GeV}/c^2$ and spanning the full $q^2$-range.

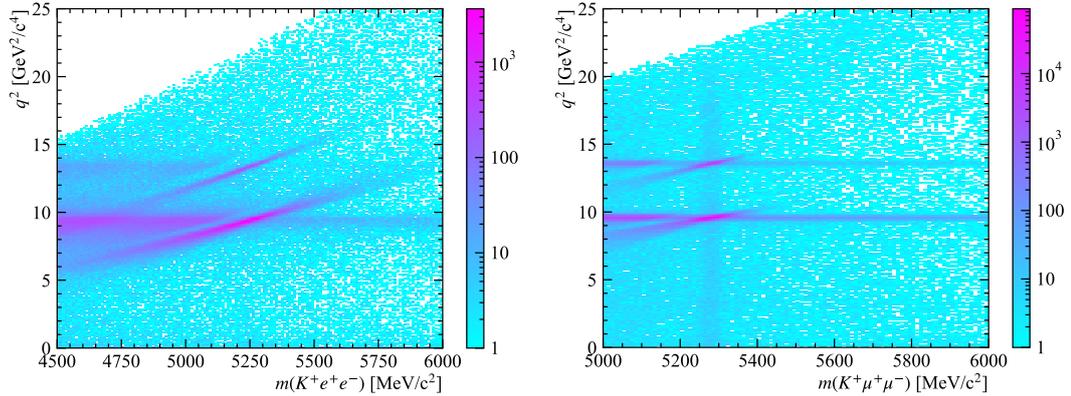

**Figure 6.1.1:** Two-dimensional distributions in $m(K^+\ell^+\ell^-)$, $q^2$ of the electron (left) and muon (right) final states. As can be seen, the muon dataset has a much better mass and $q^2$ resolution than the electron dataset, where the resonances are a lot wider.

- Finally, an empty triangle in the high $q^2$, low $m(K^+\ell^+\ell^-)$ corner, corresponding to the limits of the kinematically allowed region.

### 6.1.1.1 Trigger efficiencies

The LHCb detector selects muons at the trigger level using the muon stations at the very end of the detector, and triggers electrons using information from the electromagnetic calorimeter; both are explained in Section 4.5.1. Muons are minimum ionizing particles that travel across the LHCb detector, losing a negligible fraction of their momentum and allowing for a precise measurement and accurate simulation. For electrons, the momentum resolution and tracking efficiency depend strongly on the amount of material in the detector, which must be well known to reproduce the efficiency in simulated events. Moreover, the trigger efficiency and bremsstrahlung recovery depend on the performance of the calorimeters, which is affected by effects such as aging and the occupancy of the detector, both quantities that are poorly modeled in the simulation.

Therefore, electron reconstruction and trigger selection are significantly less efficient than the equivalent for muons, by approximately a factor of two and three, respectively. The exact differences in efficiencies are discussed in Section 6.4 and listed in Tab. 6.4.1 for electrons and in Tab. 6.4.2 for muons. As a consequence, the $B^+ \rightarrow K^+e^+e^-$ final state contains fewer events than the $B^+ \rightarrow K^+\mu^+\mu^-$ final state.

### 6.1.1.2 $q^2$ resolution and bremsstrahlung recovery

There is a crucial aspect in the differences of the two leptons for this analysis that requires a deeper investigation in the high $q^2$ region: the bremsstrahlung and its recovery. Bremsstrahlung is emitted primarily due to the interaction of a particle with the Coulomb field of the material; with a probability that is $\propto E/m^2$, electrons, due to their light mass, are highly affected by it. As there is little material *within* the magnet, nearly all the radiation happens before or after the magnet. An electron track from a $B^+ \to K^+ J/\psi(e^+e^-)$ decay typically loses $\sim 20\%$ momentum in bremsstrahlung before reaching the calorimeters. To compensate for this loss, a bremsstrahlung recovery algorithm is applied that assigns radiated photons with an energy above 75 MeV, detected in the ECAL, to electrons, as schematically depicted in Fig. 6.1.2.

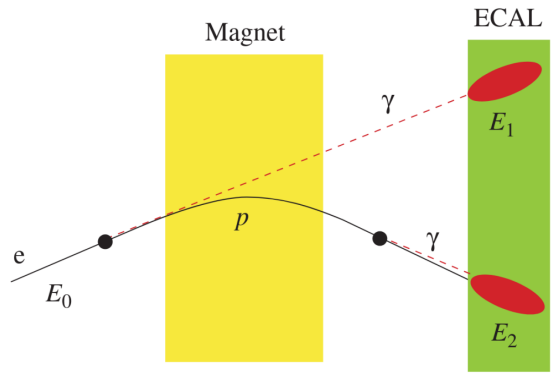

**Figure 6.1.2:** Depiction of the LHCb bremsstrahlung recovery algorithm with two possible scenarios. If the electron radiates after the magnet, the energy will be deposited in the same cluster $E_2$ as the electron itself. If the electron radiates before the magnet, the photon hits the calorimeter in a different cluster, $E_1$, that can be tracked back to the electron track before the magnet and reassigned to it.

If radiation occurs after the magnet, the photon will deposit its energy in the same cluster of the calorimeter as the electron, thereby altering neither the energy estimate from the ECAL nor the momentum measurement by the tracking system. However, if radiation occurs before, the photon cluster in the calorimeter is extrapolated back in order to match the electron track before the magnet. The recovery algorithm, matching to the correct electron track, is imperfect, and the recovered energy cluster, assumed to be a photon, can be assigned to the wrong electron track or be missed altogether.

This imperfection gives rise to two effects: First, if photons are missed, some energy loss of the electron is not recovered, thereby underestimating the electron momentum and resulting in a tail in the low $m(K^+\ell^+\ell^-)$ spectrum below the signal peak. Second, and especially important for the analysis of the high $q^2$ region, the recovered energy cluster can be assigned to the wrong electron track, overestimating the electron momentum, resulting in a tail in the high $m(K^+\ell^+\ell^-)$ spectrum above the signal peak. To illustrate the effect of bremsstrahlung, the distribution of the energy of the electrons and the recovered bremsstrahlung photons, as well as the $q^2$ resolution, are compared in Fig. 6.1.3.

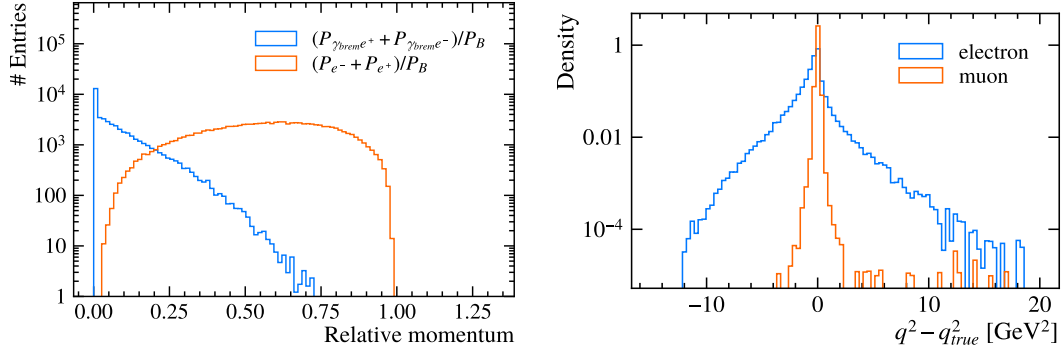

**Figure 6.1.3:** Fraction of the momentum carried away by the bremsstrahlung of the true total electron momentum (left) and a comparison of the electron and muon $q^2$ resolution (right) for simulated $B^+ \to K^+ e^+ e^-$ and $B^+ \to K^+ \mu^+ \mu^-$ events in the whole $q^2$ range.

Decays with two electrons in the final state can therefore be classified into three distinct *bremsstrahlung categories*, bremsstrahlung category zero, where neither of the two electrons has bremsstrahlung clusters added to them; bremsstrahlung category one, where one of the two electrons has bremsstrahlung energy added; and bremsstrahlung category two, where both electrons have added bremsstrahlung energy. Mass and momentum resolution, as well as the particle identification efficiencies associated with a particle, are strongly dependent on the bremsstrahlung category to which the event belongs.

Using bremsstrahlung recovery, the momentum resolution for an electron track from the $B^+ \to K^+ J/\psi(e^+e^-)$ decay is approximately 12%. This is still significantly worse than for a muon track, which typically has a momentum resolution of 0.6%. This poses four main challenges for the analysis:

- The worse resolution of $B^+ \to K^+ e^+ e^-$ final states results in a significantly lower reconstruction efficiency of these;

- Systematic uncertainties are dominated by effects related to the modeling of the background contributions in $m(K^+\ell^+\ell^-)$ of the $K^+ e^+ e^-$ final state, mainly from the combinatorial model that is heavily warped due to the limited available phasespace, the empty triangle in Fig. 6.1.1;

- The wrongly added bremsstrahlung recovery, *i.e.,* the tail to higher momenta, results in a worse $q^2$ resolution, affecting the removal of the charmonium resonant background. The leakage of $B^+ \to K^+ \psi(2S)(\ell^+\ell^-)$ events into the $B^+ \to K^+ e^+ e^-$ final state is a particular problem, as can be seen in Fig. 6.1.1;

- The high momentum of the leptons increases the misidentification probability of electrons while leaving the muon identification nearly unchanged.

The first point, together with the low trigger efficiency as discussed in Section 6.1.1.1, makes the electron mode the statistically limiting final state and its sensitivity the analysis' priority. For the second point, to get the efficiencies better under control, especially those originating from the calorimeter, corrections are derived using data events and comparing them with the control channel, as discussed later in Section 6.3.2.1.

Although the first two points have already been partially a challenge in the central $q^2$ region, the last two points, and the combinatorial background shape, are particular to the analysis in the high $q^2$ region and pose a major problem due to the large branching fraction of the charmonium resonance $\psi(2S)$ and associated partially reconstructed backgrounds, which leak into the signal region.

## 6.1.2 $q^2$ selection

The analysis aims to measure $R_K$ in the rare mode in the high $q^2$ region above the two charmonium resonances $J/\psi$ and $\psi(2S)$. Adequate selection requirements in $q^2$ are needed to select both resonances, which has been established in numerous other analyses, and, more importantly, to select the signal region *above* these two resonances. As the $\psi(2S)$ is heavier at a mass of $3686.10 \pm 0.06\,\text{MeV}/c^2$ than the $J/\psi$ at a mass of $3096.900 \pm 0.006\,\text{MeV}/c^2$ [7], the $\psi(2S)$ falloff mainly determines the $q^2$ signal selection threshold. With a mass squared of approximately $13.59\,\text{GeV}^2/c^4$ and a natural width squared of $0.35\,\text{GeV}^2/c^4$, the peak in the $q^2$ spectrum is expected to fall off at approximately $14\,\text{GeV}^2/c^4$. However, the actual width of the reconstructed invariant di-lepton mass arises from the uncertainty of the reconstructed momentum on the leptons, slightly widening the width of the $\psi(2S)$ peak.

As discussed previously in Section 6.1.1, not only do track momentum resolution effects increase the widths of the electron decay mode but also the wrongly added photons from the bremsstrahlung recovery, creating a tail in the higher energy spectrum of $q^2$. These added bremsstrahlung photons can have large energies, as seen in Fig. 6.1.3 and — improperly added to an electron — can push the $\psi(2S)$, and even the $J/\psi$, invariant mass squared well above $14\,\text{GeV}^2/c^4$. This presents significant background contributions from both the $K^+\psi(2S)(e^+e^-)$ decay and partially reconstructed decays $B^+ \to \psi(2S)(e^+e^-)X$.

Although the comparably minor effect of the track momentum resolution smearing cannot be disabled, the effect of the bremsstrahlung recovery can be disabled, and therefore the wrong association of bremsstrahlung photons to electrons. Computing $q^2$ using *only* the track information will not have bremsstrahlung recovery applied. This is referred to as the $q^2_{\text{track}}$ variable defined as

$$q^2_{\text{track}} \equiv \left( (E^{\text{track}}_{e_1} + E^{\text{track}}_{e_2})^2 - \sum_{i=x,y,z} (p^{\text{track}}_{i\,e_1} + p^{\text{track}}_{i\,e_2})^2 \right)^2 . \tag{6.2}$$

Performing a selection on the $q^2_{\text{track}}$ variable has a striking effect: it removes both charmonium resonances nearly completely, as no events below 14.3 remain, as depicted in Fig. 6.1.4. Applying a selection requirement on the $q^2_{\text{track}}$ variable instead of the $q^2$ variable is the main pillar of the analysis strategy and is, alongside a BDT based $q^2$ selection approach, further discussed in Section 6.2.4.

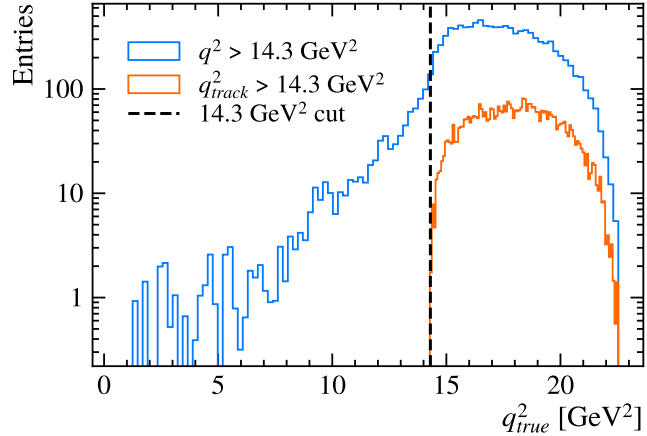

**Figure 6.1.4:** Comparison of the $q^2$ and $q^2_{\text{track}}$ distributions for the $B^+ \to K^+ e^+ e^-$ channel with a signal selection of $q^2_{\text{true}} > 14.3 \, \text{GeV}^2/c^4$. The selection requirement on the $q^2$ variable includes events well below 14.3 $\text{GeV}^2/c^4$, including events containing $J/\psi$ and $\psi(2S)$ resonances, which the $q^2_{\text{track}}$ variable does not.

The main implication for the measurement arises in the different $q^2$ dependence of the efficiency of $B^+ \to K^+ e^+ e^-$ and $B^+ \to K^+ \mu^+ \mu^-$, which is further enhanced by the $q^2_{\text{track}}$ selection requirement, as they do not cancel in the $R_K$ ratio. This implies a strong dependence on the simulation estimate of the efficiency with respect to $q^2$. Contrary to other effects in *different* variables, the difference in efficiency in $q^2$ cannot be assessed in the normalization channel, as $q^2$ serves as the channel discriminating variable itself. To correct for any non-uniformity, the same $q^2$ distribution in both modes is enforced for the $R_K$ measurement by applying a weighting on the $B^+ \to K^+ \mu^+ \mu^-$ events as further described in Section 6.5.3.

### 6.1.3 Normalization and control modes

The branching fractions of the decays $J/\psi \to e^+ e^-$ and $J/\psi \to \mu^+ \mu^-$ are measured to be equal to unity within subpercent precision [7]:

$$\frac{\mathcal{B}(J/\psi \to e^+ e^-)}{\mathcal{B}(J/\psi \to \mu^+ \mu^-)} = 0.9983 \pm 0.0078. \tag{6.3}$$

These are tree-level processes that predominantly decay through $b \to s c \bar{c}$ where no NP is expected to appear. The resonant decays $B^+ \to K^+ J/\psi(\ell^+\ell^-)$ and $B^+ \to K^+\psi(2S)(\ell^+\ell^-)$ have branching fractions that are two to three orders of magnitude larger than the expected signal branching fractions, providing comparably clean high-yield channels. Therefore, the $B^+ \to K^+ J/\psi(e^+e^-)$ and $B^+ \to K^+ J/\psi(\mu^+\mu^-)$ channels are used as normalization channels. Furthermore, the normalization channels $B^+ \to K^+ J/\psi(\ell^+\ell^-)$ and the decays $B^+ \to K^+\psi(2S)(\ell^+\ell^-)$, with $\ell\ell$ either two muons or two electrons, are also used in this analysis to derive corrections and for various cross-checks. Using the normalization channel, $R_K$ can be measured via a double ratio defined as

$$R_K = \frac{\int_{14.3\,\text{GeV}^2/c^2}^{\infty} \frac{\text{d}\mathcal{B}(B^+ \to K^+\mu^+\mu^-)}{\text{d}q^2} \text{d}q^2}{\int_{14.3\,\text{GeV}^2/c^2}^{\infty} \frac{\text{d}\mathcal{B}(B^+ \to K^+ e^+ e^-)}{\text{d}q^2} \text{d}q^2} \cdot \frac{\mathcal{B}(B^+ \to K^+ J/\psi(e^+e^-))}{\mathcal{B}(B^+ \to K^+ J/\psi(\mu^+\mu^-))} \tag{6.4}$$

$$= \frac{N(B^+ \to K^+\mu^+\mu^-)}{\varepsilon(B^+ \to K^+\mu^+\mu^-)} \cdot \frac{\varepsilon(B^+ \to K^+ e^+ e^-)}{N(B^+ \to K^+ e^+ e^-)}$$
$$\cdot \frac{\varepsilon(B^+ \to K^+ J/\psi(\mu^+\mu^-))}{N(B^+ \to K^+ J/\psi(\mu^+\mu^-))} \cdot \frac{N(B^+ \to K^+ J/\psi(e^+e^-))}{\varepsilon(B^+ \to K^+ J/\psi(e^+e^-))}, \tag{6.5}$$

where $N(X)$ indicates the yield estimate of decay mode $X$. The yield is obtained from an extended, unbinned maximum log-likelihood fit as discussed in Section 3.3.2 to the invariant mass $m(K^+\ell^+\ell^-)$ with a suitable selection requirement on $q^2$, and $\varepsilon(X)$ is the efficiency for selecting decay mode $X$, as computed from simulation samples which are corrected for various effects as described in Section 6.3.

The selection applied to the normalization modes is kept identical to that applied to the rare modes, except for the $q^2$ selection that differentiates the rare and normalization modes, as previously discussed in Section 6.1.2. The double ratio is designed in such a way that many systematic uncertainties cancel between the rare and normalization modes. The absolute size of *e.g.*, tracking, particle identification, or trigger efficiencies of one channel need not be known exactly, only the ratio of efficiencies between the rare mode and the corresponding control mode must be understood, *i.e.*,

$$\frac{\varepsilon(B^+ \to K^+ e^+ e^-)}{\varepsilon(B^+ \to K^+ J/\psi(e^+e^-))} \quad \text{and} \quad \frac{\varepsilon(B^+ \to K^+\mu^+\mu^-)}{\varepsilon(B^+ \to K^+ J/\psi(\mu^+\mu^-))} \tag{6.6}$$

are the quantities that must be controlled.

If the kinematic distributions of variables related to the rare mode were identical to those of the normalization mode, all efficiency ratios would be unity, and the measurement would be free from any efficiency-related uncertainties. In most one-dimensional variables, the distributions of the rare mode and the normalization mode are very similar. However, some variables show disagreement due to the different $q^2$ regions, as shown in Fig. 6.1.5. This means that the efficiency ratios, $\varepsilon(B^+ \to K^+ e^+ e^-)/\varepsilon(B^+ \to K^+ J/\psi(e^+e^-))$ and $\varepsilon(B^+ \to K^+\mu^+\mu^-)/\varepsilon(B^+ \to K^+ J/\psi(\mu^+\mu^-))$, will not fully cancel, and some residual sys-tematic uncertainties remain. For such variables, the dependence of the selection efficiency as a function of the variables must be well under control to correctly evaluate the effect of possible efficiency mismodeling; it would not fully cancel in the $R_K$ double ratio. The difference between the two channels is most pronounced in the two-dimensional plot of the maximum momentum of the two leptons against the angle between them, $\alpha_{K^+}$, as shown in Fig. 6.1.6.

This difference highlights one of the experimental motivations to measure the high $q^2$ region, as the kinematic differs *in the opposite way* with respect to the normalization mode than in the central $q^2$ region. The high $q^2$ region accesses kinematic distributions with a $q^2$ larger than that of the normalization channel, while the central $q^2$ region has kinematic distributions with a $q^2$ lower than the normalization channel. The kinematic differences of the two $q^2$ regions and the normalization mode are depicted in Fig. 6.1.6.

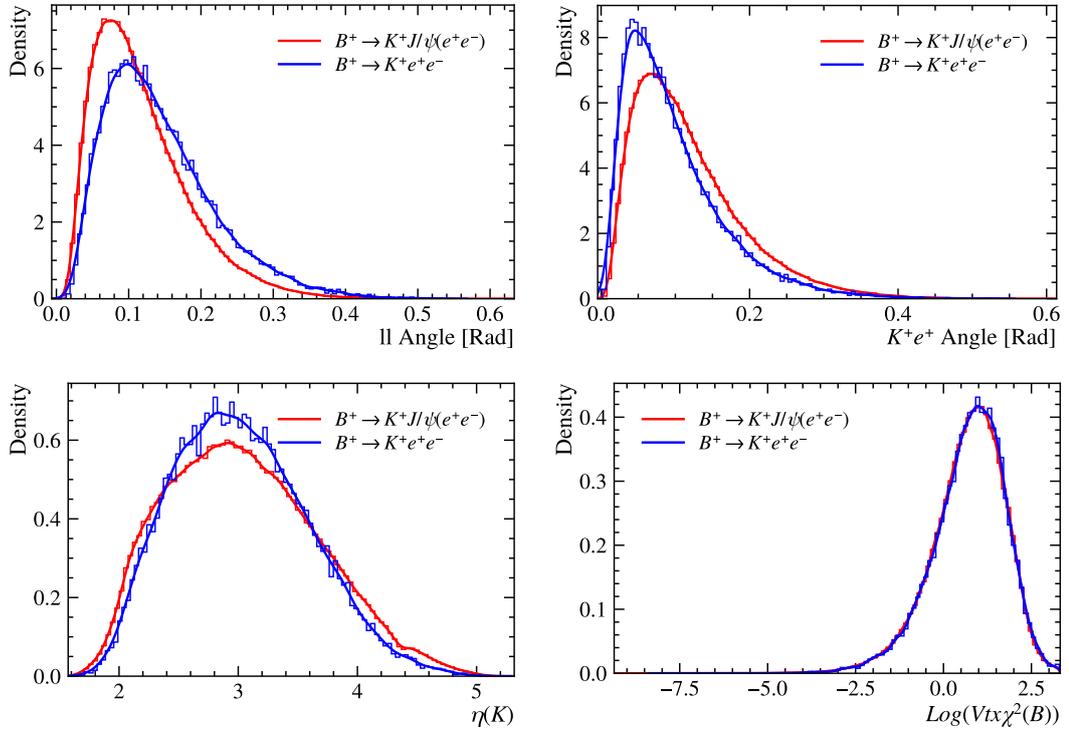

**Figure 6.1.5:** Distributions of kinematic variables for the $B^+ \to K^+ J/\psi(e^+e^-)$ mode (blue line) and the $B^+ \to K^+e^+e^-$ mode (red line). The angle between the two leptons (top left), the angle between one lepton and the kaon (top right), and the pseudorapidity of the kaon (lower left) show some expected disagreement. Contrary to these differences, most variables agree very well, such as the vertex $\chi^2$ of the $B$ (lower right).

In order to demonstrate that the efficiencies are controlled, several cross-checks are performed. The first of these cross-checks is the single ratio measurement of branching fractions between the normalization modes, called $r_{J/\psi}$, and measured in Section 6.8.1, which must be unity, in agreement with existing measurements as depicted in Eq. 6.3:

$$r_{J/\psi} = \frac{\mathcal{B}(B^+ \to K^+ J/\psi(\mu^+\mu^-))}{\mathcal{B}(B^+ \to K^+ J/\psi(e^+e^-))} \tag{6.7}$$

$$= \frac{N(B^+ \to K^+ J/\psi(\mu^+\mu^-))}{\varepsilon(B^+ \to K^+ J/\psi(\mu^+\mu^-))} \cdot \frac{\varepsilon(B^+ \to K^+ J/\psi(e^+e^-))}{N(B^+ \to K^+ J/\psi(e^+e^-))}. \tag{6.8}$$

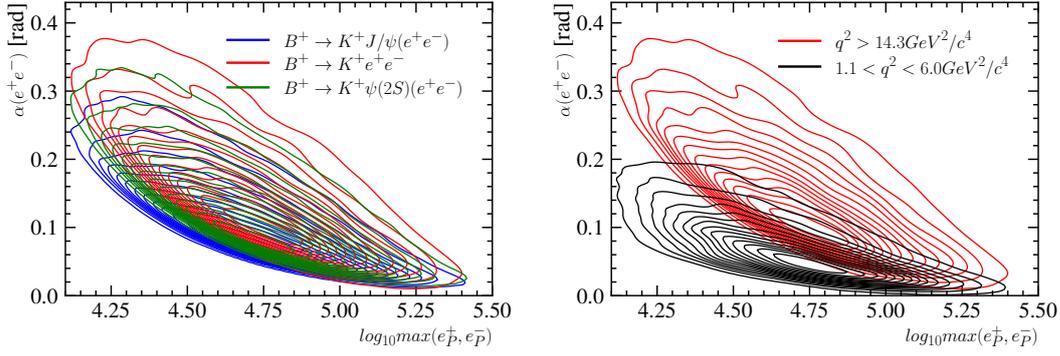

**Figure 6.1.6:** Distributions of kinematic variables for the $B^+ \to K^+ J/\psi(e^+e^-)$, $B^+ \to K^+ \psi(2S)(e^+e^-)$ and $B^+ \to K^+ e^+ e^-$ mode (left) and the comparison with the central $q^2$ region (right).

Because $r_{J/\psi}$ is a single ratio, the muon and electron efficiencies must be controlled directly with respect to each other. This is a strict cross-check, as systematic effects will not cancel as they do in the double ratio $R_K$. Moreover, because $N(B^+ \to K^+ J/\psi(\mu^+\mu^-))$ and $N(B^+ \to K^+ J/\psi(e^+e^-))$ are both relatively large, the statistical uncertainty is small and the total uncertainty is dominated by systematic effects.

As an additional cross-check, $r_{J/\psi}$ is measured differentially, *i.e.,* as a function of one kinematic variable as described in Section 6.8.2. Those cross-checks enable insight into whether the dependency of the efficiencies on the kinematic variable is properly understood and will therefore cancel well in the double ratio.

### 6.1.4 Backgrounds

The collected data samples contain not only signal events but also different background contributions coming from different decays or combinations thereof. Some of these backgrounds can be largely removed through a series of selection criteria, elaborated in Section 6.2, while some will remain partially in the data set. To extract the yields to obtain $R_K$ as defined in Eq. 6.1.3, a fit to the invariant $m(K^+ \ell^+ \ell^-)$ distribution is performed, which includes the remaining background components as part of the fit model, detailed in Section 6.5. This section introduces the different types of backgrounds, both in the rare and normalization modes, and the treatment of each of them.

### 6.1.4.1 Partially reconstructed backgrounds

The most prevalent type of physical background contribution, *i.e.,* events that originate from a specific decay, enters as partially reconstructed background, which are decays that have not been fully reconstructed and miss one or more particles, thus giving them the appearance of a different decay than they actually are. Decays with invisible particles, such as neutrinos that occur from semileptonic decays, are a special type of partially reconstructed background and are discussed separately in Section 6.1.4.2. Partially reconstructed background events that are relevant for the analysis originate from an actual *b*-hadron. This makes them peak below the *B* mass, as the energy of the non-reconstructed particles is missing. They are therefore mostly problematic in the electron mode, where the mass resolution is significantly worse than in the muon mode, bringing them closer to the signal peak.

For most contributions, the overall efficiency of partially reconstructed background is lower than for the signal decay itself after the selection requirements. Therefore, mainly decays with a much larger branching fraction play a role, such as decays with a resonance. The dominant types of partially reconstructed background that are encountered can generally be classified into two categories. The "strange" partially reconstructed background is a $b \to s\ell^+\ell^-$ decay where the *s* quark hadronizes into an excited strange resonance. A prominent example is the $B^0 \to K^{*0}\ell^+\ell^-$ decay with $K^* \to K\pi$, where the pion is not reconstructed. The other type of partially reconstructed background is labeled "charmed", which enters via a charmonium resonance in the $b \to sc\bar{c}$ transition. For example, $B^0 \to K^{*0}\psi(2S)\ell^+\ell^-$ where the pion is lost. To further reduce the number of partially reconstructed background events, a dedicated BDT was trained, called partially reconstructed background BDT, and is described in Section 6.2.5.1. The remaining partially reconstructed background contributions are modeled in the fit using simulation templates as described in Section 6.5.1.2.

### 6.1.4.2 Cascade backgrounds

Following the previous introduction of "strange" and "charm" partially reconstructed backgrounds, another prominent source of partially reconstructed background is "(semilep­tonic) cascade" decays. These decays start with a semileptonic $V_{cb}$ decay followed by a semileptonic $V_{cs}$ decay. Examples of this type of decay include $H_b \to H_c(\to K^+\ell^-\bar{\nu}_\ell X)\ell^+\nu_\ell Y$, where $H_b$ is a beauty hadron, such as $B^+$, $B^0$, $B_s^0$ or $\Lambda_b^0$, $H_c$ an open charm hadron[1], such as *D* mesons $D^0$, $D^+$, $D^*$, and *X*, *Y* denote an arbitrary number of additional missing particles. In addition to $X$ and $Y$ being missing particles, such decays contain two neutrinos, which are not reconstructed in the detector, making them peak at a lower invariant mass than the signal. These decays tend to have a large branching fraction of the order of a few percent, approximately four orders of magnitude larger than the signal decay, and are suppressed with a dedicated veto, as described in Section 6.2.3.2.

---

[1] Open charm refers to a hadron containing netto charm quarks and therefore acts charm-like; the opposite is a hidden charm consisting of $c\bar{c}$ quarks that cancel any total charm quantum number.

### 6.1.4.3 Cascade backgrounds with misidentified tracks

In addition to the previous cascade backgrounds, there are also partially reconstructed semileptonic cascade decays in which one or more of the final-state particles are misidentified. These decays are called "cascades with misidentification". A strong source of misidentified background originates from decays where a pion is misidentified as an electron.

### 6.1.4.4 Peaking backgrounds with double misidentification

Another source of background comes from fully reconstructed decays where two particles have been misidentified. The most significant of this kind comes from pions or kaons, which are misidentified as electrons, such as $B^+ \to K^+ \pi^+_{\to \ell^+} \pi^-_{\to \ell^-}$ and $B^+ \to K^+ K^+_{\to \ell^+} K^-_{\to \ell^-}$, which have branching fractions of $\mathcal{O}(10^{-5})$ [7].

Another type of double misidentified background events are the so-called "swaps", an event where two different particles are misidentified for each other, *i.e.*, where they swap their identities. An example of this is the decay $B^+ \to K^+_{\to \ell^+} J/\psi(\ell^+_{\to K^+} \ell^-)$, where the $K^+$ is misidentified as the $\ell^+$ and vice versa.

In order to control these misidentified backgrounds, strong selection requirements on PID variables are applied as described in Section 6.2.3.1. The remaining backgrounds are estimated using simulation samples and a data-driven method called "pass-fail", both used in the central $q^2$ region analysis [86].

Thereby, different samples are produced by inverting the PID selection requirement on the electrons, retrieving events where one or both show a small likelihood of actually being an electron. These constitute genuine misidentified backgrounds from which per-track weights can be derived that depend on kinematic variables and are extrapolated into the signal sample. These are applied to the data sample and serve as templates to model the remaining misidentified backgrounds.

### 6.1.4.5 Peaking backgrounds with a $\pi \to K$ misidentification

There is a single fully reconstructed background with a single misidentified particle: When a pion is falsely identified as a kaon, as in the decay $B^+ \to \ell^+\ell^-\pi^+_{\to K^+}$ and $B^+ \to J/\psi\pi^+_{\to K^+}$, which enter into the rare and resonant mode, respectively. As the pion has only a slightly lower mass than the kaon and the event is fully reconstructed, the decay peaks right around the signal peak in $m(K^+\ell^+\ell^-)$, shifted upward by the mass difference of the two mesons. Since the pion contains a $d$-quark instead of a $s$-quark, the decay is suppressed by a factor of $|V_{cd}/V_{cs}|^2$ in the normalization modes and by a factor of $|V_{td}/V_{ts}|^2$ in the case of the rare modes. This type of background can be suppressed through PID requirements on the kaon, as described in Section 6.2.3.1.

### 6.1.4.6 Combinatorial background

Events that are reconstructed from random combinations of tracks and vertices that do not originate from a specific decay are referred to as combinatorial background. Its characteristics differ from the previously discussed background types, as the tracks do not have a common physical process that generated them. This renders them nearly impossible to be well simulated and can only be reliably accessed in data. Two main sources of combinatorial background events are used throughout the analysis, both originating from collected data: One is the sideband in the mass spectrum, often in the higher mass region, since the lower mass contains partially reconstructed background. Another type are selected samples that do not contain any known physical decay - such as same-sign events $K^+e^-e^-$ or $B^+ \to K^+e^+\mu^-$ events. They are therefore expected to consist predominantly of combinatorial background events.

Combinatorial background appears in all modes and channels and has to be taken care of. To reduce the fraction of combinatorial background in the dataset, several preselection requirements are applied, as described in Section 6.2.3.1. Especially vertex quality variables greatly suppress this type of background, given that a combinatorial background event is a random combination of tracks. Furthermore, a dedicated BDT is trained on the mass sideband of the data to further reduce the combinatorial background, as described in Section 6.2.5.2.

Modeling the combinatorial background component in the high $q^2$ region greatly differs between the rare and normalization modes, contrary to the central $q^2$ region analysis. Due to the high $q^2$ selection requirement, as discussed in Section 6.2.4, the background is sculpted and an elaborate dedicated model is developed, described in Section 6.5.4.

# 6.2 Selection

The raw data samples obtained from the detector were already heavily reduced due to the geometric acceptance of the detector and the trigger requirements as described in Section 4.5. These samples are still huge and contain a large number of background events, raising the need for multiple selection criteria to improve the purity of the signals. To optimize the whole selection chain and maximize the sensitivity on $R_K$, mostly limited by the sensitivity in the electron mode, an expected signal and background yield are calculated. These are obtained by calculating the efficiencies and taking the different branching fractions [7], assuming SM physics. The efficiencies are obtained using simulation of the signal and background channels with corrections applied, as described in Section 6.3.

This section outlines all the different selection criteria applied to the data sample before performing the fit to extract $R_K$. It begins with the description of variables that are used throughout the analysis in Section 6.2.1 and then follows the experimental flow of the data. Detector-specific efficiencies and trigger requirements are introduced in Section 6.2.2. Then, the offline selection requirements, including the PID, are explained in Section 6.2.3. The $q^2$ selection is separately discussed in the consecutive Section 6.2.4, including the definition of the mass fit range. Finally, Section 6.2.5 introduces the different MVA algorithms that are used to remove specific backgrounds.

## 6.2.1 Variables

Throughout the analysis, a variety of variables refering to particles or combinations thereof are used. For a general description, a decay of type $A \rightarrow bc\dots$ is considered, where $b$ and $c$ are tracks and $A$ is a short-lived particle candidate, in this analysis usually a $B$ meson.

| | |
|---|---|
| PV | Primary Vertex: origin of $A$ where the $pp$ collision takes place; |
| SV | Secondary Vertex: $A$ decay point into $b$ and $c$; |
| $x_{ECAL}(b), y_{ECAL}(b)$ | $x$ and $y$ coordinates of track $b$ when reaching the electromagnetic calorimeter; |
| $\chi^2_{\text{DV}\leftrightarrow\text{PV}}(A)$ | Difference in $\chi^2$ obtained from two fits: either assuming all tracks come from the same point, or assuming two vertices, PV and SV. Serves as an indication of the significance of the flight distance of $A$; |
| $\text{DIRA}(A)$ | Directional angle: angle between the momentum vector of $A$, reconstructed from the decay products, and the vector that links the PV and SV; |
| $IP(A)$ | Impact Parameter: closest distance from the PV to a given track; |
| $\chi^2_{\text{IP}}(A)$ | Difference in $\chi^2$ of a given PV when reconstructed with and without the considered candidate $A$; |
| $\chi^2_{\text{DV}}(A)$ | $\chi^2$ of the fit of the decay vertex of $A$; |
| $\chi^2_{\text{DV}}(A)/\text{ndf}$ | $\chi^2$ per degree of freedom of $A$ decay vertex fit; |
| **hasRich** | Whether the particle has information from the RICH detectors that are explained in Section 4.4.1. |
| **hasCalo** | Whether the particle has information from the calorimeters that are explained in Section 4.4.2. |
| **inMuonAcc** | Whether the particle is within the muon acceptance of LHCb as explained in Section 4.4.3. |
| $\text{PID}_\beta(b)$ | Log-likelihood difference between the hypotheses that the track $b$ is a particle of type $\beta$, and that $b$ is a pion ($\beta = \mu, e, K$ or $p$). This variable uses input from the calorimeters and the RICH detectors; |
| $\text{probNN}_\beta(b)$ | Probability of the track $b$ being a particle of type $\beta$ ($\beta = \mu, e, K, \pi$ or $p$), as estimated from information from all subdetectors combined using a neural network; |
| $\text{isMuon}(b)$ | Boolean variable that indicates whether $b$ is compatible with being a muon, computed using information from the muon stations; |
| $\text{prob}_{\text{ghost}}(b)$ | Probability of $b$ being a ghost, that is, a fake track coming from a random combination of hits or noise in the tracking system; |
| $\chi^2_{\text{TrackFit}}(b)$ | $\chi^2$ of the $b$ track fit; ; |
| nSPDHits | Number of hits recorded in the scintillating pad; detector; |
| $m(b, c)$ | Invariant mass of a combination of particles; |
| $\theta_{\text{clone}}$ | Angle between kaon and the same-sign electron; |
| $m_{\text{track}}(b, c)$ | Mass of a combination of particles using only the information from the tracking systems, meaning that, in the case of electrons, the energy of recovered bremsstrahlung is not added; |
| $q^2$ | Invariant mass of the di-lepton system; |
| $q^2_{\text{track}}$ | $q^2$ computed using the mass by combining the tracks, $m_{\text{track}}(b, c)$, meaning that for the electrons, no bremsstrahlung recovery information is added; |

### 6.2.2 Trigger requirements

The trigger system of the LHCb detector is described in Section 4.5, which reduces the bandwidth of events and the amount of data stored on disk to a manageable level. The decision to persist an event can be based on different criteria, with each of them stored in a variable of the data for later determination of the trigger criteria. Of special importance are the different decisions taken in the level zero trigger stage, which can be associated with a particle that is part of the signal decay, called trigger on signal (TOS), or with a particle that is not directly part of the signal decay, called trigger independent of signal (TIS). Events in TOS are triggered by a high-momentum final-state particle, an electron, muon, or kaon, triggering the `L0Electron`, `L0Muon` or `L0Hadron` category, respectively. The TIS trigger is associated with the signal $B^+$ requiring either a muon or a hadron in the *event* — not decay — to trigger the `L0Muon` or `L0Hadron` criteria.

The leptonic trigger criteria `L0Electron` and `L0Muon` are used as the nominal trigger category in each of the respective channels $B^+ \to K^+ e^+ e^-$ and $B^+ \to K^+ \mu^+ \mu^-$. Samples that have been triggered with other criteria have an efficiency comparatively much smaller than that of the nominal sample. This difference is more pronounced compared to the central $q^2$ region analysis, since the leptons carry more energy. Therefore, other trigger categories are not used for the determination of $R_K$, but only for different cross-checks, to assess systematic uncertainties and to derive certain corrections.

All trigger requirements are listed in the following Tab. 6.2.1. A logical *or* between different lines at the same trigger stage is implied. Each event has to pass all trigger stages L0, HLT1, and HLT2. Events that pass all trigger requirements are categorized according to the level-zero trigger decision, exclusive with respect to previous lines: If `L0Electron` is triggered by one of the electrons, the event is classified as $e$TOS; as $h$TOS! if `L0Hadron` is triggered by the kaon *but the event is not classified as $e$*TOS; as TIS! if `L0TIS` is triggered through `L0Electron`, `L0Hadron`, `L0Muon`, or `L0Photon` of a particle *not contained* in the signal decay and falls neither in the $e$TOS nor the $h$TOS! category. For the muon mode, events triggered with `L0Muon` are referred to as $\mu$TOS.

The HLT1 trigger line `Hlt1TrackMVA` uses a MVA algorithm based on kinematic variables and vertex quality. In HLT2, topological trigger lines using a *bonsai* BDT (BBDT), a fast implementation of a BDT, are used. They are designed to select reconstructed events that contain a $b$ hadron, either with two or three tracks.

**Table 6.2.1:** Trigger requirements and definitions that are used throughout the analysis. A track or $B$ candidate indicated in parentheses means that a TOS requirement is applied on that candidate. An event is required to pass *one of* the trigger criterion *for each* trigger stage L0, HLT1 and HLT2. Logically, there is an *or* between lines in a trigger stage and an *and* between the trigger stages. Here, "threshold" refers to appropriate requirements to equalize the dataset in each year, where thresholds in the trigger have been varied throughout the period of data-taking.

|       | **Electron mode** | **Muon mode** |
|-------|-------------------|---------------|
| L0    | `L0Electron` $(e)$ & $E_\mathrm{T}(e) >$ threshold | `L0Muon` $(\mu)$ & $p_\mathrm{T}(\mu) > 0.8\,\mathrm{GeV}$ |
|       | `L0Hadron` $(K)$ & $E_\mathrm{T}(K) > 3.5\,\mathrm{GeV}$ | & $p_\mathrm{T}^\mathrm{L0}(\mu)\,\mathrm{GeV}$ threshold |
|       | `L0Electron` $\parallel$ `L0Hadron` $\parallel$ |  |
|       | `L0Muon` $\parallel$ `L0Photon` $\parallel$ `L0TIS` |  |
| HLT1  | `Hlt1TrackMVA`$(B)$ | `Hlt1TrackMVA`$(B)$ |
| HLT2  | `Hlt2Topo[2,3]BodyBBDT`$(B)$ | `Hlt2Topo[2,3]BodyBBDT`$(B)$ |
|       |  | `Hlt2TopoMu[2,3]BodyBBDT`$(B)$ |

The trigger variables have slight complications as the thresholds for the `L0Electron` and `L0Muon` trigger in $E_\mathrm{T}$ and $p_\mathrm{T}$, respectively, have been varied during the data taking periods. However, this variation is not directly reflected in the simulation. Both together complicate the evaluation of efficiencies. Fiducial selection requirements are applied to the data samples, bringing the selection requirements to the same minimum for each year and aligning the value with what is available in the simulation. This results in a loss of $\sim 2\%$ of $B^+ \to K^+ J/\psi (e^+ e^-)$ events in $e$TOS, $\sim 2.5\%$ of $B^+ \to K^+ e^+ e^-$ events in $e$TOS, and $\sim 6\%$ of $B^+ \to K^+ J/\psi (\mu^+ \mu^-)$ events. Therefore, this loss is small and has a negligible effect on the double ratio of $R_K$.

## 6.2.3 Preselection

### 6.2.3.1 Fiducial selection requirements

A set of preselection requirements is applied to the data, reducing the amount of data to a manageable size and performing a first selection of events. The preselection imposes loose quality and PID selection requirements to remove the vast majority of background events while maintaining high signal efficiency. Further selection requirements are applied to remove events that are outside of subdetector acceptances. The preselection also contains a requirement of ghost probability; ghost tracks are tracks that have been reconstructed from hits that are not associated with a real particle, as explained in Section 4.3. All analysis preselection requirements are listed in Tab. 6.2.2.

**Table 6.2.2:** Analysis preselection requirements for both the electron mode (left) and the muon mode (right). Units, if not given, are in MeV, MeV/$c$ and MeV/$c^2$ where appropriate. The different variables are explained in Section 6.2.1.

| Electron | | | Muon | | |
|---|---|---|---|---|---|
| **Event quality** | | | | | |
| nSPDHits | $<$ | 450 | nSPDHits | $<$ | 450 |
| $\text{prob}_{\text{ghost}}(K, e)$ | $<$ | 0.3 | $\text{prob}_{\text{ghost}}(K, \mu)$ | $<$ | 0.3 |
| $\cos(\theta_{\text{clone}})$ | $<$ | 0.9999998 | $\cos(\theta_{\text{clone}})$ | $<$ | 0.9999998 |
| **Cascade & misidentification vetoes** | | | | | |
| $m(K^+e^-)$ | $>$ | 1885 | $m(K^+\mu^-)$ | $>$ | 1885 |
| $m_{e\to\pi}^{Track}(K^+e^-)$ | $\notin$ | $m(D^0) \pm 40$ | $m_{\mu\to\pi}(K^+\mu^-)$ | $>$ | 1885 |
| | | | $m_{K\to\mu}(K^+\mu^-)$ | $\notin$ | $m(J/\psi) \pm 60$ |
| | | | $m_{K\to\mu}(K^+\mu^-)$ | $\notin$ | $m(\psi(2S)) \pm 60$ |
| **Fiducial selection requirements** | | | | | |
| $\chi^2_{to\ \text{PV}}(B^+)$ | $>$ | 100 | $\chi^2_{to\ \text{PV}}(B^+)$ | $>$ | 121 |
| $\cos(\text{DIRA})(B^+)$ | $<$ | 0.995 | $\cos(\text{DIRA})(B^+)$ | $<$ | 0.9999 |
| $\chi^2_{\text{IP}}(B^+)$ | $<$ | 25 | $\chi^2_{\text{IP}}(B^+)$ | $<$ | 16 |
| $\chi^2_{\text{DV}}/\text{ndof}(B^+)$ | $<$ | 9 | $\chi^2_{\text{DV}}/\text{ndof}(B^+)$ | $<$ | 8 |
| $\chi^2_{\text{IP}}(K)$ | $>$ | 9 | $\chi^2_{\text{IP}}(K)$ | $>$ | 6 |
| $p_{\text{T}}(K)$ | $>$ | 400 | | | |
| $\chi^2_{\text{DV}\leftrightarrow\text{PV}}(e^+e^-)$ | $>$ | 16 | $\chi^2_{\text{DV}\leftrightarrow\text{PV}}(\mu^+\mu^-)$ | $>$ | 9 |
| $\chi^2_{\text{DV}}/\text{ndof}(e^+e^-)$ | $<$ | 9 | $\chi^2_{\text{DV}}/\text{ndof}(\mu^+\mu^-)$ | $<$ | 12 |
| $\chi^2_{\text{IP}}(e)$ | $>$ | 9 | $\chi^2_{\text{IP}}(\mu)$ | $>$ | 9 |
| $p_{\text{T}}(e)$ | $>$ | 500 | $\texttt{inMuonAcc}(K, \mu)$ | $=$ | $\texttt{true}$ |
| $p(e)$ | $>$ | 3 GeV | $\texttt{hasRich}(K, \mu)$ | $=$ | $\texttt{true}$ |
| $\texttt{hasRich}(K, e)$ | $=$ | $\texttt{true}$ | $p_{\text{T}}(\mu)$ | $>$ | 0.8 GeV |
| $\texttt{hasCalo}(e)$ | $=$ | $\texttt{true}$ | | | |
| $x_{ECAL}(e)$ | $>$ | 363 mm | | | |
| $or\ y_{ECAL}(e)$ | $>$ | 282 mm | | | |
| $\chi^2_{\text{TrackFit}}$ (all) | $<$ | 3 | | | |
| **PID selection requirements** | | | | | |
| $\text{probNN}_K(K)$ | $>$ | 0.2 | $\text{probNN}_K(K)$ | $>$ | 0.2 |
| $\text{PID}_e(K)$ | $<$ | 0 | $\text{isMuon}(K)$ | $=$ | $\texttt{false}$ |
| $\text{PID}_e(e)$ | $>$ | 4 | $\text{PID}_\mu(\mu)$ | $>$ | 0 |
| | | | $\text{isMuon}(\mu)$ | $=$ | $\texttt{true}$ |

**6.2.3.2 Cascade background vetoes**

As described in Section 6.1.4.2, there is a significant contribution of cascade backgrounds in the electron mode, which mainly originate from $D^0$ decays. To suppress these backgrounds, two strategies were studied: Require a rectangular selection, a veto, or use a dedicated BDT to remove the background.

Since the cascade background is mainly caused by $D^0$ decays, a veto requirement can be applied to the invariant mass of the decay products $m(K^+\ell^-)$, requiring $m(K^+\ell^-) > m(D^0)$. For misidentified cascades, as described in Section 6.1.4.3, the same selection requirement is applied, but the lepton mass hypothesis is switched to that of the pion. For the electron mode, following similar arguments about the use of the $q^2_{\text{track}}$ variable for the $q^2$ selection in Section 6.1.2, the $D^0$ mass is computed using the electron track momenta without bremsstrahlung recovery. Although the veto allows for a clean selection of cascade backgrounds, it also removes a significant fraction of signal events, about 30%.

As an alternative approach, a BDT was trained using kinematic and track quality variables, but without the invariant mass $m(K^+\ell^-)$, the most discriminatory variable, to avoid introducing explicit correlations with $m(K^+\ell^+\ell^-)$. The BDT approach did not turn out to be as powerful as the veto in suppressing cascade events and was therefore discarded.

### 6.2.4 $q^2$ selection

Experimentally, the different decay modes $B^+ \to K^+e^+e^-$, $B^+ \to K^+J/\psi(e^+e^-)$ and $B^+ \to K^+\psi(2S)(e^+e^-)$ are distinguished by a $q^2$ selection requirement, as shown in Fig. 6.0.2. Electrons in the $K^+e^+e^-$ final state have a comparably bad momentum resolution due to their interaction with material and photon radiation as they traverse the LHCb detector, as discussed in Section 6.1.1. This complicates the correct $q^2$ selection, as wrong bremsstrahlung recovery can significantly alter the measured momentum of decays, notably by pushing events from resonances into the high $q^2$ region.

#### 6.2.4.1 $q^2$ track

To avoid resonant background contributions, the $q^2_{\text{track}}$ variable is used as defined in Eq. 6.2 instead of the $q^2$ variable. $q^2_{\text{track}}$ is not affected by bremsstrahlung recovery and does not suffer from possible erroneous addition of bremsstrahlung energy. Migration to a higher value of $q^2_{\text{track}}$ than the true $q^2$, except for track resolution effects, is therefore not possible in this variable.

The effect of this selection requirement, also in comparison to the normal $q^2$ selection requirement, on the signal as well as different backgrounds, can be seen in Fig. 6.2.1.

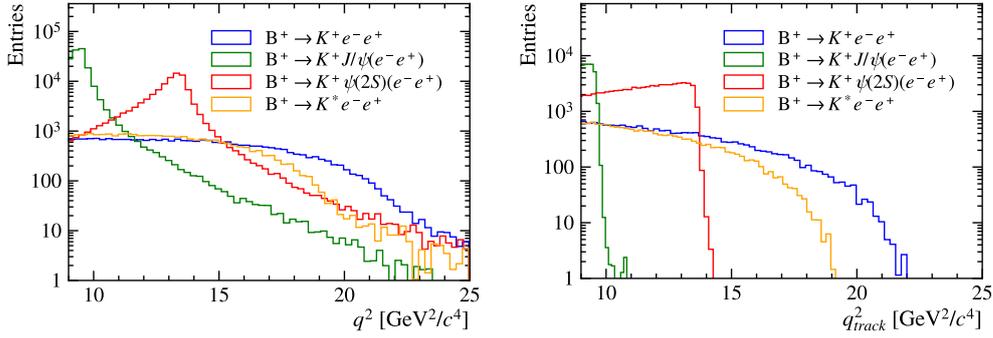

**Figure 6.2.1:** Comparison of the different channels, signal $B^+ \to K^+ e^+ e^-$ (blue), $B^0 \to K^{*0} e^+ e^-$ (orange), $B^+ \to K^+ J/\psi(e^+ e^-)$ (green) and $B^+ \to K^+ \psi(2S)(e^+ e^-)$ (red) in $q^2$ (left) and $q^2_{track}$ (right) distributions.

The $q^2_{track}$ allows for a clean removal of the charmonium resonant backgrounds containing $J/\psi$ and $\psi(2S)$ around $14\,\text{GeV}^2/c^4$. However, this selection has the disadvantage that about half of all true signal events are removed, mostly in bremsstrahlung category two and partially in bremsstrahlung category one. This change in the distribution of the bremsstrahlung categories can be seen in Fig. 6.2.2.

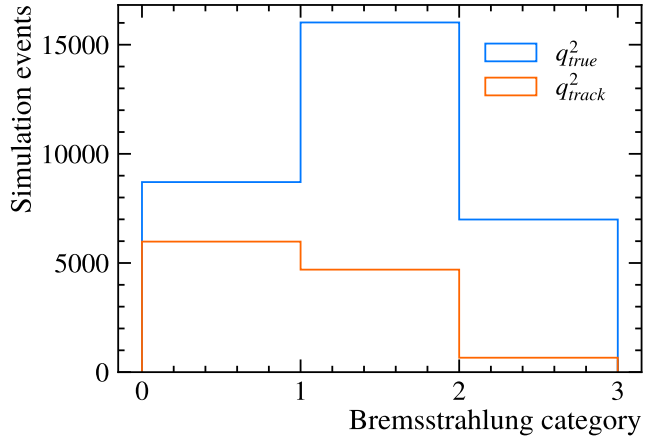

**Figure 6.2.2:** Signal events by bremsstrahlung category in $B^+ \to K^+ e^+ e^-$ simulation with a true selection of $q^2_{\text{true}} > 14.3\,\text{GeV}^2/c^4$ (blue) and a $q^2_{\text{track}}$ selection $q^2_{\text{track}} > 14.3\,\text{GeV}^2/c^4$ (red).

The selection requirement on $q^2_{\text{track}}$ is imperative, as additional background events that leak into the $q^2$ window severely affect the precision of the measurement. As demonstrated in toy studies, this is mainly due to the addition of background events that contain a $\psi(2S)$, including partially reconstructed decays of the form $B^+ \to \psi(2S)(e^+ e^-)X$. Such decays are difficult to constrain well enough, as all the exact contributions and branching fractions are required. Additionally, excellent simulation would be needed not only to model the tail of the resolution correctly, but also to obtain the correct efficiencies of these partially reconstructed $B^+ \to \psi(2S)(e^+e^-)X$ decays.

The effect of the $q^2_{\text{track}}$ selection requirement is clearly visible, as the $\psi(2S)$ resonances are basically removed, while an analysis with a selection requirement on $q^2$ would essentially be dominated by the uncertainties on $B^+ \to \psi(2S)(e^+e^-)X$ leakage. Therefore, the nominal selection requires $q^2_{\text{track}} > 14.3 \, \text{GeV}^2/c^4$ with an estimated total of $0.49 \, K^+\psi(2S)(e^+e^-)$ events remaining over all years.

### 6.2.4.2 $q^2$ BDT

The approach of using a $q^2$ selection requirement based on the $q^2_{\text{track}}$ variable leads to a loss of about 50% of signal events. To lessen the impact on the signal loss of this choice, a dedicated BDT has been studied to recover parts of the events that are rejected by the $q^2_{\text{track}}$ selection requirement but have a $q^2_{\text{true}}$ within the selection range of $q^2 > 14.3 \, \text{GeV}^2/c^4$. Selecting events through this BDT constitutes a non-overlapping data sample that can be directly combined with the existing selected data sample. Different strategies and sets of variables have been tested, which are described in the following.

**Training samples**   Training and evaluation of the $q^2$ BDT was performed in simulation with the whole selection chain, including BDT requirements and the correction weights, as described in Section 6.3, applied. Using only simulation eliminates the risk of the BDT picking up differences in simulation and data instead of the true $q^2$ difference. All events considered in the training were required to have $q^2_{\text{track}} < 14.3 \, \text{GeV}^2/c^4$ to be complementary to the nominally selected sample.

The signal sample consists of $B^+ \to K^+e^+e^-$ events with $q^2_{\text{true}} > 14.3 \, \text{GeV}^2/c^4$, resulting in events that have been "falsely" removed by the $q^2_{\text{track}}$ selection requirement, as they actually have a $q^2$ in the desired signal $q^2$ range. The choice of suitable background samples and variables to use for the BDT is not as straightforward: A BDT with a powerful discrimination between events that are inside or outside the $q^2$ range, is desired, but at the same time, its response must be understood extremely well, as any undesired effect would not cancel in the double ratio $R_K$. Most importantly, the performance estimate must be impartial with respect to any background leakage. Multiple options have been considered.

$B^+ \to \psi(2S)(e^+e^-)X$

This is the most obvious choice as it remains the main background in the nominal selection. However, since these backgrounds are not very well understood, their quality cannot be compared to actual data, and the simulation samples have comparably limited number of events. Using them as a training sample could bias the estimation of the remaining events in data. Therefore, they are only used for performance estimation.

$B^+ \to K^+\psi(2S)(e^+e^-)$

These events have the exact same decay topology as the signal. They make up for a good training sample but have the disadvantage that they have a constraint phasespace of $q^2 \approx m^2_{\psi(2S)}$, which neither the rare mode nor the $B^+ \to \psi(2S)(e^+e^-)X$ events have. Therefore, the classifier detects wrong correlations that are not present in the other channels, resulting in a different BDT response for these decays compared to $B^+ \to K^+\psi(2S)(e^+e^-)$.

$B^+ \to K^+e^+e^-$

The same decay as the signal candidates but with an inverted $q^2_{\text{true}}$ selection of $q^2_{\text{true}} < 14.3\,\text{GeV}^2/c^4$. This sample has the advantage that the classifier can only learn correlations that originate from a different $q^2_{\text{true}}$ value, no matter which variables are used as input. The disadvantage is that a classifier potentially underperforms in rejecting backgrounds in comparison to using more dedicated samples, as other correlations cannot be explored. As a BDT trained on this sample is expected not to favor any specific background contribution, it is used as the main training data sample.

The different effects and performance of using these samples for training are compared in Fig. 6.2.3. As a performance comparison, the $q^2_{\text{track}}$ variable was considered similar to the BDT score, which is not competitive at all and illustrates the need for a $q^2$ BDT approach.

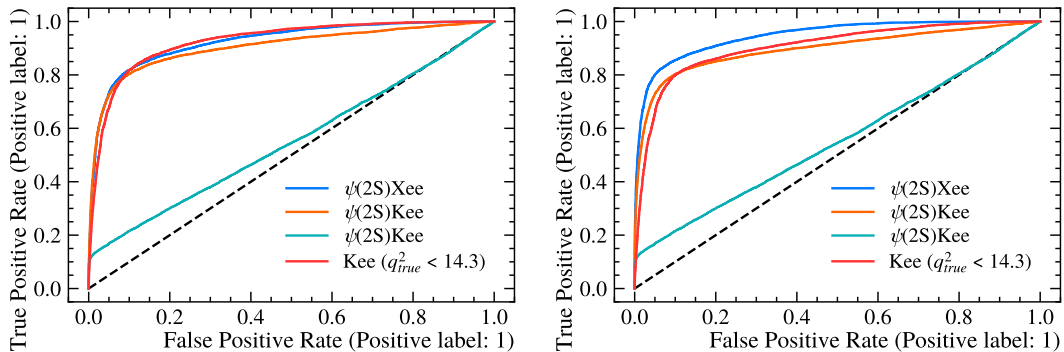

**Figure 6.2.3:** Comparison of the ROC curves for the different background samples $B^+ \to K^+e^+e^-$ with $q^2_{\text{true}} < 14.3\,\text{GeV}^2/c^4$ (left) and $B^+ \to \psi(2S)(e^+e^-)X$ (right) used in the training of the BDT with the lepton features, evaluated on $B^+ \to \psi(2S)(e^+e^-)X$ (blue), $B^+ \to K^+\psi(2S)(e^+e^-)$ (orange) and $B^+ \to K^+e^+e^-$ with $q^2_{\text{true}} < 14.3\,\text{GeV}^2/c^4$ (red). Additionally, the performance of using a $q^2_{\text{track}}$ variable by going lower than 14.3 GeV$^2/c^4$, is illustrated on the $B^+ \to K^+\psi(2S)(e^+e^-)$ sample (green).

**Variables** The selection of input features is ultimately a trade-off between the power and the biasing of the BDT to favor a specific decay. The following sets of variables were considered where each previous set is contained as a subset of the next one.

**Minimal** The smallest set of variables tested consists of the $q^2$ and $q^2_{\text{track}}$ variable. While not very powerful, it serves as a minimally biased baseline.

**Lepton features** This is used as the nominal set of variables and contains all the electron kinematic variables, with and without bremsstahlung, and reconstruction quality variables.

**Full** In addition to the lepton tracks and kinematics, track and vertex fitting quality information of both the kaon and the $B$ meson are included. This provides the most complete and powerful set of variables. However, the signature of missed photons in the reconstruction, candidates of interest to the analysis, inhibits a signature similar to the partially reconstructed background, which also misses a particle in the final state. The validation showed this behavior, namely that the efficiency in $B^+ \to \psi(2S)(e^+e^-)X$ events was significantly higher compared to fully reconstructed $B^+ \to K^+\psi(2S)(e^+e^-)$. Therefore, this set of variables was not considered feasible.

The $q^2$ and $q^2_{\text{track}}$ variable, used in the minimal set, are the strongest correlated variables among all other variables considered, as depicted in Fig. 6.2.4. The different responses of the $q^2$ BDT to the different background samples are shown in Fig. 6.2.5.

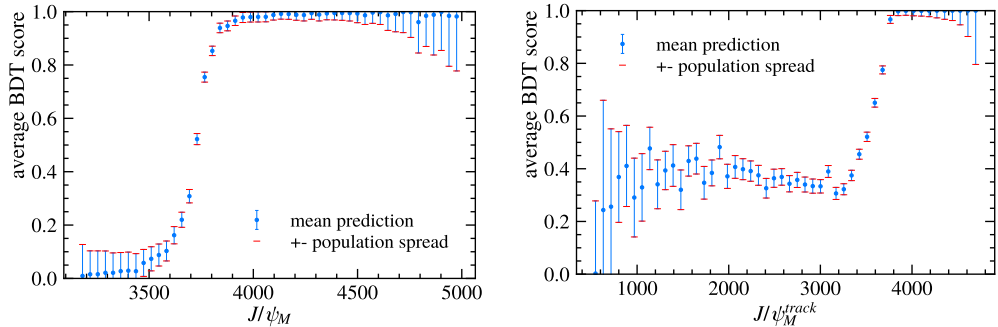

**Figure 6.2.4:** Average $q^2$ BDT prediction against the $q^2$ (left) and $q^2_{\text{track}}$ (right) variable for the $B^+ \to K^+e^+e^-$ training sample with the full variable set.

**Performance** The $q^2$ BDT performance is evaluated using the $B^+ \to \psi(2S)(e^+e^-)X$, $B^+ \to K^+\psi(2S)(e^+e^-)$ and $B^+ \to K^+e^+e^-$ samples as background. The ROC is plotted for the full feature set in Fig. 6.2.6 and for the nominal lepton set in Fig. 6.2.3. The main trade-off of using this strategy is to have more leakage from backgrounds containing a $\psi(2S)$. As previously discussed in Section 6.1.2, this is difficult to constrain and peaks under the signal in $m(K^+\ell^+\ell^-)$ as can be seen in Fig. 6.2.8 and brings back the problem that was attempted to be solved with the $q^2_{\text{track}}$ selection requirement.

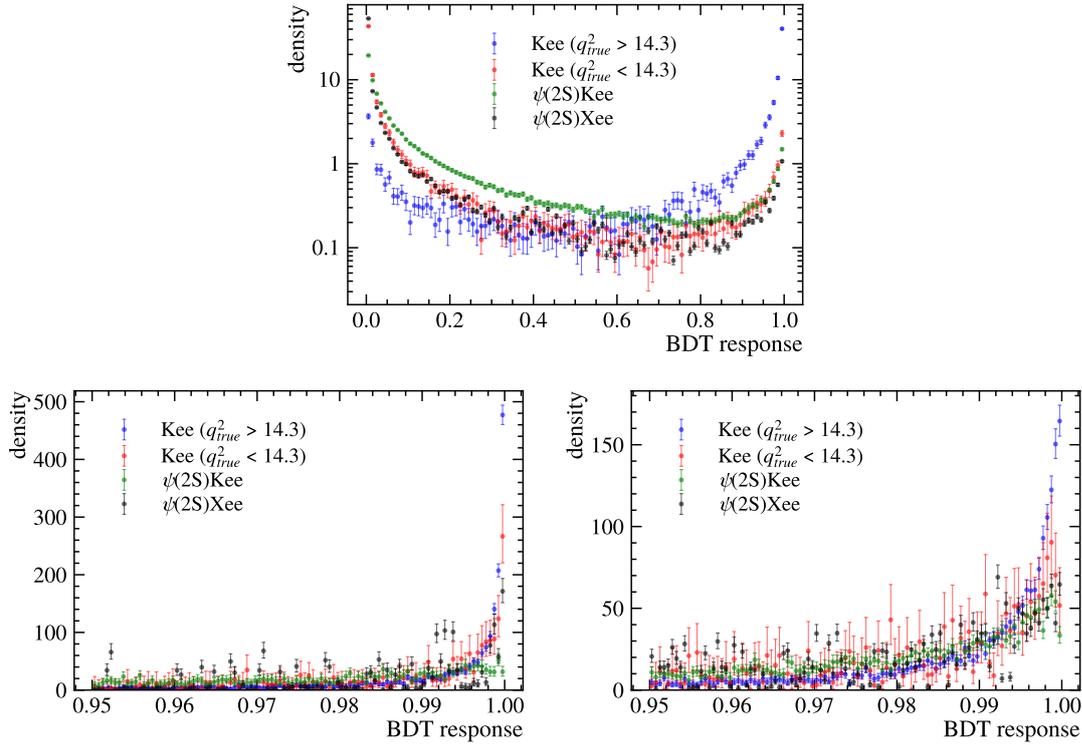

**Figure 6.2.5:** $q^2$ BDT response of the different backgrounds for the full feature set (bottom left) and the lepton feature set (bottom right), both zoomed in on the high score region, and an overview of using the lepton feature set (top).

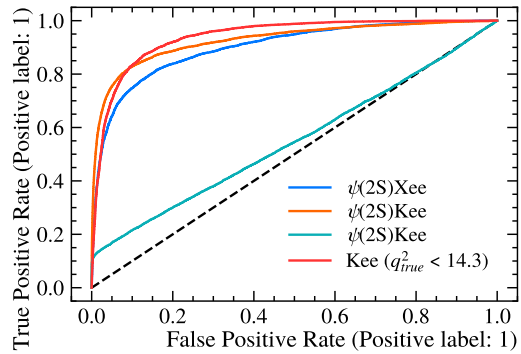

**Figure 6.2.6:** ROC curves using the full set of variables for the $B^+ \to K^+e^+e^-$ with $q^2_{\text{true}} < 14.3 \, \text{GeV}^2/c^4$, $B^+ \to \psi(2S)(e^+e^-)X$ and $B^+ \to K^+\psi(2S)(e^+e^-)$ backgrounds trained on the nominal $B^+ \to K^+e^+e^-$ with $q^2_{\text{true}} < 14.3 \, \text{GeV}^2/c^4$ sample.

In Fig. 6.2.7 the estimated gain of signal events against the leakage of $B^+ \to \psi(2S)(e^+e^-)X$ and $B^+ \to K^+\psi(2S)(e^+e^-)$ *relative* to the leakage in the nominal selection is shown for the $q^2$ BDT. As the nominal leakage is tiny, the *relative* leakage grows quickly with increasing signal gain, and is best illustrated compared to a lower selection requirement in the $q^2_{\text{track}}$ variable. The $q^2$ BDT works significantly better than the latter, yet the relative leakage is still large, too large to gain any additional sensitivity as toy studies revealed.

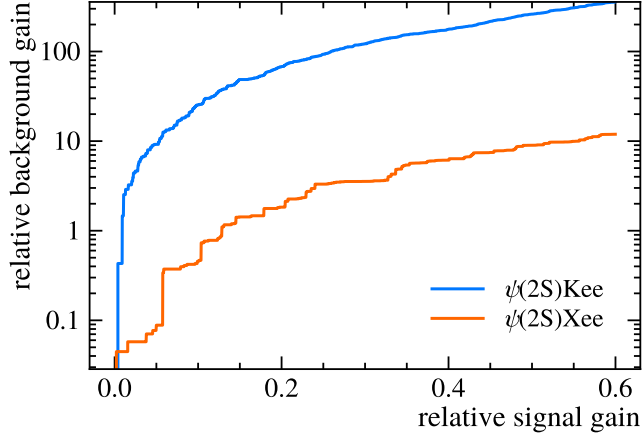

**Figure 6.2.7:** Gain in signal $B^+ \to K^+e^+e^-$ events against the leakage of $B^+ \to K^+\psi(2S)(e^+e^-)$ (blue) and $B^+ \to \psi(2S)(e^+e^-)X$ (red) *relative* to the number of events in the respective channels found using the nominal $q^2_{\text{track}}$ selection requirement.

An estimation of expected contributions in a sample that contains additional events selected by the $q^2$ BDT is shown in Fig. 6.2.8 and demonstrates the issue of additional leakage. The same arguments that have been used before motivating using the $q^2_{\text{track}}$ over the $q^2$ variable apply again. Although the $q^2$ BDT is highly effective at recovering signal events compared to an alternative loosened $q^2_{\text{track}}$ selection requirement, the purity provided by the nominal $q^2_{\text{track}}$ selection maximizes the sensitivity of $R_K$ in this measurement; the leakage of background events from $B^+ \to \psi(2S)(e^+e^-)X$ and $B^+ \to K^+\psi(2S)(e^+e^-)$ through the $q^2$ BDT is too large. Therefore, the $q^2$ BDT is not used in the nominal analysis.

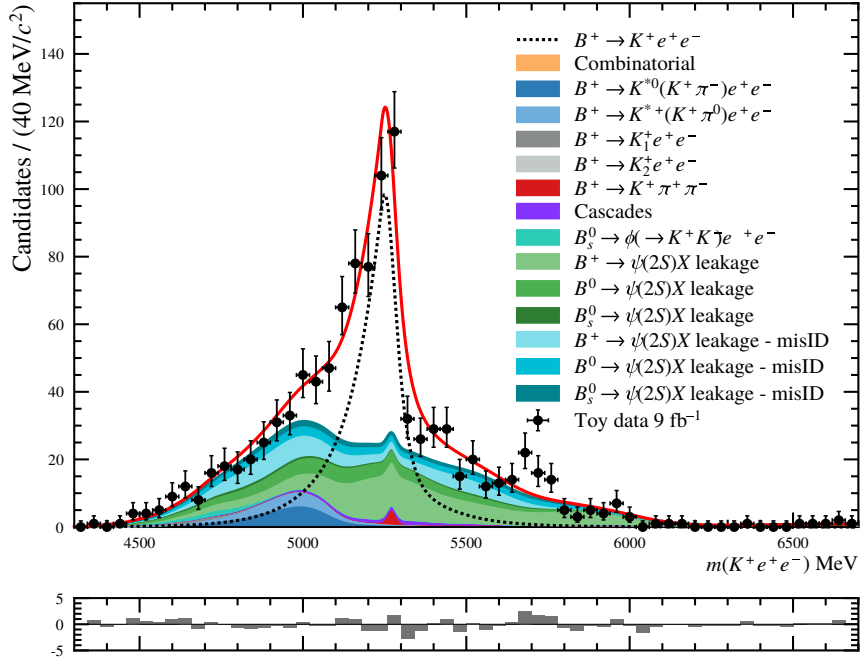

**Figure 6.2.8:** Generated example of the expected composition with an additional sample using the $q^2$ BDT selection requirement. The $\psi(2S)$ peak is clearly visible under the signal peak and becomes dominant.

### 6.2.4.3 Definition of $q^2$ and mass ranges

Throughout the analysis, the following $q^2$ and $m(K^+\ell^+\ell^-)$ range definitions are used to specify the different modes and mass ranges in the fits, respectively. They are based on the discussions throughout this section and the previous analysis.

**Table 6.2.3:** Signal, normalization and control channels and their corresponding $q^2$ or $q^2_{\text{track}}$ and mass ranges. Units are omitted due to better readability and are implied to be $\text{GeV}^2/c^4$ for $q^2$ and $\text{GeV}/c^2$ for the mass. The decay "rare" refers to the signal decay $B^+ \to K^+e^+e^-$, $J/\psi$ to the normalization channel $B^+ \to K^+J/\psi(e^+e^-)$ and $\psi(2S)$ to the control channel $B^+ \to K^+\psi(2S)(e^+e^-)$.

| Decay | electron mode | muon mode |
|---|---|---|
| rare | $q^2_{\text{track}} > 14.3\,\text{GeV}^2/c^4$ | $q^2 > 14.3\,\text{GeV}^2/c^4$ |
| | $4.30 < m(K^+e^+e^-) < 6.70$ | $5.18 < m(K^+\mu^+\mu^-) < 5.60$ |
| $J/\psi$ | $6.00 < q^2 < 12.96$ | $8.68 < q^2 < 10.09$ |
| | $5.08 < m_{\text{DTF}}{}^{J/\psi}(K^+e^+e^-) < 5.68$ | $5.18 < m_{\text{DTF}}{}^{J/\psi}(K^+\mu^+\mu^-) < 5.60$ |
| $\psi(2S)$ | $9.92 < q^2 < 16.40$ | $12.5 < q^2 < 14.2\,\text{GeV}^2/c^2$ |
| | $5.08 < m_{\text{DTF}}{}^{\psi(2S)}(K^+e^+e^-) < 5.68$ | $5.18 < m_{\text{DTF}}{}^{\psi(2S)}(K^+\mu^+\mu^-) < 5.60$ |

### 6.2.5 Multivariate selection

An MVA selection is performed in addition to the requirements established before. In comparison to rectangular selection requirements, an MVA selection can take into account higher order correlations between variables, yielding a better discriminative power. On the downside, training samples are required to learn the correlations. The techniques of ML, different algorithms, and the problems of validation and bias are explained in more detail in Section 3.2. For the implementation of the following BDTs, the XGBoost [99] library is used and a $K$-folding with $k = 10$ is used to obtain unbiased BDT predictions.

#### 6.2.5.1 Partially reconstructed background BDT

On the lower side of the $m(K^+\ell^+\ell^-)$ spectrum, there are partially reconstructed events that originate from the decay of a $B$ but have one or more tracks not assigned to the decay, as introduced in Section 6.1.4.1. For the electron mode, due to its low resolution, they leak into the signal region and must be taken into account when modeling the $m(K^+\ell^+\ell^-)$ variable, further elaborated in Section 6.5.1.2. Partially reconstructed events inhibit a wider, slightly peaking structure just below the signal peak but strongly reaching into it. Therefore, reducing this contribution greatly increases the sensitivity on $R_K$. The strategy is to use a dedicated BDT to remove these background events.

The BDT can take advantage of the systematic differences from a fully reconstructed decay to a partially reconstructed one:

- With a missing track, there will be some missing transverse momentum ($p_\mathrm{T}$) that can be determined from the position of the PV.

- The efficiency of partial reconstruction is considerably lower than for fully reconstructed decays, which is why the main background contribution is expected to come from intermediate resonances of the $KH$ system. As this resonance has a certain mass, the $q^2$ is expected to be lower than for a fully reconstructed decay.

- The missing track is close to the actual decay, as it comes from the PV. Isolation variables can therefore help to indicate the proximity of such an additional track.

The samples used to train the BDT consist of simulated samples with all selection requirements applied and weighted using the whole correction chain. For the background, a combination of three decays was used, namely

- $B^0 \to K^{*0}(\to K^+\pi^-)e^+e^-$,

- $B^+ \to K^{*+}(\to K^+\pi^0)e^+e^-$,

- $B_s^0 \to \phi(\to K^+K^-)e^+e^-$.

The datasets from the years 2016, 2017, and 2018 were merged to train the BDT to increase the number of events available for training. Using all years together showed superior performance and stability compared to training the BDT for each year separately, also in direct comparison to a BDT that was trained in a specific year, as shown for the year 2018 in Fig. 6.2.9. This clearly indicates that any possible difference between the years, which is expected to be minimal, is outweighed by the additional events added to the training set.

The relative weight of the samples follows the expected relative size at this stage of the selection. These samples only represent a part, the lighter resonances, of an array of possible partially reconstructed background decays. Heavier resonances, such as $B^+ \to H^+ (\to K^+ \pi^+ \pi^-) e^+ e^-$ with $H^+$ either $K^+ K_1^+(1270)$ or $K_2^+(1430)$, were also used for training, but overall only worsened the performance. Analogously to this choice, an independent feature optimization yielded the same set of variables as for the combinatorial background BDT that are listed in Tab. 6.2.4, with the only exception that the $\chi^2_{\mathrm{DV}}$ variable was omitted. The partially reconstructed background BDT was optimized with respect to the ROC AUC, which is used only in the electron channel, that is about 0.82 and the ROC curve shown in Fig. 6.2.9.

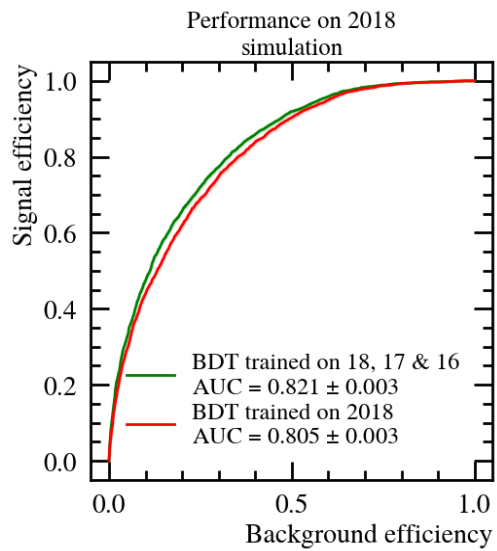

**Figure 6.2.9:** ROC curves for the partially reconstructed background BDT trained on all three years 2016, 2017 and 2018 together (green) or separately in 2018 (red).

### 6.2.5.2 Combinatorial background BDT

Combinatorial background is a random combination of tracks without a specific physical decay as discussed in Section 6.1.4.6. Preselection requirements on reconstruction quality variables can reduce this contribution, but only up to a certain point. Due to its nature of not originating from an actual decay, variables that describe the decay topology can differ a lot from signal candidates; using a MVA based selection requirement can remove a significant fraction of the remaining combinatorial background events.

To train the combinatorial background BDT, $B^+ \to K^+ e^+ e^-$ signal events with the whole correction chain applied were used to be trained against data events in the signal $q^2$ range from the upper mass sideband with $m(K^+ \ell^+ \ell^-) > 5400 \, \mathrm{MeV}/c^2$, which are expected to consist predominantly of combinatorial background events. The training was carried out using a $K$-folding technique with 10 folds as described in Section 3.2.3.2 to obtain unbiased predictions for sideband events. As for the partially reconstructed background BDT, the performance is found to improve by merging samples from different years due to the higher available statistics.

For each of the lepton modes, a separate BDT was trained and the selection requirements optimized, as some of the topological variables between the electron and muon modes differ significantly, as described in Section 6.1.1. The variables used in the training were chosen not only to provide good discrimination between the two datasets but also not to bias the efficiency estimation. They needed to be mostly uncorrelated with $m(K^+ \ell^+ \ell^-)$ to not allow the BDT to select events around the signal region and thereby sculpt the shape, possibly creating a peak of combinatorial background events under the signal peak. Furthermore, they needed to be well simulated in order for the BDT not to pick up the difference in data against simulation instead of the difference between signal and background events. Some studies on the similarity of the chosen variables are described in Section 6.3.3.3. Furthermore, the BDTs were compared with the normalization mode $B^+ \to K^+ J/\psi(e^+ e^-)$ to evaluate their performance. Both decays have the same topological signature apart from the $q^2$ and are therefore a sensitive check for $q^2$ invariance of the BDT output. In the muon mode, the BDT seems to pick up a minor correlation and performs slightly worse in the normalization channel using the same variables as for the electron mode. Therefore, some variables were dropped into the muon mode and are listed, along with the electron mode variables, in Tab. 6.2.4.

**Table 6.2.4:** List of variables used in the training of the combinatorial background BDT and the partially reconstructed background BDT for the electron and muon mode in all years 2016, 2017 and 2018 combined. A subset of the variables is used only in the electron mode; these are underlined. The $\chi^2_{\mathrm{DV}}$ is only used in the combinatorial background BDT and not in the partially reconstructed background BDT and is written in *italic*. $\ell^\pm$ represents the final-state leptons considered, $\ell^+ \ell^-$ corresponds to the di-lepton system. The variables $\min(V)$ and $\max(V)$ correspond to the lower and higher value of variable V between the two leptons, respectively.

| | |
|---|---|
| $B^+$ | $p_\mathrm{T}$, $\log \chi^2_\mathrm{IP}$, $\chi^2_\mathrm{DV}$, DIRA, $\chi^2_\mathrm{FD}$, ISO$_\mathrm{VTX}$ |
| $K^+$ | $p_\mathrm{T}$, $\log \chi^2_\mathrm{IP}$ |
| $\ell^\pm$ | $\underline{\min(p_\mathrm{T})}$, $\underline{\max(p_\mathrm{T})}$, $\min(\log \chi^2_\mathrm{IP})$, $\underline{\max(\log \chi^2_\mathrm{IP})}$ |
| $\ell^+ \ell^-$ | $p_\mathrm{T}$, $\underline{\log \chi^2_\mathrm{IP}}$, $\underline{\mathrm{N}_\mathrm{Bremsstrahlung\ photons}}$ |

The combinatorial background BDT is also used in another selection: When reconstructing tracks as described in Section 4.3.4, there are sometimes multiple possibilities that a PV can be assigned to a specific candidate. These different candidates per event, also called "multiple candidates", do not represent multiple decays, but *alternatives* to describe a specific decay. The combinatorial background BDT is used to decide, based on the highest score, which candidate is considered the most likely to describe the event and, therefore, chosen to be kept while the other alternatives are disregarded. The combinatorial background BDT was optimized with respect to the ROC AUC, which is for the electron channel 0.992 and for the muon channel 0.994, the ROC curve shown in Fig. 6.2.10

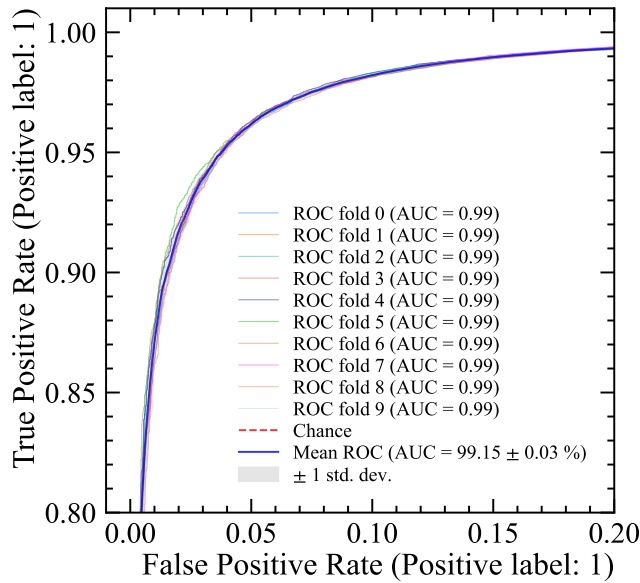

**Figure 6.2.10:** ROC curves for each of the folds separately for the combinatorial background BDT trained on the $B^+ \to K^+\mu^+\mu^-$ events of the data sample combined for the years 2016, 2017 and 2018 against $B^+ \to K^+\mu^+\mu^-$ simulation.

### 6.2.5.3 BDT selection

To determine the best selection requirement on the combinatorial background BDT for muons, the significance was chosen as a figure of merit (FoM) to be optimized. The significance is defined as

$$\mathcal{S}_{\mathcal{FoM}} = \frac{N_{sig}}{\sqrt{N_{sig} + B_{bkg}}}, \tag{6.9}$$

where $N_{sig}$ and $N_{bkg}$ are the expected number of signal and background events, respectively, in the signal region. The number of signal events is estimated from a ratio with the normalization mode, the relative efficiencies, and branching fraction. The number of background events in the signal region is estimated from an extended fit to the upper mass side band using the exponential as described in Section 6.5.4. As the muon mode is comparably clean and provides enough statistics, the approximation of the optimal value given Eq. 6.9 is considered good enough.

In contrast, the procedure for determining the selection requirement in both the combinatorial background BDT and partially reconstructed background BDT variables is different for electrons. The selection value was optimized by running toy studies on generated pseudoexperiments, which do not rely on asymptotic assumptions and require different selections on both BDT variables, using the sensitivity of $R_K$ as the target.

# 6.3 Corrections

The simulation of the LHCb detector as described in Section 4.6 is not perfect and shows disagreement with data, affecting the precision of the efficiencies. Variables that are not well modeled are mainly the PID distributions, the trigger efficiencies and the kinematics of the $B$-meson including some reconstruction quality variables. To correct for these discrepancies between data and simulation, different weights are derived from data samples and applied to simulated events. Each of the corrections is derived separately per year, the corrections for the 2018 year of data taking are used as illustration. First, Section 6.3.1 introduces the various PID calibrations. Trigger corrections are described in Section 6.3.2, followed by kinematic corrections in Section 6.3.3. Finally, the resolution of electrons in $m(K^+\ell^+\ell^-)$ and $q^2_{\mathrm{track}}$ is underestimated and corrected for by a smearing, as described in Section 6.3.4.

## 6.3.1 Particle identification corrections

The PID in LHCb involves information from multiple detector stations, including the RICH detectors, and is described in Section 4.4. Since the performance of the RICH detectors is not well simulated, the resulting PID variables are not precise enough to be used in the analysis. Instead, efficiency weights are derived using calibration samples from data as described in more detail later on. They include the efficiencies of the kaon, muon, and electron PID and kaon, muon isMuon efficiencies, which are all used in the signal selection as listed in Tab. 6.2.2. The efficiency of a misidentified pion as kaon is also needed to determine the background contribution from $B^+ \to \pi^+_{\to K^+} J\!/\psi(\mu^+\mu^-)$, as described in Section 6.1.4.5, allowing to constrain its yield in the fit to the control mode.

The calibration histograms are binned in year, the kinematic variables momentum and pseudorapidity of the particle and, apart from the electron PID, in polarity. The binning is chosen fine enough to not introduce any bias and then rebinned to avoid fluctuations by merging bins with similar enough efficiencies, within 2.5 $\sigma$ uncertainty. No binning is attempted in any occupancy proxy, where some dependency with the PID variables is expected, such as nTracks or nSPDHits, as these are notoriously badly modeled in simulation. The PID efficiencies are determined using a tag & probe method, whereby all particles in decay, apart from one, receive a tight PID selection requirement to have a clean sample, the tag particles, and the remaining particle may or may not pass a specific selection requirement, the probe particle. An event weight derived from this method is the product of all particles in it

$$w^{\mathrm{PID}} = \varepsilon^{\mathrm{PID}}_{K^+} \cdot \varepsilon^{\mathrm{PID}}_{\ell^-} \cdot \varepsilon^{\mathrm{PID}}_{\ell^+} \ , \tag{6.10}$$

where $\varepsilon^{\mathrm{PID}}_{K^+}$ and $\varepsilon^{\mathrm{PID}}_{\ell^\pm}$ describe the PID efficiency of the kaon and the leptons, respectively. These weights replace the selection requirement on PID variables for the simulation. The fits to the samples as well as the extraction of the efficiencies are done using the `PIDCalib` package [100] that is, together with the choice of the calibration samples, standard within LHCb.

In the following, the specific methods to obtain the actual numbers of passing particles and the calibration samples are described for each particle species. To ensure that the calibration histograms are meaningful when applied to the relevant channels, the same selection requirements, apart from specific exceptions explicitly mentioned, have been applied in the calibration samples as are applied in the offline selection listed in Tab. 6.2.2. An overview of the different calibration histograms is given in Tab. 6.3.1 with the corresponding figures.

**Table 6.3.1:** Overview of the different PID calibration histograms obtained with the `PIDCalib` package. The first row indicates the particle species for which calibration is performed, followed in the second row by the PID selection requirement for which the efficiency is calculated. Specific selection requirements are applied to make the calibration meaningful: each decay is required to have been within the detectors responsible for a specific PID variable. The term "$HLT1$ unbias" refers to TIS requirements applied to a set of HLT lines that prevent the efficiency of a kaon misidentified as a muon to be biased. For brevity, "cut" is used to denote a selection requirement.

| | **PID selection** | **Fiducial selection** | **Fig.** |
|---|---|---|---|
| | **For muon samples** | | |
| $K$ | isMuon $= 0$ | `hasRich` $= 1$ & `inMuonAcc` $= 1$ & $HLT1$ unbias & nSPDHits selection | 6.3.3 |
| | probNN$_K > 0.2$ | `hasRich` $= 1$ & isMuon $= 0$ & `inMuonAcc` $= 1$ & nSPDHits selection | 6.3.2 |
| $\mu$ | isMuon $= 1$ | `hasRich` $= 1$ & `inMuonAcc` $= 1$ & nSPDHits selection | 6.3.4 |
| | DLL$_\mu > 0$ | `hasRich` $= 1$ & `inMuonAcc` $= 1$ & isMuon $= 1$ & nSPDHits selection | 6.3.4 |
| $\pi$ | probNN$_K > 0.2$ | `hasRich` $= 1$ & isMuon $= 0$ & `inMuonAcc` $= 1$ & nSPDHits selection | 6.3.1 |
| | **For electron samples** | | |
| $K$ | probNN$_K > 0.2$ & DLL$_e < 0$ | `hasRich` $= 1$ & nSPDHits selection | 6.3.2 |
| $\pi$ | probNN$_K > 0.2$ & DLL$_e < 0$ | `hasRich` $= 1$ & nSPDHits selection | 6.3.1 |

#### 6.3.1.1 Kaon, muon and pion identification

Kaon PID and pion misidentification efficiencies are calibrated using $D^0 \to K^- \pi^+$ from the $D^{*+} \to D^0 (\to K^- \pi^+) \pi^-$ decay, where the latter contains a soft, bachelor[2] pion, making the determination of the charge, and therefore of $D^{*\pm}$, $D^0$ and $K^\pm$ unambiguous.

For the muon PID calibration, the decay $J/\psi \to \mu^+ \mu^-$ is used where only one muon had to pass the PID requirement, the other being used as a tag particle. Kaon, pion and muon PID variables are largely independent of the reconstructed mass distribution of the mother particle and allow for the *sPlot* technique to be used to extract *sWeight* [101]. This allows for a weighted sum of the passing events in each bin to retrieve the PID efficiency, denoted as $\varepsilon^{\text{PID}}$ and defined as

$$\varepsilon_i^{\text{PID}} = \frac{\sum_{pass_i} {}_s W}{\sum_{tot_i} {}_s W} \ ,$$ (6.11)

where ${}_s W$ is a *sWeight*, the index $i$ denotes the bin, $\sum_{pass_i}$ runs over all events in bin $i$ that satisfy the PID requirement and $\sum_{tot_i}$ runs on the same bin $i$ including all events. To avoid biasing the efficiencies, the isMuon requirement on the kaon and the muon has been dropped in the calibration sample preselection for the PID variables. For the misidentification of the pion, the requirement isMuon = 0 is kept, since the isMuon variable is well represented in simulation and its mismodeling has a minor impact, given that it is only used to constrain a background yield. The resulting PID efficiencies are shown for the kaon in Fig. 6.3.2, for the muon in Fig. 6.3.4 and for the pion in Fig. 6.3.1.

---

[2]Meaning that it comes from the $D^{*\pm}$.

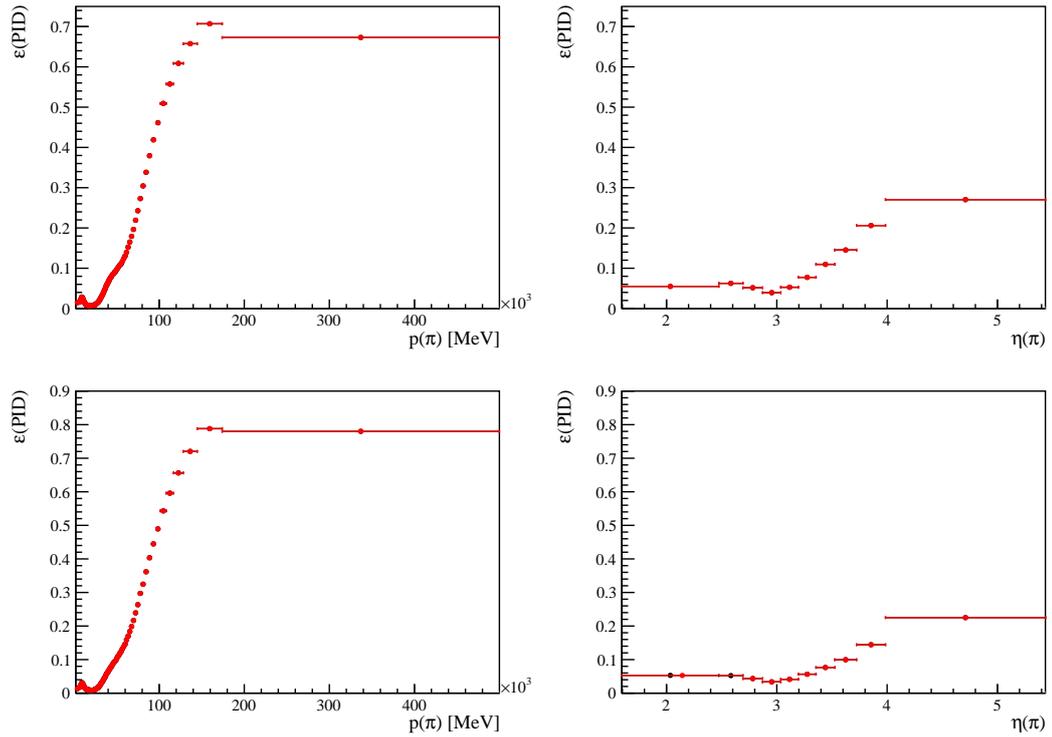

**Figure 6.3.1:** Pion misidentified as a kaon efficiencies, projected on the momentum (left) and pseudorapidity (right) axis. The top row shows the efficiencies for the electron mode, the bottom row for the muon mode. The black points show the efficiencies with the initial high number of bins, the red points after the rebinning scheme.

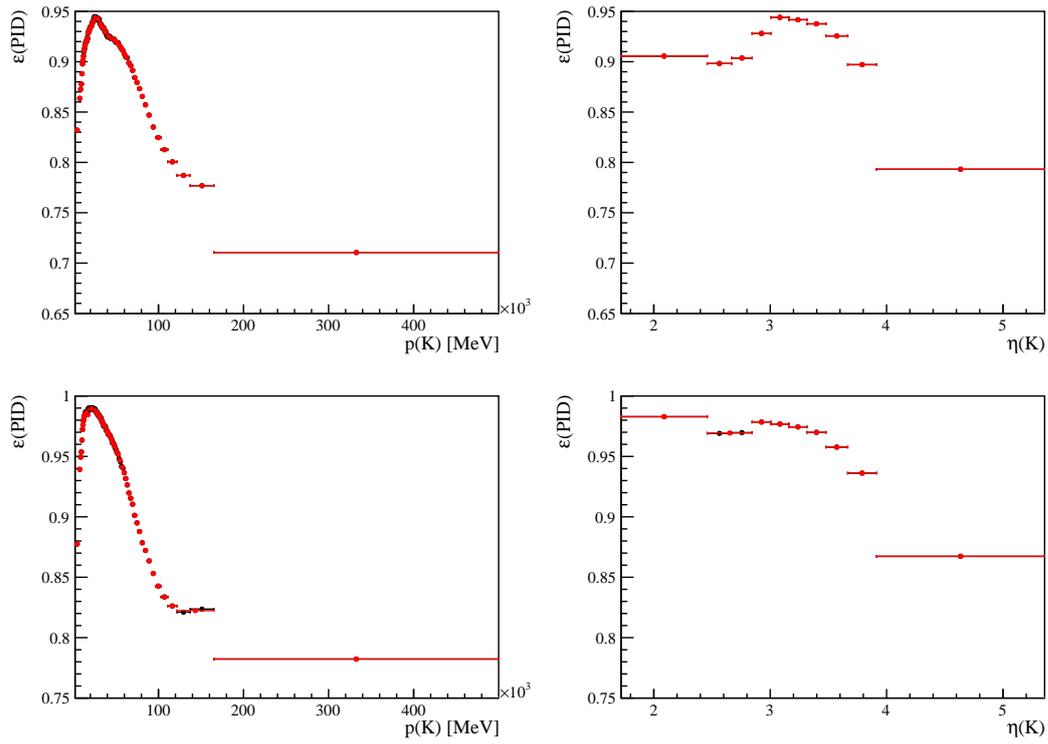

**Figure 6.3.2:** Calibrated PID efficiencies for kaons for the electron sample (top row) and muon sample (bottom row), projected on the momentum (left) and pseudorapidity (right) axis. The black points show the efficiencies with the initial high number of bins, the red points after the rebinning scheme.

### 6.3.1.2 Muon identification

The muons have an extra identification system, as introduced in Section 4.4.3, which relies on a set of additional tracking detectors. These return the isMuon variable, which is also corrected in the same manner as the PID variables before, although the correction is significantly smaller. It is significantly better modeled in the simulation, as it does not rely on information from the RICH detectors.

In the nominal selection, a isMuon = 1 requirement is applied to the muon and isMuon = 0 to the kaon, as listed in Tab. 6.2.2. To avoid biasing the PID efficiencies, separate samples from those used for the PID calibration in Section 6.3.1.1 are used. The resulting histograms are shown in Fig. 6.3.3 for the kaon and in Fig. 6.3.4 for the muon.

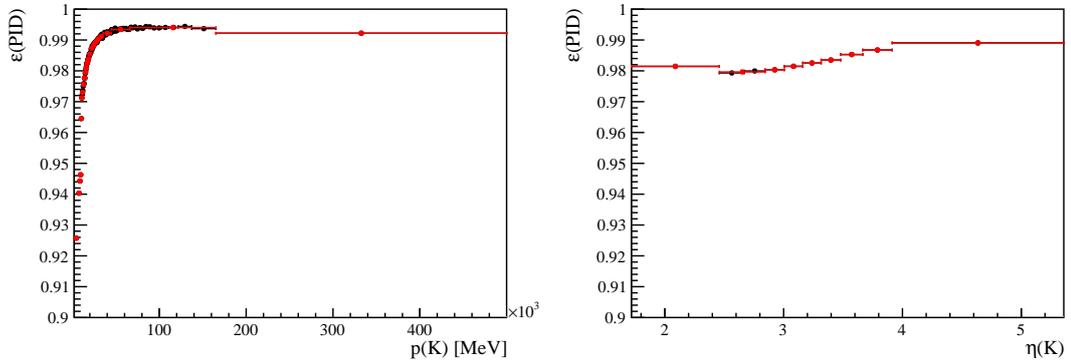

**Figure 6.3.3:** Calibration efficiencies for the isMuon variable on the kaon with the isMuon = 0 requirement in 2018 magnet-up data, projected on the momentum (left) and pseudorapidity (right) axis. The black points show the efficiencies with the initial high number of bins, the red points after the rebinning scheme.

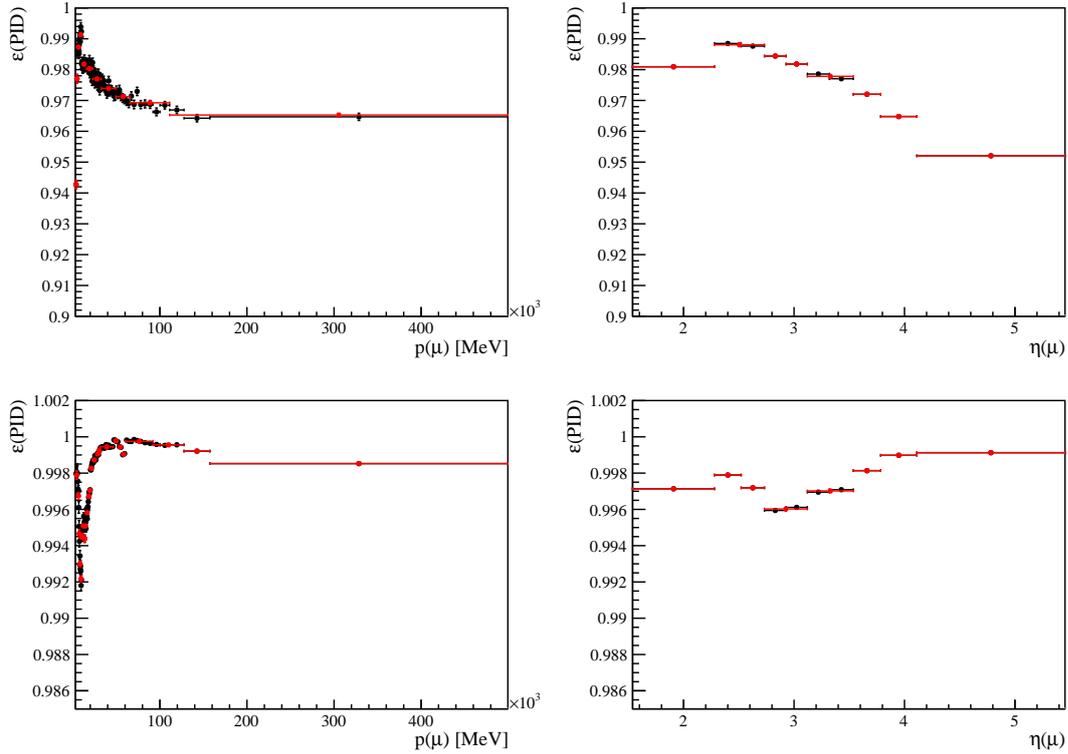

**Figure 6.3.4:** Calibration histograms for the muon in the isMuon (top) and the PID (bottom) variables, projected on the momentum (left) and pseudorapidity (right) axis. The black points show the efficiencies with the initial high number of bins, the red points after the rebinning scheme.

### 6.3.1.3 Electron identification

For the electron PID efficiencies, the normalization mode $B^+ \to K^+ J/\psi(e^+ e^-)$ is used, although in a different manner than for the other particles. The *sPlot* method from before in Eq. 6.11 cannot be applied, as the momentum, thereby also the PID, is correlated with the mass. This serves as the discriminating variable in the *sPlot* fit, thus violating the assumptions of the *sPlot* method. Even ignoring the previous correlation, a high background level in the electron mode makes it difficult to extract reliable *sWeight*s. To determine the number of passing electrons, a fit & count method is used and compared to the pure *sWeight* method, together with an alternative binning in Fig. 6.3.5.

Thus, the number of electrons remaining in a given bin for a given selection requirement is determined through an extended maximum likelihood fit to the $m(K^+\ell^+\ell^-)$. Each fit is performed in bins with merged magnet polarities and, additionally to the binning in kinematics and year, in two bins of the hasBrem variable that distinguishes between decays that radiated bremsstrahlung photons and those that did not. This extra binning in hasBrem is chosen because the electron PID variable $DLL_e$ uses hasBrem as a direct input. The PID weights are calculated as the ratio of the number of passing electrons to the total number of electrons in each bin, obtained as the yields from the fits as

$$\varepsilon_i^{\mathrm{PID}} = \frac{N_{pass_i}}{N_{tot_i}} \ , \qquad (6.12)$$

with $N_{pass_i}$ and $N_{tot_i}$ the yields of the passing and total electrons in bin $i$, respectively. The electron calibration histograms are shown in Fig. 6.3.5.

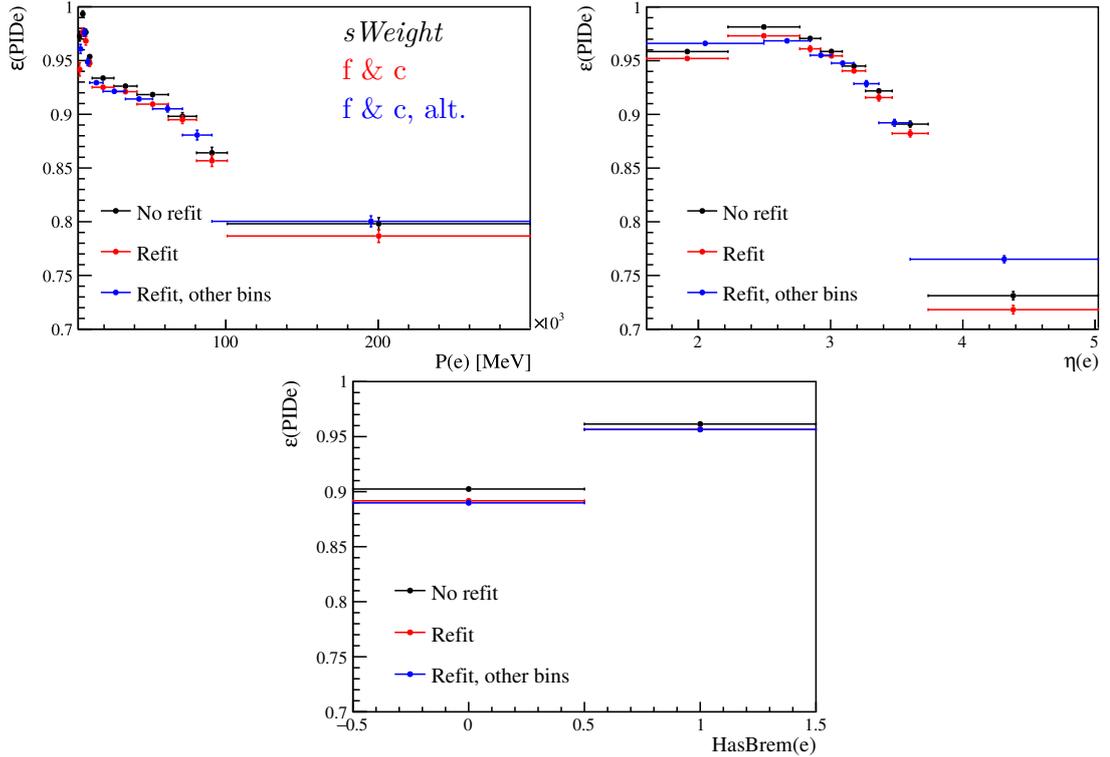

**Figure 6.3.5:** Calibration histogram for PID requirements on the electron, projected on the momentum (top left), pseudorapidity (top right) and hasBrem (bottom) axis. The black dots correspond to the efficiencies calculated with the *sWeight* method that is expected to produce biased results due to correlations, the red dots the efficiencies calculated with the fit & count (f & c) method (nominal) and the blue dots correspond to an alternative (alt.) binning scheme used to compute systematic uncertainties.

## 6.3.2 Trigger calibration

It is known that the L0 trigger variables, described in Section 4.5.1, are not perfectly modeled in the simulation. This is true for the four categories used in this analysis,

`L0Electron` and `L0Muon` for the nominal selection, as well as *h*TOS! and TIS!. Therefore, a correction is derived using the $B^+ \to K^+ J/\psi(e^+e^-)$ and $B^+ \to K^+ J/\psi(\mu^+\mu^-)$ data samples with the tag & probe method, in a similar fashion to the PID calibration, as described in Section 6.3.1. The entire preselection chain with PID selection requirements is applied to the samples and *sWeight*s are calculated. However, in contrast to the PID weights, the trigger weights are not directly obtained from the data, as the HLT selection is strongly correlated with the L0 response. Instead, trigger weights are computed relative to the simulation, as the ratio of the trigger efficiencies measured in data and simulation. Another complication arises as events that did *not* pass any L0 trigger are not stored in the data, requiring other L0 triggers to be used as a proxy sample.

The procedure to determine the corrections is similar for the electron and muon mode, in the following referred to as *lepton*:

- First, the efficiency of the `L0Lepton` requirement $\varepsilon^{\mathrm{L0}e}(\ell)$ is measured, both on simulation and data.

- A function is fit to each of these histograms as a function of transverse energy ($E_{\mathrm{T}}$) and $p_{\mathrm{T}}$ for electrons and muons, respectively, to parameterize the efficiency.

- The ratio of these two efficiencies is calculated and fitted with a function to obtain the trigger correction.

These efficiencies are measured in multiple categories:

**direct** The efficiency is measured on events that have the full offline preselection and PID requirements applied. This corresponds to the efficiency that should be measured by the definition of the L0 efficiency after the preselection. The data sample puts serious limitations on this method, as the collected events are only stored if they have passed both the L0 and HLT trigger stages and this efficiency is not accessible in data.

**TIS** The efficiency is measured on events that have passed the trigger using the TIS requirement. Since this sample is largely independent of the signal decay, it allows to measure the efficiency directly in data.

*K* **tag** This method uses a sample that is triggered on the `L0Hadron` line and is therefore also independent of the leptons in the signal decay. Compared to the TIS method, this sample is much smaller.

$\ell_{\mathrm{other}}$ The efficiency is measured on events that have the offline preselection and PID requirements applied, but the efficiency is calculated using the other lepton than the one responsible for the triggering of `L0Lepton`. This method has a drawback: The `L0Electron` uses the $E_{\mathrm{T}}$ of the electrons, which is measured by the ECAL with a low resolution that also depends on the different regions, as described in Section 4.4.2. The trigger variables for both leptons were checked for correlations and found to be correlated in the electron channel for a small lepton opening angle.

This most likely comes from the fact that the ECAL stores only the maximal energy received. Therefore, this method is not used for the electron channel.

The nominal trigger corrections are derived using the TIS sample. As this was selected slightly differently from the nominal analysis criteria based on $\ell$TOS, this was compared and found to introduce only a minor bias compared to the `direct` method.

The correction weights are calculated according to

$$w^{\ell \text{TOS}} = \frac{1 - (1 - \varepsilon_{\text{data}}^{\text{L0}\ell}(\ell^+)) \cdot (1 - \varepsilon_{\text{data}}^{\text{L0}\ell}(\ell^-))}{1 - (1 - \varepsilon_{\text{sim}}^{\text{L0}\ell}(\ell^+)) \cdot (1 - \varepsilon_{\text{sim}}^{\text{L0}\ell}(\ell^-))}. \tag{6.13}$$

This equation holds under the assumptions that the trigger efficiencies of both leptons factorize and are independent of each other. As mentioned above for the $\ell_{\text{other}}$ tag, this is not exactly true for electrons with a small opening angle.

### 6.3.2.1 $e$TOS **corrections**

The $e$TOS trigger line contains the presence of an electron with a high $E_{\text{T}}$ in the ECAL. The ECAL, as described in Section 4.4.2, consists of three different granularity regions, the inner, middle and outer region. Therefore, the $e$TOS trigger efficiency correction histograms are derived separately for each of these regions.

The corrections to the `L0Electron` efficiency are shown in Fig. 6.3.6. As can be seen, the agreement between data and simulation differs depending on the $E_{\text{T}}$ region. For high energies, they both agree well, approaching the threshold of the trigger fiducial selection requirement, and they start to diverge. This justifies, on the one hand, the need for the calibration procedure and, on the other hand, the use of the trigger fiducial selection requirement.

### 6.3.2.2 $\mu$TOS **calibration**

The `L0Muon` efficiency measured on simulated $B^+ \rightarrow K^+ J/\psi(\mu^+\mu^-)$ events is shown in Fig. 6.3.7 as a function of the transverse momentum. Similarly to the electron mode, the data and simulation efficiencies agree quite well for high transverse momenta but start to diverge for lower transverse momenta, strongly divergent for the lowest transverse momenta below the trigger fiducial selection requirement.

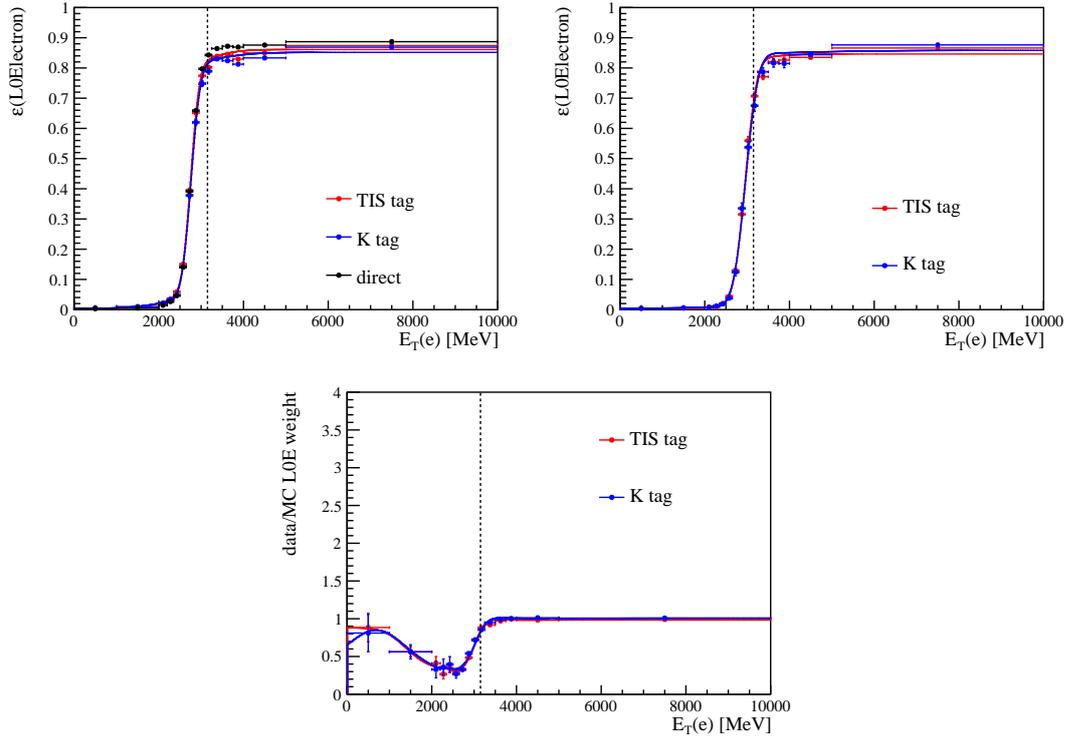

**Figure 6.3.6:** Calibration histograms of the trigger L0Electron in 2018 data taking conditions using $B^+ \to K^+ J/\psi(e^+e^-)$ events shown in the outermost ECAL region. The efficiency is calculated and fitted on simulation (left) and data (right). The ratio of data over simulation (bottom) is used as correction weights on the simulation. The solid line depicts the fit to the different histograms. The dashed, vertical line on the lower $E_T$ denotes the trigger fiducial selection requirement.

### 6.3.3 Kinematic corrections

The $B^+$ momentum and pseudorapidity spectra are imperfectly generated in the simulation. Furthermore, the reconstructed $\chi^2_{DV}$ and $\chi^2_{IP}(B^+)$ distributions are also poorly modeled. To correct for some of these reasonably well-known effects, corrections are derived and applied in the form of weights. There are two types of kinematic corrections, a first set, called `genlevel`, derived from the true variables to be applied before the trigger corrections described in Section 6.3.2, and a second one, called `reco`, using a set of reconstructed variables. The derived `reco` corrections are applied to all the samples used in the analysis, both to the electron mode and muon mode, except to generator level samples. These are only corrected with the first set of corrections, as they do not contain any reconstructed information. The kinematic weights are computed separately for each year of data taking.

The samples used to extract the correction weights are computed by comparing simulation and data distributions of $B^+ \to K^+ J/\psi(\mu^+\mu^-)$ events selected in the $\mu$TOS category. This particular sample is used because it has the largest number of events and the highest signal-to-background ratio. Moreover, it is the mode for which the PID and trigger efficiencies are best under control. Alternative weights for cross-checks have been derived using different triggers or the $B^+ \to K^+ J/\psi(e^+e^-)$ sample instead of the $B^+ \to K^+ J/\psi(\mu^+\mu^-)$ sample.

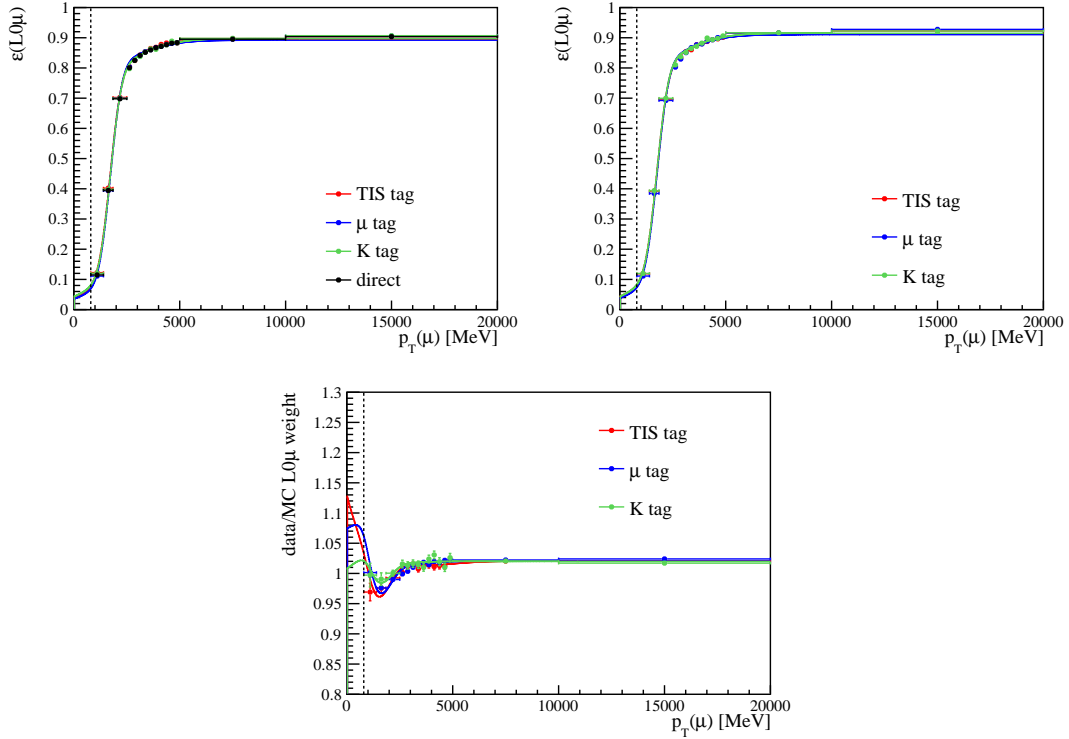

**Figure 6.3.7:** Calibration histograms of the trigger `L0Muon` in 2018 data taking conditions using $B^+ \to K^+ J/\psi(\mu^+\mu^-)$ events are shown. The efficiency is calculated and fitted on simulation (left) and data (right). The ratio of data over simulation (bottom) is used as correction weights on the simulation. The solid line depicts the fit to the different histograms. The dashed, vertical line on the lower $p_T$ denotes the trigger fiducial selection requirement.

The $K^+\mu^+\mu^-$ final state reconstruction is significantly better than the $K^+e^+e^-$ final state reconstruction, in fact, the reconstructed $K^+\mu^+\mu^-$ variables are very close to the $K^+\mu^+\mu^-$ true variables. In the $B^+ \to K^+ J/\psi(e^+e^-)$ sample, the *true* variables are closer to the *reconstructed* $B^+ \to K^+ J/\psi(\mu^+\mu^-)$ variables than to the reconstructed $B^+ \to K^+ J/\psi(e^+e^-)$ variables. Therefore, the weights are applied to the $B^+ \to K^+ \mu^+\mu^-$ sample using the variables from which the weights were derived from while for the $B^+ \to K^+ e^+ e^-$ sample, the weights are applied to true kinematic variables. Although the normalization channels contain only a small fraction of background events, in order to obtain a clean sample, *sWeight* are derived and applied, as for previous corrections of the PID in Section 6.3.1 and the trigger in Section 6.3.2.

The corrections correspond to data over simulation ratios of kinematic distributions that are estimated either via histograms or through a more elaborate, BDT based reweighting method called Gradient Boosted reweighter (GBReweighter) [102]. The two sets of weights are:

**genlevel**  A first set that corrects the two-dimensional ($p_{\mathrm{T}}(B^+); \eta(B^+)$) distribution using histograms as a density estimation to derive a correction ratio. This correction is used to compute the geometric and reconstruction efficiencies, $\varepsilon^{\mathrm{geom}}$ and $\varepsilon^{\mathrm{rec,strip}}$ as described in Section 6.4, and to improve the trigger corrections;

**reco**  A second set that corrects the four-dimensional $p_{\mathrm{T}}(B^+); \eta(B^+); \chi^2_{\mathrm{DV}}(B^+); \chi^2_{\mathrm{IP}}(B^+)$ distribution. The corrections are derived using the GBReweighter. These weights are used in the computation of the remaining efficiencies, $\varepsilon^{\mathrm{presel}}$, $\varepsilon^{\mathrm{PID}}$, $\varepsilon^{\mathrm{trig}}$ and $\varepsilon^{\mathrm{BDT}}$ as described in Section 6.4 as well as in fits to simulation as described in Section 6.5.

### 6.3.3.1 Generator level weights

For the reweighting procedure itself, a 2D histogram is used to obtain the weights. The simulation used is corrected with the previously derived PID and L0 corrections applied. A comparison of simulation and data before and after reweighting of the transverse momentum is shown in Fig. 6.3.8.

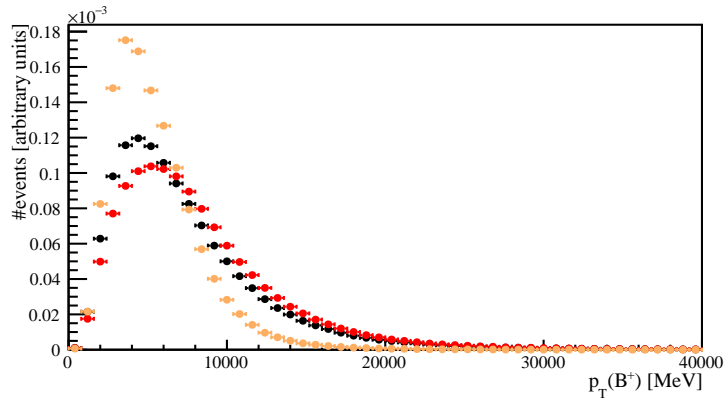

**Figure 6.3.8:** Distribution of the $B^+$ transverse momentum in the $B^+ \to K^+ J/\psi(\mu^+\mu^-)$ decay of the simulation without weights (yellow) and with generator level corrections (red) applied and the *sWeighted* data (black).

**6.3.3.2 Reconstruction variable corrections**

The weights are derived using a GBReweighter instead of histograms, since the former copes better with the curse of dimensionality. The GBReweighter method works as follows:

- First, a shallow DT with a depth of three is trained to separate the events of simulation and data to the best of its ability. This creates two bins, which have a maximal imbalance of both populations.

- Then, a weight is calculated from the remaining ratios in these two bins. The weight is chosen so that if applied to the simulation events, it would put the same number of events in both bins.

- The first step is repeated, aiming again at a maximal splitting of the events into the two categories of simulation and data, but this time with the newly obtained weights applied in addition to weights derived from previous rounds.

This closely follows the boosting technique described in Section 3.2.1. The one-dimensional projections of the distributions before and after the reweighting can be seen in Fig. 6.3.9 for the $B^+ \to K^+ J/\psi(e^+e^-)$ and in Fig. 6.3.10 for the $B^+ \to K^+ J/\psi(\mu^+\mu^-)$ decay.

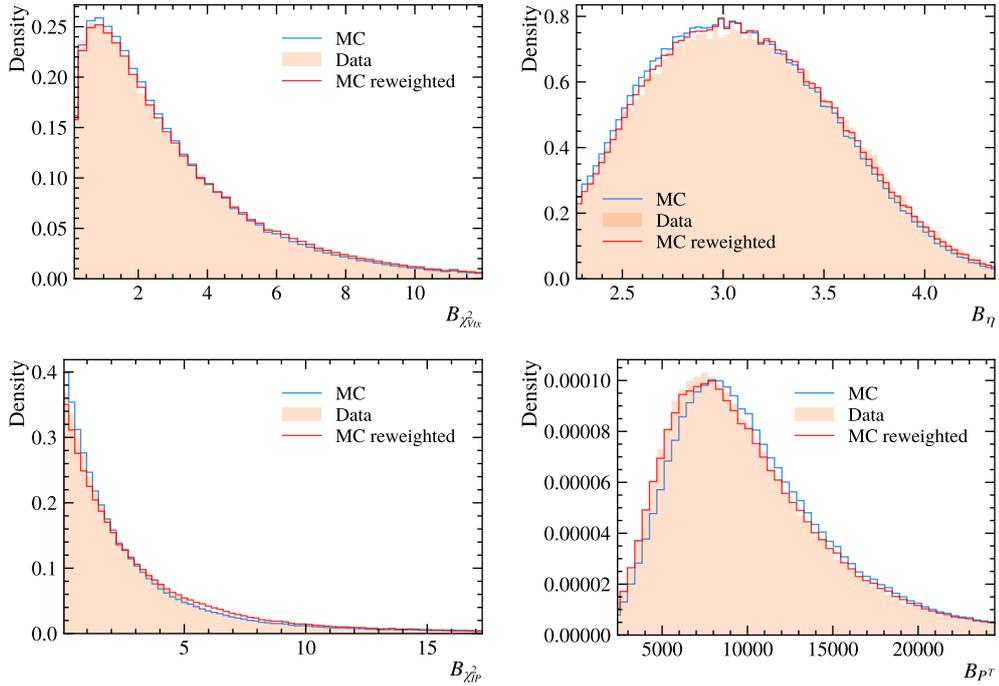

**Figure 6.3.9:** One-dimensional projections of the $B^+ \to K^+ J/\psi(e^+e^-)$ simulation (MC) distributions with (red) and without (blue) the kinematic corrections applied and the *sWeighted* data (orange).

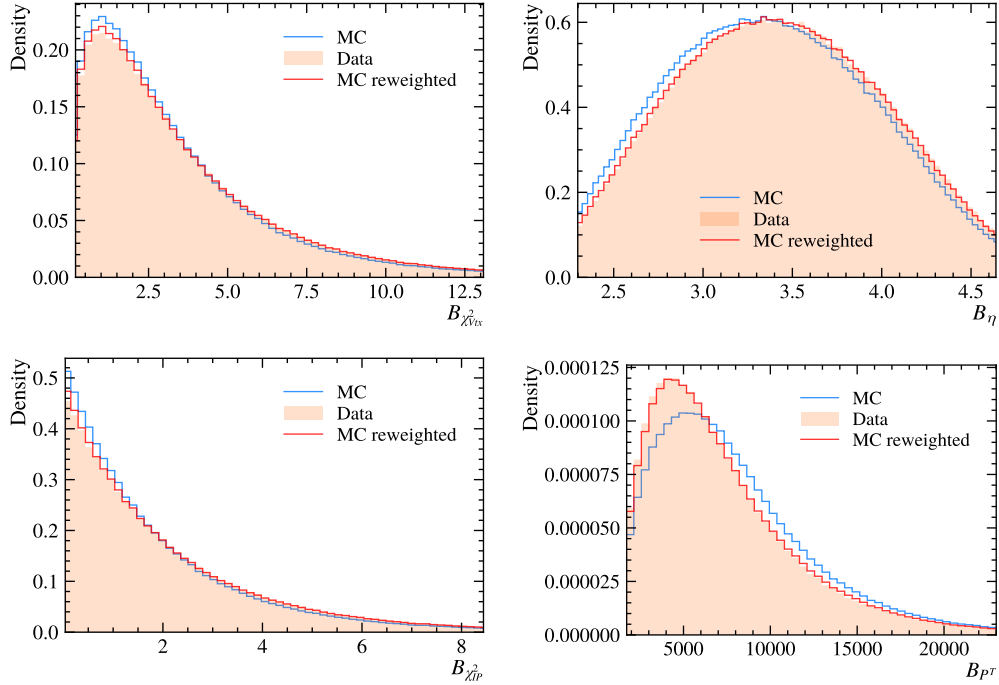

**Figure 6.3.10:** One dimensional projections of the $B^+ \to K^+ J/\psi(\mu^+\mu^-)$ simulation (MC) distributions with (red) and without (blue) the kinematic corrections applied and the *sWeighted* data (orange).

However, as with all ML algorithms, the power and flexibility come with the caveat that the response needs to be validated. Finding the right set of hyperparameters for the reweighter is not trivial [103], as the metric to quantify the success of the reweighting needs to be a distance measure between two n-dimensional distributions: This is even for one-dimensional distributions up to assumptions and there is simply no well-defined, reliable metric. To get an approximate measure of the agreement, one-dimensional projections can be used; however, they do not include all correlations and do not imply that the n-dimensional distributions are in good agreement.

This method was compared with a histogram reweighting procedure similar to the one used in Section 6.3.3.1 and showed minor improvements in the distribution comparison; this was expected as the GBReweighter should be able to better take into account correlations in multidimensional space without suffering from low statistics that a histogram approach would.

### 6.3.3.3 GBReweighter performance

In order to obtain a thorough performance estimate of the GBReweighter approach, a validation procedure using a BDT is applied, as BDTs excel at estimating the density of high-dimensional data. The validation is performed in two steps:

- The training dataset, consisting of $K^+J/\psi(\mu^+\mu^-)$ simulation and data as described previously in Section 6.3.3.2, are used to reweight the simulation sample. Therefore, a $K$-folding technique is used, as described in Section 3.2.3.2, with $K = 10$. This yields an unbiased set of weights.

- Once the $K^+J/\psi(\mu^+\mu^-)$ simulation is reweighted, a BDT is trained, also using $K$-folding, to distinguish between simulation with the newly obtained kinematic weights and data.

With approximately optimized hyperparameters, the response of the BDT serves as an indication for the similarity of the two data samples: If it performs well, the samples have little overlap; if it performs poorly, the samples are similar. For comparison of the BDT performance, a ROC curve is plotted and the AUC is calculated. The assumption is that a perfect splitting corresponds to a ROC AUC of 1, while a perfect overlap of the two distributions corresponds to a ROC AUC of 0.5, the same number that would also be obtained with a random selection. This is in general valid, but has a possible caveat.

With limited statistics, care has to be taken that the weights do not become too large, as this would bias the $K$-folding procedure. Large weights effectively reduce the available sample size. As an optimization on the lowest ROC AUC possible would incentivize the reweighting algorithm to place isolated, large weights, thereby biasing but also worsening the ROC AUC, this cannot be used without further checks. Therefore, with sufficient regularization and small enough weights, this can still be used as a proxy to quantify the agreement between two samples. The weights of both the muon and electron modes are compared with the weights produced using a histogram in Fig. 6.3.11.

This validation was performed in two different configurations. For the first one, only the variables already used for the reweighting were used in the training of the BDT. Compared to the weights derived using histograms, this showed no improvements. The other is the comparison when including variables that were not directly exposed to the reweighter. These variables are used in subsequent steps, such as in the training of different BDTs as described in Section 6.2, and showed a minor improvement, in line with expectations that the GBReweighter better picks up correlations, yet the difference overall was small.

### 6.3.4 Smearing

The last difference between the simulation and data samples that is corrected is the overestimation of the $q^2$ and $m(K^+\ell^+\ell^-)$ resolutions in the electron mode. The failure to correctly estimate the resolution mostly comes from two effects: First, any difference in the material budget between simulation and the real detector affects the scattering pattern - because of the low mass of the electrons, they suffer significantly more from this effect than other particles. Second, the resolution is affected by bremsstrahlung and the recovery procedure as previously discussed in Section 6.1.1.2.

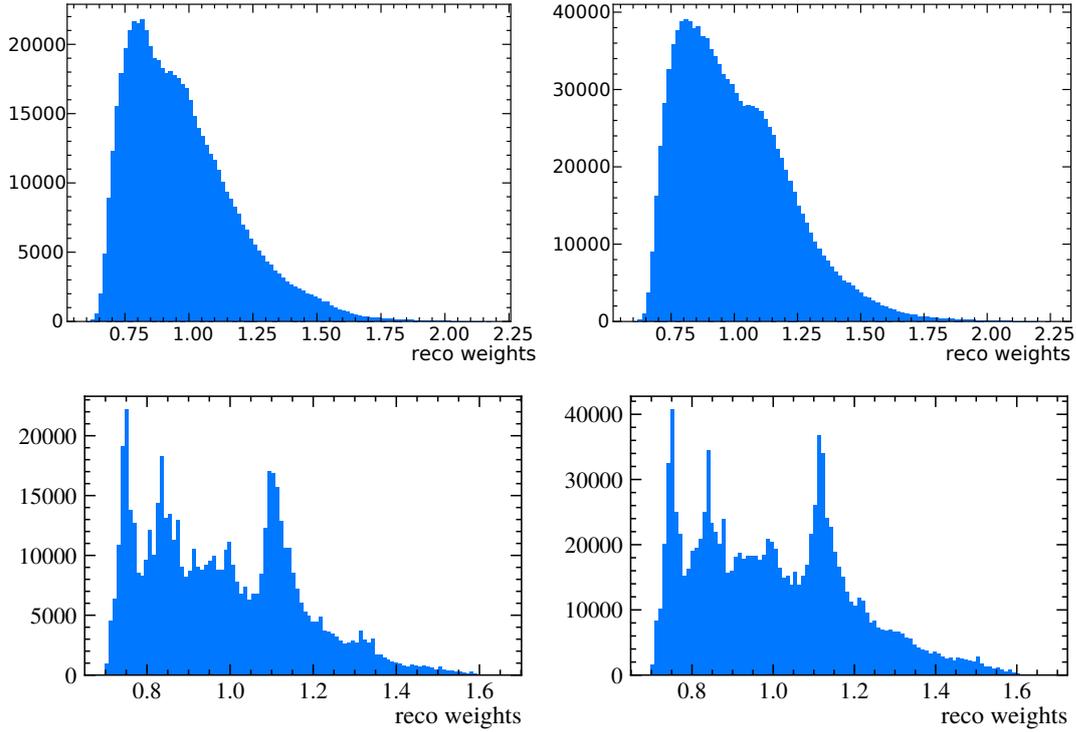

**Figure 6.3.11:** Comparison of the kinematic correction weights using histograms (top) and the GBReweighter (bottom) for the $B^+ \to K^+e^+e^-$ (left) and $B^+ \to K^+\mu^+\mu^-$ (right) samples. The shape of the distribution differs, yet the overall magnitude of the weights is comparable.

This affects primarily the calculation of the efficiency of the fit range in $m(K^+\ell^+\ell^-)$ and the $q^2_{\mathrm{track}}$ that is used in the $q^2$ selection. The basic idea of the correction is to randomly shift the values by a sampled value from a distribution encoding the increase resolution width that is determined from the difference of the data and simulation shapes. This procedure is also known as "smearing" of the variables. To obtain the smearing distribution, $B^+ \to K^+J/\psi(e^+e^-)$ simulation and data samples with the full preselection, including all BDTs and corrections, are used. Furthermore, a tighter lower fit range requirement of 5200 MeV/c² in $m_{\mathrm{DTF}}^{J/\psi}$ is applied to suppress the partially reconstructed background.

### 6.3.4.1 Mass resolution

The smearing of $m(K^+\ell^+\ell^-)$ is straightforward, as the shape is well behaved and can be described by a PDF consisting of a `Gaussian` with a mean and a width and an exponential tail to the left and one to the right, (DCB) for all bremsstrahlung categories. The mean and width have a shift and scale parameter, respectively, and are fitted first to simulation and then to data, following the same procedure and parameterization as described in Section 6.5.

Then, the variable is smeared using a `Gaussian` distribution with width and scale obtained in the fit, corrected by the shift and scale parameters. The resulting smeared $m(K^+\ell^+\ell^-)$ distributions are shown in Fig. 6.3.12.

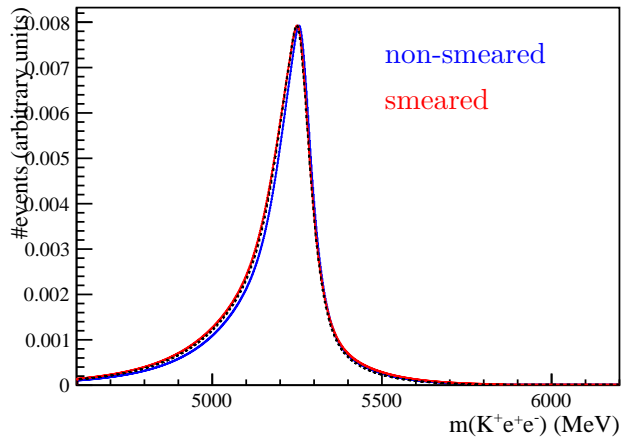

**Figure 6.3.12:** Fit in $m(K^+\ell^+\ell^-)$ to $B^+ \to K^+ J/\psi(e^+e^-)$ data (dotted line), simulation without smearing (blue) and simulation after smearing (red).

### 6.3.4.2 $q_{\text{track}}^2$ resolution

Smearing the $q_{\text{track}}^2$ variable poses a slightly larger challenge: Due to missing bremsstrahlung, the distributions in bremsstrahlung categories one and two inhibit a long tail to the lower $q_{\text{track}}^2$ values as can be seen in Fig. 6.3.13. To capture the shape of it, the fit model uses a log-normal function. The procedure to obtain the corrected variables is described below for the $q_{\text{track}}^2$ variable:

1. The samples are first divided into the different bremsstrahlung categories to perform the following steps independently for each category.

2. A fit is performed to the $q_{\text{track}}^2$ distribution in simulation. For bremsstrahlung category zero, a DCB is fitted, while for bremsstrahlung categories one and two a mixture of two Crystal Ball functions, a `Gaussian` distribution with an exponential tail, (CB) and a log-normal function are used as the model.

3. The fitted model is convoluted with a `Gaussian` distribution that is used to capture the effect of the underestimated resolution and is fit to the data sample. Thereby, the `Gaussian` width is the only parameter that is optimized, keeping the other parameters fixed to the values obtained from the fit to simulation.

4. This `Gaussian` is used to enhance the $q^2_{\mathrm{track}}$ variable in the simulation by adding a random value drawn from the `Gaussian` to each event. The resulting variable is then used to calculate the efficiencies in Section 6.4.

The resulting distributions are shown in Fig. 6.3.13. It can be seen that the agreement deteriorates with increasing bremsstrahlung category, posing, however, little concern over the procedure, as bremsstrahlung category two contributes less than 10% in the signal region.

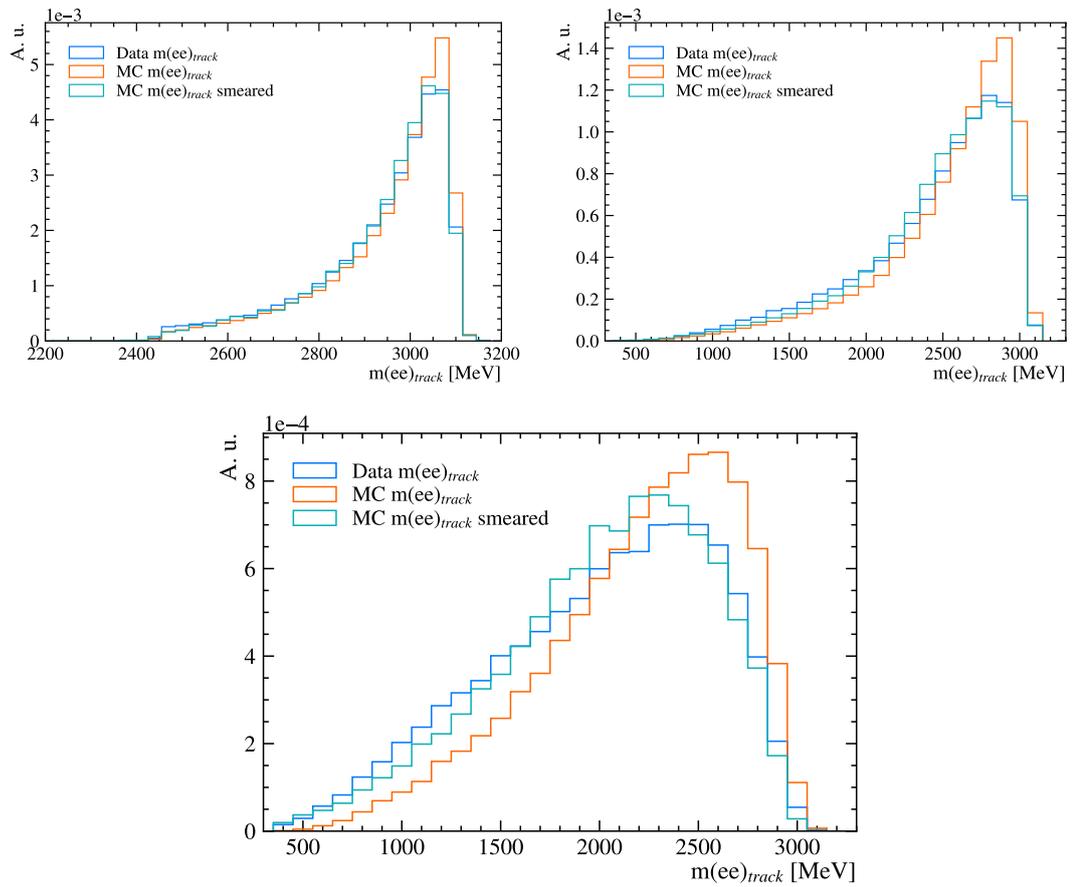

**Figure 6.3.13:** Smeared simulation for the bremsstrahlung category zero (top left), bremsstrahlung category one (top right) and bremsstrahlung category two (bottom)in $q_{\text{track}}^2$ ($m(ee)$). The original simulation (orange) and the smeared version (green) are displayed together with the data (blue).

## 6.4 Efficiencies

The efficiencies establish the relative number of events remaining in the data sample after every selection requirement. Each efficiency is defined as the ratio of the number of events that pass a given selection requirement to the number of events that pass the previous requirement. These ratios are obtained with simulation using the product of the frequency weights described in Section 6.3 as

$$\varepsilon^{sel} = \frac{\sum_i^{\text{sel}} w_i}{\sum_i^{\text{prev}} w_i} \,, \tag{6.14}$$

with $w_i$ the product of the corrections for the event $i$.

The total efficiency is the product of the individual efficiencies and is given by

$$\varepsilon^{\text{tot}} = \varepsilon^{\text{geom}} \cdot \varepsilon^{\text{rec,strip}} \cdot \varepsilon^{\text{presel}} \cdot \varepsilon^{\text{PID}} \cdot \varepsilon^{\text{trig}} \cdot \varepsilon^{\text{BDT}} \cdot \varepsilon^{\text{fitrange}} \tag{6.15}$$

The individual efficiencies are defined as follows:

$\varepsilon^{\text{geom}}$    denotes the efficiency of selecting events fully within the LHCb detector acceptance as described in Section 4.2;

$\varepsilon^{\text{rec,strip}}$    accounts for the tracking, vertex reconstruction and stripping requirements, including the global event selection requirement on nSPDHits, but omitting the stripping PID selection requirements;

$\varepsilon^{\text{presel}}$    denotes the efficiency of the preselection requirements as described in Section 6.2.3.1, yet omitting the PID selection requirements, and the $q^2$ selection requirement described in Section 6.2.4.3;

$\varepsilon^{\text{PID}}$    denotes the PID selection efficiencies of all particles in the final state as defined in Eq. 6.10;

$\varepsilon^{\text{trig}}$    denotes the trigger efficiency of L0 and HLT, as described in Section 6.2.2;

$\varepsilon^{\text{BDT}}$    is the selection efficiency of the BDT, as described in Section 6.2.5;

$\varepsilon^{\text{fitrange}}$    is the efficiency of retaining events in the fit range, as described in Section 6.2.4.3.

Using Eq. 6.14 and Eq. 6.15, the total efficiency can be expressed as

$$\varepsilon^{\text{tot}} = \underbrace{\frac{\sum_{rec} w_{\text{gen}}^{\text{kin}}}{\sum_{gen} w_{\text{gen}}^{\text{kin}}}}_{\varepsilon^{\text{geom}} \cdot \varepsilon^{\text{rec,strip}}} \cdot \underbrace{\frac{\sum_{sel} w_{\text{rec}}^{\text{kin}} \cdot w^{\text{PID}} \cdot w^{\text{trig}}}{\sum_{rec} w_{\text{rec}}^{\text{kin}}}}_{\varepsilon^{\text{presel}} \cdot \varepsilon^{\text{PID}} \cdot \varepsilon^{\text{trig}} \cdot \varepsilon^{\text{BDT}} \cdot \varepsilon^{\text{fitrange}}} \,, \tag{6.16}$$

Here, the subscripts denote a sum over specific events, where *gen* stands for generator level events, *rec* for reconstructed events and *sel* for fully selected events. Note that $w^{\text{PID}}$ and $w^{\text{trig}}$ are weights that emulate a selection requirement, therefore, only appearing on top of the fraction, while $w^{\text{kin}}$ denote genuine corrections to the overall distribution.

Individual efficiencies are summarized in Tab. 6.4.1 for the electron mode and in Tab. 6.4.2 for the muon mode under the 2018 data taking conditions; other years are listed in Appendix A.1. Together with the observed yields in the fits, the efficiencies are used to compute the efficiency corrected yields $\mathcal{N}$ as

$$\mathcal{N} = \frac{N^{\text{sel}}}{\varepsilon^{\text{tot}}} \ , \tag{6.17}$$

where $N^{\text{sel}}$ is the observed yield from the fit with the full selection applied. The efficiency-corrected yield for the rare mode needs to take into account the $q^2$ selection of the analysis and refers to the fully selected rare decay events in the range $q^2 > 14.3\,\text{GeV}^2/c^4$, given by

$$N^{\text{sel}}(B^+ \to K^+ \ell^+ \ell^-) = \mathcal{N}_{\text{all}}(B^+ \to K^+ \ell^+ \ell^-) \cdot \varepsilon^{\text{tot}}(B^+ \to K^+ \ell^+ \ell^-) \tag{6.18}$$

$$= \mathcal{N}_{\text{in}}(B^+ \to K^+ \ell^+ \ell^-) \cdot \frac{1}{f^{q^2}} \cdot \varepsilon^{\text{tot}}(B^+ \to K^+ \ell^+ \ell^-), \tag{6.19}$$

where $\mathcal{N}_{\text{all}}$ is the total number of events in the whole $q^2$ range, and $f^{q^2}$ is the fraction of $B^+ \to K^+ \ell^+ \ell^-$ events in the true range $q^2{}_{\text{true}} > 14.3\,\text{GeV}^2/c^4$. Thereby, $f^{q^2}$ is computed from simulated events at generator level as

$$f^{q^2} = \frac{\sum_{\text{gen},q^2{}_{\text{true}}>14.3} w^{\text{kin}}_{\text{gen}}}{\sum_{\text{gen},q^2{}_{\text{all}}} w^{\text{kin}}_{\text{gen}}} \ . \tag{6.20}$$

The efficiency corrected yields can be used to express the ratio in the signal mode and the normalisation mode as

$$\frac{\mathcal{B}(B^+ \to K^+ \ell^+ \ell^-)_{\text{in}}}{\mathcal{B}(B^+ \to K^+ J/\psi(\ell^+ \ell^-))} = \frac{\mathcal{B}(B^+ \to K^+ \ell^+ \ell^-)_{\text{in}}}{\mathcal{B}(B^+ \to K^+ J/\psi(\ell^+ \ell^-))}$$

$$= \frac{\mathcal{N}_{\text{in}}(B^+ \to K^+ \ell^+ \ell^-)}{\mathcal{N}(B^+ \to K^+ J/\psi(\ell^+ \ell^-))}$$

$$= \frac{\mathcal{N}_{\text{all}}(B^+ \to K^+ \ell^+ \ell^-)}{\mathcal{N}(B^+ \to K^+ J/\psi(\ell^+ \ell^-))} \cdot f^{q^2}$$

$$= \frac{N^{\text{sel}}(B^+ \to K^+ \ell^+ \ell^-)}{N^{\text{sel}}(B^+ \to K^+ J/\psi(\ell^+ \ell^-))} \cdot \frac{\varepsilon^{\text{tot}}(B^+ \to K^+ J/\psi(\ell^+ \ell^-))}{\varepsilon^{\text{tot}}(B^+ \to K^+ \ell^+ \ell^-)} \cdot f^{q^2}$$

The efficiencies in Tab. 6.4.1 and Tab. 6.4.1 are obtained using the definition in Eq. 6.14, either with all corrections derived in Section 6.3 applied or without corrections at all. Ghost events are only included at the BDT and fit range levels and remain at the 2% level.

**Table 6.4.1:** Efficiencies for $B^+ \to K^+ J/\psi(e^+e^-)$ and $B^+ \to K^+ e^+ e^-$ decays, as well as efficiency ratios $\varepsilon(B^+ \to K^+ e^+ e^-)/\varepsilon(B^+ \to K^+ J/\psi(e^+e^-))$. All efficiencies and ratios are displayed in percent, and the uncertainty quoted is the statistical uncertainty related to the finite size of the simulation samples. The efficiencies are computed either without corrections to the simulation or with all nominal correction weights.

| | electron modes | | |
|---|---|---|---|
| $\varepsilon[\%]$ | $J/\psi$ | rare | ratio |
| | **No corrections** | | |
| $\varepsilon^{\text{geom}}$ | $17.31 \pm 0.03$ | $17.58 \pm 0.03$ | $101.59 \pm 0.25$ |
| $\varepsilon^{\text{rec,strip}}$ | $14.32 \pm 0.01$ | $13.74 \pm 0.02$ | $95.93 \pm 0.18$ |
| $\varepsilon^{\text{presel}}$ | $75.51 \pm 0.04$ | $3.88 \pm 0.04$ | $5.13 \pm 0.05$ |
| $\varepsilon^{\text{PID}}$ | $77.49 \pm 0.04$ | $74.0 \pm 0.42$ | $95.5 \pm 0.55$ |
| $\varepsilon^{\text{trig}}$ | $45.41 \pm 0.05$ | $85.18 \pm 0.4$ | $187.56 \pm 0.91$ |
| $\varepsilon^{\text{BDT}}$ | $103.27 \pm 0.03$ | $100.19 \pm 0.05$ | $98.9 \pm 0.06$ |
| $\varepsilon^{\text{fitrange}}$ | $101.31 \pm 0.02$ | $104.51 \pm 0.26$ | $101.2 \pm 0.26$ |
| $\varepsilon^{\text{BDT}} \& \varepsilon^{\text{fitrange}}$ | $100.35 \pm 0.01$ | $100.16 \pm 0.05$ | $99.81 \pm 0.05$ |
| $\varepsilon^{\text{tot}}$ | $0.66 \pm 0.0$ | $0.06 \pm 0.0$ | $8.94 \pm 0.11$ |
| | **PID, trigger and kinematic corrections (nominal)** | | |
| $\varepsilon^{\text{geom}}$ | $17.66 \pm 0.03$ | $17.94 \pm 0.03$ | $101.58 \pm 0.25$ |
| $\varepsilon^{\text{rec,strip}}$ | $13.87 \pm 0.01$ | $13.27 \pm 0.02$ | $95.72 \pm 0.19$ |
| $\varepsilon^{\text{presel}}$ | $73.44 \pm 0.04$ | $3.82 \pm 0.04$ | $5.2 \pm 0.05$ |
| $\varepsilon^{\text{PID}}$ | $68.01 \pm 0.05$ | $66.09 \pm 0.53$ | $97.17 \pm 0.78$ |
| $\varepsilon^{\text{trig}}$ | $31.09 \pm 0.05$ | $56.94 \pm 0.51$ | $183.14 \pm 1.67$ |
| $\varepsilon^{\text{BDT}}$ | $103.14 \pm 0.03$ | $99.2 \pm 0.17$ | $98.61 \pm 0.17$ |
| $\varepsilon^{\text{fitrange}}$ | $100.6 \pm 0.01$ | $104.53 \pm 0.28$ | $101.35 \pm 0.28$ |
| $\varepsilon^{\text{BDT}} \& \varepsilon^{\text{fitrange}}$ | $99.69 \pm 0.02$ | $99.18 \pm 0.17$ | $99.48 \pm 0.17$ |
| $\varepsilon^{\text{tot}}$ | $0.38 \pm 0.0$ | $0.03 \pm 0.0$ | $8.95 \pm 0.12$ |

**Table 6.4.2:** Efficiencies for $B^+ \to K^+ J/\psi(\mu^+\mu^-)$ and $B^+ \to K^+\mu^+\mu^-$ decays, as well as efficiency ratios $\varepsilon(B^+ \to K^+\mu^+\mu^-)/\varepsilon(B^+ \to K^+ J/\psi(\mu^+\mu^-))$. All efficiencies and ratios are displayed in percent, and the uncertainty quoted is the statistical uncertainty related to the finite size of the simulation samples. The efficiencies are computed either without corrections to the simulation or with all nominal correction weights.

| $\varepsilon[\%]$ | muon modes | | |
| --- | --- | --- | --- |
| | $J/\psi$ | rare | ratio |
| | No corrections | | |
| $\varepsilon^{\text{geom}}$ | $17.4 \pm 0.03$ | $17.68 \pm 0.03$ | $101.66 \pm 0.25$ |
| $\varepsilon^{\text{rec,strip}}$ | $27.68 \pm 0.01$ | $26.93 \pm 0.03$ | $97.27 \pm 0.12$ |
| $\varepsilon^{\text{presel}}$ | $52.38 \pm 0.03$ | $13.44 \pm 0.05$ | $25.65 \pm 0.09$ |
| $\varepsilon^{\text{PID}}$ | $93.9 \pm 0.02$ | $93.16 \pm 0.09$ | $99.21 \pm 0.1$ |
| $\varepsilon^{\text{trig}}$ | $82.82 \pm 0.03$ | $87.88 \pm 0.13$ | $106.1 \pm 0.16$ |
| $\varepsilon^{\text{BDT}}$ | $101.31 \pm 0.01$ | $100.52 \pm 0.03$ | $99.94 \pm 0.03$ |
| $\varepsilon^{\text{fitrange}}$ | $100.58 \pm 0.01$ | $100.1 \pm 0.01$ | $98.81 \pm 0.02$ |
| $\varepsilon^{\text{BDT}} \&\varepsilon^{\text{fitrange}}$ | $99.88 \pm 0.0$ | $97.59 \pm 0.06$ | $97.7 \pm 0.06$ |
| $\varepsilon^{\text{tot}}$ | $1.96 \pm 0.0$ | $0.51 \pm 0.0$ | $26.09 \pm 0.13$ |
| | PID, trigger and kinematic corrections (nominal) | | |
| $\varepsilon^{\text{geom}}$ | $17.75 \pm 0.03$ | $18.04 \pm 0.03$ | $101.65 \pm 0.25$ |
| $\varepsilon^{\text{rec,strip}}$ | $27.08 \pm 0.01$ | $26.31 \pm 0.03$ | $97.14 \pm 0.13$ |
| $\varepsilon^{\text{presel}}$ | $51.92 \pm 0.03$ | $13.35 \pm 0.05$ | $25.71 \pm 0.09$ |
| $\varepsilon^{\text{PID}}$ | $91.08 \pm 0.03$ | $89.79 \pm 0.15$ | $98.58 \pm 0.17$ |
| $\varepsilon^{\text{trig}}$ | $75.33 \pm 0.04$ | $80.26 \pm 0.16$ | $106.54 \pm 0.21$ |
| $\varepsilon^{\text{BDT}}$ | $101.24 \pm 0.01$ | $99.66 \pm 0.04$ | $99.53 \pm 0.04$ |
| $\varepsilon^{\text{fitrange}}$ | $100.13 \pm 0.01$ | $99.52 \pm 0.03$ | $98.3 \pm 0.03$ |
| $\varepsilon^{\text{BDT}} \&\varepsilon^{\text{fitrange}}$ | $99.43 \pm 0.01$ | $96.83 \pm 0.08$ | $97.38 \pm 0.08$ |
| $\varepsilon^{\text{tot}}$ | $1.7 \pm 0.0$ | $0.44 \pm 0.0$ | $25.97 \pm 0.13$ |

## 6.5 Mass model

The yields of the different channels of interest are determined by unbinned, extended maximum likelihood fits to the $m(K^+\ell^+\ell^-)$ distributions of $K^+e^+e^-$ and $K^+\mu^+\mu^-$ final states. Different background components are added in a mixture to reflect the actual population in the sample. The fits and mass modeling are performed mainly using the ZFIT package [98] that is described in Chapter 7 and cross-checked in some fits with the ROOFIT package [88]. In order to nominally obtain the yields of the resonant modes $B^+ \to K^+ J/\psi(\ell^+\ell^-)$ and $B^+ \to K^+\psi(2S)(\ell^+\ell^-)$, a fit is performed to the masses $m_{\mathrm{DTF}}{}^{J/\psi}(K^+\ell^+\ell^-)$ and $m_{\mathrm{DTF}}{}^{\psi(2S)}(K^+\ell^+\ell^-)$ where DTF, *i.e.,* the decay tree fit, is performed to the constrained nominal $B$ mass. Hereby, the $B$ mass is calculated while the di-lepton mass is required to be equal to that of the $J/\psi$ mass and $\psi(2S)$ mass respectively. While the model follows closely that of the previous analysis [104] in the central $q^2$ region, the fit strategy in the $B^+ \to K^+ J/\psi(e^+e^-)$ channel has been adapted to improve the overall fit quality, resulting in a change in signal yield of around one sigma. Both fit strategies are retained however, as the previous strategy offers better insight into the behavior of different bremsstrahlung categories, while the new one is more robust and allows for a better fit-quality.

In the signal and background modeling, simulation samples are used in most channels to infer the shape, often parameters, of the distributions, "pre-fitting" them. These fits are performed using an unbinned weighted maximum likelihood to determine the best fit of the parameters to simulation, a procedure that is validated in Section 6.5.1.5. The uncertainties displayed are calculated using the asymptotically correct method for weighted likelihoods based on a Hessian approximation as described in Ref. [105] and as implemented in ZFIT. The weights used in the simulation samples are described in Section 6.3.

Although the simulation samples are corrected for most differences between data and simulation, there is still a known discrepancy in the mass resolution that is effectively manifest in a shift of the mean and a scaling of the width. To take this effect into account, two additional parameters are introduced, which effectively shift or scale the mean or width of the different models between the simulation and data models, respectively. Formalized, a parameter $P_i$ annotated with *shifted* or *scaled* is understood as a parameter that has been shifted or scaled by the parameter $P_{shift}$ or $P_{scale}$, respectively. The definitions are

$$P_i^{shifted} \equiv P_i + P_{shift}\,, \tag{6.21}$$

$$P_i^{scaled} \equiv P_i \cdot P_{scale}\,, \tag{6.22}$$

where $P_{shift}$ and $P_{scale}$ are fixed to 0 and 1, respectively, for the fits to simulation and are left free to be optimized in the minimization in the fits to data. The parameter $P_i$ that was allowed to vary during the fit to simulation is later fixed in the fit to data to the value obtained from the fit to simulation. Both parameters, $P_{shift}$ and $P_{scale}$, are shared between all means and widths of different models in the fit, as the discrepancy in the resolution is expected to be the same for all components.

To extract $R_K$ as defined in Eq. 6.1.3, fits are performed to the $m(K^+\ell^+\ell^-)$ distribution of the $K^+e^+e^-$ and $K^+\mu^+\mu^-$ final states both in the rare mode and in the normalization mode. In the following, the fit models are described, starting with the normalization channels $B^+ \to K^+ J/\psi(e^+e^-)$ in Section 6.5.1 and $B^+ \to K^+ J/\psi(\mu^+\mu^-)$ in Section 6.5.2. Next, the dependence of $R_K$ on $q^2$ and a strategy to prevent effects from possible mismodeling in the rare mode simulation is described in Section 6.5.3. Then, the combinatorial background model of the rare mode is covered in Section 6.5.4. Finally, the entire model for $B^+ \to K^+e^+e^-$ is described in Section 6.5.5, followed by the model of $B^+ \to K^+\mu^+\mu^-$ in Section 6.5.6.

### 6.5.1 $B^+ \to K^+ J/\psi(e^+e^-)$ **fit model**

In this section, the components of the fit model for the $B^+ \to K^+ J/\psi(e^+e^-)$ channel are described and derived. The fits to data and the results are shown in Section 6.6.1.

#### 6.5.1.1 Signal shape

The signal model is composed of a mixture of PDFs with individual shapes for each of the bremsstrahlung categories. The model for each category consists of a DCB. All three shapes have shared mean and width parameters. The parameters of these shapes are fixed from a weighted fit, as shown in Fig. 6.5.1, to the simulated $B^+ \to K^+ J/\psi(e^+e^-)$ sample.

These distributions are then combined in a mixture, where the fraction of each distribution is constrained to the relative fraction found in the simulation, *i.e.*, the value obtained from the weighted number of events found in each bremsstrahlung category. Additionally, a `Gaussian` distribution is added to the mixture of the three distributions, which contributes the same fraction in each distribution to the bremsstrahlung category model. The addition of the `Gaussian` distribution is new to this analysis with respect to the central $q^2$ region analysis and was introduced to improve the fit quality and the pulls. The final model is given by

$$f_{signal} = \sum_{i=1}^{3} frac_i \cdot (f_{brem}^i(\theta_i, \mu_{signal}^{shifted}, \sigma_{signal}^{scaled}) + f_{Gauss}(\mu_{signal}^{shifted}, \sigma_{gauss}^{scaled})) \qquad (6.23)$$

As can be seen from the pulls of the individual components in Fig. 6.5.1 of the different bremsstrahlung categories, the model in Eq. 6.23 does not fully capture the simulation distribution, notably with biases in the same place for the three categories. Summing up the three categories, it is not surprising to find that the full model does not capture the merged components, *i.e.*, the full signal $B^+ \rightarrow K^+ J/\psi(e^+e^-)$ simulation sample, as shown in Fig. 6.5.2.

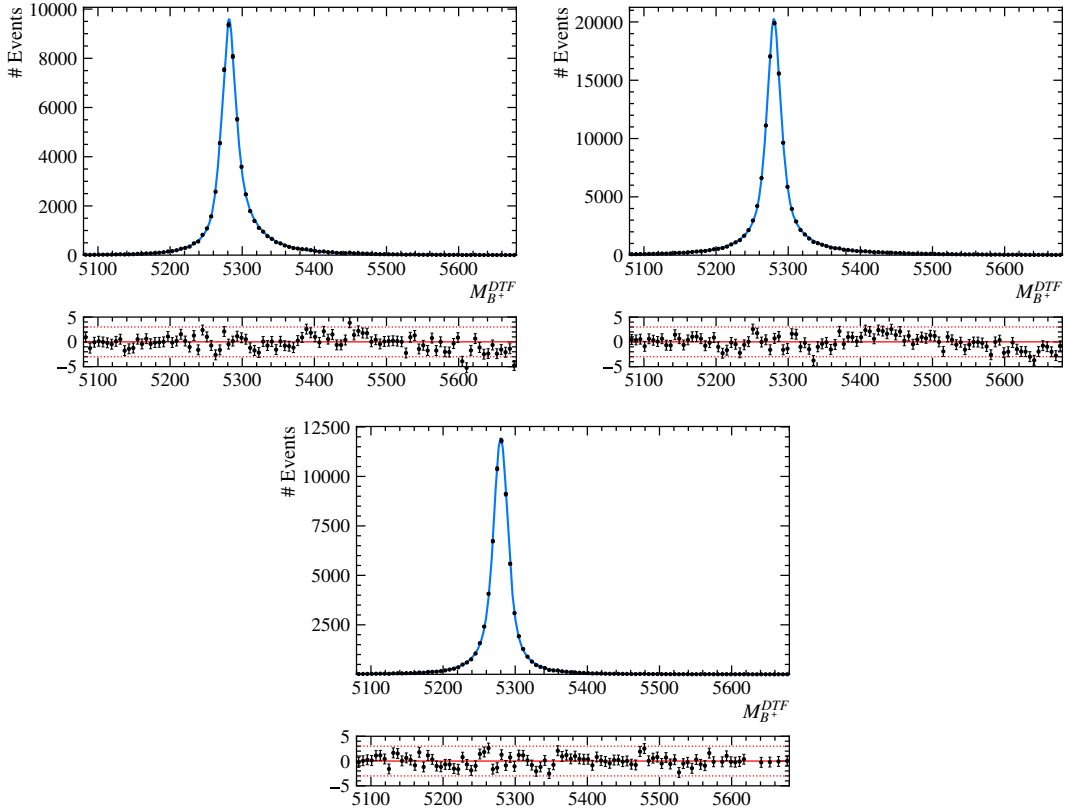

**Figure 6.5.1:** Fit to $B^+ \rightarrow K^+ J/\psi(e^+e^-)$ simulation, split in the different bremsstrahlung categories zero (top left), one (top right) and two (bottom). The data points are in black and the PDF is drawn in blue.

Therefore, an improved model was derived and is nominally used. The improved model consists of the same basic shapes, but differs in the values of the parameters. The entire model, the sum of all brem contributions, the three DCB and the `Gaussian`, is fitted in one model to the entire simulation sample, without splitting in bremsstrahlung categories. This results in a more flexible model that is able to capture the full signal shape significantly better, as can be seen in Fig. 6.5.2. The downside of this approach is that the components of the signal lose their meaning of representing the different bremsstrahlung categories, which is of no concern as the shapes are, in general, an ad-hoc choice to model the signal.

Such an approach, using a more flexible model, could introduce a bias: When parameters are fixed from simulation, the assumption is that the uncertainty on them is small enough to fix them directly, as explained in Section 3.3. To ensure that this is a valid assumption and not just overfitting the sample, a simultaneous fit in data and simulation was performed, without fixing any parameters, and the two approaches were compared in Section 6.5.1.5. This guarantees that the whole likelihood is taken into account and resulted in uncertainties reduced by about one-third using the new approach; the signal yield shifted by about one sigma. Although this is not significant for the extraction of $R_K$, it is important to note that the improved shape is overall more robust and allows for better fit quality.

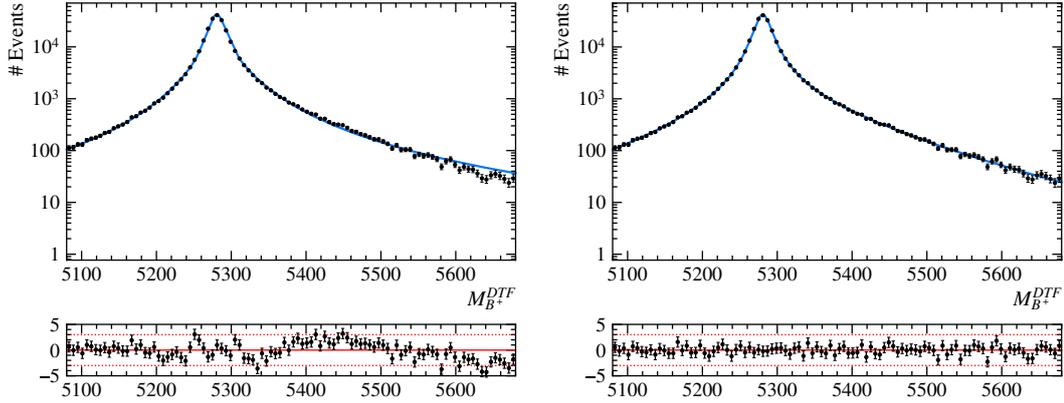

**Figure 6.5.2:** Comparison of the fit methods, where both plots show a fit to $B^+ \to K^+ J/\psi(e^+e^-)$ signal, either with pre-fitted shapes from individual bremsstrahlung categories(left) or with an additionally fit of the sum to the entire simulation sample (right). The data points are in black and the PDF is drawn in blue.

### 6.5.1.2 Partially reconstructed background model

Due to the worse resolution for electrons, partially reconstructed background as explained in Section 6.1.4.1 is leaking into the signal mass window. The partially reconstructed background component comes predominantly from $B^0 \to K^{*0} J/\psi(e^+e^-)$ decays and is significant enough to be modeled as a background in the fit. The shape of the contribution is obtained using a template, namely a kernel density estimation (KDE) function that uses the improved Sheather-Jones (ISJ) algorithm, determined from $B \to J/\psi(e^+e^-)X$ simulation samples. The components are divided into two categories, charm and strange, as explained in Section 6.1.4.1, and enter as two separate templates in the fit. Both components are shown in Fig. 6.5.3.

The fraction of strange to charm partially reconstructed events is `Gaussian` constrained to the weighted fraction of events observed in simulation. It is to note that the uncertainty on the fraction is negligible in the determination of $R_K$: A stronger constraint would not improve the precision of the control mode yield by any notable amount, which is a quantity that in any case does not limit the precision of the $R_K$ measurement.

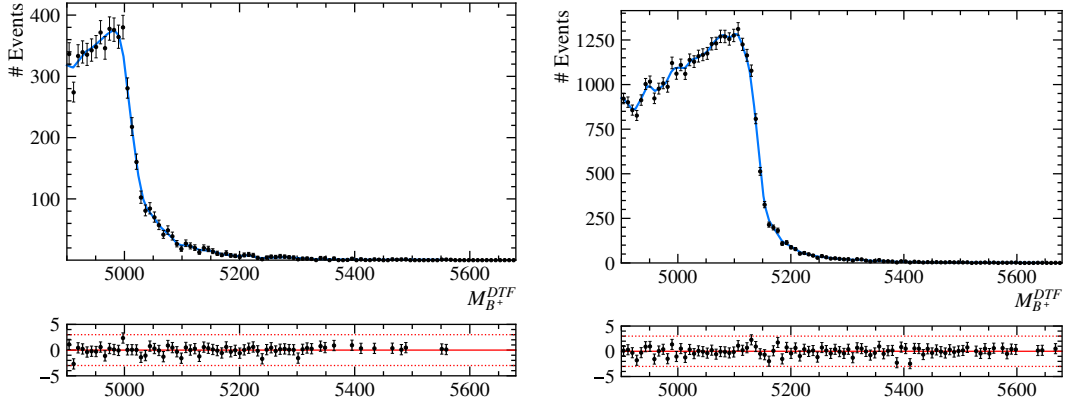

**Figure 6.5.3:** Partially reconstructed $B \to J/\psi(e^+e^-)X$ for the charm (left) and strange (right) background. The data points are in black and the PDF is drawn in blue. The fit range extends further on the left to avoid any boundary effects that can appear in a KDE.

### 6.5.1.3 Single misidentified background

Another source of background is $B^+ \to \pi^+ J/\psi(e^+e^-)$ events, where the pion is wrongly identified as a kaon, as explained in Section 6.1.4.5. As it also falls into the fitting range of the electron mode, it is modeled as a background using a DCB. A dedicated simulated sample is used, and a fit to the mass distribution is performed to fix the tail parameters of the distribution as shown in Fig. 6.5.4. As for the signal component, it has a scale and width parameter that is shifted and scaled by the common shift and scale parameters as described in Eq. 6.22.

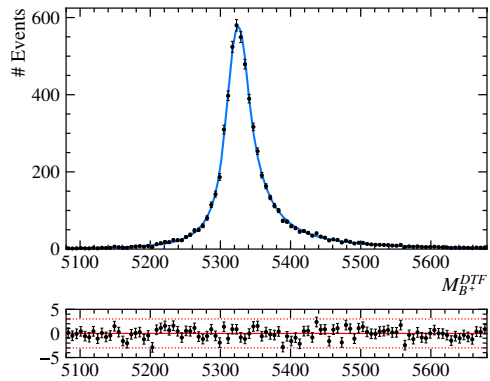

**Figure 6.5.4:** Fit to the single misidentified background channel $B^+ \to \pi^+ J/\psi(e^+e^-)$. The data points are in black and the PDF is drawn in blue.

#### 6.5.1.4 Combinatorial background model

The combinatorial background, as explained in Section 6.1.4.6, is modeled using an exponential shape function. Its free parameter, $\lambda$, is determined from data and therefore allowed to float in the data fit. The fit on the right side band of the data is shown in Fig. 6.5.5 and serves as an illustrative purpose, validating the selection of the background model.

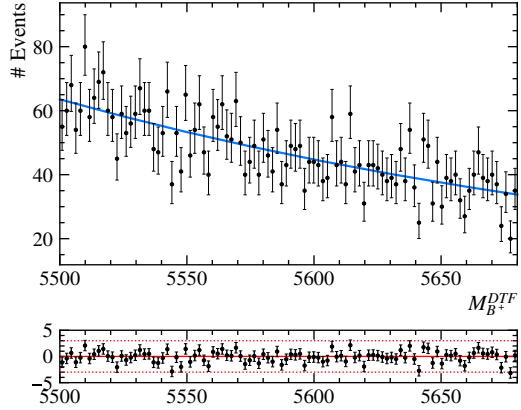

**Figure 6.5.5:** Fit to the combinatorial background on the right side band of the $B^+ \to K^+ e^+ e^-$ data. The data points are in black and the PDF is drawn in blue.

#### 6.5.1.5 Model shape validation

Most of the components in the fit are obtained by fitting to simulated events and then fixing most of the obtained nuisance parameters to their best-fit values, often the tail parameters but also the widths and means. While the assumption goes that the simulation is correct with all the corrections applied, the samples that are used to infer the nuisance parameters are finite. This implies that the parameters can only be known up to a certain precision. Usually, this is assumed to be negligible as the sample size is large and the number of nuisance parameters comparably small. However, in the case of the improved model in the $B^+ \to K^+ J/\psi(e^+ e^-)$ model described in Section 6.5.1.1, the model fit to the whole simulation has about three times more free parameters than the previous approach, making it necessary to check that there is no overfitting. The final shape fitted to the $B^+ \to K^+ J/\psi(e^+ e^-)$ data using the method of splitting the bremsstrahlung categories in comparison with the merged approach is shown in Fig. 6.5.6. This highlights the difference between the old and the new method for the signal model; the new method shows improved pulls on the data fit. This was expected, as the new method captures the full sample shape significantly better in the fit to simulation. To properly encode this uncertainty about the nuisance parameter, the same fit is performed without fixed parameters but simultaneously to data and simulation. The results show that the total uncertainty of parameters increases by about 10% on average, which is small enough to make the assumption that the nuisance parameters are known to infinite precision from the fit to simulation.

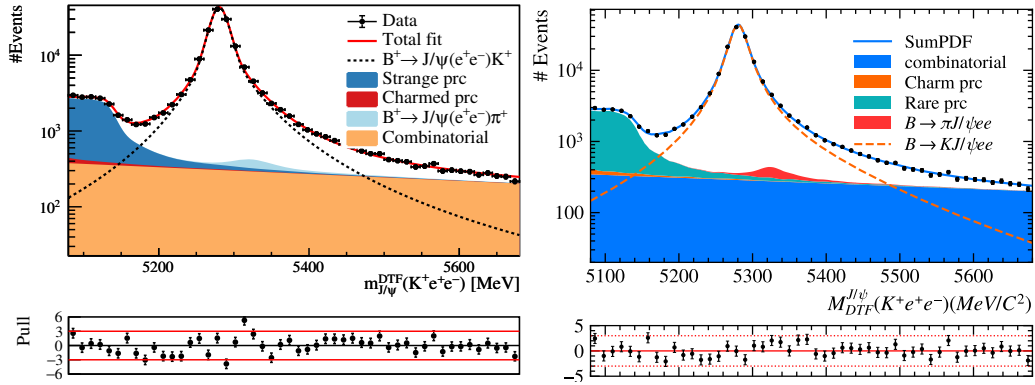

### 6.5.2 $B^+ \to K^+ J/\psi(\mu^+ \mu^-)$ model

In comparison to the electron mode, the muon mode with its better mass resolution fits in a narrower window, leaving fewer background sources polluting the signal region. Therefore, the fit model is simplified in comparison to the electron mode. In the following, the different components of the fit model are discussed. The fit to data and the results are shown in Section 6.6.2.

#### 6.5.2.1 Signal model

The $B^+ \to K^+ J/\psi(\mu^+ \mu^-)$ signal component is modeled using a mixture of a `Gaussian` and a Hypatia function [106], with one power-law tail to the left and another to the right and a wide `Gaussian` distribution, (Hypatia). The latter helps capture the shape of the tails and makes up only about $\sim 2\%$ of the total signal PDF. The different nuisance parameters of the PDF are obtained from a fit to simulated $B^+ \to K^+ J/\psi(\mu^+ \mu^-)$ events, as shown in Fig. 6.5.7. As in the electron mode, the width and mean of the PDF are shifted and scaled using the same procedure with $P_{shift}$ and $P_{scale}$ as denoted in Eq. 6.22.

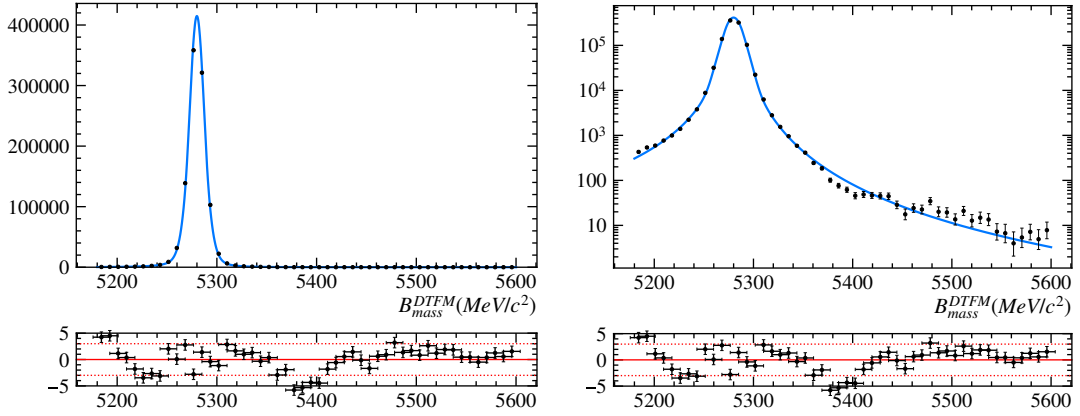

**Figure 6.5.7:** Fit to the $B^+ \to K^+ J/\psi(\mu^+\mu^-)$ signal simulation in 2018 data-taking conditions. The left plot is in linear scale, and the right plot is in logarithmic scale to highlight the agreement in the tails. The data points are in black and the PDF is drawn in blue.

### 6.5.2.2 Single misidentified background model

In the mass range considered, the only background contributions that come from a physical decay entering the sample are single misidentified events, where a pion is reconstructed as a kaon as described in Section 6.1.4.5. These are from the $B^+ \to \pi^+ J/\psi(\mu^+\mu^-)$ decay, which is modeled with a DCB function. Both the width and the mean of the DCB are shifted and scaled by the same parameters $P_{shift}$ and $P_{scale}$, as defined in Eq. 6.22, as the signal model in Section 6.5.2.1. The tail parameters are fixed from a fit to simulated $B^+ \to \pi^+ J/\psi(\mu^+\mu^-)$ events, as shown in Fig. 6.5.8. The yield ratio of the $B^+ \to \pi^+ J/\psi(\mu^+\mu^-)$ component to the $B^+ \to K^+ J/\psi(\mu^+\mu^-)$ signal is `Gaussian` constrained to the ratio of branching fractions [7] multiplied by the ratio of efficiencies:

$$\frac{N(B^+ \to \pi^+ J/\psi(\mu^+\mu^-))}{N(B^+ \to K^+ J/\psi(\mu^+\mu^-))} = \frac{\mathcal{B}(B^+ \to \pi^+ J/\psi(\mu^+\mu^-))}{\mathcal{B}(B^+ \to K^+ J/\psi(\mu^+\mu^-))} \cdot \frac{\varepsilon_{J/\psi K}}{\varepsilon_{J/\psi \pi}}. \tag{6.24}$$

The width of this constraint is set to the uncertainty derived from the branching fraction measurements, which is about 10%. This is much larger than the uncertainty expected on the ratio of efficiencies.

### 6.5.2.3 Combinatorial background

The second background component comes from the combinatorial background, as described in Section 6.1.4.6. It is modeled by an exponential shape function, whose parameter $\lambda$ is left free to be optimized in the fit. This is the same procedure as in the $B^+ \to K^+ J/\psi(e^+e^-)$ case and is shown in Fig. 6.5.8.

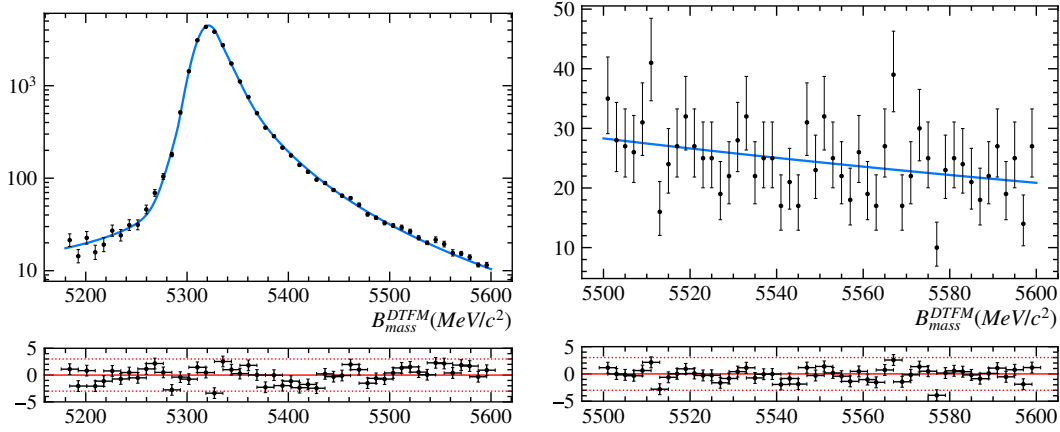

**Figure 6.5.8:** Fit to the $B^+ \to \pi^+ J/\psi(\mu^+\mu^-)$ misidentified background (left) and fit to the combinatorial background of the $B^+ \to K^+ J/\psi(\mu^+\mu^-)$ in the right side band in data (right). The data points are in black and the PDF is drawn in blue.

#### 6.5.2.4 Data model

The resulting fit is illustrated in Fig. 6.5.9 for 2018 data and with the BDTs requirements applied, in Section 6.6.2 the fits and results for each year and both with and without the BDTs applied are shown.

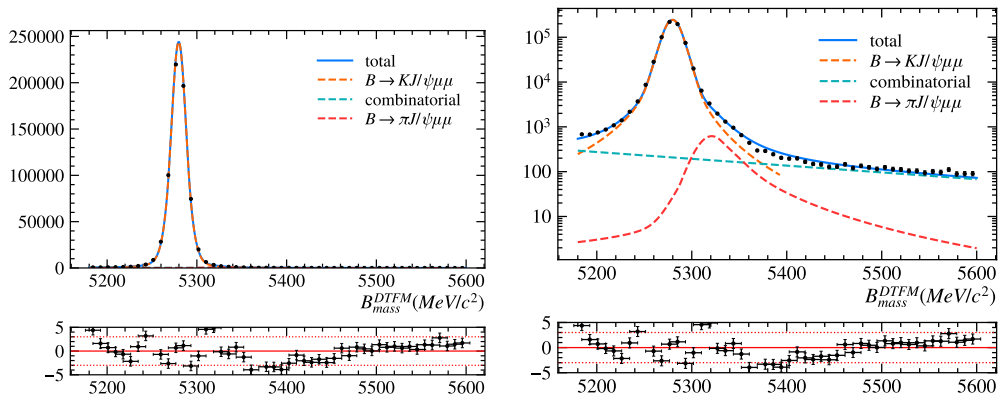

**Figure 6.5.9:** Fit to the $B^+ \to K^+ J/\psi(\mu^+\mu^-)$ $m_{\mathrm{DTF}}^{J/\psi}(K^+\ell^+\ell^-)$ distribution in 2018 data. The left plot is in linear scale, the right plot in log scale to highlight the agreement in the tails. The data points are in black and the PDF is drawn in blue. The different PDF components are displayed as dashed lines in different colors.

### 6.5.3 Extraction of $R_K$

The observable $R_K$ is defined in Eq. 6.1 as the integral over a given $q^2$ region. Experimentally, most of the events are not observed directly, but are removed as described in Section 6.2, and the efficiency of the remaining events is estimated, as described in Section 6.4. These efficiencies are not uniform across all variables, and for most cross-checks, simulation resembles the data well enough to take the dependency from the corrected simulation. Furthermore, for many variables, the efficiency dependence is similar in both the electron and the muon modes, canceling to a high degree in the ratio; this is one of the major experimental advantages of a ratio. The $q^2$ dependence of the efficiency is more complicated, especially in the high $q^2$ region, where the shape in the electron mode is heavily sculpted due to the $q^2_{\text{track}}$ selection requirement as depicted in Fig. 6.5.10.

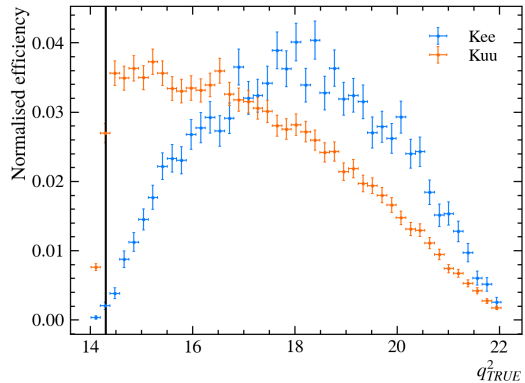

**Figure 6.5.10:** Comparison of normalized lepton efficiencies differential in $q^2$ of $B^+ \to K^+ e^+ e^-$ (blue) and $B^+ \to K^+ \mu^+ \mu^-$ (red).

The high $q^2$ region contains contributions from multiple wide resonances above the open-charm threshold, which are expected to not violate LFU [107]. This additional component should be taken into account equally in both lepton modes, as it would otherwise dilute any deviation from the SM that could be measured. Fig. 6.5.11 illustrates the higher $q^2$ spectrum including all resonances expected in the rare mode. To remove any dependence of the model in the $q^2$ spectra of the signal modes, the efficiencies of both modes are aligned. The muon mode, being the larger sample and having a better reconstructed $q^2$ variable, is weighted to match the efficiency of the electron mode.

The weights obtained for the muons and the shape of the new efficiency are depicted in Fig. 6.5.12. This procedure has a major drawback: The combinatorial background in the muon mode is no longer exponential and therefore cannot be modeled with an exponential function, complicating the fit to obtain the yields. Therefore, a different approach is chosen, where the muon mode is fit separately and *sWeight*s are obtained. These are multiplied directly by the efficiency, yielding the $q^2$ efficiency corrected yield that compares directly with the electron mode.

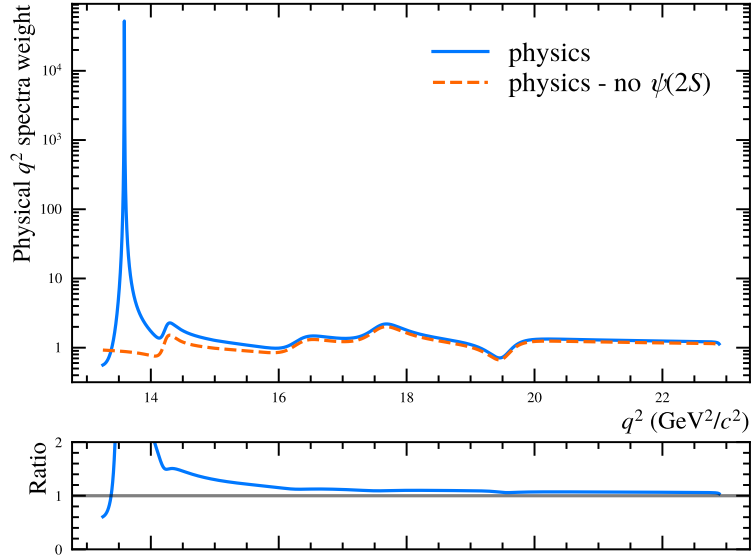

**Figure 6.5.11:** $q^2$ spectrum llustration including all charmonium resonances in the high $q^2$ regions with (blue, solid) and without (orange, dashed) the $\psi(2S)$.

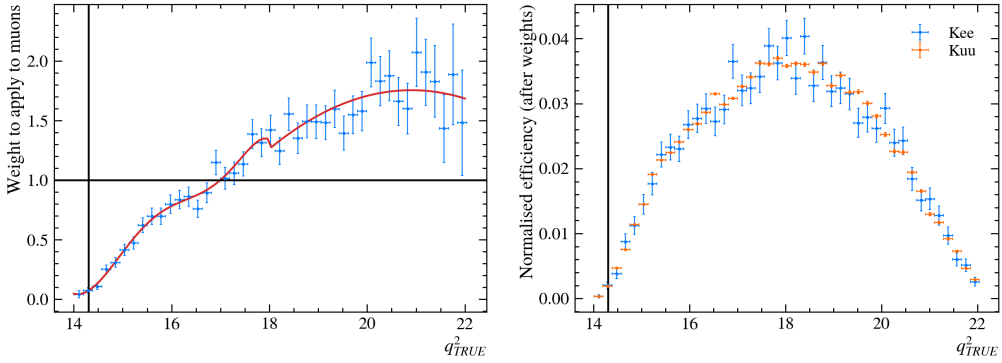

**Figure 6.5.12:** Muon mode weights to align the lepton efficiencies with the weights (blue) and the weighting function (red line) in dependence of $q^2_{\text{true}}$ (left) and the efficiencies in the electron mode (blue) and in the muon mode with the new weights applied (blue), normalized (right).

In order to capture the likelihood of this procedure and allow it to be used in the direct determination of $R_K$, the data is bootstrapped and repeatedly fitted. The obtained yield values are approximately normally distributed and the shape is described with a `Gaussian` distribution. This is added to the likelihood of fit $R_K$, which effectively incorporates the fit to $B^+ \to K^+\mu^+\mu^-$.

### 6.5.4 High $q^2$ combinatorial background model

The combinatorial background differs significantly with respect to the central $q^2$ region and the normalization mode. The background shape is affected by a phasespace effect, since the $B$ mass has to be above the minimum energy carried by the three-body constituents in $K^+e^+e^-$, where the $q^2$ selection requirement poses a high lower limit. This induces a shoulder-like shape on the lower side of the $B$ mass as depicted in Fig. 6.5.13. As the shoulder peaks approximately under the signal region, understanding and verifying the model is crucial as any uncertainties or biases through mis-modeling will directly affect the signal yield. Due to the large fit range in $m(K^+\ell^+\ell^-)$ of the electron mode, this effect is only relevant in this mode, whereas the combinatorial model for the muon mode remains an exponential, as good resolution allows for a tight selection requirement around the $B$ mass.

The strategy to obtain a suitable model starts with a discussion of proxy samples in Section 6.5.4.1. Then, the theoretical model is developed using a combination of a phasespace and an exponential in Section 6.5.4.2. Further corrections to the selection efficiencies and reconstructed variables are introduced in Section 6.5.4.3. The model is validated by fitting to proxy samples in Section 6.5.4.4.

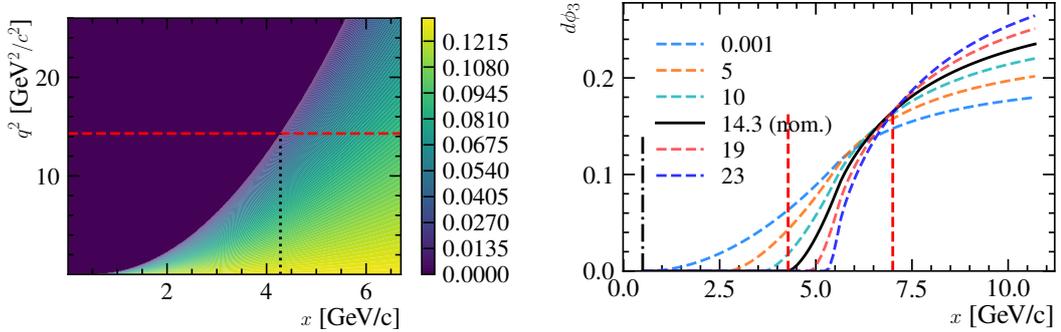

**Figure 6.5.13:** Illustration of the two-dimensional phasespace in $x$, that is $\sqrt{p_{tot}^2}$, and $q^2$ for a three-body final state $K^+e^+e^-$ (left) and the projection on $x$ with the effect of different $q^2$ selection requirements (right). In the left plot, the black dashed line denotes the $q^2$ selection requirement used in the analysis. The right plot illustrates this effect on the projection of $m(K^+\ell^+\ell^-)$, integrating over $q^2$, for different $q^2$ selection requirements, as denoted in the legend. The solid black line shows the $q^2$ selection requirement used in the analysis, the two vertical red-dashed lines indicate the $x$ range used in the mass fit. The vertical, dashed black line at the lower end of the plot indicates the minimum $x$.

### 6.5.4.1 Combinatorial control samples

To better understand and model the combinatorial background shape in the high $q^2$ region, a number of control samples that function as a proxy of the actual combinatorial background are used to infer and validate the shape of the background model. A comparison of the different samples is shown in Fig. 6.5.21.

$B^+ \to K^+ e^+ \mu^-$

This sample consists of an electron and a muon in the final state instead of two leptons of the same family. In the absence of any notable LFU violation, this decay is forbidden, and this sample is therefore assumed to consist of random combinations of tracks, apart from an additional semi-leptonic contribution, as explained in Section 6.1.4.2. In the same way as for the $B^+ \to K^+ e^+ e^-$ channel, this can be removed with a selection requirement on the $m(K^+\ell^-)$ mass. With the phasespace dominating the shape of this sample, the effect is strong enough to infer the shape. One inherent downside to this sample is that the $\mu$ does not contain any bremsstrahlung information, making the sample fall between the $B^+ \to K^+ e^+ e^-$ and $B^+ \to K^+ \mu^+ \mu^-$ samples in terms of corrections and BDT requirements applied. It also does not contain any isolation information, deteriorating the performance of the multivariate selection.

$K^+ e^- e^-$ **same sign**

This sample is ideal in many ways because it mimics the rare mode exactly with the only difference that both electrons have the same electrical charge; additionally, the charge is opposite to that of the kaon. Such a decay is forbidden under the assumption of lepton flavor conservation and therefore mainly consists of combinatorial background events. However, it is not a completely clean sample, as it can contain semileptonic backgrounds: the default veto applies a selection requirement on the $m(K^+\ell^-)$ targeting the $D^0$ mass using the electron with the *opposite* sign. Since both electrons have the same charge, there are now two possibilities to build $m(K^\pm e^\mp)$, which would require an additional selection, distorting the $m(K^+\ell^+\ell^-)$ distribution, and further reducing the already small sample. To better preserve events in this sample, the combinatorial background BDT selection requirement is only applied as mass-dependent efficiency weights extracted from the $K^+ e^+ \mu^-$ sample.

**Mixed** $B^+ \to K^+ e^+ e^-$

A proxy of combinatorial background can be created by simply mixing random tracks in existing decays to constitute a new event. The advantage of this sample is that realistic tracks are used, while practically unlimited amounts of data can be generated. The downside is that not all the variables can be created correctly such as isolation variables or variables based on vertex constraints, namely any variable that is computed from other information of the decay.

**inverse BDT** $B^+ \rightarrow K^+ e^+ e^-$

The combinatorial background BDT is trained to distinguish signal events from combinatorial background events. Instead of using this to select the former and remove the latter, the selection requirement can be inverted, selecting combinatorial background instead. This provides the "true" background that *is* actually in the signal sample, or at least some with a lower combinatorial background BDT score. As the signal is blinded and this sample also contains signal events, a cross-check was performed to ensure that the region is clean and the potential selection requirements were tested on the whole $q^2$ spectrum, including the $J/\psi$ and $\psi(2S)$ resonances. Only a small peak is observed, corresponding to $\approx 0.1\%$ of the $B^+ \rightarrow K^+ J/\psi(e^+ e^-)$ yield, implying a signal efficiency of $< 0.1\%$ and therefore an expected number of signal events in this sample of less than 1. Another possible effect could arise if the shape of the sample is not independent of the combinatorial background BDT selection requirement applied. This would imply that samples chosen this way, with a low combinatorial background BDT selection requirement, look different in the $m(K^+ \ell^+ \ell^-)$ than in the signal, which is chosen with a high selection requirement. Different bins of combinatorial background BDT selection requirements are shown in Fig. 6.5.14 and inhibit only minimal changes in shape.

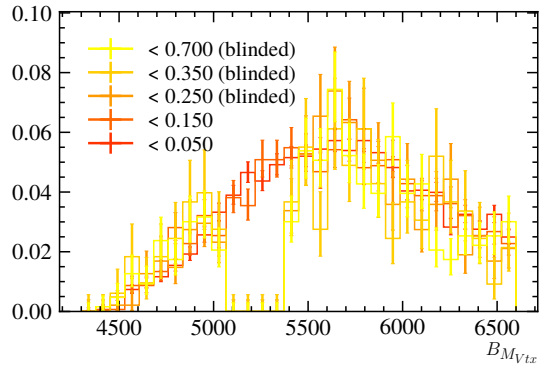

**Figure 6.5.14:** Different selection requirements using the combinatorial background BDT on the $B^+ \rightarrow K^+ e^+ e^-$ sample, where the chosen selection requirements are exclusive between combBDT$_{previous}$ < combBDT < combBDT$_{sel}$, with combBDT$_{sel}$ listed in the legend. The combinatorial background BDT selection requirements either are low enough so that none of these samples contains any signal but only combinatorial background or the signal region has been blinded.

### 6.5.4.2 Phasespace model

To model the combinatorial background, a few approaches were studied. Using toy studies, it is clear that the signal yield, and thus $R_K$, depends strongly on the combinatorial background shape. Therefore, the objective was to reduce the number of free parameters in the background model to keep the highest significance possible. This justified the use of an elaborate model that incorporates as much information as possible and emulates the effects of the selection on the combinatorial background.

First, the basic model for the combinatorial background needs to capture the phasespace effect, with other effects studied later in Section 6.5.4.3. The model includes two assumptions: For one, the shape of the combinatorial background on the *right* side of the signal peak in the $m(K^+\ell^+\ell^-)$ can be described by an exponential. This is well-founded when looking at the right side band, has also been observed in the muon mode, and in countless other similar analyses. In terms of differential decay width, this corresponds to the modulus of the matrix element of the combinatorial component, which follows an exponential as

$$|\mathcal{M}|^2 \propto e^{-\lambda x}$$

with $x = |p_\text{tot}| = \sqrt{p_\text{tot}^2}$ and $\lambda$ the combinatorial slope parameter. The total momentum is given by the sum of the kaon and the two lepton momenta:

$$p_\text{tot} = p_K + q = p_K + p_{e^+} + p_{e^-}$$

with $|q| = \sqrt{q^2}$.

Second, going to lower masses, the phasespace is warped due to the high selection requirement in $q^2$, effectively disallowing a reconstructed $B$ mass below $q$ plus the kaon mass, as depicted in Fig. 6.5.13. This 3-body phasespace with a lower $q^2$ selection requirement is then used to describe the lower limit of the $B$ mass. The complexity of the phasespace can be reduced by the assumption that the two-body phasespace of the two leptons, denoted as $d\phi_2(q; p_{e^+}, p_{e^-})$, is equal to one. Its value is close to one at very high $q^2$, typically $> 0.999$,

$$\sqrt{1 - 4m_\ell^2/q^2} \overset{q^2 > 14.3\,\text{GeV}^2/c^2}{\longrightarrow} 1 \tag{6.25}$$

Therefore, the double differential decay width can be expressed as a function of $x$ and $|q|$ by building the product of the matrix element squared and the simplified three-body phasespace as

$$\begin{aligned}
\frac{d\Gamma}{dx d|q|} &= \frac{1}{2x} e^{-\lambda x} d\phi_3(x; p_K, q) \\
&= \frac{1}{4\,x\,|q|} e^{-\lambda x} \frac{\sqrt{x^2 - (m_K + |q|)^2}}{x} \frac{\sqrt{x^2 - (m_K - |q|)^2}}{x} \ .
\end{aligned} \tag{6.26}$$

For the fit model, the differential decay width as a function of $x$, that is actually $m(K^+\ell^+\ell^-)$, is needed, which is obtained by integrating over $|q|$:

$$\Gamma(x, |q|, \lambda) = \int \frac{d\Gamma}{dx d|q|} d|q| \tag{6.27}$$

This can be written as a function of $x$, the $q^2$ cutoff value, $q^2_{\text{min}}$, and the exponential slope parameter ,$\lambda$, and the definite integral becomes

$$\gamma(x, \sqrt{q^2_{\text{min}}}, \lambda) = \Gamma(x, x - m_K, \lambda) - \Gamma(x, \sqrt{q^2_{\text{min}}}, \lambda) \qquad (6.28)$$

This expression is implemented as a PDF, normalized over $x$, with the other two parameters left to be inferred from data samples as

$$f_{comb} = \frac{\gamma(x; q_{\text{min}}, \lambda)}{\int_{m^{\text{fit}}_{\text{lower}}}^{m^{\text{fit}}_{\text{upper}}}} \ . \qquad (6.29)$$

The shape for $\lambda = 1.1$ and different $q^2$ selection requirements is shown in Fig. 6.5.15.

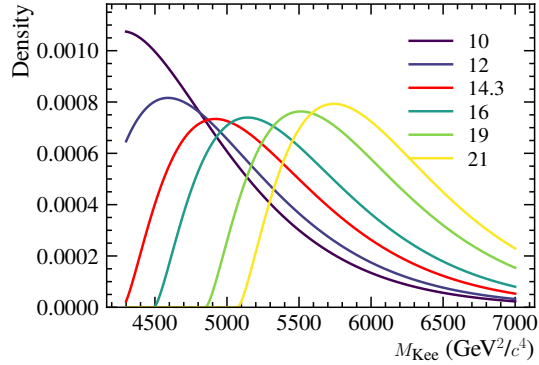

**Figure 6.5.15:** Combinatorial background model, combining a three-body phase-space and an exponential for different selection of $q^2_{\text{min}}$ displayed in the legend in GeV$^2$/$c^4$. The nominal $q^2$ selection requirement is indicated by the red line.

As can be seen, a $m(K^+\ell^+\ell^-)$ window 4800 to 6700 MeV/$c^2$ is not sufficient to fit the $q^2_{\text{min}}$, instead widening the window from 4300 to 6700 MeV/$c^2$ includes the entire phasespace of the combinatorial background. A comparison of the fit of the model to the $K^+e^-e^-$ sample, once with a fixed $q^2$ selection, once with a floating one, is shown in Fig. 6.5.16, where the only other free parameter is the $\lambda$ of the exponential.

The model is not quite able to capture the $q^2$ increase correctly, which is improved by floating $q^2_{\text{min}}$ freely, resulting in a small loss in sensitivity. At the same time, the fits also show that the *overall shape* of the background is well enough described by the model and motivates the use of it as a basis, with further refinements to be incorporated.

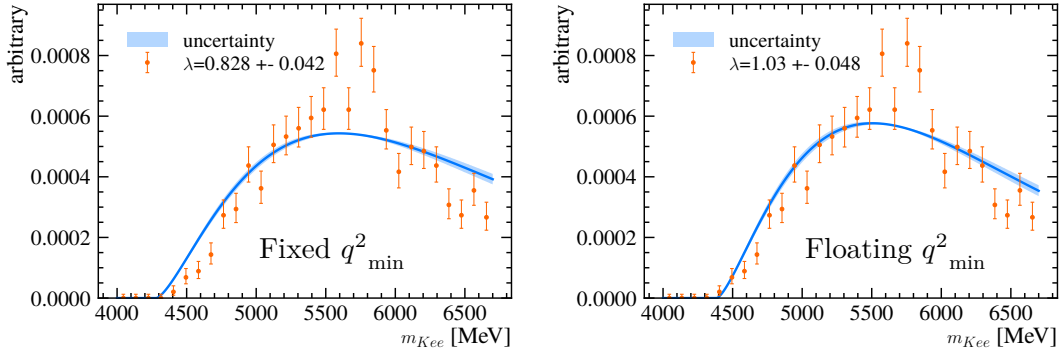

**Figure 6.5.16:** Phasespace model with exponential fit to $K^+e^+e^-$ with a $q^2_{min}$ fixed to the actual $q^2$ selection requirement (left) and a floating $q^2_{min}$ (right).

### 6.5.4.3 Efficiency and reconstruction effects

The phasespace model derived in Eq. 6.26 is a model for the combinatorial background assuming true variables. Some corrections need to be made as the expected $m(K^+\ell^+\ell^-)$ is affected by selection requirements and the reconstruction of the variable.

The first correction concerns the reconstructed $m(K^+\ell^+\ell^-)$ variable that contains resolution effects and is spread more widely than the true $m(K^+\ell^+\ell^-)$. As the underlying, true $m(K^+\ell^+\ell^-)$ is not known, the smearing is obtained from the signal simulation in $m(K^+\ell^-)$ of $B^+ \to K^+e^+e^-$ by fitting the sample with a CB in bremsstrahlung category zero and a DCB in both bremsstrahlung categories one and two, then summing the three contributions according to their relative fractions. The resolution depends, as expected, highly on the bremsstrahlung category, with bremsstrahlung category two having the worst resolution. The $B^+ \to K^+e^+e^-$ differs from the combinatorial background, as it constraints a degree of freedom, the $m(K^+\ell^+\ell^-)$ to the $B$ mass, yet it is valid to assume that the resolution effect is approximately independent of the $m(K^+\ell^+\ell^-)$ itself, at least in the region of interest 4300 to 6700 MeV/$c^2$. This is convolved with the phasespace model in Eq. 6.29 and has a minor impact on the shape, as shown in Fig. 6.5.17.

Using the $q^2_{track}$ variable instead of $q^2$ for the selection requirement introduces some shaping effects, which are correlated with $m(K^+\ell^+\ell^-)$. The efficiency of the $q^2_{track}$ selection requirement compared to a $q^2$ selection is obtained with the mixed high-statistics sample and shown, together with other samples that show the same trend, in Fig. 6.5.18.

The selection requirement of the cascade veto, as described in Section 6.2.3.2, is correlated with $m(K^+\ell^+\ell^-)$. Both effects are therefore measured and folded into the combinatorial background model through an efficiency derived from a proxy sample. For this, the mixed sample is used due to its high statistics. The derived efficiency corrections for the cascade veto are shown in Fig. 6.5.19.

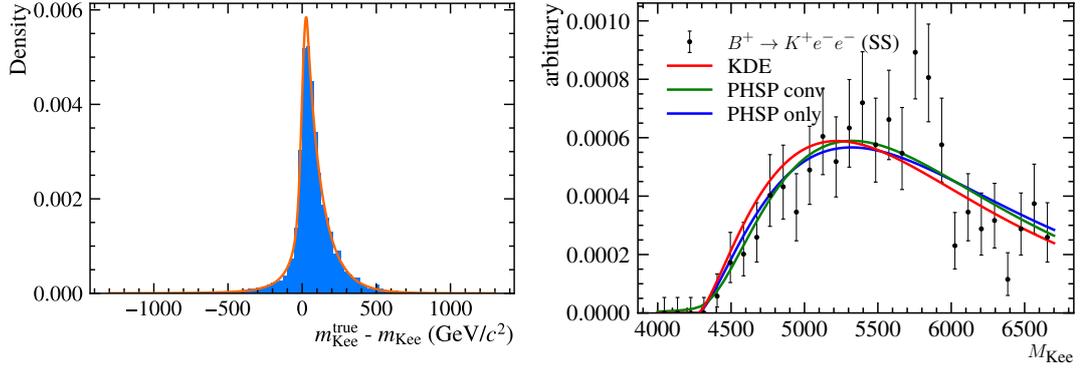

**Figure 6.5.17:** Fit of the kernel to the difference of the $m(K^+\ell^+\ell^-)$ that contains the smearing of the reconstruction effect in $B^+ \to K^+e^+e^-$ to the true variable (left). The effect of the smearing correction on the $K^+e^-e^-$ sample (right) is shown without (blue) and with (green) the convolution and a KDE (red).

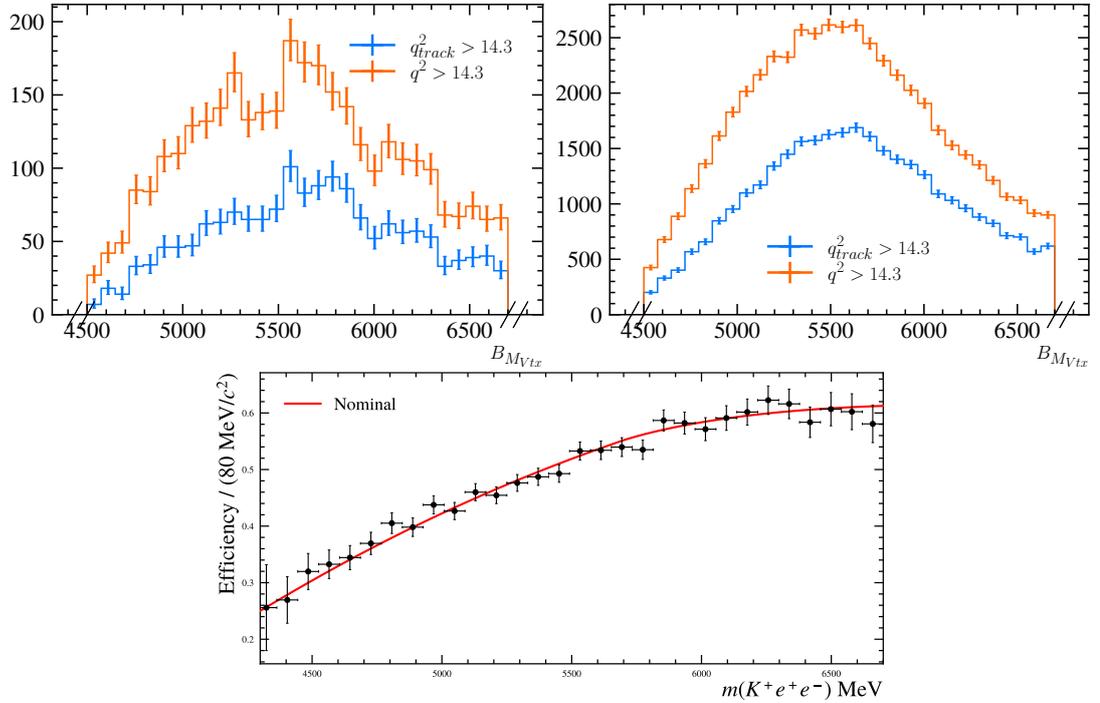

**Figure 6.5.18:** Comparison of a selection requirement on $q^2_{\mathrm{track}}$ versus $q^2$ as a function of $m(K^+\ell^+\ell^-)$ in the $K^+e^-e^-$ (top left) and $K^+e^+\mu^-$ (top right) samples and the efficiency obtained from a mixed high-statistics sample (bottom).

Similarly, the effect of the BDTs, described in Section 6.2.5.2, is correlated with $m(K^+\ell^+\ell^-)$. In a similar fashion, the combinatorial background BDT efficiency is measured in the mixed sample and folded into the combinatorial background model. The efficiency corrections are shown in Fig. 6.5.20.

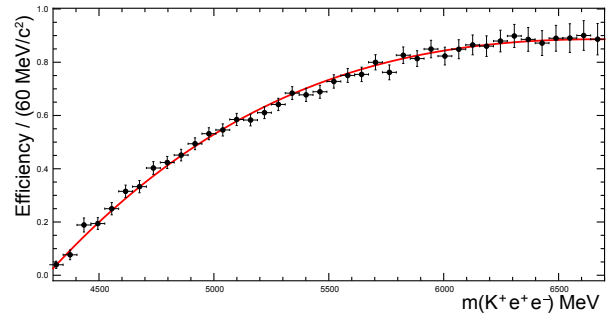

**Figure 6.5.19:** Efficiency correction for the cascade veto as a function of $m(K^+\ell^+\ell^-)$ that is folded into the combinatorial background model. The corrections are derived from the mixed sample.

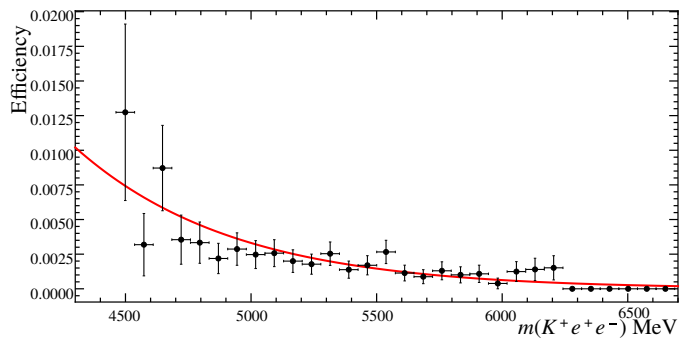

**Figure 6.5.20:** Efficiency correction for the combinatorial background BDT as a function of $m(K^+\ell^+\ell^-)$ that is folded into the combinatorial background model. The corrections are derived from the mixed sample.

#### 6.5.4.4 Fitting control samples

The complete model is fit to the different control samples, as shown in Fig. 6.5.21. To understand the quality of the model and possible biases caused in the fit, toy studies are performed. Each control sample is fitted with a KDE and used to generate a toy sample. Fitting back with the model, the bias on $R_K$ is obtained from the best-fit value. The biases with and without $q^2_{min}$ floating are shown in Fig. 6.5.22 and are found to be $-0.007$ and $-0.013$, respectively.

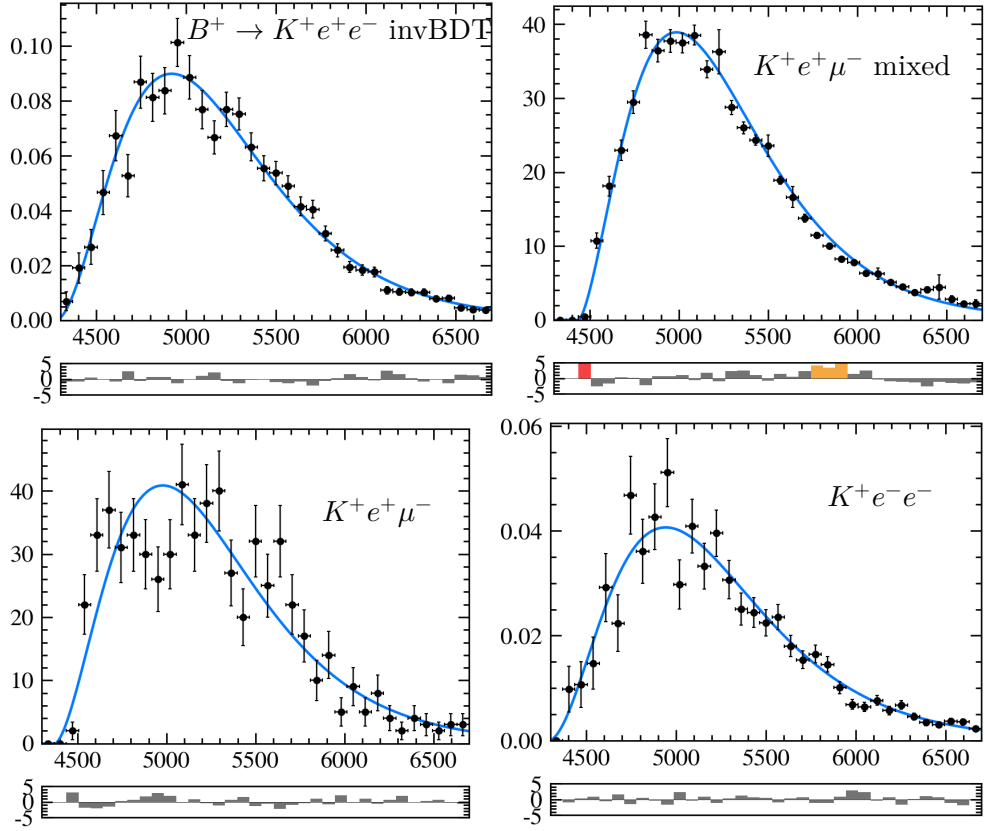

**Figure 6.5.21:** Fit of the full combinatorial background model to different proxies in $m(K^+\ell^+\ell^-)$; the $B^+ \to K^+e^+e^-$ invBDT (top left), the $K^+e^+\mu^-$ mixed (top right), the $K^+e^+\mu^-$ (bottom left) and the $K^+e^-e^-$ (bottom right).

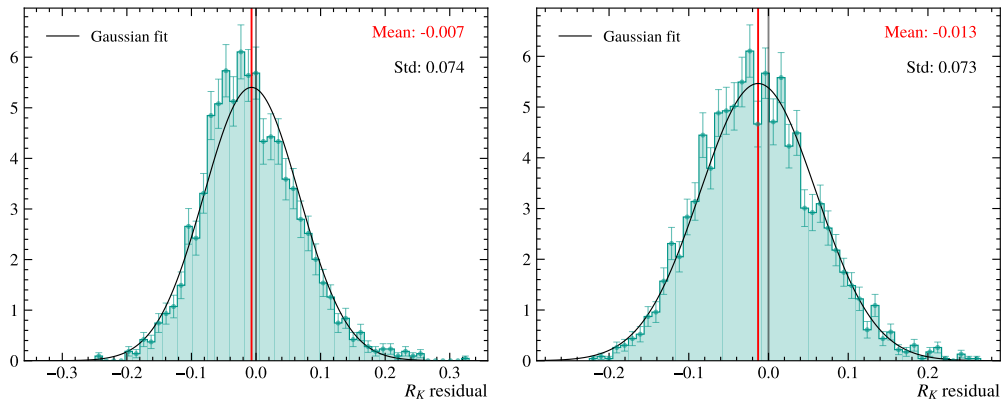

**Figure 6.5.22:** $R_K$ values obtained from fitting the KDE generator models using the complete combinatorial background model with the parameter $q^2_{\min}$ free in the fit (left) and fixed to the selection requirement value (right).

### 6.5.5 $B^+ \rightarrow K^+ e^+ e^-$ **fit model**

The fit model to the rare mode consists of components similar to the normalization mode described in Section 6.5.1 with the following differences:

- The signal model consists also of three bremsstrahlung categories, but they are modeled slightly differently. Bremsstrahlung category zero is modeled with a DCB while both bremsstrahlung categories one and two are each modeled with a mixture of a DCB and a `Gaussian`.

- Cascade backgrounds as described in Section 6.1.4.2 are incorporated into the fit model. The cascade background is modeled with a KDE and each channel is weighted according to its expected branching fraction in the template. The fit to simulation is shown in Fig. 6.5.23.

- The "charm" partially reconstructed background coming from a $J/\psi$ does not enter the rare mode fit, but instead leakage from decays containing the heavier charmonium resonance $\psi(2S)$. The contribution is modeled using a KDE from a dedicated sample and is shown in Fig. 6.5.23.

- Some additional backgrounds enter the rare mode, namely $B^+ \rightarrow K^+\pi^+\pi^-e^+e^-$, $B^\pm \rightarrow KKK$ and $B_s^{0^+} \rightarrow K^+\phi(\rightarrow e^+e^-)$. The former two are both double misidentified background and are modeled with a DCB while the latter enters as a KDE obtained from simulation.

- The combinatorial background model is used as previously described in Section 6.5.4.

The fit to data is discussed in Section 6.9.1 and the fits to the signal simulation as well as the background components $B^+ \to K^+\pi^+\pi^-e^+e^-$, $B^\pm \to KKK$ and $B_s^{0+} \to K^+\phi(\to e^+e^-)$ are shown in Appendix A.2.

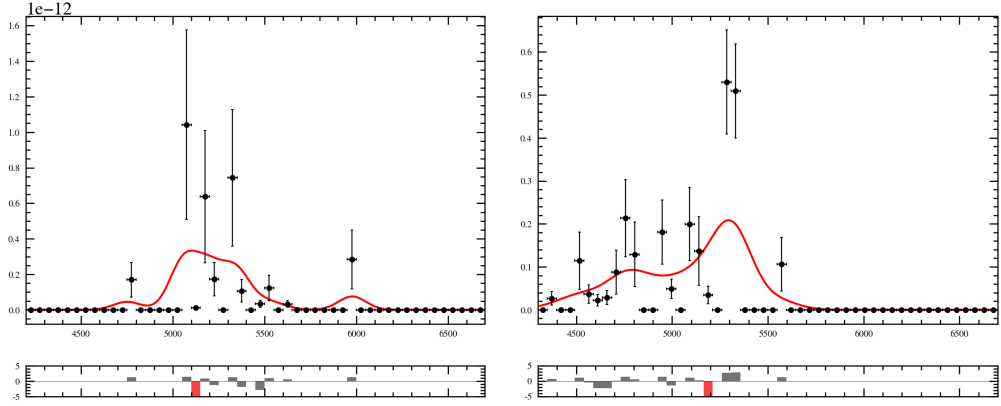

**Figure 6.5.23:** Fit to different cascade backgrounds (left) and leakage of different partially reconstructed background of $B^+ \to \psi(2S)(e^+e^-)X$ (right). The samples are comparably small and the effect of this is taken into account in the evaluation of the systematic uncertainties as described in Section 6.7. The data points are shown in black and the PDF is drawn in red.

### 6.5.6 $B^+ \to K^+\mu^+\mu^-$ mass model

The model for the $B^+ \to K^+\mu^+\mu^-$ decay in $m(K^+\ell^+\ell^-)$ closely follows the normalization mode in Section 6.5.2, with the main difference that there is no misidentified background contribution but instead leakage of $B^+ \to K^+\psi(2S)(\mu^+\mu^-)$. The signal is modeled with a sum of a DCB and a `Gaussian` that is fit to the simulation, which is shown in Fig. 6.5.24. For the fit to data, the mean and width have a free floating shift and a scale parameter, respectively, as defined in Eq. 6.22.

Apart from the combinatorial background that is modeled by an exponential, as mentioned in Section 6.5.4, there is another contribution from the leakage of $B^+ \to K^+\psi(2S)(\mu^+\mu^-)$. This is modeled by a sum of a CB, with the tail to the upper mass side, and a `Gaussian`, where all parameters are obtained from a fit to simulation. The same shift and scale parameters are applied as above for the signal. The fit to simulation is shown in Fig. 6.5.24.

The model can be written as

$$\mathcal{P}_{rare}(m) = N_{rare} \cdot P_{sig}(m) + N_{comb} \cdot P_{comb}(m; \lambda) + N_{\psi(2S)} \cdot B_{\psi(2S)}(m) \qquad (6.30)$$

where the yield of the $\psi(2S)$ leakage $N_{\psi(2S)}$ is constrained by the $J/\psi$ yield $N_{J/\psi}$, the branching fractions and the efficiencies as

$$N_{\psi(2S)} \;=\; N_{J/\psi} \cdot \frac{\mathcal{B}(B^+ \to K^+ \psi(2S)(\mu^+\mu^-))}{\mathcal{B}(B^+ \to K^+ J/\psi(\mu^+\mu^-))} \cdot \frac{\varepsilon_{\psi(2S)}}{\varepsilon_{J/\psi}} \;. \tag{6.31}$$

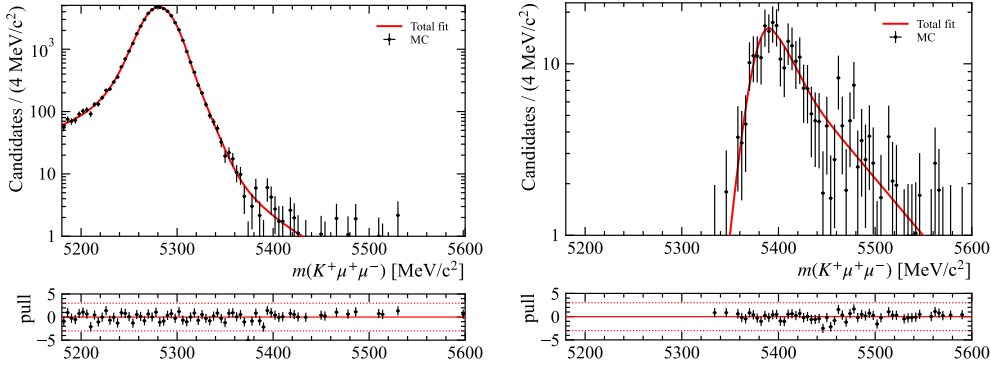

**Figure 6.5.24:** Fit to $m(K^+\mu^+\mu^-)$ for fully selected $B^+ \to K^+\mu^+\mu^-$ (left) and $B^+ \to K^+\psi(2S)(\mu^+\mu^-)$ (right). The data points are shown in black and the PDF is drawn in red.

# 6.6 Normalization fits

In the following sections, the fits to the normalization modes in 2018 data are shown, the years 2016 and 2017 are shown in Appendix A.3. The fits were performed using an extended unbinned maximum likelihood to data. The $q^2$ and $m(K^+\ell^+\ell^-)$ selection requirements that select the resonances and specify the fit range, respectively, are defined in Section 6.2.4.3.

### 6.6.1 Fits to $B^+ \to K^+ J/\psi(e^+e^-)$

The models and components are described in Section 6.5.1. The total PDF is the mixture of the signal, the partially reconstructed background, the misidentified background and the combinatorial background PDFs. In the fit to data, the yields of the different components are allowed to vary in the optimization. Furthermore, shape parameters like the fraction of the two partially reconstructed background components, the $\lambda$ parameter of the combinatorial PDF and the shift and scale parameters are also floating. The fits to the $B^+ \to K^+ J/\psi(e^+e^-)$ data in 2018 are shown in Fig. 6.6.1 with and without the BDTs selection requirement and the results of the fits are summarized in Tab. 6.6.1.

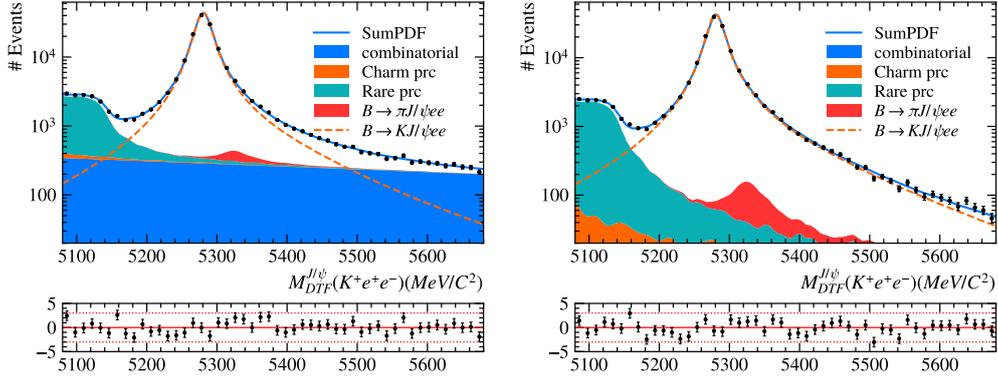

**Figure 6.6.1:** Normalization mode fits to $B^+ \to K^+ J/\psi(e^+e^-)$ data in the year 2018, without (left) and with (right) the combinatorial background BDT applied. The data points are shown in black and the PDF is drawn in red. The different PDF components are filled with different colors and stacked.

**Table 6.6.1:** Yields from the fit to the $B^+ \to K^+ J/\psi(e^+e^-)$ decay in 2018 data, with and without the BDTs applied. The uncertainties on the yields are statistical only. Individual models for the different contributions are described in Section 6.5.1. The abbreviation *prc* stands for *partially reconstructed background*.

| **channel** | With BDTs | No BDTs |
|---|---|---|
| | 2018 | |
| $B^+ \to K^+ J/\psi(e^+e^-)$ | $146847 \pm 410$ | $150747 \pm 510$ |
| $B^+ \to \pi^+ J/\psi(e^+e^-)$ | $666 \pm 67$ | $779 \pm 68$ |
| combinatorial | $550 \pm 130$ | $13276 \pm 490$ |
| charm prc | $377 \pm 55$ | $412 \pm 62$ |
| strange prc | $13198 \pm 140$ | $13559 \pm 220$ |

### 6.6.2 Fits to $B^+ \to K^+ J/\psi(\mu^+\mu^-)$

The components included in the fit to $B^+ \to K^+ J/\psi(\mu^+\mu^-)$ are described in Section 6.5.2. The model to fit the data is a mixture of all components. Following the same procedure as for the electron mode described in Section 6.5.1, the same corrective parameters $P_{shift}$ and $P_{scale}$ from Eq. 6.22 are left floating in the fit to data. In addition to these two parameters, each of the components yields and the $\lambda$ of the combinatorial background PDF are free to be optimized; the rest of the shape parameters are taken from the fits to simulation. The fits to the $B^+ \to K^+ J/\psi(\mu^+\mu^-)$ data in 2018 are shown in Fig. 6.6.2 with and without the combinatorial background BDT selection requirement, the fits to other years are shown in Appendix A.3, and the results of the fits are summarized in Tab. 6.6.2.

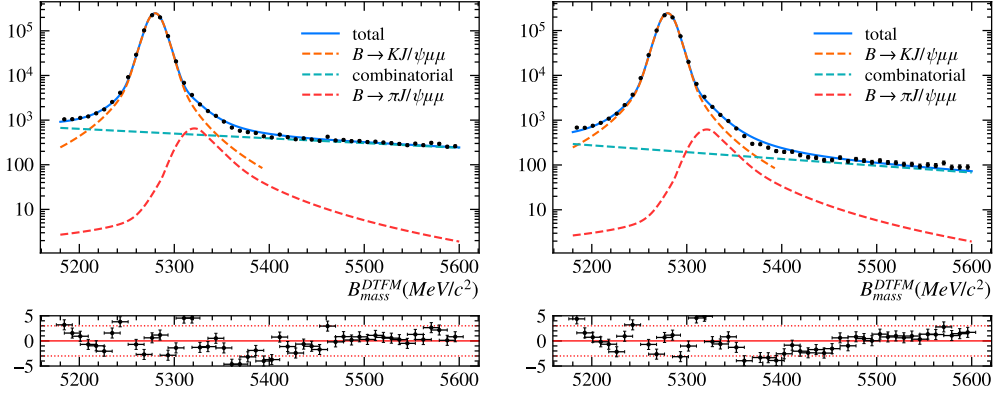

**Figure 6.6.2:** Normalization mode fits to $B^+ \to K^+ J/\psi(\mu^+\mu^-)$ data of the year 2018 with the full selection chain, without (left) and with (right) the combinatorial background BDT requirement applied. The data points are shown in black and the PDF is drawn in red. The different PDF components are displayed as dashed lines in different colors.

**Table 6.6.2:** Yields from the fit to the $B^+ \to K^+ J/\psi(\mu^+\mu^-)$ decay in 2018 data, with and without the BDTs applied. The uncertainties on the yields are statistical only. Individual models for the different contributions are described in Section 6.5.2.

| **channel** | With BDTs | No BDTs |
|---|---|---|
| | \multicolumn{2}{c}{2018} | |
| $B^+ \to K^+ J/\psi(\mu^+\mu^-)$ | $736005 \pm 880$ | $735713 \pm 880$ |
| $B^+ \to \pi^+ J/\psi(\mu^+\mu^-)$ | $4055 \pm 180$ | $4085 \pm 180$ |
| combinatorial | $22173 \pm 250$ | $22434 \pm 250$ |

## 6.7 Uncertainties

The uncertainties that enter the $R_K$ measurement are divided into two categories: statistical and systematic uncertainties. Statistical uncertainties represent the uncertainty that originates *directly* from the limited size of the data sample and can be reduced by collecting more data. Systematic uncertainties refer to the rest of uncertainties, which can arise as well from limited sample sizes such as simulation, but, more importantly, not directly from the signal data. The measurement of $R_K$ is expected to be dominated by the statistical uncertainty, which is estimated to be 7.26%, while the combined systematic uncertainty including systematic effects is estimated to be 7.89%. In the following, the most prominent sources of systematic uncertainties are discussed; the values are given in Tab. 6.7.1.

| $R_K$ statistical error | 0.0726 |
|---|---|

| Systematic | Size (in units of $R_K$ statistical error) |
|---|---|
| Combinatorial - model choice | 0.30 |
| Combinatorial - BDT efficiencies | 0.13 |
| Other sources | 0.28 |

| syst/stat | 0.43 |
|---|---|

| $R_K$ combined error | 0.079 |
|---|---|

**Table 6.7.1:** All uncertainties on $R_K$. Systematic uncertainties are reported in units of $R_K$ statistical uncertainty as obtained in the fit toys.

### 6.7.1 Combinatorial background modeling

The background model described in Section 6.5.4 is the major source of systematic uncertainties, as its shape is inferred from a variety of proxies and it peaks right under the signal. To obtain an estimate of the uncertainty, toy studies are performed in which the background sample is created using a KDE. The fit model then consists of either the same KDE or the nominal phasespace-based model. A deviation of approximately 30 % of the statistical uncertainty is observed when fitting with the latter. Furthermore, an additional contribution arises from the BDT efficiency folding, explained in Section 6.5.4.3, which is observed to be 0.125. The rest of the systematic uncertainties are, all together, estimated to be 0.277, which arises mainly from the pass-fail method.

## 6.8 Cross-checks

Multiple cross-checks have been performed to verify that the corrections to the efficiencies and shapes are under control. A cross-check differs from assigning a systematical uncertainty in that it is only a check with pass or fail as an outcome. If it does not pass, this indicates that parts of the procedure to obtain $R_K$ are most likely flawed, yet passing *does not* directly imply that the procedure is correct, rather that parts of it do not look wrong[3].

In the following, $r_{J/\psi}$, which is well measured to be unity is used as a cross-check. The cross-check ratio is performed separately by year and either "integrated", using the whole dataset of a year, or "differential" in bins of specific variables as described below in Section 6.8.2.

### 6.8.1 Integrated $r_{J/\psi}$

The check is performed on data samples pre-selected using the full selection chain, with and without BDTs. The $q^2$ regions and the fit ranges as a selection in $m(K^+\ell^+\ell^-)$ are used as defined in Section 6.2.4.3. Simulation is used to determine the efficiencies, which are computed with all corrections applied as discussed in Section 6.3. The yields are extracted using the fits as described in Section 6.6.1 for the electron mode and Section 6.6.2 for the muon mode. The integrated $r_{J/\psi}$ single ratio, as defined in Eq. 6.8.

### 6.8.2 Differential $r_{J/\psi}$

As an additional cross-check and to specifically assess the agreement between data and simulation in kinematic distributions of $K^+J/\psi(\ell^+\ell^-)$, the differential $r_{J/\psi}$ is measured in bins of different variables. The check is performed as in the integrated ratios described above in Section 6.8.1 with an additional selection requirement for events to be inside a specific bin. As before, the yields are extracted using fits to count events in specific bins. This eliminates any potential bias caused by correlations between kinematic variables and the invariant $B$ mass variable used in the fit.

The simulation, particularly for the electron mode, is imperfect, meaning that *some* differential $r_{J/\psi}$ histograms are expected to have variations across the bins. To quantify

---

[3]For example, a small, additive constant would not be detected in the normalization mode, but would comparably strongly affect the rare mode.

**Table 6.8.1:** Integrated values of $r_{J/\psi}$, with and without BDTs, computed for the three years 2016, 2017 and 2018 separately. The muon modes are measured in the $\mu$TOS category, and the electron modes are measured in the three categories $e$TOS, $h$TOS! and TIS! where the latter two are used as additional cross-checks. The uncertainties are statistical only.

| | $r_{J/\psi}$ | |
| | With BDTs | No BDTs |
| **2016** | | |
| $e$TOS | $1.049 \pm 0.005$ | $1.054 \pm 0.005$ |
| $h$TOS! | $1.101 \pm 0.013$ | $1.051 \pm 0.011$ |
| TIS! | $1.1388 \pm 0.0097$ | $1.117 \pm 0.010$ |
| **2017** | | |
| $e$TOS | $1.010 \pm 0.005$ | $1.001 \pm 0.005$ |
| $h$TOS! | $0.975 \pm 0.009$ | $0.956 \pm 0.010$ |
| TIS! | $0.989 \pm 0.008$ | $0.978 \pm 0.008$ |
| **2018** | | |
| $e$TOS | $1.013 \pm 0.005$ | $1.008 \pm 0.005$ |
| $h$TOS! | $0.966 \pm 0.009$ | $0.945 \pm 0.009$ |
| TIS! | $0.980 \pm 0.007$ | $0.970 \pm 0.007$ |

these variations, the flatness parameter $d_f$ is introduced, defined as

$$d_f \equiv \frac{\sum\limits_{i=1}^{n_{\text{bins}}} \varepsilon_\mu^{\text{rare},i} \cdot y_\mu^i}{\sum\limits_{i=1}^{n_{\text{bins}}} \varepsilon_\mu^i \cdot y_\mu^i} \cdot \frac{\sum\limits_{i=1}^{n_{\text{bins}}} \varepsilon_\mu^i}{\sum\limits_{i=1}^{n_{\text{bins}}} \varepsilon_\mu^{\text{rare},i}} \left/ \frac{\sum\limits_{i=1}^{n_{\text{bins}}} \varepsilon_e^{\text{rare},i} \cdot y_e^i}{\sum\limits_{i=1}^{n_{\text{bins}}} \varepsilon_e^i \cdot y_e^i} \cdot \frac{\sum\limits_{i=1}^{n_{\text{bins}}} \varepsilon_e^i}{\sum\limits_{i=1}^{n_{\text{bins}}} \varepsilon_e^{\text{rare},i}} \right. - 1 \qquad (6.32)$$

with $n_{\text{bins}}$ the number of bins in the variable of interest, $\varepsilon_\mu^{\text{rare},i}$ the efficiency of the rare decay, $y_\ell^i$ the efficiency-corrected yield and $\varepsilon_\ell^i$ the efficiency of the $K^+ J/\psi(\ell^+\ell^-)$ decay. The superscript $i$ denotes that this quantity is in the $i$-th bin, subscript $\ell$ refers to either the electron or muon mode. The flatness $d_f$ value can be interpreted as follows: If the deviations from flatness across bins of the efficiency-corrected yield histogram are assumed to be genuine variations of the efficiency from a perfectly flat distribution, given the distribution of events in the rare decay, what would be the resulting deviation in the double ratio $R_K$ be? The size is expected to be on the magnitude of the total systematic uncertainty of $R_K$, where a larger flatness parameter indicates possibly an additional source of systematic uncertainty. This would not cancel in the ratio $\varepsilon(B^+ \to K^+\ell^+\ell^-)/\varepsilon(B^+ \to K^+ J/\psi(\ell^+\ell^-))$ as expected in $R_K$.

The differential $r_{J/\psi}$ has been measured in data taken during 2018, to which the entire selection chain *including the BDT* requirement has been applied. Flatness parameters $d_f$ are computed for all the different variables. Most of the flatness parameters are less than one permille, with the exception of a few:

- The flatness of the kinematic variables $B$ and kaon transverse momentum are at the 1 % level. Both are expected to differ between the normalization and rare mode.

- Similarly, the flatness of the lepton kinematic variables, including bremsstrahlung related variables, are between 0.5 to 3%, with most around 1 %.

- The flatness of a BDT that was trained to distinguish between the rare and normalization mode is at the 1 % level, picking up the differences in the kinematic distributions; importantly, the actually used combinatorial background BDT is at half a permille.

The plots of $r_{J/\psi}$ in four variables that highlight the kinematic differences between the normalization and the rare mode, the kaon angle and the lepton angle are shown in Fig. 6.8.1; the $B$ transverse momenta and a dedicated BDT that distinguishes between $B^+ \to K^+ J/\psi(e^+ e^-)$ and $B^+ \to K^+ e^+ e^-$, are shown in Fig. 6.8.2. The plots show the $r_{J/\psi}$ in the $e$TOS category for the $B^+ \to K^+ J/\psi(e^+ e^-)$ mode and the $\mu$TOS category for the $B^+ \to K^+ J/\psi(\mu^+ \mu^-)$ mode with the flatness parameters for the electron and muon mode, respectively. The other plots and flatness parameters of 2018 data are shown for $e$TOS in Appendix A.4.0.1 with plots to the alternative trigger category $h$TOS! in Appendix A.4.0.2 and TIS! in Appendix A.4.0.3.

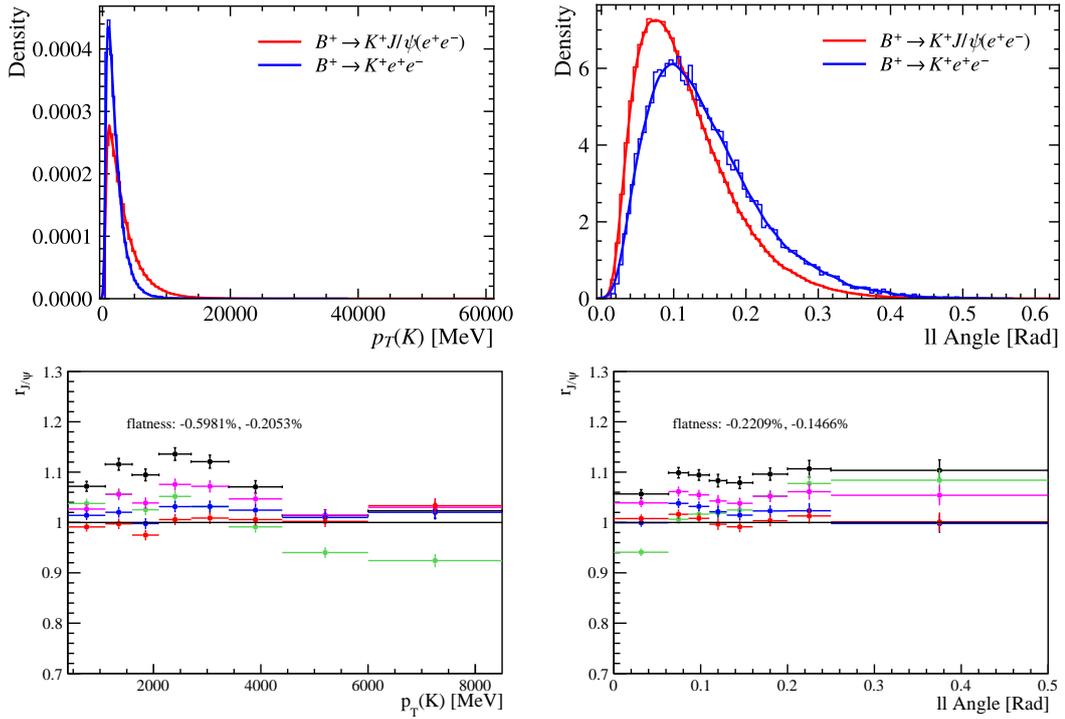

**Figure 6.8.1:** $r_{J/\psi}$ in the transverse kaon momentum (left) and the lepton angle (right). The black points are computed with no kinematic weights.
The red points are computed applying the nominal kinematic weights. The other colors are alternative kinematic weights, derived using different samples instead of the $\mu$TOS muon mode: The blue points use the electron $e$TOS, the pink points the TIS muon sample, and the green the TIS electron sample. At the top of each plot, the kinematic distributions for $B^+ \to K^+ e^+ e^-$ (blue) and $B^+ \to K^+ J/\psi(e^+ e^-)$ (red) are shown, a histogram overlaid with a KDE.

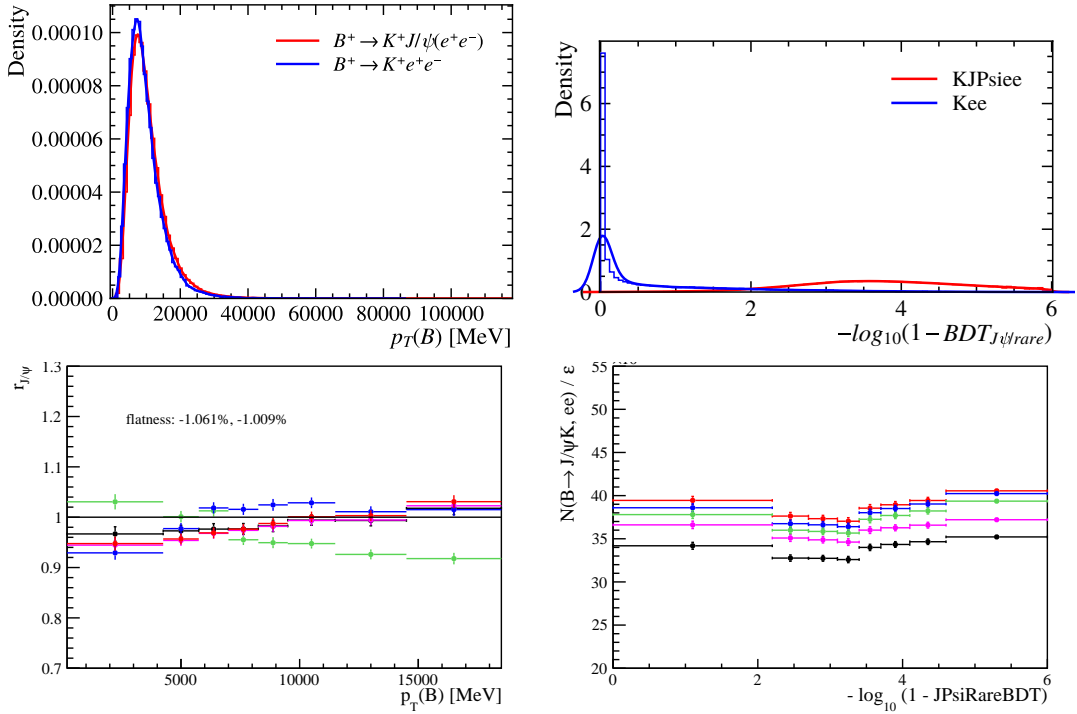

**Figure 6.8.2:** $r_{J/\psi}$ in the transverse $B$ momentum (left) and a dedicated BDT (right). The black points are computed with no kinematic weights. The red points are computed applying the nominal kinematic weights. The other colors are alternative kinematic weights, derived using different samples instead of the $\mu$TOS muon mode: The blue points use the electron $e$TOS, the pink points the TIS muon sample and the green the TIS electron sample. At the top of each plot, the kinematic distributions for $B^+ \to K^+ e^+ e^-$ (blue) and $B^+ \to K^+ J/\psi(e^+ e^-)$ (red) are shown, a histogram overlaid with a KDE.

## 6.9 Results

With all the fit models at hand from Section 6.5, the selection defined from Section 6.2 and the efficiencies obtained in Section 6.4, the fit to data in the rare mode can be performed and the results can be obtained. As the analysis is still in a blinded stage, awaiting further cross-checks to pass, the fits to $B^+ \rightarrow K^+ e^+ e^-$ have not yet been performed. Instead, toy studies are performed to assess the sensitivity of the measurement.

### 6.9.1 Rare mode

The muon mode has already been unblinded due to its comparably clean nature. In Fig. 6.9.1, the fit to the $B^+ \rightarrow K^+ \mu^+ \mu^-$ data for the entire Run 2, including the small dataset of 2015, is shown, with the model and the fits to simulation described in Section 6.5.6. The yields obtained from the fits to the muon mode are of no direct relevance for the measurement of $R_K$ as explained in Section 6.5.3, but are used for a branching fraction measurement of the decay channel.

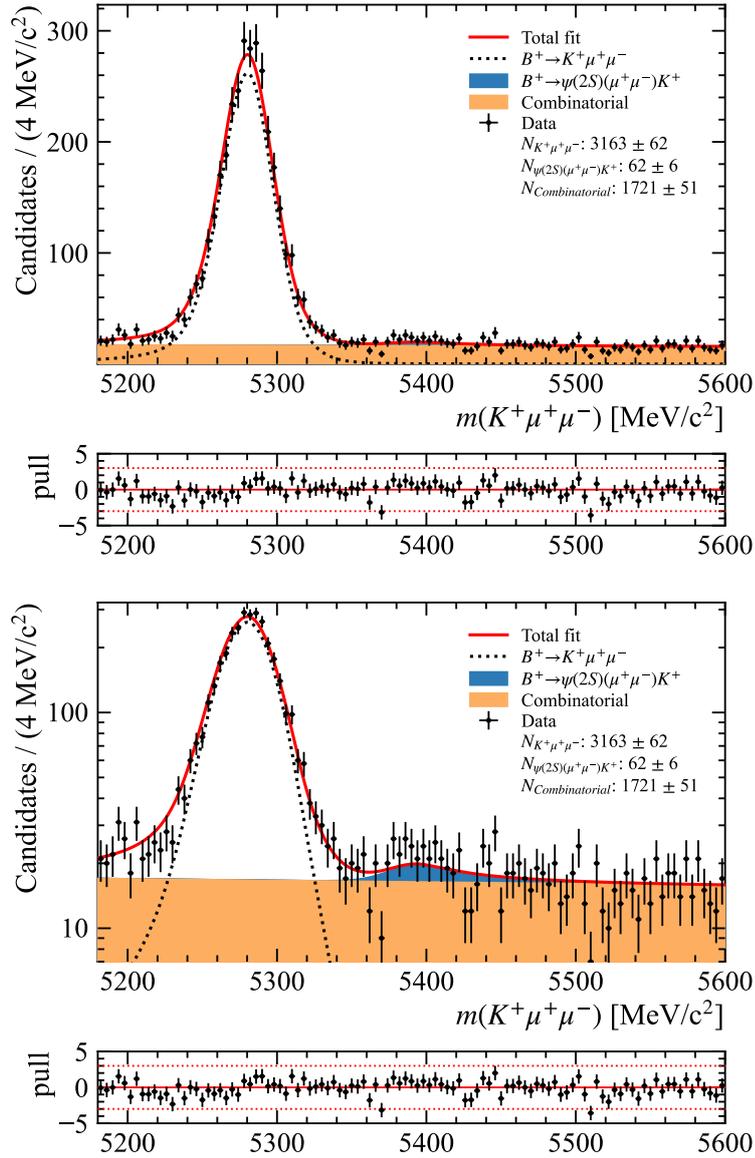

**Figure 6.9.1:** Fit to the $B^+ \to K^+\mu^+\mu^-$ data in the rare mode for the full Run 2 dataset in linear scale (top) and log scale (bottom). The data points are shown in black and the PDF is drawn in red. The different PDF components are filled with different colors and stacked.

## 6.9.2 Expected sensitivity on $R_K$

To assess the expected sensitivity of the measurement and study different scenarios, toy studies were performed using the model established in Section 6.5.5.

A sample is generated and fit back to test different scenarios of selection requirements and determine the expected precision of the measurement. The different contributions were modeled with KDEs, where no apparent signal model is present, which was otherwise used. The relative contribution was estimated using the efficiencies calculated in Section 6.4 and the respective branching fractions, with the electron yield as a relative baseline taken from the measurement of the muon mode, assuming LFU. The uncertainty of the measurement can be estimated from the width of the distribution of the fitted $R_K$ values, which is shown in Fig. 6.9.3.

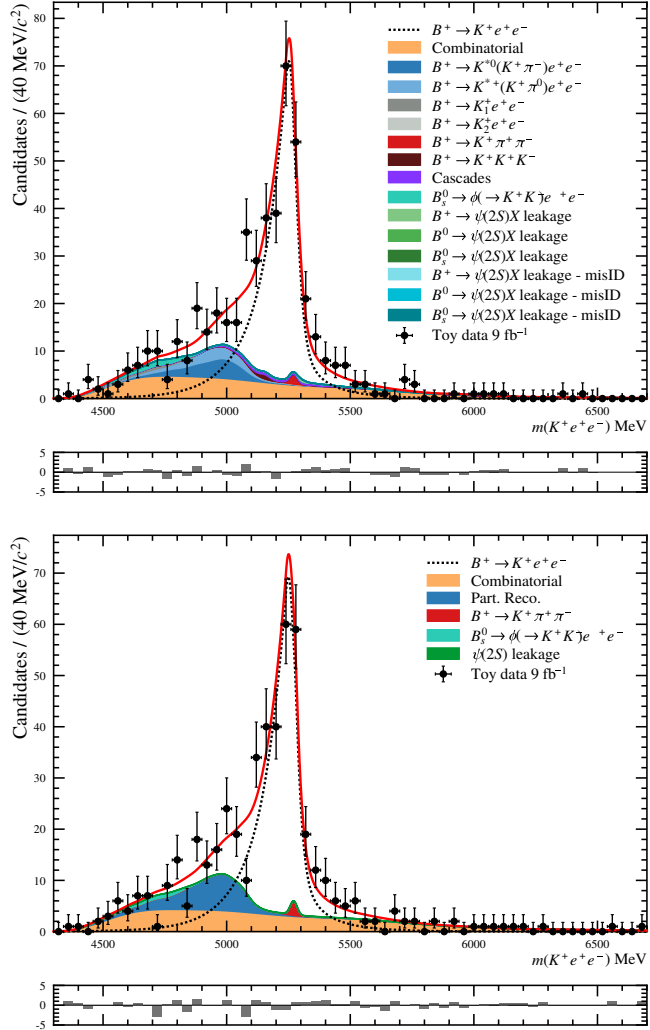

**Figure 6.9.2:** Toy data generated at the optimized selection overlaid with the PDFs used in generation (top) and fit with the nominal model (bottom).

The statistical uncertainty is estimated to be $\sigma_{R_K}^{(\text{stat})} = 0.0726$, which compares with the statistical uncertainty of the measurement in the central $q^2$ region [86] of $\sigma_{R_K}^{(\text{stat})} = +0.042/-0.041$. The expected number of events in both $q^2$ regions is similar and most efficiencies are similar, too, but due to the $q^2_{\text{track}}$ selection, about half of the events are lost in the high $q^2$ region, resulting in a larger statistical uncertainty. The systematic uncertainty, as discussed in Section 6.7, is estimated to be $\sigma_{R_K}^{(\text{syst})} = 0.031$, compared to the systematic uncertainty of $\sigma_{R_K}^{(\text{syst})} = \pm 0.022$ in the central $q^2$ region. Here, the high $q^2$ region is strongly dominated by the uncertainty on the combinatorial background, which is a minor contribution in the central $q^2$ region. Besides this difference, both analyses have comparable systematic uncertainties, with the misidentified backgrounds being the largest contribution in the central $q^2$ regions, respectively the second largest contribution in the high $q^2$ region. Combined, the total uncertainty is estimated to be $\sigma_{R_K} = 0.079$, compared to the total uncertainty of $\sigma_{R_K} = 0.047$ in the central $q^2$ region, resulting in a reduced sensitivity in the high $q^2$ region that is still comparable.

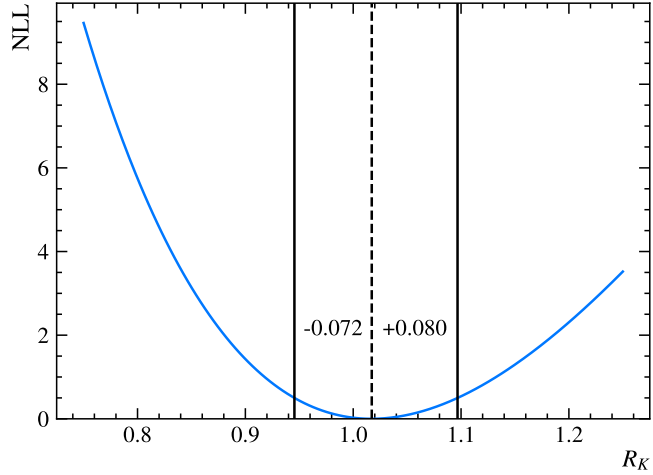

**Figure 6.9.3:** Example of a likelihood profile for $R_K$ obtained from a toy study.

# 7

# Scalable Pythonic Fitting

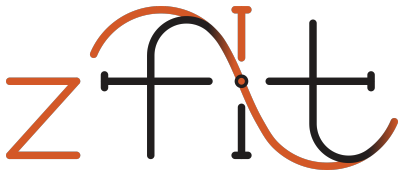

The last step of an analysis of LHCb data, and in HEP in general, often involves statistical inference. The approach most frequently used is to fit a probabilistic model to the data using the MLE described in Section 3.3.1. Likelihood-based fitting in HEP distinguishes itself substantially from other fields in the complexity and diversity of the fit models. This renders the use of generally available libraries impossible, resulting in fitting frameworks specific to HEP.

A few years ago, the most commonly used frameworks were RooFit, a general fitting framework, and, built on top of it, HistFactory. Both are part of Root, *the* large, monolithic HEP analysis toolbox. They are written in C++, the main programming language used by the LHC experiments. Due to its simpler syntax and the readily available scientific data analysis ecosystem, including deep learning libraries, Python has gained popularity globally and has slowly overtaken C++ as the main language used in HEP analyses. A major advantage of a Python- over a Root-centric analysis is that data scientists outside academia favor Python, resulting in a wide range of useful implementations. As a consequence, the HEP community can rely on industry-developed tools and focus its resources on adding tools for specific HEP needs. The organization Scikit-HEP [89] was founded in order to unite the effort to build a HEP specific Python ecosystem by coordinating the packages. For a long time, this ecosystem lacked a native Python tool for likelihood model fitting with advanced features commonly used in HEP, such as the performance to handle the large amounts of data from HEP experiments. zfit [98] was created and developed, primarily in Zurich and mainly by me, into a full-fledged library with the aim of filling this gap.

This chapter provides an overview of the library and its features. Section 7.1 introduces the needs of HEP likelihood fitting and the goals of zfit followed by a discussion of choosing the TensorFlow library as backend in Section 7.2. Section 7.3 describes the zfit application programming interface (API) and workflow design choices. The entire set of features of zfit is presented in Section 7.4 with selected example applications presented in Section 7.5. Finally, Section 7.6 summarizes the current state and outlines future plans for the library.

# 7.1 Introduction

Creating a model fitting library for HEP like ZFIT requires thorough planning and embedding into the existing ecosystem. First, the use-cases that the library is meant to handle need to be understood. With the wide variety of fits in HEP, the use cases can be categorized into different types, which is explained in Section 7.1.1. Second, existing libraries need to be examined and the place in the ecosystem established. These considerations are discussed in Section 7.1.2. And third, the guiding principles and the design philosophy that drive various decisions and ultimately characterize the library are elaborated in Section 7.1.3.

## 7.1.1 Types of fits

Fitting in HEP involves a variety of different applications that have their own complications and bottlenecks. All have in common that they use a likelihood estimator as introduced in Section 3.3. The main distinction appears in the model building stage with three relevant questions. Is the data binned or unbinned? Is the shape obtained from an analytic expression or taken as a template from a sample? And is the PDF normalized numerically or analytically?

The data stored on disk from the detector represents unbinned data. The size of these datasets can accumulate to petabytes, processed for end-user analysis in experiment-specific workflows. If the sample analyzed by the end-user is large enough, the data can also be represented by histograms. The PDFs that enter the likelihood are evaluated at data points that represent a measurement, with no addition for unbinned data. When the data is binned, *i.e.,* a histogram, the likelihood must also be binned: Instead of measuring a *single* data point, the number of data points that fall into a bin is measured. The counts of a bin follow a `Poisson` distribution, hence every bin count is treated as a single measurement being drawn from a `Poisson` distribution. The expected value of the `Poisson` distribution is the integral of the PDF over the bin.

The function representing the distribution or histogram is the unnormalized PDF. Finding a model that describes a specific population, or the whole dataset at once, is a problem-specific task. This shape may be an analytic expression derived from the expected outcome of the measurement. Sometimes, an analytical expression is however not available and instead templates can be obtained from non-tractable distributions, available through Monte-Carlo method based simulation sample (MC) generators, as described in Section 4.6. These templates can be used with nonparametric density estimation, such as a KDE or histograms, to create a PDF.

A PDF is a function with a normalization that is the integral over the whole domain of possible outcomes, see Def. 2. For the case of a binned template, the normalization is trivially obtained by summing over all the bins. For an analytic expression, the evaluation of the integral is sometimes analytically possible; otherwise the normalization has to be obtained numerically. The latter is a computationally expensive task, especially for multidimensional functions.

## 7.1.2 Fitting ecosystem

The PYTHON ecosystem, and especially SCIKIT-HEP, contain a variety of libraries used by ZFIT. ZFIT is associated with SCIKIT-HEP and developed in the same spirit ensuring maximum compatibility. In addition to ZFIT, the SCIKIT-HEP project provides other statistics packages frequently used in HEP. Their main synergies and differences are summarized in this section.

The IMINUIT [97] package mainly provides bindings to the C++ minimization library Minuit [108]. This thoroughly tested implementation is available in ZFIT, see Section 7.4.4. IMINUIT itself contains a limited amount of basic fitting tools, such as different cost functions. It lacks any form of models or parameters restricting its usage to pre-defined statistical inference functions. The library is very general, allowing great flexibility in terms of model building.

The PYHF [109, 110] library provides a PYTHON implementation of the HISTFACTORY [111] model, which previously existed only as a C++ implementation in the ROOT ecosystem. This library is restricted to fits of templated PDFs with no analytical model to binned data and only provides a very limited set of analytic PDFs for auxiliary measurements.

The HEPSTATS [95] library relies on models, minimizers, and loss functions provided by ZFIT for higher-level statistical inference. Its features include the provision of statistical tests, such as hypothesis testing, and the computation of limits, and implements an asymptotic likelihood-based approximation as well as toy generation.

## 7.1.3 Design philosophy

The design of ZFIT is based on two pillars: the API and workflow, and the backend. The API, the interface used by the user and other libraries to employ ZFIT, aims for an intuitive design that is stable over time, consistent within the framework, and flexible enough to fit into the ecosystem. To cover the wide variety of fits, the general philosophy is that "things should not be easy or hard, they should be consistent". This means that the API sometimes requires an additional one or two lines of code, but preserves a dozen lines in other places. The backend is the computational engine that performs the calculations. It must run fast, run on GPUs and CPUs, and be easy to use. Moreover, the goal is scalability: initial overhead in seconds and minutes is not important; hours and days of runtime are.

In order to fulfill its purpose as a HEP fitting library, ZFIT needs the ability to handle arbitrarily complex fits. Basic functionality has to be provided with the library, but it has to be built around the flexibility to allow for extensions and customization. This necessary flexibility is deeply incorporated into the design of ZFIT by providing a base class for every step of the workflow. These base classes are designed with inheritance in mind and provide well-defined hooks that allow an easy implementation of complicated objects by the user, where the heavy lifting and preprocessing is done by the base class.

## 7.2 Computational backend

PYTHON is an interpreted language that is flexible and easy to use. This comes at a great cost of speed for numerical algorithms that are *purely* implemented in PYTHON, contrary to a compiled language like C++. Merging the best of both worlds can be achieved using AOT compiled libraries and JIT compilation, as explained in Section 3.5.2. To achieve the goal of being on pair with C++ frameworks in terms of speed, ZFIT uses TENSORFLOW, a framework that is also widely used for deep learning tasks, as its computational backend. The complexity of fits covered by ZFIT needs advanced elements, including control flow and loops, which are generally not supported across other alternative backends. Implementing a common API for different backends is not feasible and would require large efforts and force the usage of the *minimal set of supported features* by all supported libraries. As a consequence, ZFIT is limited to TENSORFLOW for *optimized* computations, yet allows the usage of other libraries and even arbitrary functions that can be directly incorporated by TENSORFLOW.

TENSORFLOW [112] is an open source software library for numerical computations using data flow graphs developed by Google for deep learning applications. The two key techniques that significantly reduce the computation time in TENSORFLOW are *JIT compilation* and *automatic differentiation*; both are described in Section 3.5.2. In its versions 0.x and 1.x, when ZFIT first started as a library, TENSORFLOW relied on the definition of a computation graph, which is then compiled and executed. In addition to improved user-friendliness, the current version 2.x allows a more dynamic approach with JIT compilation instead of graphs, entirely fulfilling the needs for ZFIT.

In comparison with other deep learning frameworks, TENSORFLOW offers a variety of features that are useful for the HEP use-case:

- support of complex numbers necessary for analyses involving the computation of amplitudes in the model;

- possible JIT compilation on unknown shapes occurring in numerical methods;

- global variables to store model parameters or data;

- an existing rich ecosystem of libraries based on statistical methods;

- and the possibility for execution and inclusion of arbitrary code, including C++.

The TensorFlow function `numpy_function` allows to execute arbitrary Python code including functions. The execution is thus out of the hands of TensorFlow as the functions are called as a black-box. This prevents computational optimizations as well as the automatic calculation of gradients. The first is an expected price to pay[1] that affects performance, but the latter is conceptually problematic. As discussed in Section 3.4, gradients are usually needed in the maximization of likelihoods, with many minimizing algorithms relying on the gradient being provided as a function, as discussed in Section 7.3.5. To overcome this limitation, zfit provides a global switch, the `zfit.run.set_autograd_mode()` function, that disables the automatic gradient computation and instead uses a numerical method to compute the gradient and the Hessian matrix.

## 7.3 Workflow and API

The workflow of zfit, illustrated in Fig. 7.3.1, consists of five steps that are maximally independent and can be replaced by other parts from the Scikit-HEP ecosystem. This section introduces the workflow and provides an overview of the API design choices. This includes the variables, Section 7.3.1, the data object, Section 7.3.2, and PDF classes, Section 7.3.3. Using these building blocks, a loss function can be created as described in Section 7.3.4. The implementation of loss minimization is explained in Section 7.3.5. Finally, the fit result is obtained and introduced in Section 7.3.6.

### 7.3.1 Variables

In zfit, variables in the form of parameters, $\theta$, and observations, $x$, are handled in fundamentally different ways.

Parameters, representing $\theta$ in Eq. 3.3.1, are global objects with a unique name containing a scalar value and are designed to be modified by a minimizer. Parameters may have lower and upper bounds or a step size that can be used in the minimization, as described in Section 7.3.5. Parameters can also be composed from other parameters, as it is common to shift, scale, or sum parameters. Composed parameters cannot be minimized directly as they are a function of other parameters.

zfit observables, called *spaces*, are objects designed to specify one or multiple variables appearing in data and have no intrinsic value. *Spaces* have a – not necessarily unique – name and a lower and upper limit per dimension defining the range. A space may have a *binning* in order to represent a binned observable. The API concerning binned components follows closely the Universal Histogram Interface, a protocol containing a general definition of histograms and their axes, contained in a package that is part of Scikit-HEP, (UHI), adhering to its API definition. A binning can be defined by a lower and upper limit in combination with a number of bins or the explicit bin edges. Wherever a space is expected, a binning can be passed instead, which is converted to a space with the binning itself attached as an attribute.

---

[1] Custom code can principally also run faster than TensorFlow or be on pair with it.

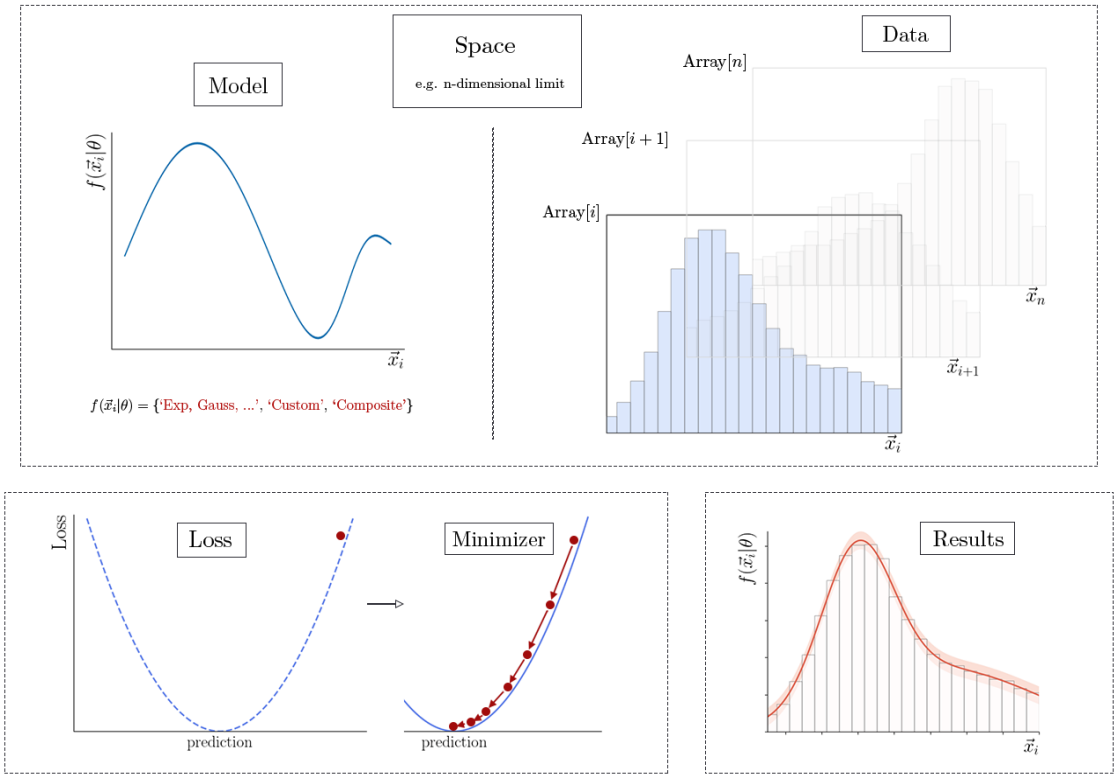

**Figure 7.3.1:** Five-stage workflow implemented in ZFIT. The first two steps represent the model building (top left) and data loading (top right); both communicate their axes via *Space* objects. Together, they form a loss function (bottom left), which is then minimized (bottom middle) to obtain the result (bottom right).

### 7.3.2 Data

Data handling in ZFIT is comparably light, thanks to existing libraries such as PANDAS [113], a widely used PYTHON tool for columnar data analysis. By design, no preprocessing is included, except for rectangular selection requirements implied by the limits of *Space*. The API for *Data* closely follows the corresponding libraries, PANDAS for unbinned data, and HIST [91] for binned data, including easy conversion to and from these libraries.

The need for an actual *Data* object arises from the JIT compilation feature of the computational backend, see Section 3.5.2 and Section 7.2. On the one hand, the JIT compilation prefers a pure TENSORFLOW array over an arbitrary PYTHON object. On the other hand, using PANDAS or HIST with their rich API is a much more user-friendly approach. The *Data* objects in ZFIT resolve these conflicting demands by providing an API that resembles the corresponding libraries but converts to a TENSORFLOW array internally, allowing them to be passed into the JIT compilation.

The *Data* class `Data` handles unbinned data by tightly wrapping around the TENSORFLOW array and providing a name for each of the axes. This is achieved through a *Space*, described in Section 7.3.1, that is passed to the constructor of the `Data` object, thus assigning a name to each axis. The lower and upper limits of the *Space* serve as rectangular selections. Additionally, the `Data` object can be weighted by providing an array of weights, which can be used in the loss function described in Section 7.3.4.

Binned data is handled by the *Data* class `BinnedData` and implements the UHI methods. It provides a constructor that takes a HIST object that is internally converted to a TENSORFLOW array, a constructor that takes the raw values of the counts and the edges of the bins, and a constructor that takes a `Data` object and a binning. The latter produces a histogram of the data in a TENSORFLOW JIT compilable way, allowing to convert the `Data` object to a `BinnedData` object *internal* of a PDF.

### 7.3.3 Model building

The largest part of ZFIT concerns the model building with PDFs. To cover all use cases, ZFIT provides three different types of PDFs: out-of-the-box PDFs that represent mostly basic shapes, PDFs that are built from other PDFs and custom PDFs that can be defined by the user. They exist as binned or unbinned PDFs, which can be converted into each other. A PDF can be extended, in which case it has a *yield* parameter that provides an absolute scaling and *enables additional functionality*. The contract is that any extended PDF can be used instead of a nonextended PDF while the opposite is not guaranteed to hold.

The base class of all PDFs is `BasePDF`, which provides preprocessing, fallback defaults and a convenient interface to the user. In general, a PDF requires for its instantiation the following.

**obs** The observables as a *Space* of the PDF. They communicate the axes of the PDF *by order*, which has to be defined in the documentation of the PDF. By default, the `obs` are also taken as the normalization range of the PDF.

**params** The parameters of the PDF, which are `Parameter`-like objects or scalars. Each pdf is assumed to have a parameterization by name, *i.e.,* a `Gaussian` has a *mu* and a *sigma* parameter, and *params* provides a mapping from the name to the actual parameter. In general, the idea of a subclass is to request the name of the parameter in the constructor as a pure PYTHON argument, but there can be cases, *i.e.,* sums of PDFs, where the parameters are naturally provided differently.

**pdfs** Optional argument for composite PDFs that depend on other PDFs.

**extended** Special parameter that *can* be `Parameter`-like, but can also be a boolean: If `True`, the PDF is automatically extended defined by the normalization, if `False`, the PDF is not extended.

**norm** Normalization range of the PDF, provided as a *Space*.

A ᴢꜰɪᴛ PDF provides the following methods.

**pdf** The probability density function, which is the main method of a PDF. It takes the data as the first argument and returns the value of the PDF at the data points. As an additional argument, the normalization range can be given, which can be `False` to disable normalization.

**sample** Samples from the PDF and returns a ᴢꜰɪᴛ *Data* object. The base class provides a default implementation for numerical sampling in case no analytic inverse integral is provided, which would allow for analytical, fast sampling.

**integrate** Integrates the PDF over the given range. Either the base class uses an analytic integral, if provided, or uses numerical integration as a fallback.

**get_yield** Returns the yield of the PDF, an expected number of events. It is used in extended likelihoods, as explained in Section 3.3.2

**get_params** Returns the parameters of the PDF as a *set*. By default, this extracts all free parameters from the PDF.

Both methods **pdf** and **integrate** have an extended version starting with `ext_` that returns values scaled by the yield.

Binned PDFs have additional methods that follow closely the UHI API.

**counts** Returns the counts of the PDF in the bins of the given *Space* and is only available for extended PDFs.

**rel_counts** Returns the relative counts of the PDF, that is, the counts normalized to the total counts.

### 7.3.4 Loss

The general loss function in ᴢꜰɪᴛ takes the PDF and the data, encapsulating the computational aspects of the evaluation and likelihood computation. As previously explained in Section 3.3, there are multiple advanced requirements for HEP fitting, including constraints and simultaneous fits. A key aspect of the design in ᴢꜰɪᴛ is to make the loss its own part of the workflow to incorporate these features. The constructor of a loss function takes the following arguments.

**pdf** The PDF to be fitted to the data or a list thereof for simultaneous fits.

**data** The data to be fitted to the PDF or a list thereof for simultaneous fits.

**constraints** Optional constraint terms to be added to the loss function. Can be a list of multiple *constraints*, a special class of PDFs in ZFIT.

Giving a collection of data and PDFs, both of equal length, to the constructor seamlessly incorporates the ability to perform simultaneous fits. Equivalently, a sum of multiple losses also results in a simultaneous likelihood.

The design decision to make the loss an essential part of the workflow has some explicit advantages, as opposed to using a product of uncorrelated, multidimensional PDFs. First, the loss function can be used to perform statistical tests independently of other objects, as it contains the data and the PDF. It even allows to have a loss function without a PDF. Second, simultaneous fits are a natural extension of the loss function and do not require a special dataset: In general, the dataset fit to different PDFs have a different number of points or can even be different in their nature, *e.g.,* a binned and an unbinned dataset. Third, this allows for optimizations of the loss function, which can be determined by the loss itself. The downside of this approach compared to using a product of PDFs is the generation of data: Sampling from a PDF is a natural operation in a PDF, while sampling from a loss function is not.

A loss has the following methods.

**value** This evaluates the loss function and returns the value, a scalar. Optionally, parameter values can be given as a dictionary or a list of values.

**gradient** Calculates the gradient of the loss function with respect to the parameters. The implementation used depends on the backend and current mode as explained in Section 7.2.

**hessian** Calculates the Hessian matrix of the loss function with respect to the parameters. As for the gradient, the implementation depends on the backend and mode.

**Combinations** A loss also contains the methods `value_gradient` and `value_gradient_hessian`, which combine the evaluation of the loss function with the computation of the gradient and the Hessian matrix. These methods are more efficient than calling the various methods separately.

### 7.3.5 Minimization

The minimization in ZFIT relies on algorithms provided by other libraries and provides a unified interface to the user. The chosen minimizer API agrees closely with some libraries, while fundamentally disagrees with others.

The design is driven by three requirements.

**Stateless** The minimizer should be stateless, meaning that it should not store any information about the minimization that happened.

**Minimization is universal** The API of the minimization *must not depend* on the method of minimization itself. A single method `minimize` with a signature that depends solely on the function and the information associated with it should suffice[2].

**Minimizers are not universal** There are a plethora of minimization algorithms, some use gradients, some are based on trustregion and so forth. Therefore, the configuration between minimizer algorithms is fundamentally different. A distinct API for the configuration of each algorithm is needed, as the combinatorial possibilities of the configuration are too large to be handled by a single API.

Technically, the last two points *could* be merged into a single API using composition, *e.g.,* by passing a minimizer or configuration object to the `minimize` function, which practically would result in a similar API. The ZFIT design of minimizers incorporates these requirements by making each minimizer a class with its own constructor and a `minimize` method with the same signature for all minimizers. Changing the minimizing algorithm is as simple as changing the class of the minimizer object *without* any changes at the place where the minimization occurs. This is also essential for libraries that build on top of ZFIT, which can use any minimizer without having to know the details of the minimizer algorithm.

The API of the minimizers constructor is as arbitrary as the minimizer algorithm itself. However, some keywords that are common in different minimizers have been standardized:

**tol** The tolerance of the minimizer. The minimization process stops once the convergence criterion is smaller than *tol*.

**verbosity** The amount of information that is printed during the minimization. Values range from 0 to 10, where 0 is no information and 10 is the most verbose setting, printing every step of the minimization.

**maxiter** The maximum number of iterations that the minimizer is allowed to perform. This is an approximate number and serves as a measure to prevent minimizers from running too long.

Additionally, minimizers can take a strategy and a convergence criterion. A strategy is a minimizer-agnostic object that serves as a callback, *i.e.,* it is called at every step of the minimization, thus allowing to change the minimization process arbitrarily. Furthermore, it provides builtin ways to catch and handle issues during minimization, such as the occurrence of *NaN*s. The convergence criterion is evaluated whenever the minimizer checks for convergence.

The `minimize` method is the only method that is required to be implemented by a minimizer. It takes the following three arguments when called.

---

[2]This is not as clean as it seems, as local minimizers need initial values, which global minimizers, although rarely ever used in likelihood fitting, do not.

**loss** The objective function to minimize. This should be a ZFIT loss but can also be an arbitrary function, as long as it returns a scalar.

**params** An optional argument when a ZFIT loss is used that specifies the parameters to minimize. If none are provided, all parameters with floating set to true and on which the loss depends are minimized.

**init** Initial information for the minimization, in the form of a `FitResult`. This serves as an option to warm-start the minimization, as the minimizer can use all the information contained in the `FitResult` object, which will be discussed in Section 7.3.6.

As a shortcut, a `FitResult` can be given as a single first argument, in which case the loss, parameters and any other information are taken from it. This allows to chain minimizers by passing the `FitResult` of the previous minimization as the first argument to the next minimizer.

```
globalminimizer = GlobalMinimizer(tol=1)
localminimizer = LocalMinimizer(tol=0.001)
resultglob = globalminimizer.minimize(loss)
resultloc = localminimizer.minimize(resultglob)
```

The chain of minimization can be wrapped within a single call to `minimize`, thereby having the best of both, the stateless and the stateful, worlds.

The parameters, whether taken from the loss, from the `FitResult` or explicitly given, are used to determine the starting values of the minimization. If limits are given, they are used to constrain the minimization within the boundaries if the minimizer supports it, otherwise they are enforced through clipping. As most minimizers rely on an initial step size, which can help determine the order of magnitude of change to expect, this is taken from the parameters *if available*, otherwise a default value is used.

### 7.3.6 Result

The result of a minimization is a `FitResult` object, which contains information about the minimization and the result. Its main purpose is to store the values of the parameters at the minimum and to evaluate some basic uncertainties.

The `FitResult` object contains the following information as attributes.

**converged** A boolean that indicates whether the minimization converged.

**valid** A boolean that indicates whether the result is valid or not. A valid fit implies convergence, yet it also covers a variety of other sanity checks, amongst others, whether parameters are at their limits.

**fmin** The value of the loss function at the minimum.

**edm** The estimated distance to the minimum, the stopping criterion of the minimizer.

**params_at_limit** A boolean that indicates whether any of the parameters are at their limits.

**params** Mapping with the parameter or their names as keys and all the information belonging to it, from the optimum to uncertainties, as values.

Furthermore, the `FitResult` provides access to the loss function and the minimizer. As the information can greatly vary from minimizer to minimizer, the `FitResult` object unifies some of the information but also provides a `info` attribute that contains additional, algorithm-specific information from the minimizer.

The `FitResult` provides two methods to evaluate the uncertainties of the parameters, `hesse` and `errors`. The `hesse` method uses the Hessian matrix to approximate the local curvature of the loss function, which is usually quick, especially if the Hessian matrix is already available from the minimization itself. It provides symmetric uncertainties that are, in general, valid for large enough datasets. The `errors` method uses a profile likelihood scan to estimate the uncertainties, thus providing asymmetric uncertainties. This method is more accurate for nonhyperbolic minima, but also more expensive, as it requires multiple minimizations; it returns two objects, where the second value *can* be a new `FitResult` object if a new minimum was found.

## 7.4 Features

The API definition and backend choice enable the actually implemented features of ZFIT. This section will give an overview of the most important features and outline the rationale behind them. Many more features are hidden in details or as additional options and are available in the ZFIT documentation. Usage examples of a variety of features can be found in Section 7.5.

### 7.4.1 Data handling

Data objects in ZFIT are by design minimal. One feature of the `Data` class is the possibility of converting it to a `BinnedData` object through the `to_binned` method. This takes either a binning or a number of bins as arguments, allowing for a quick conversion. The `BinnedData` class, following the UHI, can directly be plotted with MPLHEP [94] using the `histplot` method. `BinnedData` objects can also be converted to a `Hist` class instance from HIST [91], which contains multiple plotting methods, through the `to_hist` method, .

Access to the unbinned values in a numpy-like array is provided in two ways. Once through the `value` method, which returns a $(N, d)$ shaped object, where $N$ is the number of events and $d$ the dimension of the data. Alternatively, the columns of the `Data` object can be accessed in the same manner as a Pandas `DataFrame` object, indexing with the column name.

There is a special implementation of `Data`, a `Sampler`. This class is returned from the `create_sampler` method from a PDF and has the same interface as the `Data` class. The `Sampler` contains a reference to the PDF and carries an internal state, the sampled values, which can be updated through the `resample` method. This allows for a quick resampling *inplace* without the need to rebuild a JIT compiled loss function. The main use case appears in toy studies, where data is repeatedly sampled from a PDF and then fitted with the same PDF, possibly under different sampling conditions.

### 7.4.2 PDFs

The first type of PDFs are basic, analytic functions. They are the backbone of ZFIT, where each PDF represents a certain shape without depending on other PDFs. They include `Gaussian`, `Poisson`, `exponential`, CB PDFs and more. A variety of polynomial PDFs are provided as well, including the `Legendre`, `Chebyshev` and `Hermite` polynomials among others. ZFIT also provides a Wrapper class `WrapDistribution` that allows wrapping any TensorFlow-Probability distribution, converting it into a ZFIT PDF.

Nonparametric, unbinned models are implemented through a KDE implementation, where every point corresponds to the mean of a kernel. The selection of KDEs is extended with a binned versions, a version based on fast Fourier transform (FFT) and an implementation based on ISJ, all can handle large template datasets with ease and are explained in more detail in Ref. [114]. The KDEs support different bandwidth estimators, with global and adaptive bandwidths, different kernels, mirroring at the boundaries, and weighted samples. Binned, nonparametric models are implemented through a `HistogramPDF` class, which mimics a histogram. They can also handle unbinned data and evaluate the PDF at any point.

The second type of PDFs in ZFIT are a composition of one or more other PDFs.

**Mixture** The `SumPDF` and its binned version `BinnedSumPDF` are a weighted sum of PDFs. If all PDFs that are summed are extended, the sum is also extended.

**Product** The `ProductPDF` calculates the product of PDFs. These can be in the same or in different observables. If the observables are different, the PDFs are assumed to factorize, allowing for internal optimizations.

**Convolution** The `FFTConvPDFV1` is a convolution of two PDFs, assuming that the kernel PDF is *independent* of the variable that is convolved. This performs a numerical convolution using FFT and a linear interpolation of the resulting grid.

Additionally, a variety of binned composite PDFs are implemented.

**Morphing** The `SplineMorphingPDF` is a binned PDF that interpolates between a set of templates using a parameter $\alpha$. This procedure is also known as morphing and is used to model systematic uncertainties.

**Scale modifier** The `BinwiseScaleModifier` is a binned PDF that multiplies every bin in a histogram PDF with a parameter. Together with a constraint obtained from a template, these scale parameters can encode systematic uncertainties; this procedure is also known as the Barlow-Beeston or Barlow-Beeston light method [115], depending on whether the constraint is per template and bin, or per bin only.

The conversion from binned to unbinned models and vice versa are also composite models. Binned models can be obtained from any unbinned model through the `to_binned` method, which takes a binned *Space* as input. The inverse is possible through the `to_unbinned` method, which takes a binned PDF as input. Converting a binned PDF to an unbinned PDF can also be done through spline interpolation between the bin centers, which is implemented in the `SplinePDF` class.

The third type of PDFs is made from the `CustomPDF` class, which allows to create a custom PDF with an arbitrary shape, dimensions and parameters. All that is needed, as a minimum, is to implement the `_pdf` method, which takes the data as input and returns the unnormalized PDF value. The method has to be decorated with the `supports` decorator, which specifies, through the keyword `norm`, whether the PDF is already normalized. The default case is `False`, which means that ZFIT takes care of the normalization behind the scenes. Using `True` means that ZFIT will not normalize the PDF and the user has to take care of it, also allowing for special cases where the normalization is done in unconventional ways. The function can be implemented in TENSORFLOW or PYTHON, allowing for flexibility, as explained in Section 7.2.

Additionally, `register_analytic_integral` can be used to register an analytic integral for the normalization; if no analytic integral is registered, the normalization is performed numerically by ZFIT. Further customizations are possible by implementing the `_sample` or `_integrate` methods, which are used for sampling and integration, respectively.

For a simplified use case and to lower the entry barrier, ZFIT provides the `ZPDF` class, which also takes care of the constructor that was described in Section 7.3.3 The list of parameters is defined via the `_PARAMS` class attribute and the number of observables in the `_N_OBS` class attribute. An example of a custom PDF is shown in Section 7.5.5.

### 7.4.3 Loss functions

The implemented loss functions are all likelihood-based and differ in the way in which they are constructed by requiring binned or unbinned data, and a corresponding PDF. This enables to keep the PDF and data as attributes accessible, which can be used by libraries like HEPSTATS to perform statistical tests. Furthermore, they exist in two variants, one that is *extended* and one that is not. An *extended* loss takes into account the total number of events in the data and the yield of the PDF, as described in Section 3.3, by multiplying the likelihood with a `Poisson` term containing the total number of events and the yield.

**UnbinnedNLL, ExtendedUnbinnedNLL** The basic negative log-likelihood as introduced in Section 3.3 for unbinned data and an unbinned PDF. If the dataset is weighted, the probability density of each event is taken to the power of the weight.

**BinnedNLL, ExtendedBinnedNLL** The negative log-likelihood for binned data and a binned PDF, which is implemented using a `Poisson` distribution. The uncertainties on the frequency weights from the histograms are taken into account using scaled `Poisson` distributions from [116].

**BinnedChi2, ExtendedBinnedChi2** The $\chi^2$ loss function for binned data and a binned PDF. It uses either the uncertainty of the histogram or a `Poisson` uncertainty from the PDF as the denominator of the $\chi^2$. The $\chi^2$ is a special case of a NLL with the assumption that each measurement is independent and normally distributed with a mean of the binned PDF value and a standard deviation given by the uncertainty.

These likelihoods use internal optimizations to enhance numerical stability and speed.

To provide the user with the possibility to create custom loss functions, ZFIT has the `SimpleLoss` class. This takes a function and the corresponding parameters that it depends on. The cost of flexibility given by this approach is that no PDF or data attributes are available, allowing only for a limited set of statistical tests by HEPSTATS.

### 7.4.4 Minimization

ZFIT offers a variety of implemented minimizers that use different algorithms. A minimizer has to decide what the parameter values are and only interacts with the loss function via the `value` method and, depending on the algorithm, the `gradient` method. This decouples the minimizer from the computational backend TENSORFLOW. An issue arises regarding the stopping criteria, which differs between minimizers algorithms. Typical criteria include an absolute or relative tolerance of the gradient, the value, or the parameters, or combinations of these tolerances. In a variety of tested examples, it was found that the most reliable metric is the estimated distance to minimum (EDM), the default stopping criterion in IMINUIT.

The EDM is defined as

$$\text{EDM} = g^T \cdot H^{-1} \cdot g \tag{7.1}$$

where $g$ is the gradient and $H$ the Hessian matrix. It estimates the remaining vertical distance to the minimum, assuming a quadratic behavior of the function around the minimum as explained in Section 3.4.2. The problem with this stopping criteria is that the Hessian matrix is extremely expensive to compute exactly. IMINUIT and similar quasi-Newton methods carry an internal approximation of the Hessian matrix, making the calculation of the EDM computationally feasible. However, not all minimizers rely on a Hessian matrix or the estimated Hessian matrix cannot be accessed in the implementation. The advantage of using this criterion uniformly accros all minimizers, despite its possible high computational cost, is that minimizers can be directly compared in terms of steps needed. As explained in Section 7.3.5, an arbitrary stopping criteria can be given instead of the EDM. In the following, the different minimizers libraries are described. TENSORFLOW offers minimizers, which are well suited for the task of deep learning training but not suitable for the task of MLE, as explained in Section 3.4, and are therefore not included in ZFIT.

**iminuit** The IMINUIT library [97] provides bindings to a C++ library for minimization and wrapped within ZFIT as `Minuit`. Its strength lies in its robustness, especially for the case of low statistics.

**SciPy** The SCIPY package offers a variety of algorithms for minimization, including quasi-Newton methods, conjugate gradient methods, and trustregion minimizers. They are implemented in ZFIT as `ScipyName`, with "Name" the name of the minimizer algorithm.

**NLopt** The NLOPT is a library for nonlinear optimization with many bindings to other languages. Like SCIPY, it offers a variety of algorithms that are available as `NLoptName`.

**ipyopt** The IPYOPT PYTHON package is a binding to the `Ipopt` library [117], a large-scale nonlinear optimization library. This is usually used to optimize differential equations for finite element calculations, yet shows promising results to minimize problems with many, well-behaved parameters.

### 7.4.5 FitResult

The result of a minimization is returned as a `FitResult` object, which was introduced in Section 7.3.6. Most of the information that the `FitResult` holds is taken from the output of the minimizer, with a few calculated by ZFIT.

The `FitResult` provides two methods to evaluate the uncertainties of the parameters, `hesse` and `errors`. Both methods can either use the methods provided by the IMINUIT library, which can be used independently of the minimizer that performed the minimization, or use custom implementations in ZFIT. A third option for `hesse` is an approximate version that uses the internal Hessian matrix of the minimizer. The `hesse` method implements the *asymptotically correct* form of uncertainty for unbinned weighted datasets [105]. This requires the Jacobian matrix of the loss function *w.r.t.* the parameters, which is calculated using the automatic differentiation of TENSORFLOW or the *jacobi* [118] library for numerical differentiation, as explained in Section 7.2.

The `FitResult` object has a beautified representation for enhanced human readability. An example is shown in Fig. 7.4.1. It starts at the top with the information about the loss function, here an unbinned negative log-likelihood, and the minimizer used, here `Minuit`. Below is the information about the minimum, including the overall validity of the result, the convergence status, the EDM, whether any of the parameters is at its limit and the approximate value of the loss function at the minimum, once the full loss and once the optimized, internal version. Finally, the parameters listed at the bottom with

```
FitResult of
<UnbinnedNLL model=[<zfit.<class 'zfit.models.functor.SumPDF'>
params=[Composed_autoparam_0, fraction]] data=[<zfit.core.data.Sampler object at
0x7fd40c0b4790>] constraints=[]>
with
<Minuit Minuit tol=0.001>
```

| valid | converged | param at limit | edm | approx. fmin (full | internal) |
|-------|-----------|----------------|---------|---------------------|
| True | True | False | 8.3e-06 | 2782.91 \| 9885.221 |

```
Parameters
name          value  (rounded)        hesse        approx                     errors    at limit
--------  ------------------  -----------  -----------  ------------------  ----------
fraction            0.334286  +/-   0.022  +/-   0.022  -   0.022  +  0.022       False
mu                  0.942148  +/-   0.081  +/-   0.081  -   0.081  +  0.081       False
sigma                 1.0542  +/-    0.08  +/-    0.08  -   0.076  +  0.085       False
lambda            -0.0734167  +/-  0.0076  +/-  0.0076  -  0.0077  +  0.0076      False
```

**Figure 7.4.1:** Representation of a `FitResult` object.

the name, the value at the minimum and the different uncertainties. In this example, the `hesse`, `approx` and `errors` were calculated. The latter shows asymmetric uncertainties, which deviate from the `hesse` estimate in the parameter *sigma*. In the last column, an indication is given of whether the parameter is at its limit.

## 7.5 Fitting examples

Together, all features allow ZFIT to perform a variety of different fits. Through the possibility of various compositions of PDFs and simultaneous likelihoods, arbitrarily complicated models can be created. In the following, a few examples are shown that depict the typical usage of ZFIT.

### 7.5.1 Gaussian

A simple example is a fit to a Gaussian distribution. It serves as a minimal example to show the basic usage of ZFIT and demonstrate the API and workflow. First, the observables are created, a `Space` with the name `'x'` and the limits (`-5, 5`). Then, the parameters are instantiated, `mu` and `sigma`, with their initial values and limits. The `Gauss` model is created with the parameters and the space. The data is loaded from a NumPy array `data_np`, and the `obs` specifies the observable. With the model and the data, the NLL is created using the `UnbinnedNLL`. To perform an MLE, the `Minuit` minimizer is used. Finally, an uncertainty estimate is performed with the `hesse` method.

```
obs = zfit.Space('x', limits=(-5, 5))
mu = zfit.Parameter('mu', 0.5, -4, 4)
sigma = zfit.Parameter('sigma', 1.2, 0, 10)
model = zfit.pdf.Gauss(mu=mu, sigma=sigma, obs=obs)

data = zfit.Data.from_numpy(obs=obs, array=data_np)

nll = zfit.loss.UnbinnedNLL(model=model, data=data)

minimizer = zfit.minimize.Minuit()
result = minimizer.minimize(nll)

result.hesse()
```

For more elaborate examples, the workflow remains the same, some steps just become more complex.

### 7.5.2 Unbinned mass

A typical example in HEP is a fit to a mass distribution, *e.g.,* the invariant mass of a combination of particles. The first difference from the previous example is the model, which consists of a signal and a background component, a `Gaussian` and an exponential, respectively.

```
lam = zfit.Parameter('lambda', -0.02, -1, -0.001)
n_sig = zfit.Parameter('n_sig', 1000, 0, 10000)
n_bkg = zfit.Parameter('n_bkg', 300, 0, 4000)

signal = zfit.pdf.Gauss(mu=mu, sigma=sigma, obs=obs, extended=n_sig)
background = zfit.pdf.Exponential(lam, obs=obs, extended=n_bkg)
model = zfit.pdf.SumPDF([signal, background])
```

Furthermore, both PDFs are extended by a yield, and a `ExtendedUnbinnedNLL` is used to take the number of events into account.

```
nll = zfit.loss.ExtendedUnbinnedNLL(model=model, data=data)
```

The remaining part of the workflow is the same as in the previous example. This highlights the strength of the ZFIT workflow, since the only change was with respect to the PDF and the loss.

A plot of the result using native PYTHON plotting libraries is shown in Fig. 7.5.1. For a more elaborate example, mass fits from the $R_K$ analysis in Chapter 6 show fits to simulation and data from the LHCb experiment in Section 6.5 and in Section 6.6, respectively.

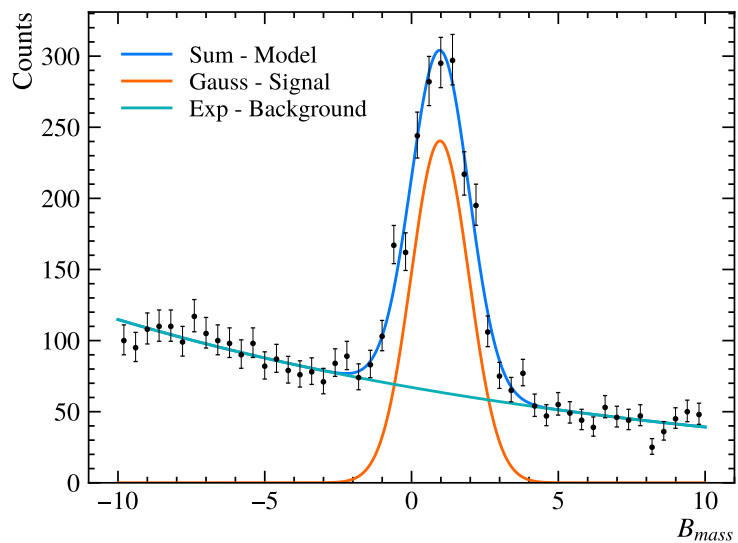

**Figure 7.5.1:** Example plot of an unbinned fit to the invariant mass of a $B$ particle. The fit model is a mixture (blue) of a `Gaussian` (orange) and an exponential PDF (green). The data is shown in black.

### 7.5.3 Binned mass

Starting from the previous example, only minor changes are needed to perform a binned fit: the data and the model need to be binned, and the loss needs to be changed to an `ExtendedBinnedNLL`. The binned observable is created with the same limits as the unbinned one, but with a number of bins.

```
binned_obs = obs.with_bins(100)
```

The data and model are created with the binned observable.

```
bin_data = data.to_binned(binned_obs)
bin_model = model.to_binned(binned_obs)
```

The loss takes the binned model and data.

```
bin_nll = zfit.loss.ExtendedBinnedNLL(model=bin_model, data=bin_data)
```

The rest of the workflow remains the same, explicitly shown again for illustrative purposes.

```
minimizer = zfit.minimize.Minuit()
result = minimizer.minimize(bin_nll)
result.hesse()
```

The result of the fit is shown in Fig. 7.5.2 and can be compared with the previous, unbinned example in Fig. 7.5.1.

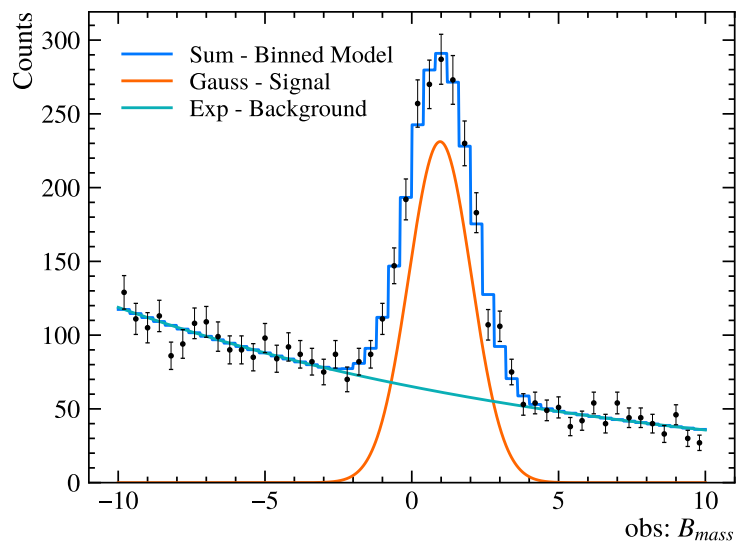

**Figure 7.5.2:** Example plot of a binned mass fit, where the mixture of the unbinned PDFs has been binned. The fit model is a binned mixture (blue) of a `Gaussian` (orange) and an exponential PDF (green). The data is shown in black.

### 7.5.4 Binned template

This example uses templates from simulation instead of analytic shapes as before. The template PDF is created from a ʜɪsᴛ histogram `template_h` together with two more templates, `template_p1s_h` and `template_m1s_h`, which represent the positive, $+1\sigma$ and negative $-1\sigma$ systematic uncertainty, arising from an uncertainty in the simulation. The parameter that causes the variations is encoded as `alpha` parameter and is `Gaussian` constrained.

Note that the `HistogramPDF` can take any ʜɪsᴛ histogram *directly* as input, without the need to convert it to a ᴢғɪᴛ `BinnedData` object first. It also automatically creates the corresponding `Space` that is retrieved in the example with the `space` attribute.

```
nominal = zfit.pdf.HistogramPDF(template_h)
obs_binned = nominal.space
p1s = zfit.pdf.HistogramPDF(template_p1s_h)
m1s = zfit.pdf.HistogramPDF(template_m1s_h)

alpha = zfit.Parameter('alpha', 0, -10, 10)
```

The histogram PDFs are then morphed with the `SplineMorphingPDF` class, which expects a mapping of the uncertainty to the corresponding template. A `Gaussian` constraint is created using the `alpha` parameter, which is then passed to the `ExtendedUnbinnedNLL`.

```
morphed = zfit.pdf.SplineMorphingPDF({0: nominal, 1: p1s, -1: m1s}, alpha)
constraints = zfit.constraint.GaussianConstraint(mu=alpha, sigma=1)
nll = zfit.loss.ExtendedUnbinnedNLL(model=morphed, data=data,
                    constraints=constraints)
```

The rest of the workflow is the same as in the previous example.

### 7.5.5 Angular

So far, the examples used PDFs provided by ZFIT. As discussed in Section 7.4.2, a core feature of ZFIT is the ability to create custom PDFs. This is demonstrated in the following example with a fit to an angular distribution of the decay $B^0 \to K^{*0}\ell^+\ell^-$, described by three angles $\theta_l$, $\theta_k$ and $\phi$. The mathematical formulation is taken from Ref. [119], Eq. (4). Inside the PDF the ZFIT backend is used through `zfit.z.numpy`, aliased as `znp`. A shortened, conceptualized version is shown here; the full PDF implementation can be found in Section A.5.

```python
import zfit.z.numpy as znp

class AngularPDF(zfit.pdf.ZPDF):
    _PARAMS = ['FL', 'S3', 'S4', 'S5', 'AFB', 'S7', 'S8', 'S9']
    _N_OBS = 3

    @zfit.supports()
    def _pdf(self, x):
        FL = self.params['FL']
        S3 = self.params['S3']
        ...

        costheta_l, costheta_k, phi = z.unstack_x(x)

        sintheta_k = znp.sqrt(1.0 - costheta_k * costheta_k)
        sintheta_l = znp.sqrt(1.0 - costheta_l * costheta_l)
        ...

        pdf = ((3.0 / 4.0) * (1.0 - FL) * sintheta_2k +
        FL * costheta_k * costheta_k +
        (1.0 / 4.0) * (1.0 - FL) * sintheta_2k * cos2theta_l +
        ...
        S9 * sintheta_2k * sintheta_2l * znp.sin(2.0 * phi))

        return pdf
```

Using this PDF is already sufficient to perform a full fit. To assess the sensitivity of fit parameters under different scenarios, toy studies can be performed, whereby the PDF is sampled and fitted to the resulting data. Sampling is performed with the `sample` method, which takes the number of samples as argument, in the example 1000.

```python
data = pdf.sample(1_000)
```

With the sampled data, a fit can be performed as before. The fitted PDF is shown in Fig. 7.5.3.

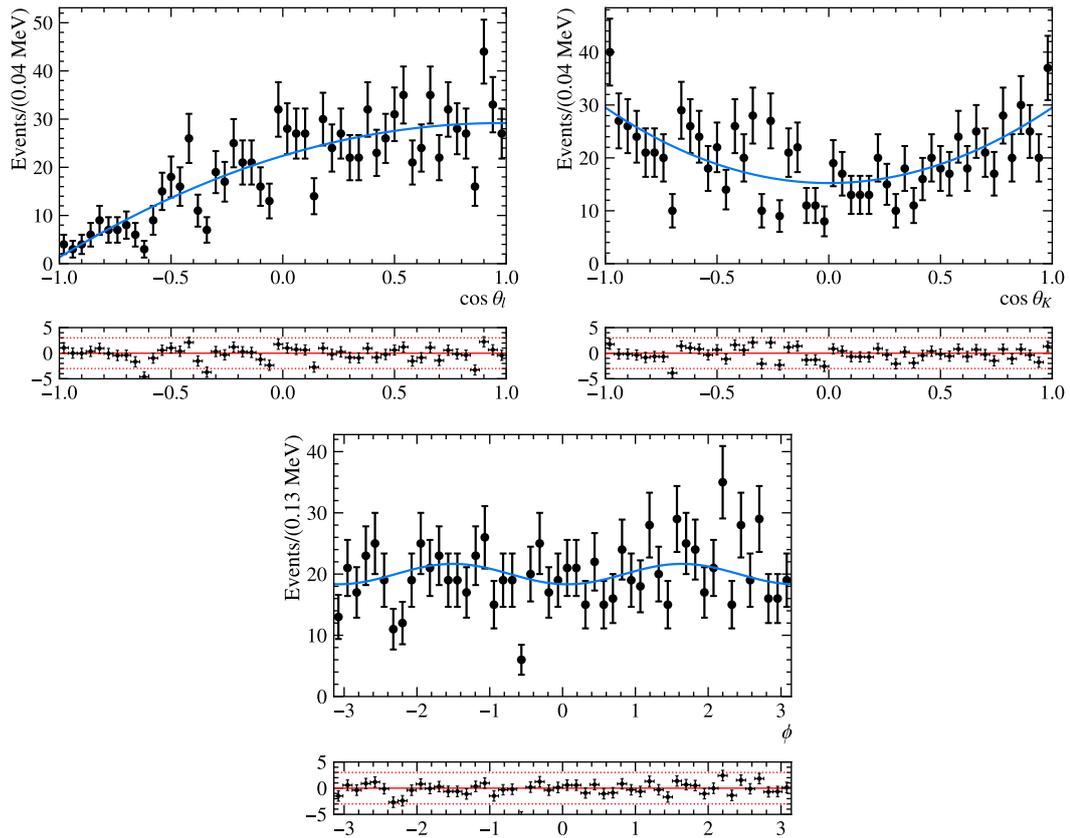

**Figure 7.5.3:** Example plots of an angular toy fit to $B^0 \to K^{*0}\ell^+\ell^-$, shown in the angular variables $cos(\theta_l)$ (top left), $cos(\theta_k)$ (top right) and $\phi$ (bottom).

### 7.5.6 Partial wave analysis

The last example concerns a partial wave analysis, probably the most elaborate type of fit that is performed in HEP in terms of the PDF expression. These fits can take hours or days to perform and are excellent candidates for parallelization on a GPU. Building amplitude models itself is a complex task, which is not covered by zfit. Instead, the CoMPWA project [120] is specialized to build the function, which can be automatically converted to a TensorFlow function. Using the `BasePDF`, similarly to the example in Section 7.5.5, the PDF can be implemented in zfit. As an example to illustrate interoperability, an amplitude model of the decay $J/\psi \to \gamma\pi^0\pi^0$ with five intermediate resonances was built with CoMPWA and converted to a zfit PDF, which was then fit to generated data. The full example is available online in CoMPWA; the resulting PDF fit to data is shown in Fig. 7.5.4.

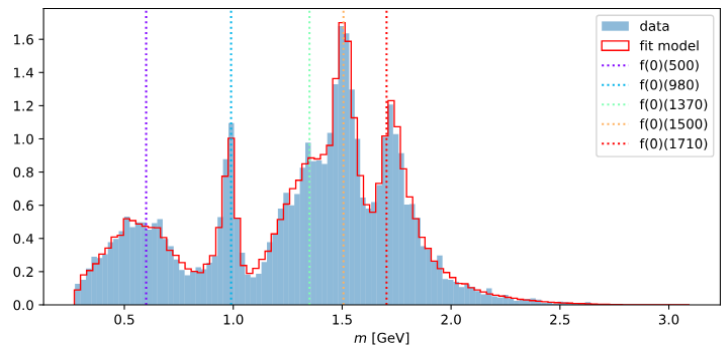

**Figure 7.5.4:** Example of an amplitude fit to $J/\psi \to \gamma\pi^0\pi^0$. The fit model (red) contains five resonances that are recognizable as peaks in the data (dashed, vertical line). The model was fitted to the generated data (blue).

## 7.6 Discussion

The goal of ᴢꜰɪᴛ is to fill the gap of a flexible, modular and performant fitting library that interoperates well with the Sᴄɪᴋɪᴛ-HEP libraries and the rest of the Pʏᴛʜᴏɴ ecosystem. Today, ᴢꜰɪᴛ is used in different communities[3], mostly in HEP collider experiments like LHCb, Belle II, ATLAS and CMS, but also in the PandaX neutrino experiment and even in industry applications. The use cases range from small fits with a few dozen events to large-scale fits with millions of events. When distilling user feedback, ᴢꜰɪᴛ is mostly liked for simple fits, due to its ease of use and interoperability with the Pʏᴛʜᴏɴ ecosystem, or for large-scale, often unconventional, fits due to its customization ability and performance.

Over the years, ᴢꜰɪᴛ improved greatly. Behind the scenes, it went from TᴇɴsᴏʀFʟᴏw 1 to TᴇɴsᴏʀFʟᴏw 2, with the internals of the library having been completely rewritten. On the user-facing side, the ᴢꜰɪᴛ project was extended to include binned PDFs, data and loss as well as a completely new minimizer interface. At the same time, legacy code and issues started to pile up through initial assumptions that are not valid anymore and make certain features difficult, if not impossible, to implement. Furthermore, the backend, TᴇɴsᴏʀFʟᴏw, underwent many major changes and shows similar signs of age, most notably with bugs that arise from newer design decisions interfering with old ones.

In the next year, a completely new version is planned, which will include major changes in the backend to a more general approach, reducing the reliance on TᴇɴsᴏʀFʟᴏw. Furthermore, a native integration of a human-readable serialization format is planned, called HS[3], the *HEP Statistics Serialization Standard* [121]. The format, still under development, is a common effort between different similar libraries in HEP, notably ᴘʏʜꜰ and RᴏᴏFɪᴛ. On a design level, a new `Func` object will be introduced to handle the requirements of different backends and allow for more automatic optimization and interoperability.

In summary, ᴢꜰɪᴛ is a flexible, modular and performant, fitting library. It is designed to be flexible and performant enough to meet the requirements of a variety of HEP analyses and is well integrated into the Sᴄɪᴋɪᴛ-HEP ecosystem. This makes it ideal to go from prototyping in an analysis to large-scale fits running on a cluster or GPUs without the need to change the code.

---

[3]Assessing a user base is inherently difficult, estimates here are conservatively based on confirmed cases.

*"Nach dem Spiel ist vor dem Spiel."*

Sepp Herberger

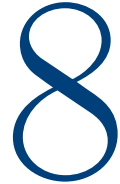

# Conclusion

The presented analysis of $R_K$ used the full Run 1 and Run 2 dataset of proton-proton collisions collected by the LHCb experiment. Despite the agreement with the SM of similar experimental results, the measurement of $R_K$ provides necessary complementary information including a test of various analysis procedures in the high $q^2$ region. Furthermore, the studies and techniques developed in this thesis bear insights for similar analyses in this kinematic regime. Nevertheless, more data is needed to improve the sensitivity of the presented, as well as other measurements. The available dataset will increase drastically during Run 3 and Run 4 of the LHC with a five times higher instantaneous luminosity provided at the LHCb interaction point.

The great increase in luminosity invokes challenges for efficient real-time data processing, which can partially be addressed through topological event reconstruction with DFEI. The performance of the prototype is remarkable in comparison with conventional trigger and reconstruction methods, but also demonstrates the potential for similar algorithms. Due to its entirely CPU and GPU based trigger system enabling the use of such sophisticated algorithms, the LHCb experiment is well prepared for the future.

Rapidly increasing data rates also require more efficient analysis tools that are user-friendly, flexible, and fast, using state-of-the-art hardware. The approach of ZFIT, to be built on top of a framework like TENSORFLOW, provides a future-proof toolbox with the ability to handle and analyze the large LHCb datasets.

# Appendices

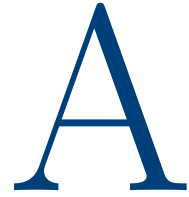

# **Appendix**

## A.1 Efficiency tables

The following efficiencies are obtained using the definition as in Eq. 6.14 and either with all corrections derived in Section 6.3 or without for the years 2016 and 2017. Ghost events are only included at the BDT and fit range level and remain at the two-percent level.

**Table A.1.1:** Efficiencies for $B^+ \to K^+ J/\psi(e^+e^-)$, $B^+ \to K^+e^+e^-$, $B^+ \to K^+ J/\psi(\mu^+\mu^-)$ and $B^+ \to K^+\mu^+\mu^-$ decays, as well as efficiency ratios $\varepsilon(B^+ \to K^+e^+e^-)/\varepsilon(B^+ \to K^+ J/\psi(e^+e^-))$ and $\varepsilon(B^+ \to K^+\mu^+\mu^-)/\varepsilon(B^+ \to K^+ J/\psi(\mu^+\mu^-))$, in 2016 data taking conditions. All efficiencies and ratios are displayed in percent, and the uncertainty quoted is the statistical uncertainty related to the finite size of the simulation samples. The efficiencies are computed in two ways: first using no correction to the simulation, second by including the nominal corrections as described in Section 6.3.

| $\varepsilon[\%]$ | electron modes | | | muon modes | | |
|---|---|---|---|---|---|---|
| | $J/\psi$ | rare | ratio | $J/\psi$ | rare | ratio |
| | No corrections | | | | | |
| $\varepsilon^{\text{geom}}$ | $17.32 \pm 0.03$ | $17.59 \pm 0.03$ | $101.54 \pm 0.25$ | $17.41 \pm 0.03$ | $17.7 \pm 0.03$ | $101.66 \pm 0.25$ |
| $\varepsilon^{\text{rec,strip}}$ | $14.25 \pm 0.01$ | $13.87 \pm 0.01$ | $97.33 \pm 0.11$ | $29.46 \pm 0.01$ | $28.58 \pm 0.02$ | $97.01 \pm 0.07$ |
| $\varepsilon^{\text{presel}}$ | $75.81 \pm 0.04$ | $3.98 \pm 0.02$ | $5.25 \pm 0.02$ | $51.82 \pm 0.03$ | $13.11 \pm 0.02$ | $25.31 \pm 0.04$ |
| $\varepsilon^{\text{PID}}$ | $77.36 \pm 0.04$ | $73.72 \pm 0.21$ | $95.29 \pm 0.27$ | $93.96 \pm 0.02$ | $93.26 \pm 0.04$ | $99.25 \pm 0.05$ |
| $\varepsilon^{\text{trig}}$ | $54.43 \pm 0.05$ | $92.24 \pm 0.15$ | $169.46 \pm 0.32$ | $77.29 \pm 0.03$ | $82.9 \pm 0.07$ | $107.26 \pm 0.1$ |
| $\varepsilon^{\text{BDT}}$ | $103.3 \pm 0.03$ | $100.15 \pm 0.02$ | $98.76 \pm 0.03$ | $101.23 \pm 0.01$ | $100.74 \pm 0.02$ | $100.1 \pm 0.02$ |
| $\varepsilon^{\text{fitrange}}$ | $101.41 \pm 0.02$ | $104.53 \pm 0.12$ | $101.19 \pm 0.12$ | $100.64 \pm 0.01$ | $100.07 \pm 0.01$ | $98.85 \pm 0.01$ |
| $\varepsilon^{\text{BDT}} \&\varepsilon^{\text{fitrange}}$ | $100.45 \pm 0.01$ | $100.15 \pm 0.02$ | $99.71 \pm 0.02$ | $99.82 \pm 0.0$ | $97.57 \pm 0.03$ | $97.75 \pm 0.03$ |
| $\varepsilon^{\text{tot}}$ | $0.79 \pm 0.0$ | $0.07 \pm 0.0$ | $8.36 \pm 0.05$ | $1.93 \pm 0.0$ | $0.5 \pm 0.0$ | $25.97 \pm 0.09$ |
| | PID, trigger and kinematic corrections (nominal) | | | | | |
| $\varepsilon^{\text{geom}}$ | $17.7 \pm 0.03$ | $17.59 \pm 0.03$ | $99.36 \pm 0.25$ | $17.79 \pm 0.03$ | $18.09 \pm 0.03$ | $101.67 \pm 0.25$ |
| $\varepsilon^{\text{rec,strip}}$ | $13.78 \pm 0.01$ | $14.19 \pm 0.01$ | $102.98 \pm 0.12$ | $28.84 \pm 0.01$ | $27.91 \pm 0.02$ | $96.79 \pm 0.07$ |
| $\varepsilon^{\text{presel}}$ | $73.74 \pm 0.04$ | $3.91 \pm 0.02$ | $5.3 \pm 0.03$ | $51.34 \pm 0.03$ | $13.02 \pm 0.02$ | $25.36 \pm 0.04$ |
| $\varepsilon^{\text{PID}}$ | $72.64 \pm 0.05$ | $70.41 \pm 0.26$ | $96.94 \pm 0.36$ | $90.09 \pm 0.03$ | $88.2 \pm 0.08$ | $97.89 \pm 0.09$ |
| $\varepsilon^{\text{trig}}$ | $36.18 \pm 0.05$ | $61.56 \pm 0.24$ | $170.14 \pm 0.7$ | $69.74 \pm 0.04$ | $75.15 \pm 0.08$ | $107.75 \pm 0.13$ |
| $\varepsilon^{\text{BDT}}$ | $103.23 \pm 0.03$ | $99.38 \pm 0.07$ | $98.61 \pm 0.07$ | $101.15 \pm 0.01$ | $99.98 \pm 0.01$ | $99.73 \pm 0.01$ |
| $\varepsilon^{\text{fitrange}}$ | $100.78 \pm 0.0$ | $104.68 \pm 0.13$ | $101.41 \pm 0.13$ | $100.25 \pm 0.01$ | $99.36 \pm 0.02$ | $98.23 \pm 0.02$ |
| $\varepsilon^{\text{BDT}} \&\varepsilon^{\text{fitrange}}$ | $99.86 \pm 0.02$ | $99.37 \pm 0.07$ | $99.51 \pm 0.07$ | $99.44 \pm 0.01$ | $96.94 \pm 0.04$ | $97.49 \pm 0.04$ |
| $\varepsilon^{\text{tot}}$ | $0.47 \pm 0.0$ | $0.04 \pm 0.0$ | $8.89 \pm 0.06$ | $1.65 \pm 0.0$ | $0.42 \pm 0.0$ | $25.67 \pm 0.09$ |

**Table A.1.2:** Efficiencies for $B^+ \rightarrow K^+ J/\psi(e^+ e^-)$, $B^+ \rightarrow K^+ e^+ e^-$, $B^+ \rightarrow K^+ J/\psi(\mu^+\mu^-)$ and $B^+ \rightarrow K^+ \mu^+ \mu^-$ decays, as well as efficiency ratios $\varepsilon(B^+ \rightarrow K^+ e^+ e^-)/\varepsilon(B^+ \rightarrow K^+ J/\psi(e^+ e^-))$ and $\varepsilon(B^+ \rightarrow K^+ \mu^+ \mu^-)/\varepsilon(B^+ \rightarrow K^+ J/\psi(\mu^+\mu^-))$, in 2017 data taking conditions. All efficiencies and ratios are displayed in percent, and the uncertainty quoted is the statistical uncertainty related to the finite size of the simulation samples. The efficiencies are computed in two ways: first using no correction whatsoever to the simulation, second by including the L0, PID corrections and the kinematic correction weights, which is the nominal way to compute the efficiencies, as described in Section 6.3.

| $\varepsilon[\%]$ | electron modes | | | muon modes | | |
|---|---|---|---|---|---|---|
| | $J/\psi$ | rare | ratio | $J/\psi$ | rare | ratio |
| | No corrections | | | | | |
| $\varepsilon^{\text{geom}}$ | $17.31 \pm 0.03$ | $17.59 \pm 0.03$ | $101.61 \pm 0.25$ | $17.4 \pm 0.03$ | $17.69 \pm 0.03$ | $101.66 \pm 0.25$ |
| $\varepsilon^{\text{rec,strip}}$ | $14.25 \pm 0.01$ | $13.67 \pm 0.02$ | $95.94 \pm 0.18$ | $29.5 \pm 0.01$ | $28.61 \pm 0.03$ | $96.98 \pm 0.12$ |
| $\varepsilon^{\text{presel}}$ | $75.81 \pm 0.04$ | $3.99 \pm 0.04$ | $5.26 \pm 0.05$ | $51.79 \pm 0.03$ | $13.08 \pm 0.04$ | $25.26 \pm 0.09$ |
| $\varepsilon^{\text{PID}}$ | $77.35 \pm 0.04$ | $74.53 \pm 0.41$ | $96.35 \pm 0.54$ | $93.94 \pm 0.02$ | $93.28 \pm 0.09$ | $99.3 \pm 0.1$ |
| $\varepsilon^{\text{trig}}$ | $49.61 \pm 0.05$ | $88.38 \pm 0.35$ | $178.16 \pm 0.74$ | $82.5 \pm 0.03$ | $86.46 \pm 0.13$ | $104.8 \pm 0.16$ |
| $\varepsilon^{\text{BDT}}$ | $103.31 \pm 0.03$ | $100.14 \pm 0.04$ | $98.77 \pm 0.05$ | $101.23 \pm 0.01$ | $100.57 \pm 0.03$ | $100.12 \pm 0.03$ |
| $\varepsilon^{\text{fitrange}}$ | $101.38 \pm 0.02$ | $104.55 \pm 0.26$ | $101.2 \pm 0.25$ | $100.45 \pm 0.01$ | $99.92 \pm 0.01$ | $98.71 \pm 0.01$ |
| $\varepsilon^{\text{BDT}} \&\varepsilon^{\text{fitrange}}$ | $100.39 \pm 0.01$ | $100.14 \pm 0.04$ | $99.75 \pm 0.04$ | $99.64 \pm 0.01$ | $97.37 \pm 0.06$ | $97.72 \pm 0.07$ |
| $\varepsilon^{\text{tot}}$ | $0.72 \pm 0.0$ | $0.06 \pm 0.0$ | $8.79 \pm 0.11$ | $2.05 \pm 0.0$ | $0.52 \pm 0.0$ | $25.33 \pm 0.12$ |
| | PID, trigger and kinematic corrections (nominal) | | | | | |
| $\varepsilon^{\text{geom}}$ | $17.65 \pm 0.03$ | $17.59 \pm 0.03$ | $99.65 \pm 0.25$ | $17.74 \pm 0.03$ | $18.04 \pm 0.03$ | $101.67 \pm 0.25$ |
| $\varepsilon^{\text{rec,strip}}$ | $13.77 \pm 0.01$ | $13.9 \pm 0.02$ | $100.96 \pm 0.2$ | $28.88 \pm 0.01$ | $27.93 \pm 0.03$ | $96.7 \pm 0.12$ |
| $\varepsilon^{\text{presel}}$ | $73.77 \pm 0.04$ | $3.92 \pm 0.04$ | $5.31 \pm 0.05$ | $51.26 \pm 0.03$ | $12.98 \pm 0.04$ | $25.31 \pm 0.09$ |
| $\varepsilon^{\text{PID}}$ | $68.4 \pm 0.05$ | $66.05 \pm 0.52$ | $96.56 \pm 0.76$ | $90.51 \pm 0.03$ | $88.97 \pm 0.15$ | $98.3 \pm 0.17$ |
| $\varepsilon^{\text{trig}}$ | $33.88 \pm 0.05$ | $60.03 \pm 0.5$ | $177.19 \pm 1.51$ | $75.26 \pm 0.04$ | $79.13 \pm 0.16$ | $105.14 \pm 0.21$ |
| $\varepsilon^{\text{BDT}}$ | $103.17 \pm 0.03$ | $99.16 \pm 0.16$ | $98.61 \pm 0.16$ | $101.15 \pm 0.01$ | $99.61 \pm 0.04$ | $99.68 \pm 0.04$ |
| $\varepsilon^{\text{fitrange}}$ | $100.56 \pm 0.01$ | $104.6 \pm 0.28$ | $101.38 \pm 0.27$ | $99.93 \pm 0.01$ | $99.23 \pm 0.04$ | $98.1 \pm 0.04$ |
| $\varepsilon^{\text{BDT}} \&\varepsilon^{\text{fitrange}}$ | $99.58 \pm 0.02$ | $99.16 \pm 0.16$ | $99.58 \pm 0.16$ | $99.1 \pm 0.01$ | $96.48 \pm 0.08$ | $97.35 \pm 0.09$ |
| $\varepsilon^{\text{tot}}$ | $0.41 \pm 0.0$ | $0.04 \pm 0.0$ | $9.11 \pm 0.12$ | $1.77 \pm 0.0$ | $0.44 \pm 0.0$ | $25.04 \pm 0.12$ |

## A.2 $B^+ \to K^+ e^+ e^-$ fit model

The fit model to the rare mode consists of components that have been described in Section 6.5.5. In the following, the plots to simulation are shown for the three bremsstrahlung categories in Fig. A.2.1 and the $B^+ \to K^+ \pi^+ \pi^- e^+ e^-$, $B^\pm \to KKK$ and $B_s^{0^+} \to K^+ \phi(\to e^+ e^-)$ backgrounds in Fig. A.2.2.

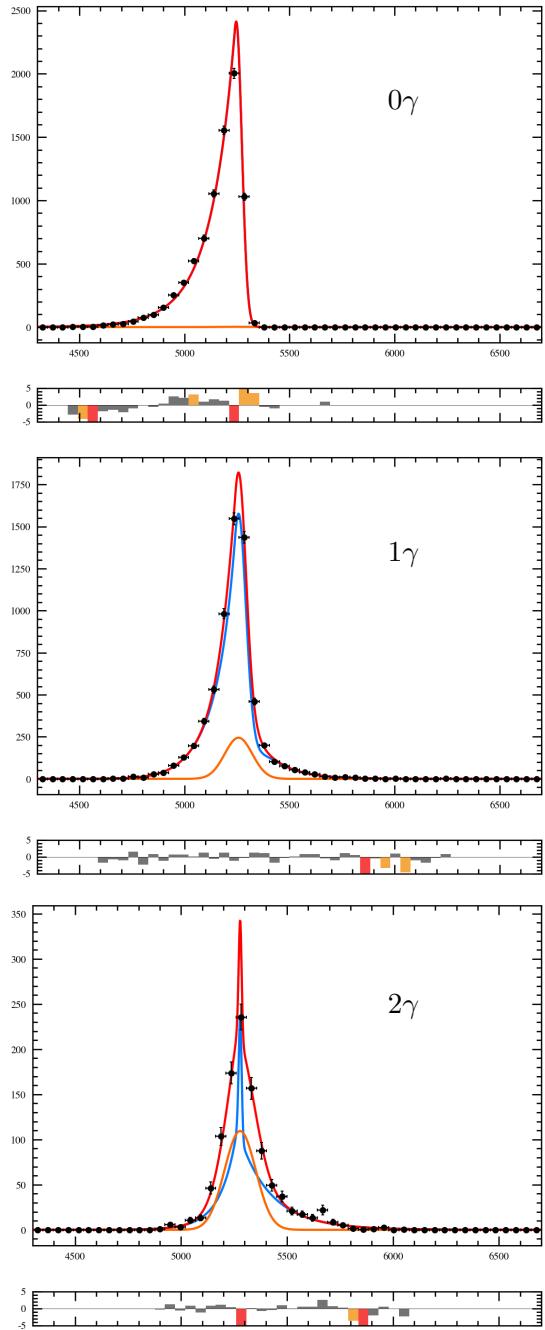

**Figure A.2.1:** Fit to the signal bremsstrahlung components in simulation, merged 2016, 2017 and 2018 data condition samples.bremsstrahlung category zero (top) is modeled with a single DCB, while bremsstrahlung category one (middle) and bremsstrahlung category two (bottom) are a mixture of a DCB and a `Gaussian`. The data points are shown in black and the PDF is drawn in red.

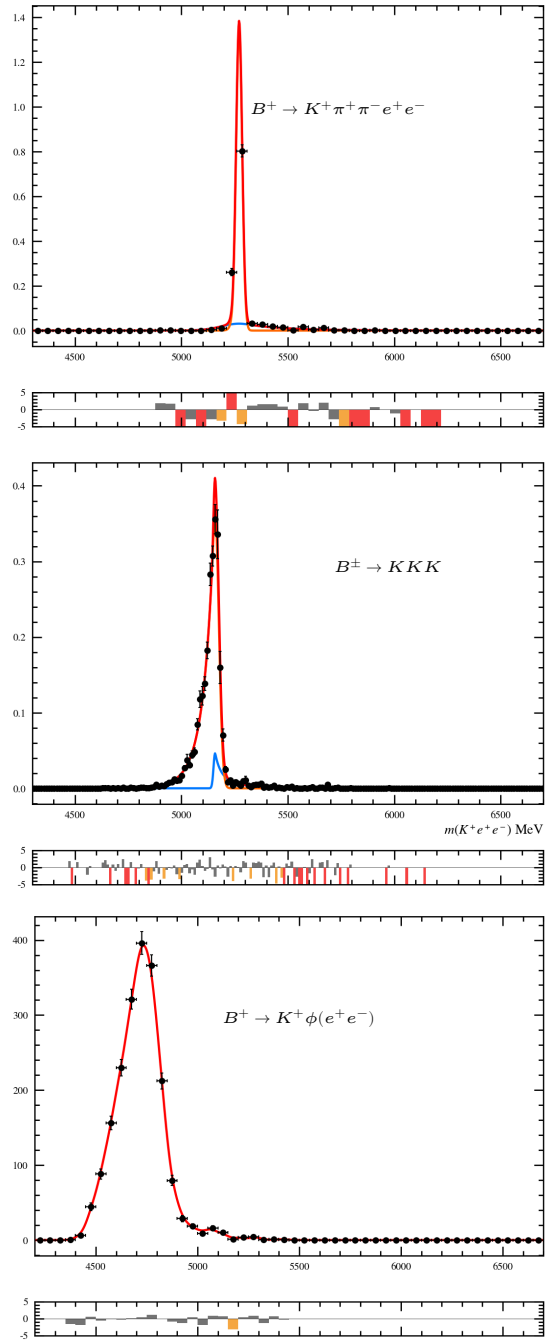

**Figure A.2.2:** Fit to the $B^+ \to K^+\pi^+\pi^-$ (top), $B^\pm \to KKK$ (middle) and $B_s^{0^+} \to K^+\phi(\to e^+e^-)$ (bottom) background components in simulation, merged 2016, 2017 and 2018 data condition samples. The data points are shown in black and the PDF is drawn in red.

# A.3 Normalization mode fits all years

In the following sections, the fits to the normalization modes are shown for all years. The fits were performed using an extended unbinned maximum likelihood to data. The $q^2$ and $m(K^+\ell^+\ell^-)$ selection requirements that select the resonances and specify the fit range, respectively, are defined in Section 6.2.4.3.

### A.3.0.1 Fits to $B^+ \to K^+ J/\psi(e^+e^-)$

The models and components are described in Section 6.5.1. The fits to the $B^+ \to K^+ J/\psi(e^+e^-)$ data in 2016, 2017 and 2018 separately are shown in Fig. A.3.1 and the results of the fits are summarized in Tab. A.3.1.

**Table A.3.1:** Yields from the fit to the $B^+ \to K^+ J/\psi(e^+e^-)$ decay in 2016, 2017 and 2018 data, with and without the BDTs applied. The uncertainties on the yields are statistical only. Individual models for the different contributions are described in Section 6.5.1. The abbreviation *prc* stands for *partially reconstructed background*.

| **channel** | With BDTs | No BDTs |
|---|---|---|
| | 2018 | |
| $B^+ \to K^+ J/\psi(e^+e^-)$ | $146847 \pm 410$ | $150747 \pm 510$ |
| $B^+ \to \pi^+ J/\psi(e^+e^-)$ | $666 \pm 67$ | $779 \pm 68$ |
| combinatorial | $550 \pm 130$ | $13276 \pm 490$ |
| charm prc | $377 \pm 55$ | $412 \pm 62$ |
| strange prc | $13198 \pm 140$ | $13559 \pm 220$ |
| | 2017 | |
| $B^+ \to K^+ J/\psi(e^+e^-)$ | $126742 \pm 380$ | $130504 \pm 460$ |
| $B^+ \to \pi^+ J/\psi(e^+e^-)$ | $584 \pm 58$ | $645 \pm 59$ |
| combinatorial | $428 \pm 120$ | $10940 \pm 450$ |
| charm prc | $393 \pm 59$ | $396 \pm 59$ |
| strange prc | $11546 \pm 130$ | $11866 \pm 200$ |
| | 2016 | |
| $B^+ \to K^+ J/\psi(e^+e^-)$ | $139543 \pm 400$ | $144177 \pm 500$ |
| $B^+ \to \pi^+ J/\psi(e^+e^-)$ | $650 \pm 63$ | $676 \pm 66$ |
| combinatorial | $379 \pm 130$ | $11713 \pm 480$ |
| charm prc | $423 \pm 64$ | $463 \pm 67$ |
| strange prc | $12462 \pm 140$ | $13169 \pm 220$ |

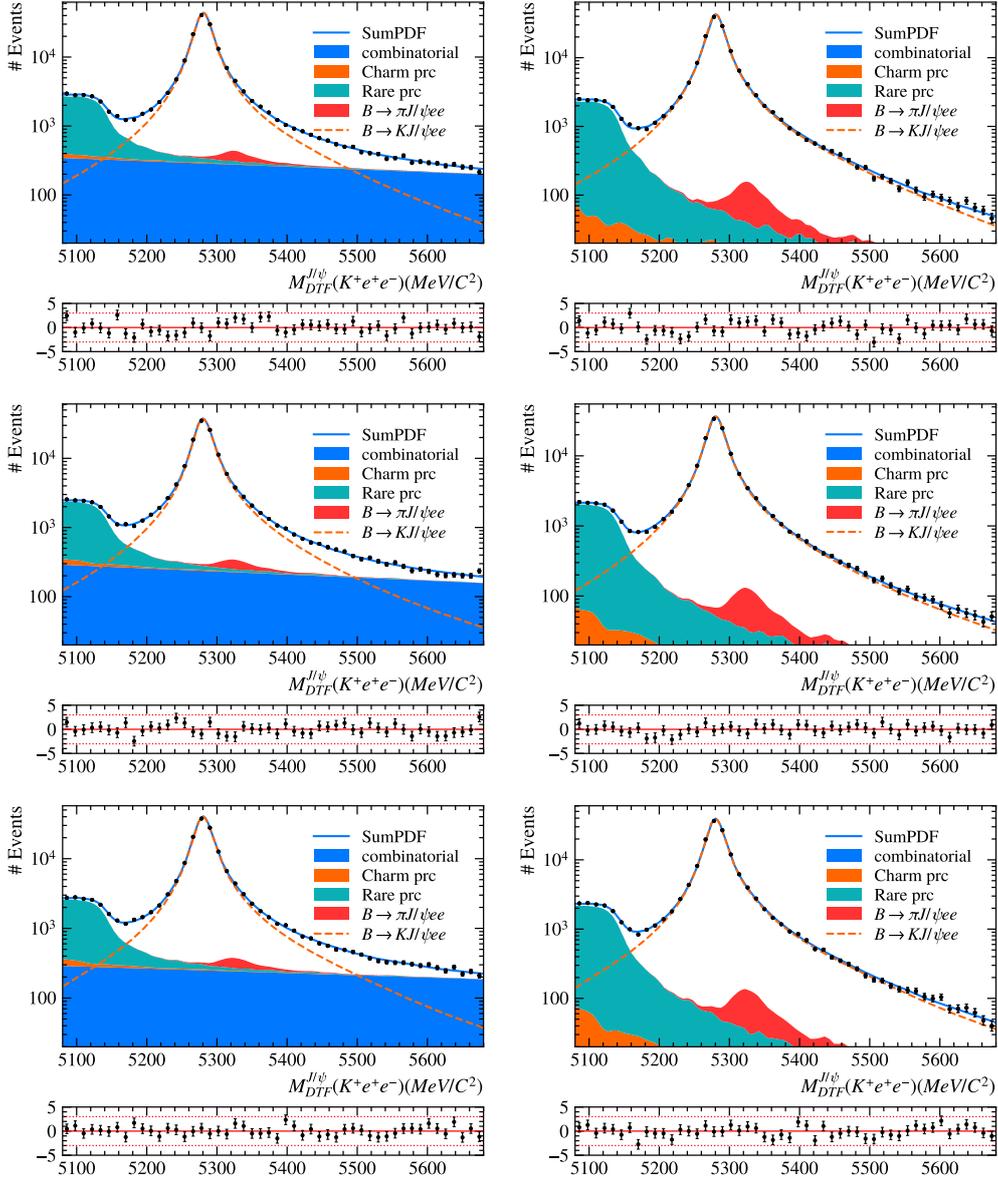

**Figure A.3.1:** Normalization mode fits to $K^+e^+e^-$ data in the $J/\psi$ $q^2$ range in the year 2018 (top), 2017 (middle) and 2016 (bottom) with the full selection chain, the nominal trigger and both without (left) and with (right) the BDTs applied. The data points are shown in black and the PDF is drawn in red. The different PDF components are filled with different colors and stacked.

### A.3.1 Fits to $B^+ \to K^+ J/\psi(\mu^+\mu^-)$

The models and components of the fit to $B^+ \to K^+ J/\psi(\mu^+\mu^-)$ are described in Section 6.5.2.

**Table A.3.2:** Yields from the fit to the $B^+ \to K^+ J/\psi(\mu^+\mu^-)$ decay in 2016, 2017 and 2018 data, with and without the BDTs applied. The uncertainties on the yields are statistical only. Individual models for the different contributions are described in Section 6.5.2.

| **channel** | With BDTs | No BDTs |
|---|---|---|
| | 2018 | |
| $B^+ \to K^+ J/\psi(\mu^+\mu^-)$ | $736005 \pm 880$ | $735713 \pm 880$ |
| $B^+ \to \pi^+ J/\psi(\mu^+\mu^-)$ | $4055 \pm 180$ | $4085 \pm 180$ |
| combinatorial | $22173 \pm 250$ | $22434 \pm 250$ |
| | 2017 | |
| $B^+ \to K^+ J/\psi(\mu^+\mu^-)$ | $599303 \pm 790$ | $598997 \pm 790$ |
| $B^+ \to \pi^+ J/\psi(\mu^+\mu^-)$ | $3533 \pm 160$ | $3541 \pm 160$ |
| combinatorial | $19199 \pm 230$ | $19497 \pm 230$ |
| | 2016 | |
| $B^+ \to K^+ J/\psi(\mu^+\mu^-)$ | $567617 \pm 770$ | $567379 \pm 770$ |
| $B^+ \to \pi^+ J/\psi(\mu^+\mu^-)$ | $2627 \pm 160$ | $2652 \pm 160$ |
| combinatorial | $17222 \pm 220$ | $17433 \pm 220$ |

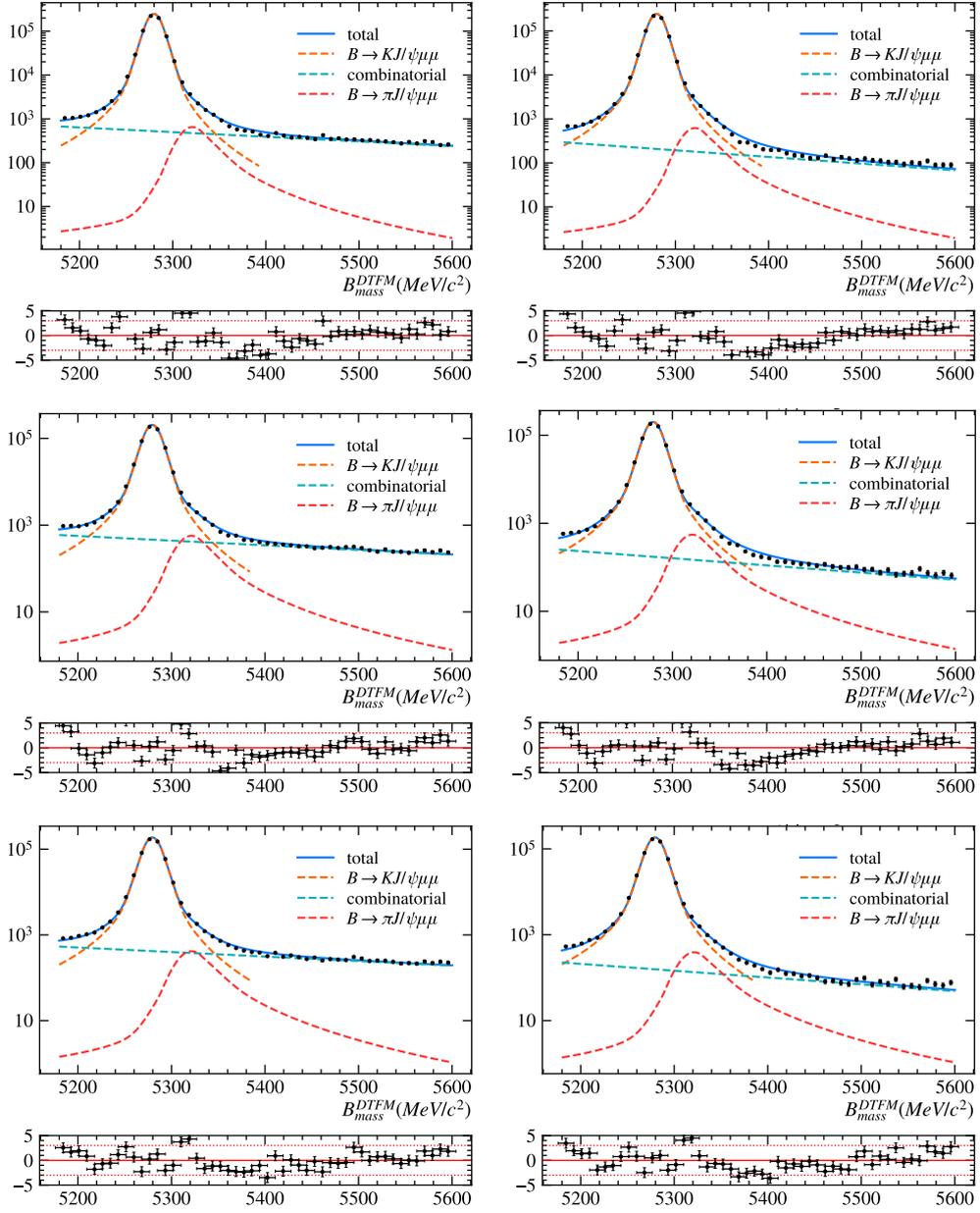

**Figure A.3.2:** Normalization mode fits to $B^+ \to K^+ J/\psi(\mu^+\mu^-)$ data of the year 2018 (top), 2017 (middle) and 2016 (bottom) with the full selection chain and both without (left) and with (right) the BDTs applied. The data points are shown in black and the PDF is drawn in red. The different PDF components are displayed as dashed lines in different colors.

## A.4 Differential $r_{J/\psi}$ for alternative triggers

As an additional cross-check and mainly to assess the agreement between data and simulation in kinematic distributions of $K^+J/\psi(\ell^+\ell^-)$, the differential $r_{J/\psi}$ is measured in bins of specific variables, as discussed in Section 6.8.2.

In the following, all plots for the year 2018 in different trigger selections are shown, the nominal $e$TOS in Section A.4.0.1 $h$TOS! in Section A.4.0.2 and TIS! in Section A.4.0.3.

### A.4.0.1 Differential $r_{J/\psi}$ for 2018

The $r_{J/\psi}$ plots obtained from data taken during 2018, to which the entire selection chain *including the BDT* has been applied, are shown in Fig. A.4.1 - Fig. A.4.5.

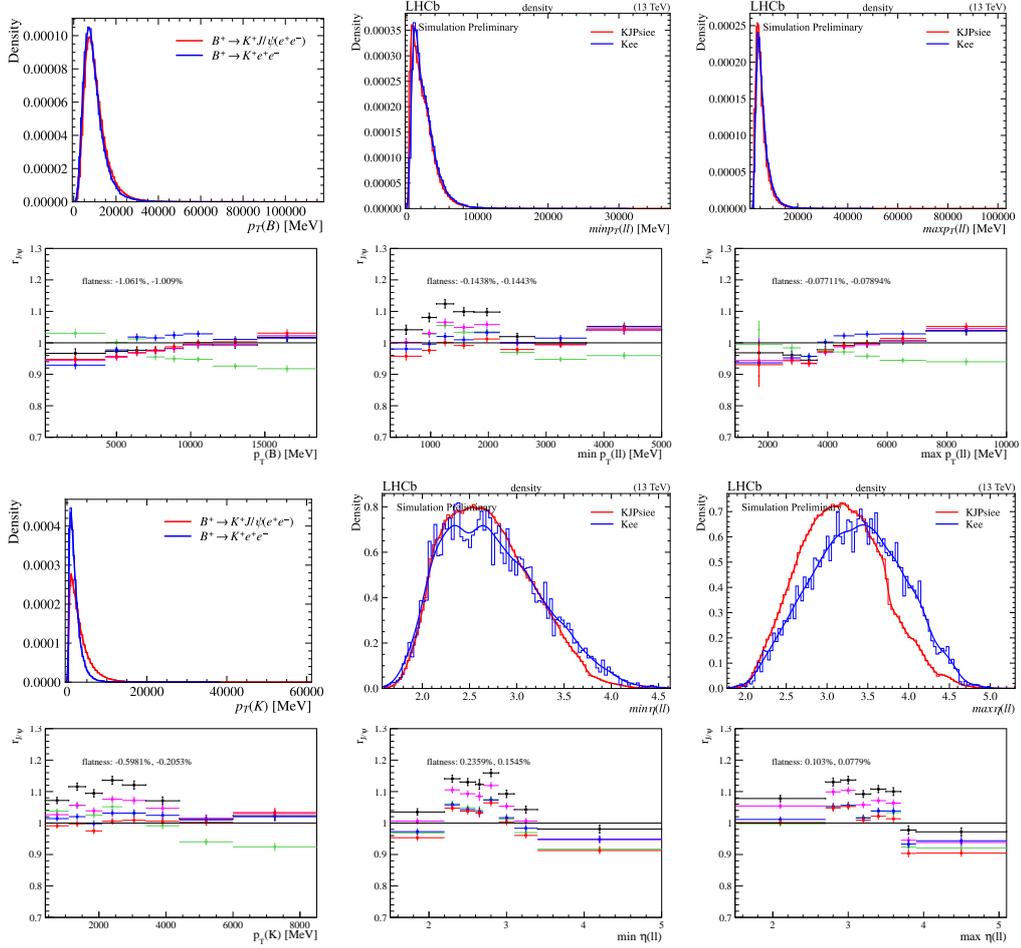

**Figure A.4.1:** $r_{J/\psi}$ for the $B^+ \to K^+ J/\psi(e^+e^-)$ mode in the $e$TOS category , in 2018 data. The black points are computed with no kinematic weights. The red points are computed applying the nominal kinematic weights. The other colors are alternative kinematic weights, derived using different samples instead of the $\mu$TOS muon mode: The blue points use the electron $e$TOS, the pink points the TIS muon sample and the green the TIS electron sample. At the top of each plot, the kinematic distributions for $B^+ \to K^+ e^+ e^-$ (blue) and $B^+ \to K^+ J/\psi(e^+e^-)$ (red) are shown, a histogram overlaid with a KDE.

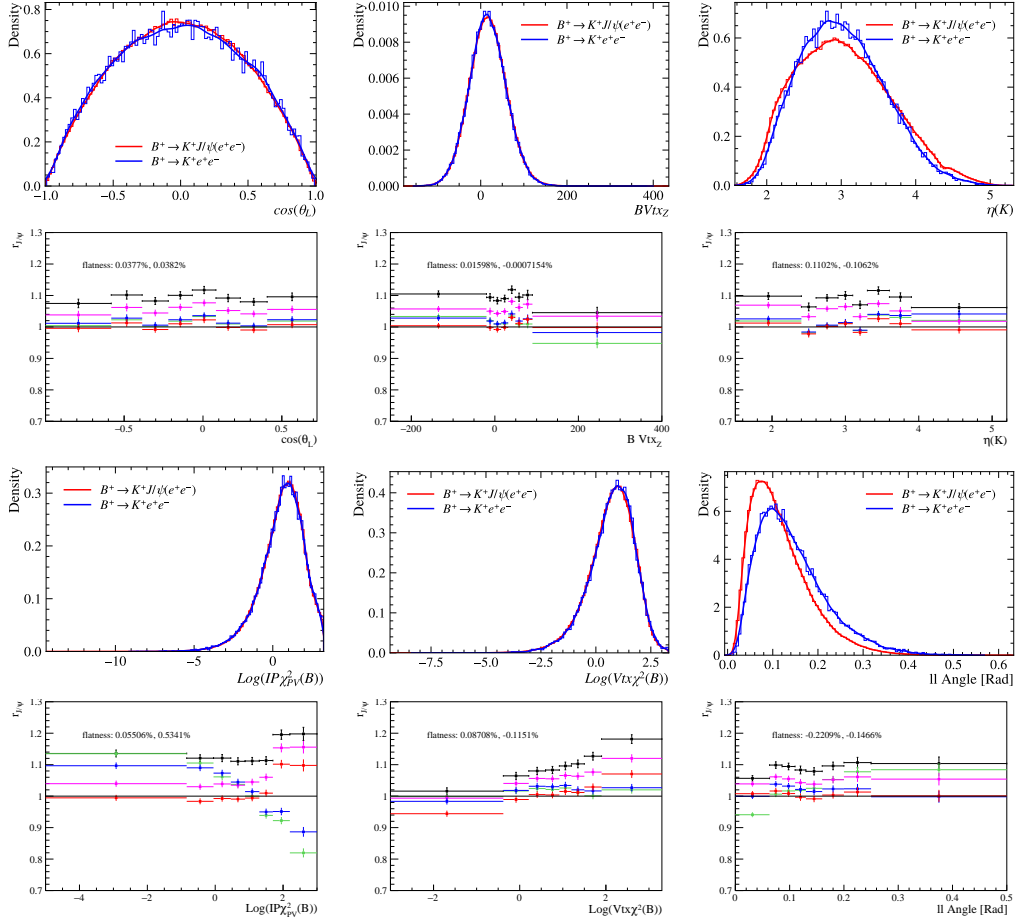

**Figure A.4.2:** $r_{J/\psi}$ for the $B^+ \rightarrow K^+ J/\psi(e^+e^-)$ mode in the $e$TOS category, in 2018 data. The black points are computed with no kinematic weights. The red points are computed applying the nominal kinematic weights. The other colors are alternative kinematic weights, derived using different samples instead of the $\mu$TOS muon mode: The blue points use the electron $e$TOS, the pink points the TIS muon sample and the green the TIS electron sample. At the top of each plot, the kinematic distributions for $B^+ \rightarrow K^+e^+e^-$ (blue) and $B^+ \rightarrow K^+J/\psi(e^+e^-)$ (red) are shown, a histogram overlaid with a KDE.

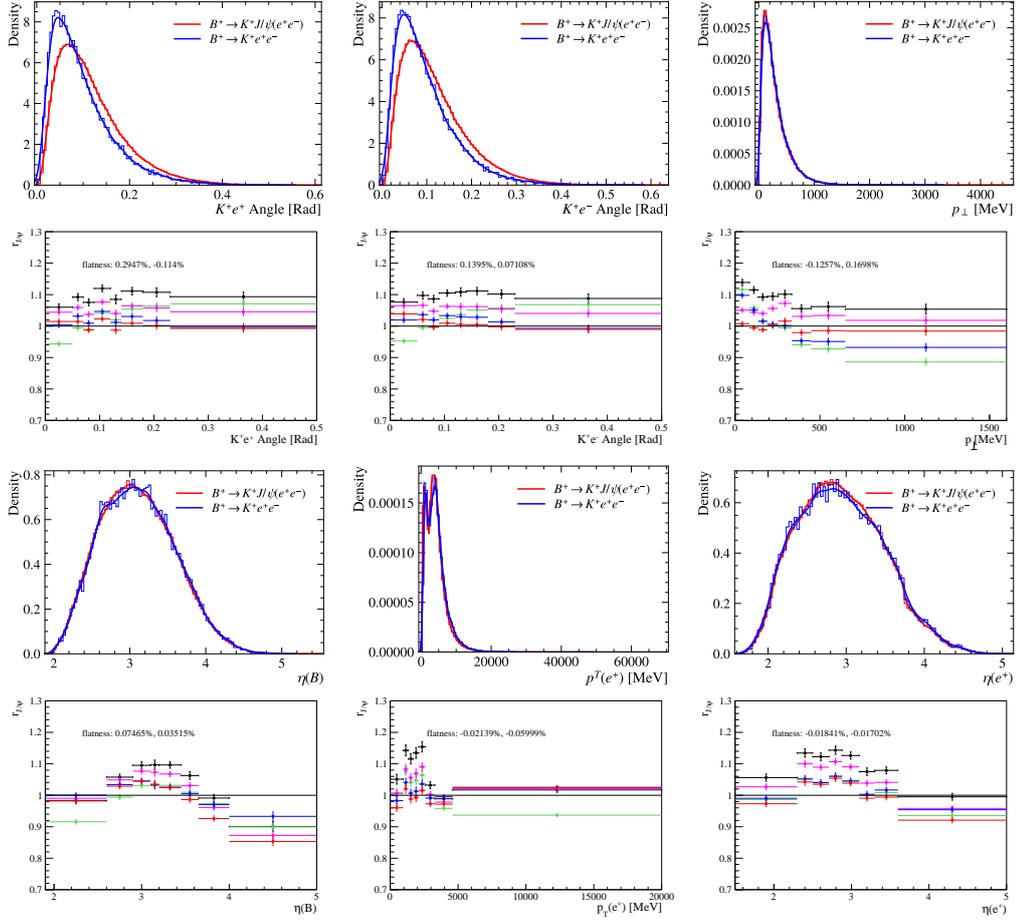

**Figure A.4.3:** $r_{J/\psi}$ for the $B^+ \to K^+ J/\psi(e^+e^-)$ mode in the $e$TOS category , in 2018 data. The black points are computed with no kinematic weights. The red points are computed applying the nominal kinematic weights. The other colors are alternative kinematic weights, derived using different samples instead of the $\mu$TOS muon mode: The blue points use the electron $e$TOS, the pink points the TIS muon sample and the green the TIS electron sample. At the top of each plot, the kinematic distributions for $B^+ \to K^+ e^+ e^-$ (blue) and $B^+ \to K^+ J/\psi(e^+ e^-)$ (red) are shown, a histogram overlaid with a KDE.

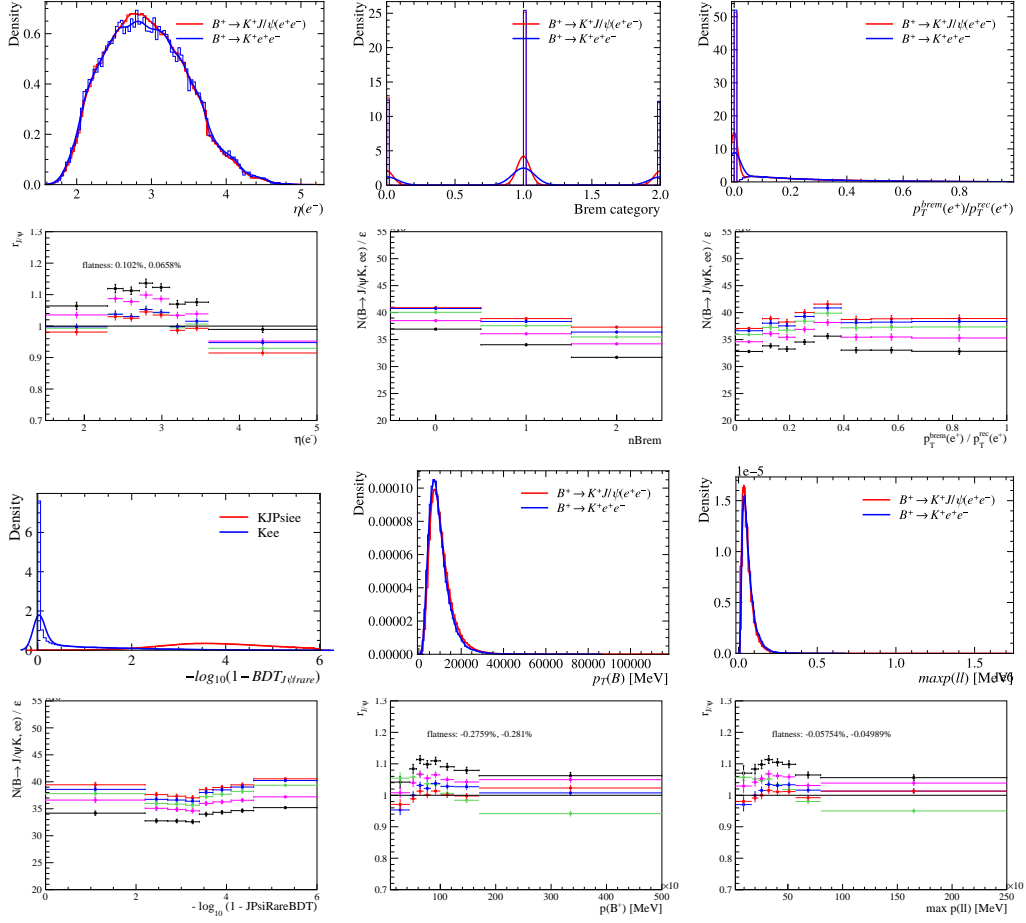

**Figure A.4.4:** $r_{J/\psi}$ for the $B^+ \rightarrow K^+ J/\psi(e^+e^-)$ mode in the $e$TOS category, in 2018 data. The black points are computed with no kinematic weights. The red points are computed applying the nominal kinematic weights. The other colors are alternative kinematic weights, derived using different samples instead of the $\mu$TOS muon mode: The blue points use the electron $e$TOS, the pink points the TIS muon sample and the green the TIS electron sample. At the top of each plot, the kinematic distributions for $B^+ \rightarrow K^+e^+e^-$ (blue) and $B^+ \rightarrow K^+J/\psi(e^+e^-)$ (red) are shown, a histogram overlaid with a KDE.

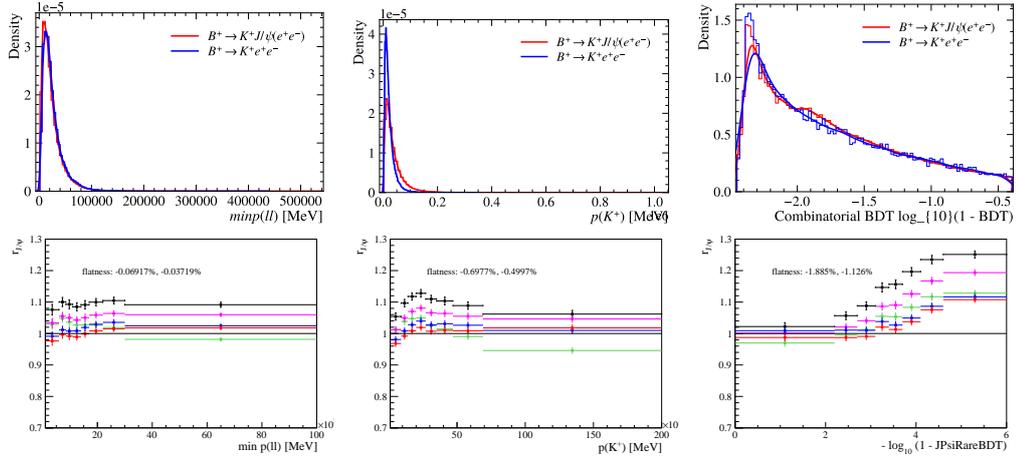

**Figure A.4.5:** $r_{J/\psi}$ for the $B^+ \to K^+ J/\psi(e^+e^-)$ mode in the $e$TOS category, in 2018 data. The L0 and PID corrections are applied to both modes, as well as the BDT selection requirement. The black points are computed with no kinematic weights. The red points are computed applying the kinematic weights to both the electron and muon modes. The blue ones are computed applying the extra pseudorapidity to the electron mode. The flatness parameters $d_f$ associated to the red and blue histogram are displayed in the plot. For the variables related to bremsstrahlung defined only for the electron mode, the efficiency corrected yield $\mathcal{Y}^i_e$ is displayed instead of $r_{J/\psi}$. At the top of each plot, the kinematic distributions for $B^+ \to K^+ \mu^+\mu^-$ (dotted red), $B^+ \to K^+ J/\psi(\mu^+\mu^-)$ (solid red), $B^+ \to K^+\psi(2S)(\mu^+\mu^-)$ (dashed red), $B^+ \to K^+ e^+e^-$ (dotted blue), $B^+ \to K^+ J/\psi(e^+e^-)$ (solid blue), and $B^+ \to K^+\psi(2S)(e^+e^-)$ (dashed blue) are shown.

### A.4.0.2  Differential $r_{J/\psi}$ for 2018 $h$TOS!

The $r_{J/\psi}$ plots obtained from data taken during 2018 and triggered on $h$TOS!, to which the entire selection chain *including the BDT* has been applied, are shown in Fig. A.4.6 - Fig. A.4.10.

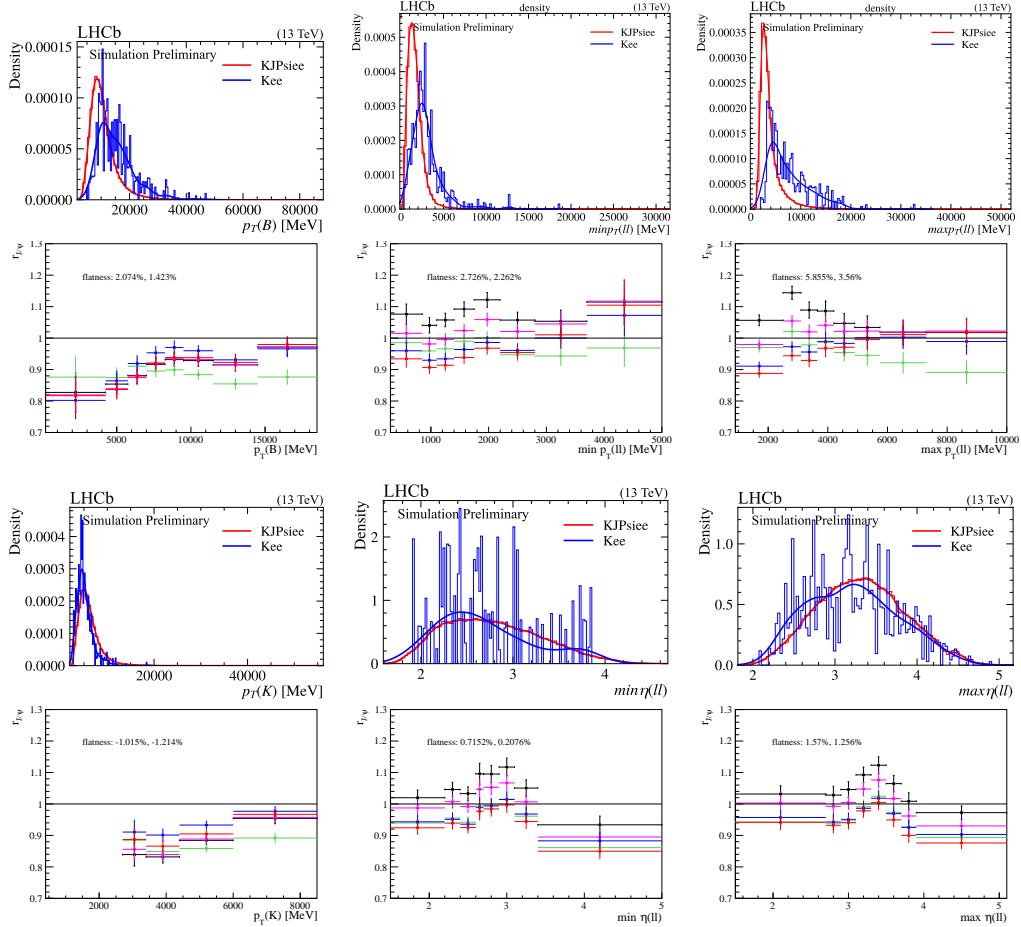

**Figure A.4.6:** $r_{J/\psi}$ for the $B^+ \to K^+ J/\psi(e^+e^-)$ mode in the $h$TOS! category , in 2018 data. The black points are computed with no kinematic weights. The red points are computed applying the nominal kinematic weights. The other colors are alternative kinematic weights, derived using different samples instead of the $\mu$TOS muon mode: The blue points use the electron $e$TOS, the pink points the TIS muon sample and the green the TIS electron sample. At the top of each plot, the kinematic distributions for $B^+ \to K^+ e^+ e^-$ (blue) and $B^+ \to K^+ J/\psi(e^+ e^-)$ (red) are shown, a histogram overlaid with a KDE.

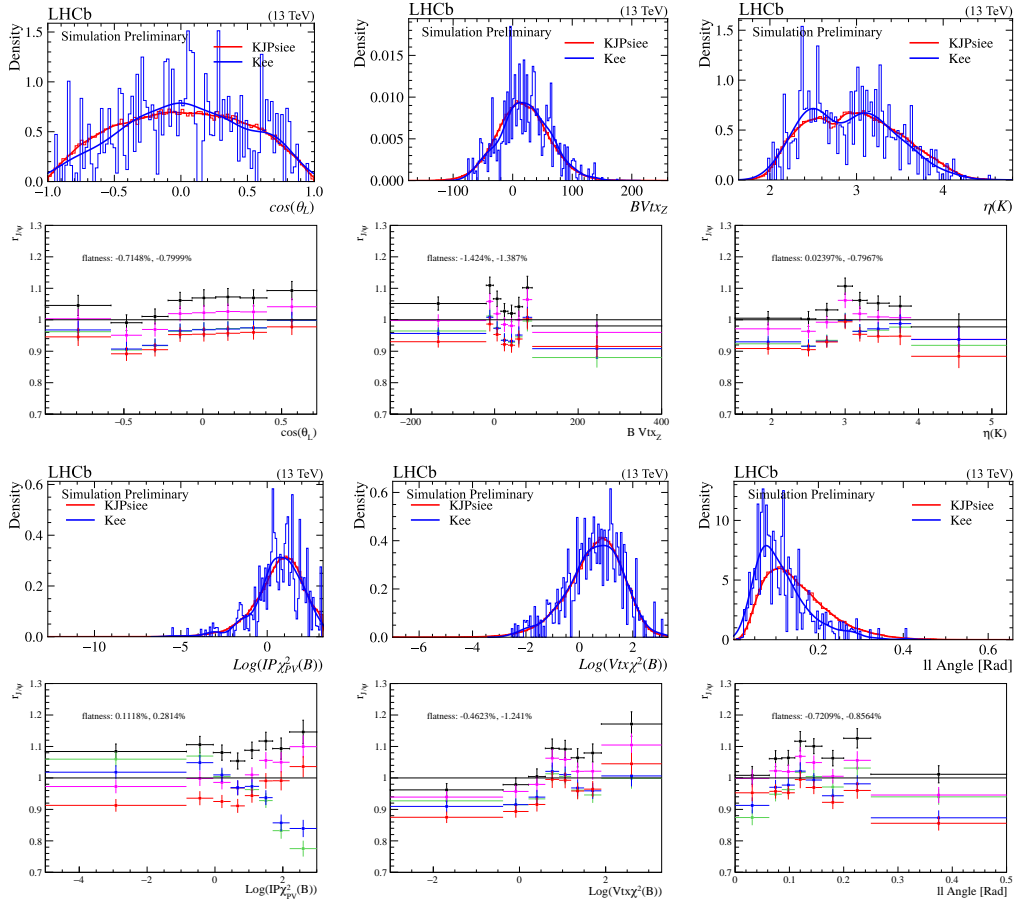

**Figure A.4.7:** $r_{J/\psi}$ for the $B^+ \to K^+ J/\psi(e^+e^-)$ mode in the $h$TOS! category , in 2018 data. The black points are computed with no kinematic weights. The red points are computed applying the nominal kinematic weights. The other colors are alternative kinematic weights, derived using different samples instead of the $\mu$TOS muon mode: The blue points use the electron $e$TOS, the pink points the TIS muon sample and the green the TIS electron sample. At the top of each plot, the kinematic distributions for $B^+ \to K^+ e^+e^-$ (blue) and $B^+ \to K^+ J/\psi(e^+e^-)$ (red) are shown, a histogram overlaid with a KDE.

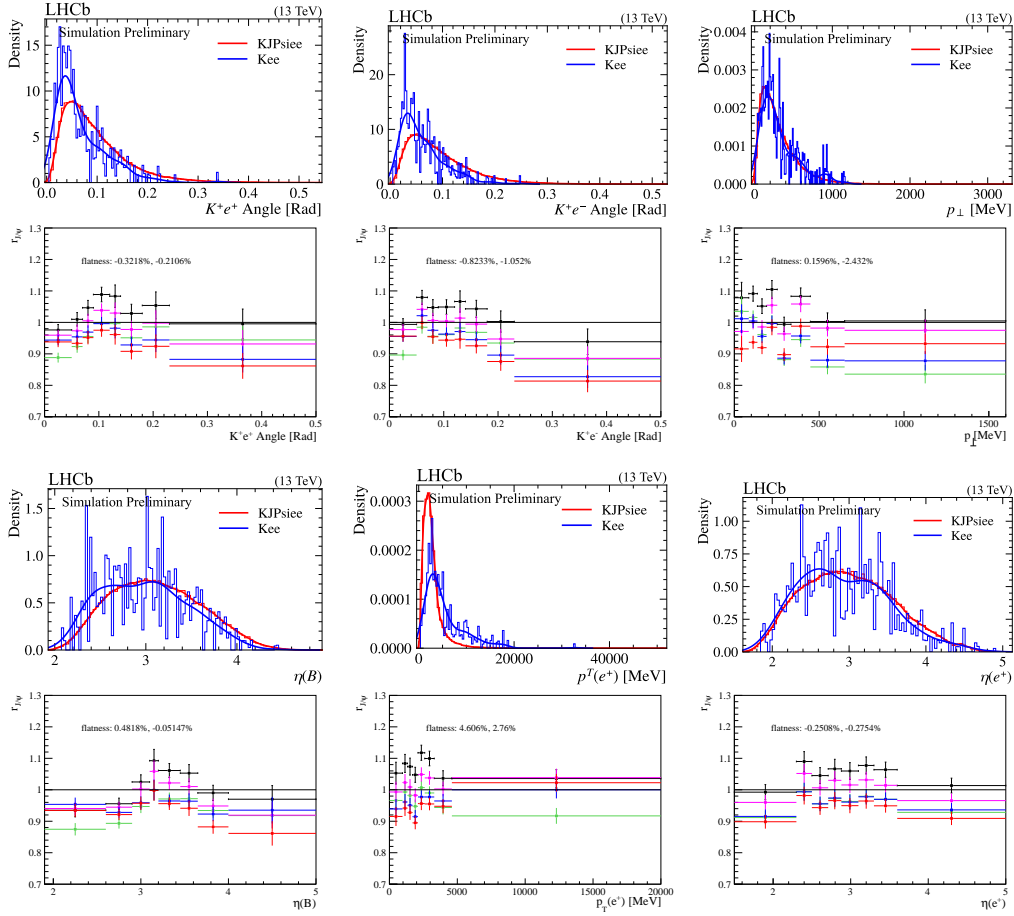

**Figure A.4.8:** $r_{J/\psi}$ for the $B^+ \to K^+ J/\psi(e^+e^-)$ mode in the $h$TOS! category , in 2018 data. The black points are computed with no kinematic weights. The red points are computed applying the nominal kinematic weights. The other colors are alternative kinematic weights, derived using different samples instead of the $\mu$TOS muon mode: The blue points use the electron $e$TOS, the pink points the TIS muon sample and the green the TIS electron sample. At the top of each plot, the kinematic distributions for $B^+ \to K^+ e^+e^-$ (blue) and $B^+ \to K^+ J/\psi(e^+e^-)$ (red) are shown, a histogram overlaid with a KDE.

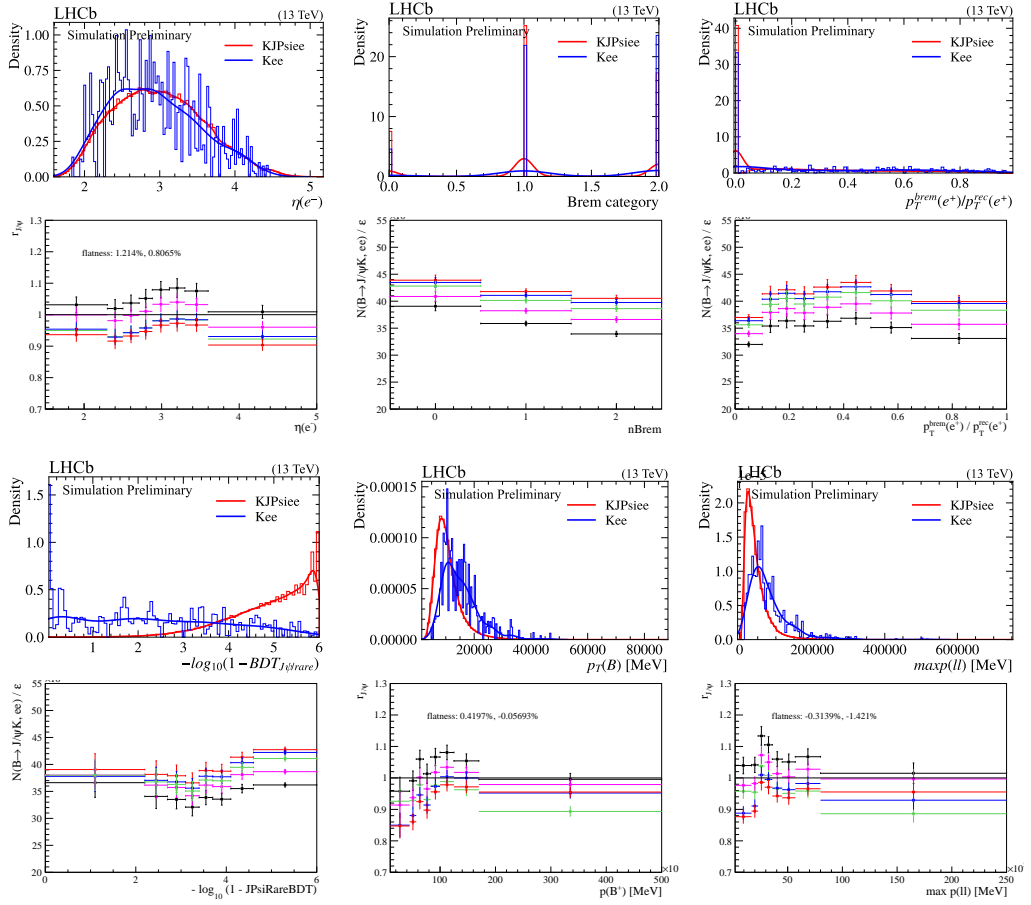

**Figure A.4.9:** $r_{J/\psi}$ for the $B^+ \to K^+ J/\psi(e^+e^-)$ mode in the $h$TOS! category , in 2018 data. The black points are computed with no kinematic weights. The red points are computed applying the nominal kinematic weights. The other colors are alternative kinematic weights, derived using different samples instead of the $\mu$TOS muon mode: The blue points use the electron $e$TOS, the pink points the TIS muon sample and the green the TIS electron sample. At the top of each plot, the kinematic distributions for $B^+ \to K^+e^+e^-$ (blue) and $B^+ \to K^+J/\psi(e^+e^-)$ (red) are shown, a histogram overlaid with a KDE.

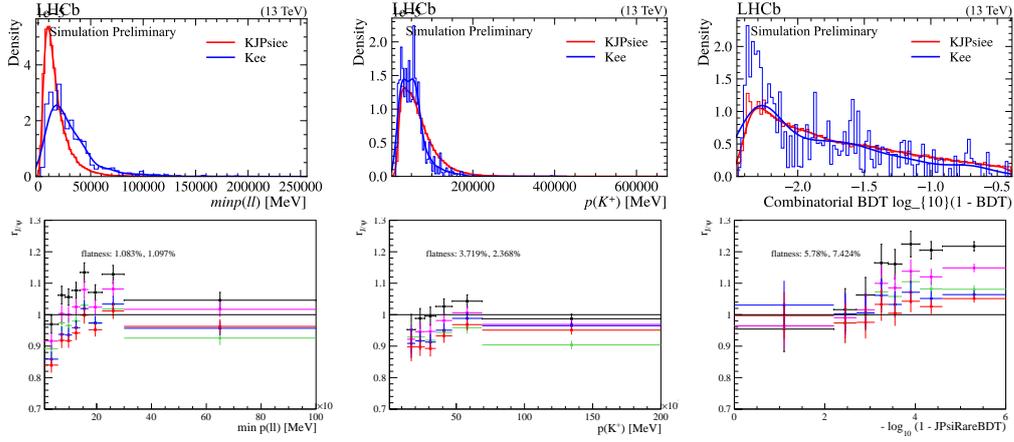

**Figure A.4.10:** $r_{J/\psi}$ for the $B^+ \to K^+ J/\psi(e^+e^-)$ mode in the $h$TOS! category , in 2018 data. The black points are computed with no kinematic weights. The red points are computed applying the nominal kinematic weights. The other colors are alternative kinematic weights, derived using different samples instead of the $\mu$TOS muon mode: The blue points use the electron $e$TOS, the pink points the TIS muon sample and the green the TIS electron sample. At the top of each plot, the kinematic distributions for $B^+ \to K^+ e^+e^-$ (blue) and $B^+ \to K^+ J/\psi(e^+e^-)$ (red) are shown, a histogram overlaid with a KDE.

### A.4.0.3 Differential $r_{J/\psi}$ for 2018 TIS!

The $r_{J/\psi}$ plots obtained from data taken during 2018 and triggered on TIS!, to which the entire selection chain *including the BDT* has been applied, are shown in Fig. A.4.11 - Fig. A.4.15.

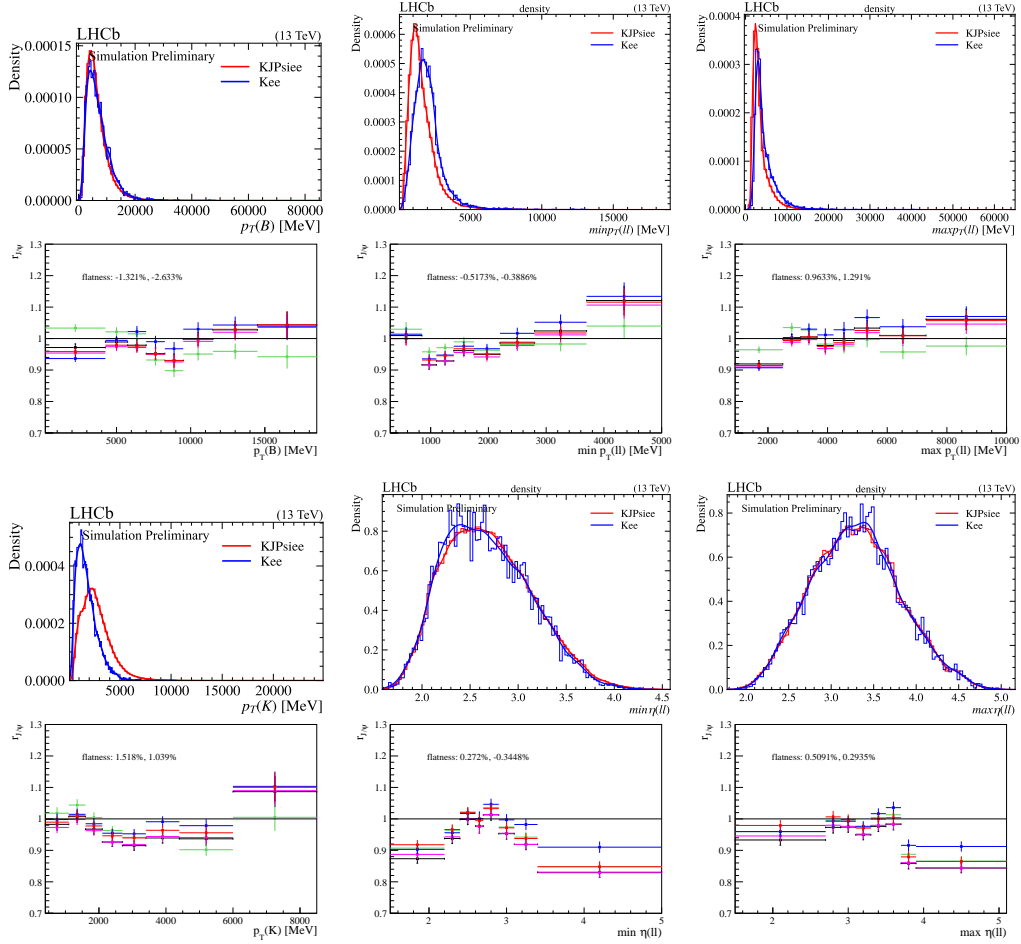

**Figure A.4.11:** $r_{J/\psi}$ for the $B^+ \to K^+ J/\psi(e^+e^-)$ mode in the TIS! category, in 2018 data. The black points are computed with no kinematic weights. The red points are computed applying the nominal kinematic weights. The other colors are alternative kinematic weights, derived using different samples instead of the $\mu$TOS muon mode: The blue points use the electron $e$TOS, the pink points the TIS muon sample and the green the TIS electron sample. At the top of each plot, the kinematic distributions for $B^+ \to K^+ e^+e^-$ (blue) and $B^+ \to K^+ J/\psi(e^+e^-)$ (red) are shown, a histogram overlaid with a KDE.

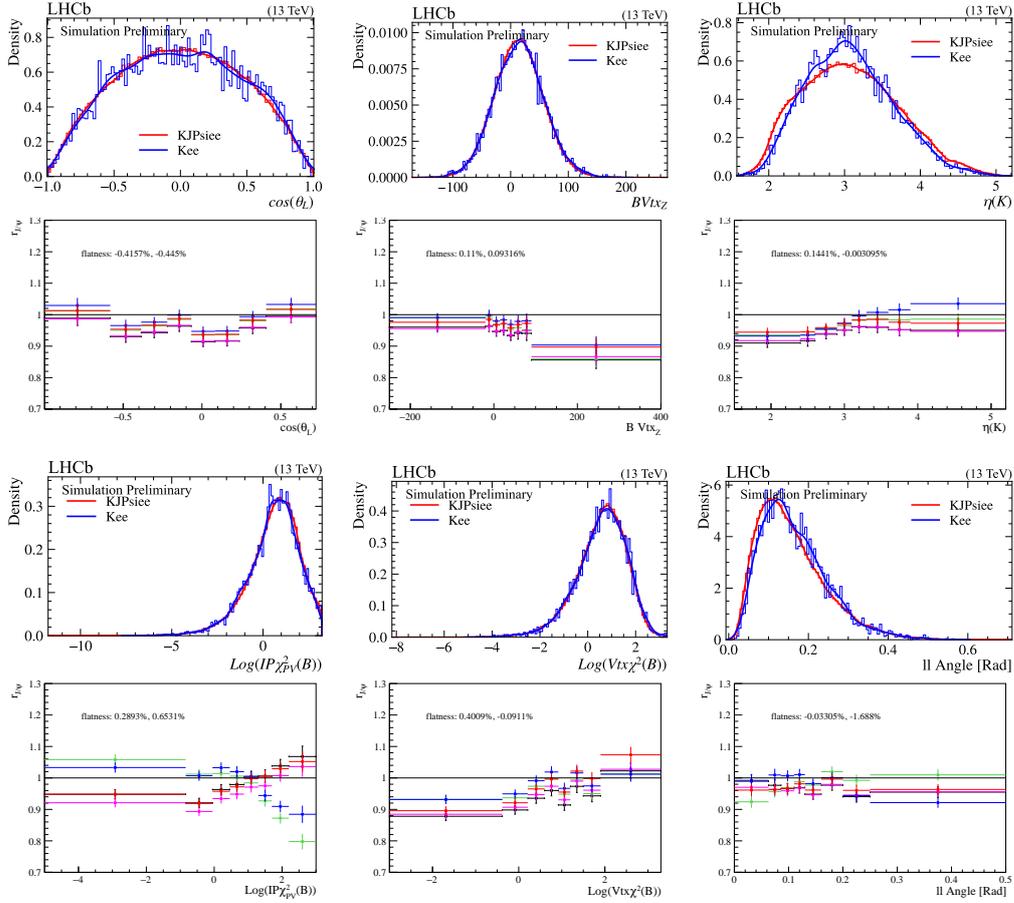

**Figure A.4.12:** $r_{J/\psi}$ for the $B^+ \to K^+ J/\psi(e^+e^-)$ mode in the TIS! category, in 2018 data. The black points are computed with no kinematic weights. The red points are computed applying the nominal kinematic weights. The other colors are alternative kinematic weights, derived using different samples instead of the $\mu$TOS muon mode: The blue points use the electron $e$TOS, the pink points the TIS muon sample and the green the TIS electron sample. At the top of each plot, the kinematic distributions for $B^+ \to K^+ e^+ e^-$ (blue) and $B^+ \to K^+ J/\psi(e^+e^-)$ (red) are shown, a histogram overlaid with a KDE.

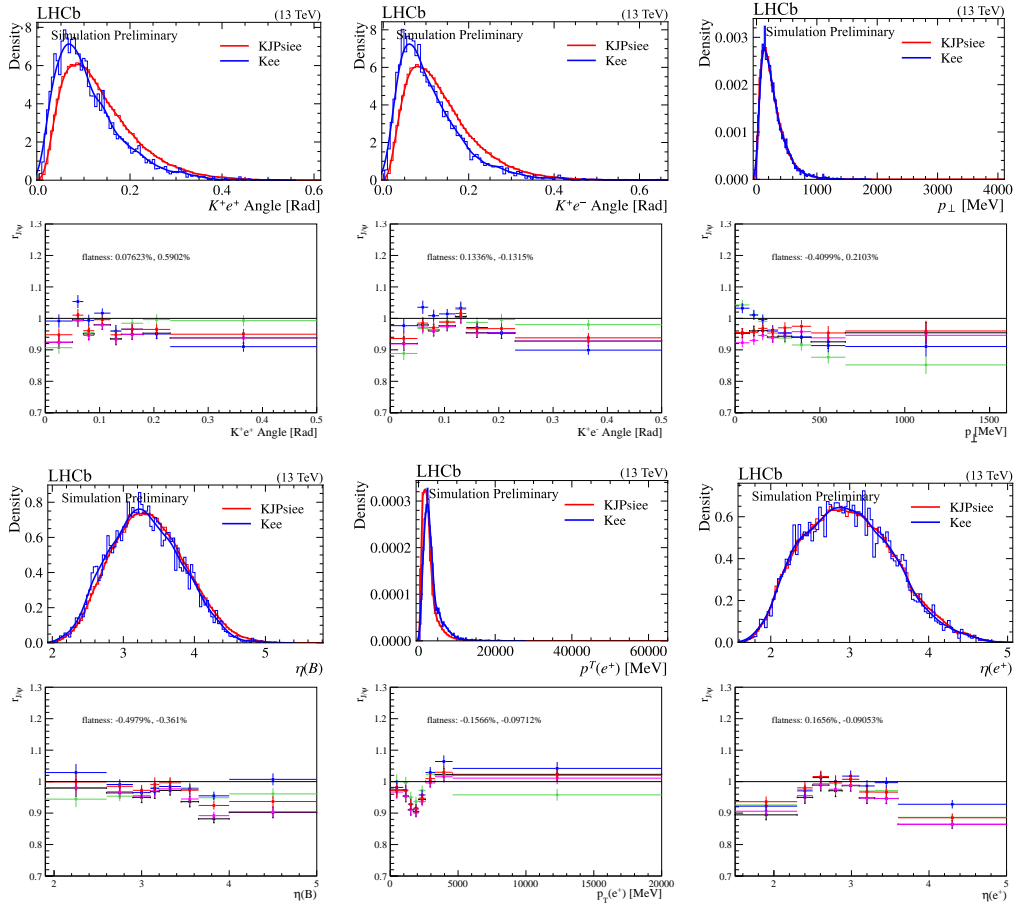

**Figure A.4.13:** $r_{J/\psi}$ for the $B^+ \to K^+ J/\psi(e^+e^-)$ mode in the TIS! category, in 2018 data. The black points are computed with no kinematic weights. The red points are computed applying the nominal kinematic weights. The other colors are alternative kinematic weights, derived using different samples instead of the $\mu$TOS muon mode: The blue points use the electron $e$TOS, the pink points the TIS muon sample and the green the TIS electron sample. At the top of each plot, the kinematic distributions for $B^+ \to K^+ e^+e^-$ (blue) and $B^+ \to K^+ J/\psi(e^+e^-)$ (red) are shown, a histogram overlaid with a KDE.

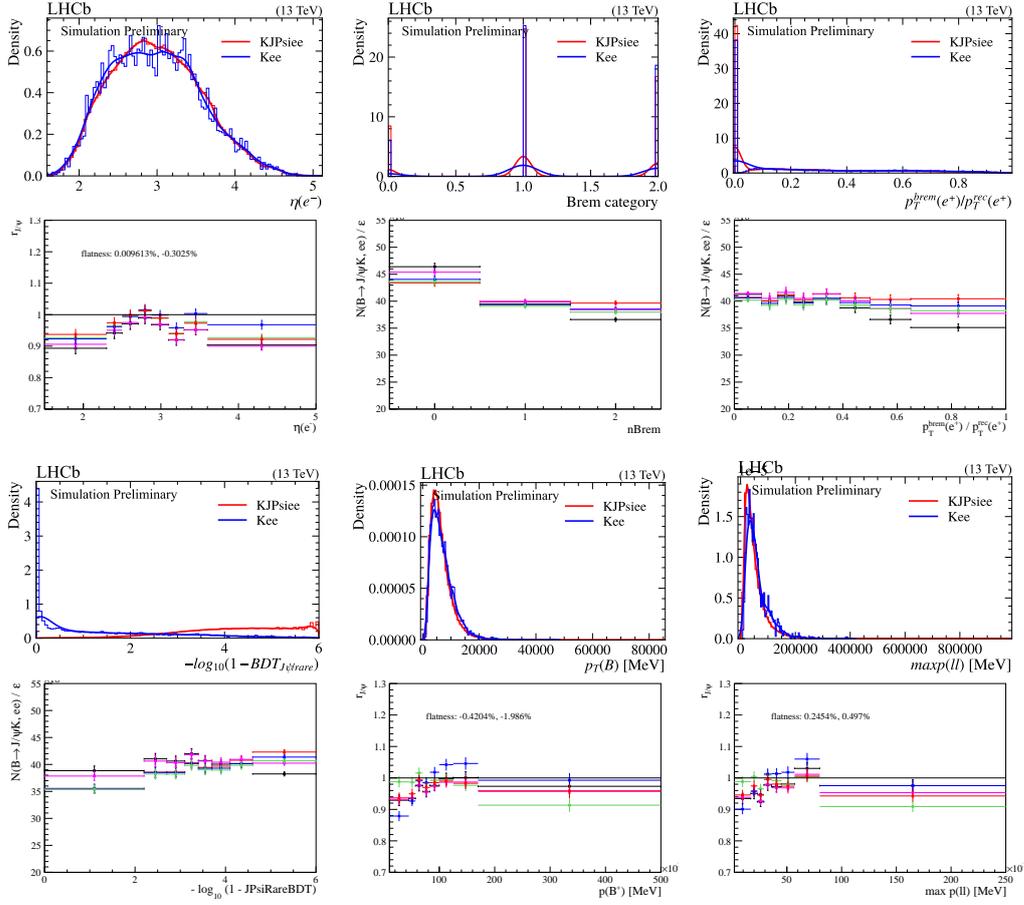

**Figure A.4.14:** $r_{J/\psi}$ for the $B^+ \to K^+ J/\psi(e^+e^-)$ mode in the TIS! category, in 2018 data. The black points are computed with no kinematic weights. The red points are computed applying the nominal kinematic weights. The other colors are alternative kinematic weights, derived using different samples instead of the $\mu$TOS muon mode: The blue points use the electron $e$TOS, the pink points the TIS muon sample and the green the TIS electron sample. At the top of each plot, the kinematic distributions for $B^+ \to K^+ e^+e^-$ (blue) and $B^+ \to K^+ J/\psi(e^+e^-)$ (red) are shown, a histogram overlaid with a KDE.

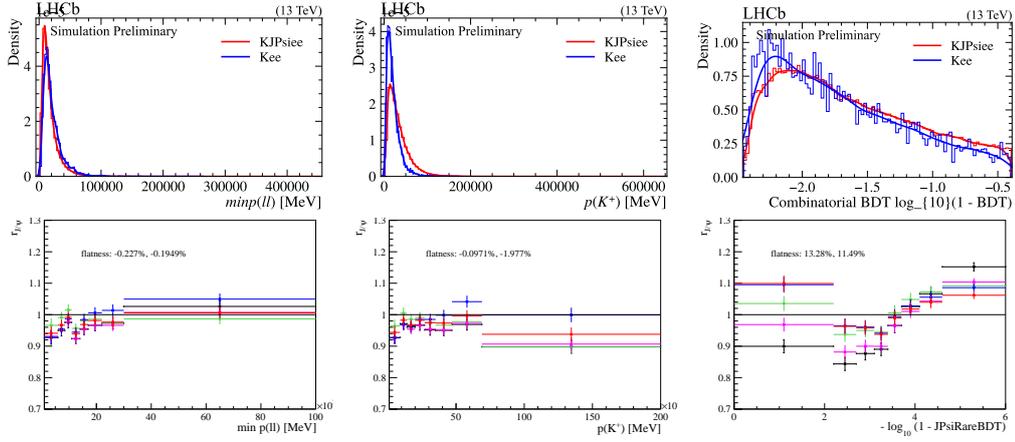

**Figure A.4.15:** $r_{J/\psi}$ for the $B^+ \to K^+ J/\psi(e^+e^-)$ mode in the TIS! category, in 2018 data. The black points are computed with no kinematic weights. The red points are computed applying the nominal kinematic weights. The other colors are alternative kinematic weights, derived using different samples instead of the $\mu$TOS muon mode: The blue points use the electron $e$TOS, the pink points the TIS muon sample and the green the TIS electron sample. At the top of each plot, the kinematic distributions for $B^+ \to K^+ e^+ e^-$ (blue) and $B^+ \to K^+ J/\psi(e^+e^-)$ (red) are shown, a histogram overlaid with a KDE.

## A.5 Angular PDF

In the following, the full implementation of the angular PDF of the decay $B^0 \to K^{*0}\ell^+\ell^-$, as described in Section 7.5.5, is shown.

```python
import zfit.z.numpy as znp

class AngularPDF(zfit.pdf.ZPDF):
    _PARAMS = ['FL', 'S3', 'S4', 'S5', 'AFB', 'S7', 'S8', 'S9']
    _N_OBS = 3

    @zfit.supports()
    def _pdf(self, x):
        FL = self.params['FL']
        S3 = self.params['S3']
        S4 = self.params['S4']
        S5 = self.params['S5']
        AFB = self.params['AFB']
        S7 = self.params['S7']
        S8 = self.params['S8']
        S9 = self.params['S9']

        costheta_l, costheta_k, phi = z.unstack_x(x)

        sintheta_k = znp.sqrt(1.0 - costheta_k * costheta_k)
        sintheta_l = znp.sqrt(1.0 - costheta_l * costheta_l)
        sintheta_2k = (1.0 - costheta_k * costheta_k)
        sintheta_2l = (1.0 - costheta_l * costheta_l)
        sin2theta_k = (2.0 * sintheta_k * costheta_k)
        cos2theta_l = (2.0 * costheta_l * costheta_l - 1.0)
        sin2theta_l = (2.0 * sintheta_l * costheta_l)

        pdf = ((3.0 / 4.0) * (1.0 - FL) * sintheta_2k +
        FL * costheta_k * costheta_k +
        (1.0 / 4.0) * (1.0 - FL) * sintheta_2k * cos2theta_l +
        -1.0 * FL * costheta_k * costheta_k * cos2theta_l +
        S3 * sintheta_2k * sintheta_2l * znp.cos(2.0 * phi) +
        S4 * sin2theta_k * sin2theta_l * znp.cos(phi) +
        S5 * sin2theta_k * sintheta_l * znp.cos(phi) +
        (4.0 / 3.0) * AFB * sintheta_2k * costheta_l +
        S7 * sin2theta_k * sintheta_l * znp.sin(phi) +
        S8 * sin2theta_k * sin2theta_l * znp.sin(phi) +
        S9 * sintheta_2k * sintheta_2l * znp.sin(2.0 * phi))

        return pdf
```

# List of Acronyms

**SPD**   Scintillating Pad Detector. 60

TIS   trigger independent of signal. 109, 126, 133

TOS   trigger on signal. 109

**TT**   Tracker Turicensis. 45, 49, 50

**VELO**   Vertex Locator. 45–47, 49, 50, 56, 60, 73

**AOT**   ahead-of-time. 41, 189

**API**   application programming interface. 186, 188–195, 197, 203

**AUC**   Area Under the Curve. 82, 140

**BDT**   boosted decision tree. 31, 32, 66, 100, 104, 106, 109, 112, 114–116, 120–124, 137, 139–141, 145, 146, 157, 161, 166, 171–177, 179

**BSM**   beyond the Standard Model. 22

**CB**   Crystal Ball functions, a `Gaussian` distribution with an exponential tail,. 143, 165, 170, 198

**central $q^2$ region**   $q^2$ region between 1.1 and 6.0 GeV$^2/c^4$. 21, 23, 26, 92–94, 102, 105, 106, 109, 149, 150, 160, 183

**CKM**   Cabibbo-Kobayashi-Maskawa. 6, 17, 19–21

**CNN**   convolutional neural network. 32, 33, 68, 70

**CoM**   center-of-mass. 1, 44, 59, 72

**CPU**   central processing unit. 40, 61, 188, 211

**DCB**   PDF consisting of a `Gaussian` with a mean and a width and an exponential tail to the left and one to the right,. 142, 150, 151, 153, 156, 165, 169, 170

**DFEI**   Deep Full Event Interpretation. 2, 61, 64, 66–68, 70–72, 74–76, 78–81, 83–89

**DL**   deep learning. 32, 62

**DNN**   deep neural network. 33, 68, 70, 81, 82

**DT**   decision tree. 31, 32, 67, 138

**EDM**   estimated distance to minimum. 200–202

**EFT**   effective field theory. 12, 19, 22

**EP**   Edge Pruning. 75, 77, 78, 80, 85, 88

**EW** electroweak. 6, 21

**FCNC** flavour changing neutral current. 18, 21, 22, 92

**FEI** Full Event Interpretation. 66, 88, 89

**FFT** fast Fourier transform. 198

**FoM** figure of merit. 123

**FSR** final state radiation. 24

**GBReweighter** Gradient Boosted reweighter. 137–141

**GIM** Glashow-Iliopoulos-Maiani. 18, 21

**GN** message passing blocks, called Graph Network. 70, 76, 80, 81, 83

**GNN** graph neural network. 64, 66–68, 70, 71, 74–77, 81, 82, 88

**GPU** graphics processing unit. 40, 56, 61, 188, 209–211

**GR** general relativity. 19

**HEP** High Energy Physics. 1, 2, 28, 32, 36, 37, 39, 41, 58, 88, 186–189, 193, 204, 209, 210

**HIE** **Wrong hierarchy**. 86, 87

**high** $q^2$ **region** $q^2$ region between 14.3 and 23.0 GeV$^2/c^4$. 26, 92–94, 97, 99, 102, 106, 112, 158, 159, 183, 211

**HL-LHC** High Luminosity LHC. 59

**Hypatia** Hypatia function [106], with one power-law tail to the left and another to the right and a wide `Gaussian` distribution,. 155

**ISJ** improved Sheather-Jones. 152, 198

**JIT** just-in-time. 41, 189, 191, 192, 198

**KDE** kernel density estimation. 152, 153, 166, 168, 169, 174, 178, 179, 182, 187, 198

**LCA** Lowest Common Ancestor. 78, 79

**LCAI** Lowest Common Ancestor Inference. 76, 78–80, 85

**LFU** lepton flavour universality. 18, 23–26, 93, 158, 161, 182

**low** $q^2$ **region** $q^2$ region between 0.1 and 1.1 GeV$^2/c^4$. 26, 93, 94

**MC** Monte-Carlo method based simulation sample. 187

**ML** machine learning. 31, 66–68, 72, 81, 83, 120, 139

**MLE** maximum likelihood estimate. 36, 186, 201, 203

**MLP** multilayer perceptron. 70, 71, 76

**MVA** multivariate analysis. 93, 107, 109, 120, 122

**NLL** negative log-likelihood. 36, 200, 203

**NOISO Not isolated**. 86, 87

**non-Abelian** non-Abelian, *i.e.,* a group that does not commute $[A, B] \neq 0$,. 11

**NP** Node Pruning. 75, 78, 80, 81, 83, 85, 88

**NP** New Physics. 20, 21, 92, 101

**PER Perfectly reconstructed**. 86, 87

**PMNS** Pontecorvo-Maki-Nakagawa-Sakata. 19

**PRC Partially reconstructed**. 86, 87

**QCD** quantum chromodynamics. 6, 10–13

**QED** quantum electrodynamics. 6, 7, 9, 10, 12–14

**QFT** quantum field theory. 4, 6, 7

**QM** quantum mechanics. 4, 6, 20

**ReLU** rectified linear unit. 32

**RNN** recurrent neural network. 70

**ROC** Receiver Operator Characteristic. 82, 83, 115–117, 121, 123, 140

**ROC AUC** ROC AUC. 82, 83, 121, 123, 140

**RV** random variable. 29

**SciFi** Scintillating Fibre Tracker. 60

**SGD** stochastic gradient descent. 38

**SI** Silicon Tracker. 49

**SIMD** single instruction, multiple data. 40

**SM** Standard Model of particle physics. 1, 4–7, 12, 17–25, 93, 94, 107, 158, 211

**SR** special relativity. 6

**SSB** Spontaneous Symmetry Breaking. 6, 15–18, 24

**SUSY** supersymmetry, a theory that suggests a symmetry between fermions and bosons,. 20

**UHI** Universal Histogram Interface, a protocol containing a general definition of histograms and their axes, contained in a package that is part of Scikit-HEP,. 190, 192, 193, 197

**UT** Upstream Tracker. 60

**WC** Wilson coefficient. 22